\documentclass[12pt]{cernart}

\usepackage{graphicx}
\usepackage{amsmath}
\usepackage{amssymb}
\usepackage{times}
\usepackage{cite}
\usepackage{multirow}
\usepackage[table]{xcolor}

\begin{document}
\setcounter{topnumber}{3}
\renewcommand{\topfraction}{0.999}
\renewcommand{\bottomfraction}{0.99}
\renewcommand{\textfraction}{0.0}
\setcounter{totalnumber}{6}
\renewcommand{\thefootnote}{\alph{footnote}}

\begin{titlepage}
\flushright{\includegraphics[width=2cm]{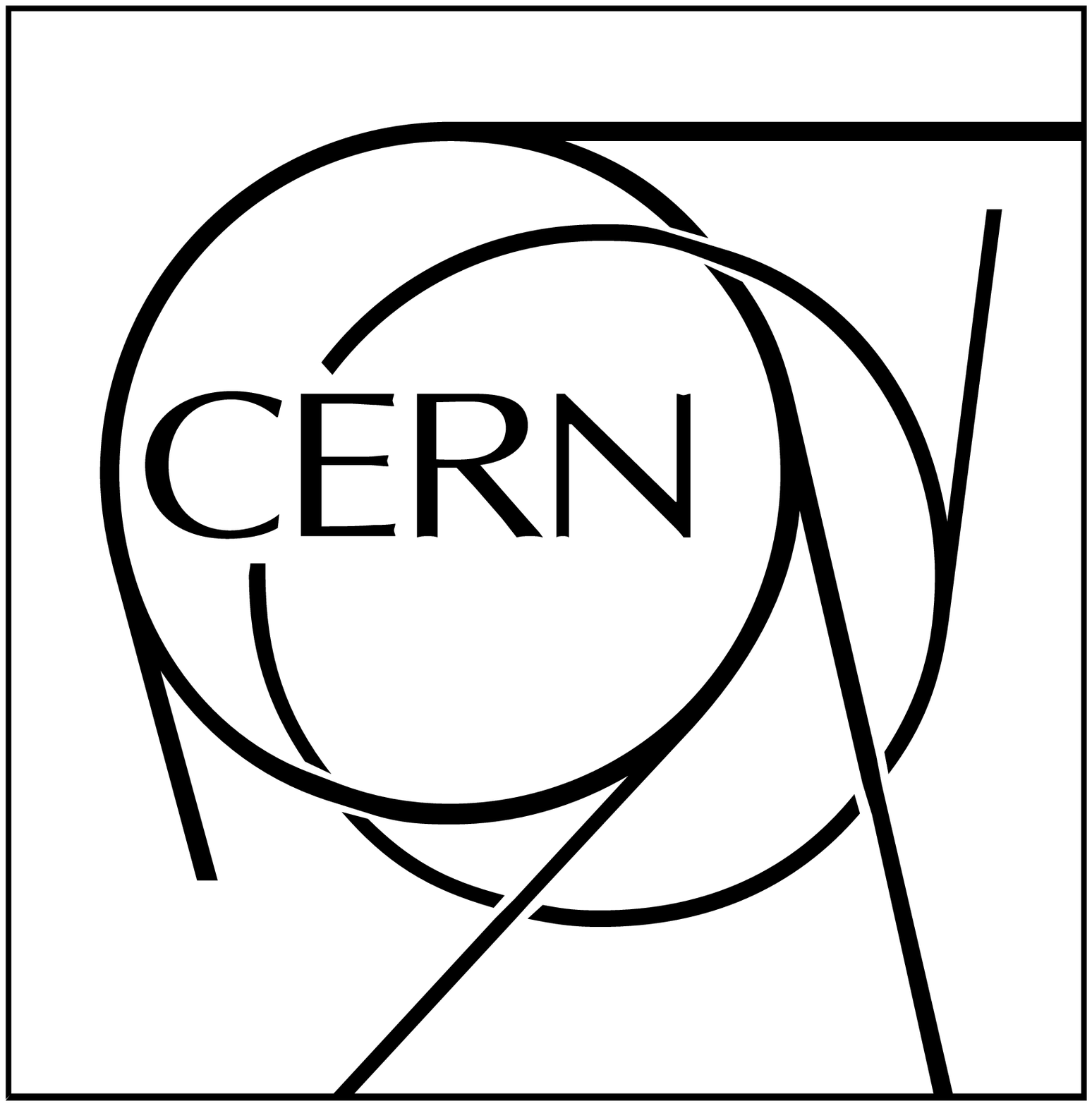}}
\docnum{CERN-PH-EP-2012-153}
\date{4 December 2012}
%\docnum{draft 1.0}
%\date{}
%
%\vspace{2mm}
\title{\large{Inclusive production of protons, anti-protons, neutrons, deuterons
and tritons in p+C collisions at 158 GeV/c beam momentum}}
\flushleft{\footnotesize{\it{This paper is dedicated to the memory of Prof. Matey Mateev}}}
\begin{Authlist}
\vspace{2mm}
\noindent
B.~Baatar$^{4}$, G.~Barr$^{8}$, J.~Bartke$^{3}$, L.~Betev$^{6}$, 
%C.~Blume$^{7}$, 
O.~Chvala$^{9,12}$, J.~Dolejsi$^{9}$, V.~Eckardt$^{7}$,
H.~G.~Fischer$^{6,}$\footnote{e-mail: Hans.Gerhard.Fischer@cern.ch}, Z.~Fodor$^{2}$, 
%P.~Foka$^{5}$, V.~Friese$^{5}$, M.~Ga\'zdzicki$^{7}$, C.~H\"{o}hne$^{5}$, 
A.~Karev$^{6}$, V.~Kolesnikov$^{4}$, M.~Kowalski$^{3}$, 
M.~Makariev$^{11}$, A.~Malakhov$^{4}$, M.~Mateev$^{10,13}$, G.~Melkumov$^{4}$, 
%M.~Mitrovski$^{7}$, S.~Mr\'owczy\'nski$^{9}$, R.~Renfordt$^{7}$, M. Rybczy\'nski$^{9}$, 
A.~Rybicki$^{3}$,
%A.~Sandoval$^{5}$, 
N.~Schmitz$^{7}$, P.~Seyboth$^{7}$, 
%G.~Stefanek$^{9}$, 
R.~Stock$^{5}$,
% H.~Str\"{o}bele$^{7}$,
G.~Tinti$^{8}$,
D.~Varga$^{1}$, G.~Vesztergombi$^{2}$, 
%D.~Vrani\'{c}$^{5}$,
S.~Wenig$^{6}$
%Z.~W{\l}odarczyk$^{9}$, A.~Wojtaszek$^{9}$
\vspace*{2mm} 

\noindent
{\it (The NA49 Collaboration)}  \\
\vspace*{2mm}
\noindent
$^{1}$E\"otv\"os Lor\'and University, Budapest, Hungary \\
$^{2}$KFKI Research Institute for Particle and Nuclear Physics, Budapest, Hungary\\
$^{3}$H. Niewodnicza\'nski Institute of Nuclear Physics, Polish Academy of Sciences, Cracow, Poland \\
%$^{5}$Gesellschaft f\"{u}r Schwerionenforschung (GSI), Darmstadt, Germany.\\
$^{4}$Joint Institute for Nuclear Research, Dubna, Russia.\\
$^{5}$Fachbereich Physik der Universit\"{a}t, Frankfurt, Germany.\\
$^{6}$CERN, Geneva, Switzerland\\
%$^{9}$Institute of Physics \'Swi\c{e}tokrzyska Academy, Kielce, Poland.\\
$^{7}$Max-Planck-Institut f\"{u}r Physik, Munich, Germany.\\
$^{8}$Oxford University, Oxford, UK \\
$^{9}$Charles University, Faculty of Mathematics and Physics, Institute of
             Particle and Nuclear Physics, Prague, Czech Republic \\
$^{10}$Atomic Physics Department, Sofia University St. Kliment Ohridski, Sofia, Bulgaria\\
$^{11}$Institute for Nuclear Research and Nuclear Energy, BAS, Sofia, Bulgaria\\
$^{12}$now at University of Tennessee, Knoxville, TN, USA\\
$^{13}$deceased
\end{Authlist}
%
%\vspace{10mm}%\vspace*{5cm}
\begin{center}
{\small{\it to be published in EPJC }}
\end{center}
\vspace*{2mm} 
\clearpage

\begin{abstract}
\vspace{-3mm}

The production of protons, anti-protons, neutrons, deuterons and
tritons in minimum bias p+C interactions is studied using a sample of
385 734 inelastic events obtained with the NA49 detector at the CERN 
SPS at 158~GeV/c beam momentum. The data cover a phase space area 
ranging from 0 to 1.9~GeV/c in transverse momentum and in Feynman 
$x$ from -0.8 to 0.95 for protons, from -0.2 to 0.3 for anti-protons 
and from 0.1 to 0.95 for neutrons. Existing data in the far backward 
hemisphere are used to extend the coverage for protons and light
nuclear fragments into the region of intra-nuclear cascading. The 
use of corresponding data sets obtained in hadron-proton collisions with 
the same detector allows for the detailed analysis and model-independent 
separation of the three principle components of hadronization in 
p+C interactions, namely projectile fragmentation, target fragmentation 
of participant nucleons and intra-nuclear cascading. 

\end{abstract}
%\vspace*{5mm}
 
\clearpage
\end{titlepage}

%
% ****************************** Section 1 ****************************
%
\section{Introduction} 
\vspace{3mm}
\label{sec:intro}

Baryon and light ion production in proton-nucleus collisions
has in the past drawn considerable interest, resulting
in an impressive amount of data from a variety of experiments.
This interest concentrated in forward direction on the
evident transfer of baryon number towards the central region,
known under the misleading label of "stopping", and in the
far backward region on the fact that the laboratory momentum
distributions of baryons and light fragments reach far beyond
the limits expected from the nuclear binding energy alone.
A general experimental study covering the complete phase
space from the limit of projectile diffraction to the detailed
scrutiny of nuclear effects in the target frame is, however, still
missing. More recently, renewed interest has been created by
the necessity of providing precision reference data for the 
control of systematic effects in neutrino physics.

In addition to and beyond the motivations mentioned above, the 
present study is part of a very general survey of elementary and 
nuclear interactions at the CERN SPS using the NA49 detector, aiming at
a straight-forward connection between the different reactions
in a purely experiment-based way. After a detailed inspection 
of pion \cite{pp_pion}, kaon \cite{pp_kaon} and baryon \cite{pp_proton} production in 
p+p interactions, a similar in-depth approach is being carried out 
for p+C collisions. This has led to the recent publication of two papers 
concerning pion production \cite{pc_pion,pc_discus} and this aim is here being extended to 
baryons and light ions.

The use of the light, iso-scalar Carbon nucleus is to be regarded
as a first step towards the study of proton collisions with heavy nuclei 
using data with controlled centrality available from NA49.
It allows the control of the transition from elementary to nuclear interactions 
for a small number of intra-nuclear collisions, thus providing an important 
link between elementary and multiple
hadronic reactions. It also allows for the clean-cut separation
of the three basic components of hadronization in p+A collisions,
namely projectile fragmentation, fragmentation of the target
nucleons hit by the projectile, and intra-nuclear cascading. 
The detailed study of the superposition of these components in a
model-independent way is the main aim of this paper. For this
end the possibility of defining net proton densities by measuring
anti-protons and thereby getting access to the yield of pair
produced baryons, will be essential. As the acceptance of the
NA49 detector does not cover the far backward region, the
combination of the NA49 results with measurements from other
experiments dedicated to this phase space area is mandatory.
A survey of the $s$-dependence of backward hadron production in 
p+C collisions has therefore been carried out and is published
in an accompanying paper \cite{pc_survey}. This allows the extension of the
NA49 data set to full phase space.

As the extraction of hadronic cross sections has been described
in detail in the preceding publications \cite{pp_pion,pp_kaon,pp_proton,pc_pion,pc_discus}, 
the present paper will concentrate on those aspects which are specific to baryon
and light ion production, particularly in the exploitation of the
NA49 acceptance into the backward hemisphere. After a short comment
on existing double differential data in the SPS energy range in
Sect.~\ref{sec:exp_sit}, a few experimental details will be given in Sect.~\ref{sec:setup} together
with the binning scheme adopted for protons, anti-protons and
neutrons. Section~\ref{sec:pid} will present a comprehensive description of 
particle identification in the backward hemisphere which is an
important new ingredient of the optimized use of the NA49 detector
in particular for the asymmetric p+A collisions. Section~\ref{sec:corr} deals with 
the extraction of the inclusive cross sections and with the
applied corrections. Section~\ref{sec:results} contains the data tables and plots
of the invariant cross sections as well as some particle ratios
and a comparison to the few available double differential yields
at SPS energy for comparison. Section~\ref{sec:backext} describes the use of the extensive
complementary data set from the Fermilab experiment \cite{bayukov} for the
data extension into the far backward direction together with
an interpolation scheme allowing for the first time the complete
inspection of the production phase space for protons in the
range -2~$< x_F <$~+0.95. This combined study is extended to deuterons
and tritons in Sect.~\ref{sec:ions}. Baryon ratios are shown in Sect.~\ref{sec:ratios}. 
Quantities integrated over $p_T$ are given in Sect.~\ref{sec:ptint} both for minimum bias trigger 
conditions and for the dependence on the number of
measured "grey" protons \cite{pc_pion,nim}. In addition, the measured $p_T$ integrated
neutron yields are presented in Sect.~\ref{sec:ptint} together with a comparison
to other integrated data in the SPS energy range. Section~\ref{sec:two_comp} contains 
a detailed discussion of the two-component mechanism
of baryon and baryon pair production in p+p collisions, thus covering the first
two components of the hadronization process defined above,
including a comment on resonance decay and a comparison to a recent microscopic simulation code.
Section~\ref{sec:feedover} contains the corresponding experimental results from p+C interactions. 
Section~\ref{sec:hadr} gives a detailed discussion of anti-proton production
including the application of the two-component mechanism
introduced in Sect.~\ref{sec:two_comp} and a study of the  $p_T$ dependence.
The discussion of $p_T$ integrated proton and net proton yields 
is presented in Sect.~\ref{sec:hadr_net} followed by the exploitation of 
double differential proton and net proton cross sections in Sect.~\ref{sec:pt_prot}.
The paper is closed by a summary of conclusions in Sect.~\ref{sec:conclusion}.

%
% ****************************** Section 2 ****************************
%
\section{The Experimental Situation}
\vspace{3mm}
\label{sec:exp_sit}

\begin{figure}[h]
  \begin{center}
  	\includegraphics[width=14.5cm]{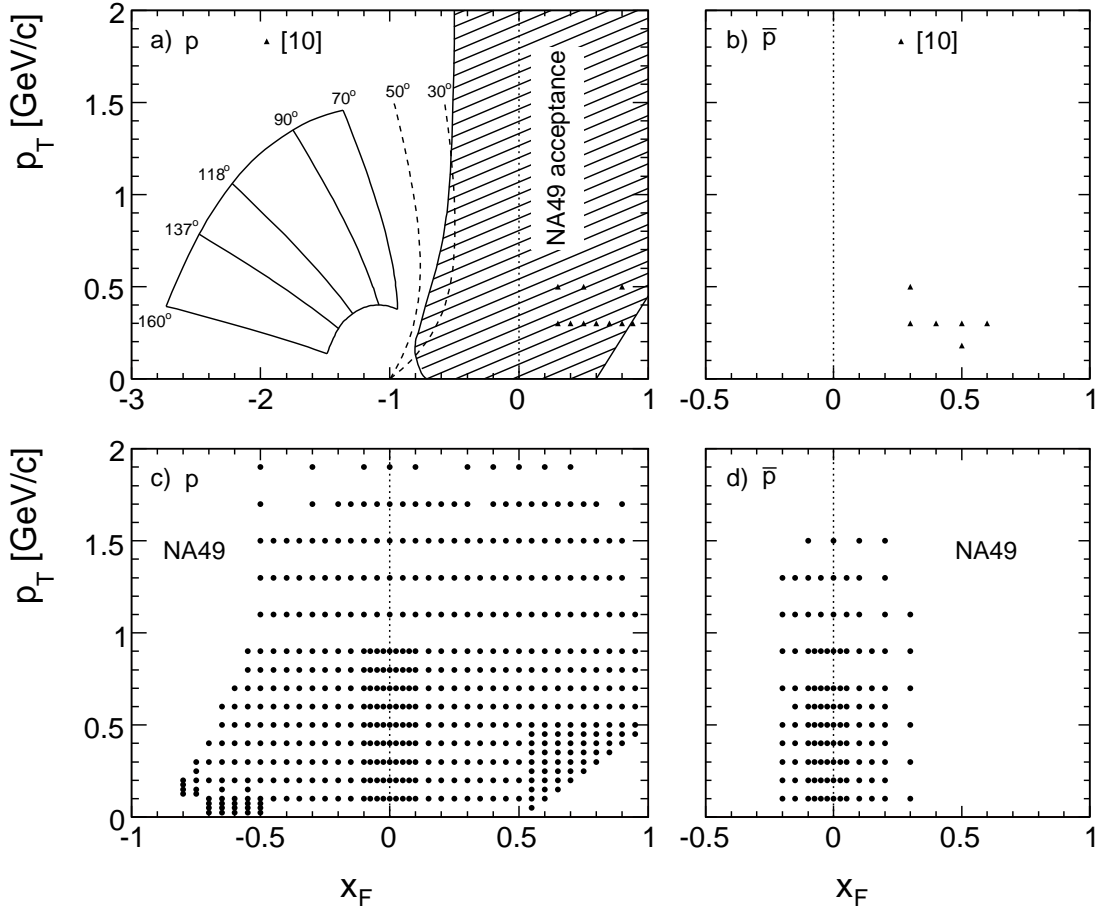}
  	\caption{Phase space coverage of existing data: a) p data from \cite{bayukov} (full lines) and \cite{barton}. 
  	         Here with the shaded area is shown the NA49 acceptance range; 
  	         b) $\overline{\textrm{p}}$ data from \cite{barton}; c) p data from NA49; and 
  	         d) $\overline{\textrm{p}}$ data from NA49. Note the extended abscissa in panel c)}
  	\label{fig:cov}
  \end{center}
\end{figure}

As already pointed out for pions in \cite{pc_pion} there are only two sets
of double differential inclusive data, for identified baryons 
and light fragments in p+C collisions in the SPS energy range.
The differential inclusive cross sections are presented in this 
paper as:

\begin{equation}
  \frac{d^2\sigma}{dx_Fdp_T^2}   ,
\end{equation}
with $x_F = 2p_L/\sqrt{s}$ defined in the nucleon-nucleon center-of-mass system (cms). A first
data set \cite{bayukov,frankel} covers the far backward direction for protons and
light ions at five fixed laboratory angles between 70 and 160~degrees for total
lab momenta between 0.4 and 1.4~GeV/c at a projectile
momentum of 400~GeV/c. A second set \cite{barton} has been obtained 
in forward direction for 0.3~$< x_F <$~0.88 and 0.15~$< p_T <$~0.5 with
100~GeV/c beam momentum. The respective phase space coverage
in $x_F$ and $p_T$ is shown in Fig.~\ref{fig:cov}a,b for protons and 
anti-protons, respectively, with a superposition of the NA49 coverage for protons. 
This coverage is presented in more detail in Fig.~\ref{fig:cov}c and for the anti-protons in
Fig.~\ref{fig:cov}d.

%    Fig.1
% Phase space coverage plot           

With the NA49 data covering lab angles of up to 40~degrees the 
combination with \cite{bayukov,frankel} into a consistent data set becomes 
possible. This allows for the first time the complete scrutiny 
of the proton phase space in the range -2~$< x_F <$~+0.95, with only 
minor inter- and extrapolation.

%
% ****************************** Section 3 ****************************
%
\section{Experimental information and binning scheme}
\vspace{3mm}
\label{sec:setup}

As a detailed description of the NA49 detector and the extraction
of inclusive cross sections has been given in \cite{pc_pion,pp_proton,pp_pion,pp_kaon,nim},
only some basic informations are repeated here for convenience.

%
% ****************************** Section 3.1 ****************************
%
\subsection{Target, "grey" proton detection, trigger cross section and event sample}
\vspace{3mm}
\label{sec:target}

The NA49 experiment is using a secondary proton beam of 158~GeV/c
momentum at the CERN SPS. A graphite target of 1.5\% interaction 
length is placed inside a "grey" proton detector \cite{pc_pion,nim} which measures 
low energy protons in the momentum range up to 1.5~GeV/c originating from
intra-nuclear cascading in the carbon target. This detector 
covers a range from 45 to 315 degrees in polar angle with a granularity of 
256 readout pads placed on the inner surface of a cylindrical 
proportional counter. An interaction trigger is defined by a small 
scintillator 380~cm downstream of the target in anti-coincidence with 
the beam. This yields a trigger cross section of 210.1 $\pm$ 2.1~mb corresponding 
to 91\% of the measured inelastic cross section of 226.3 $\pm$ 4.5~mb. This is in 
good agreement with the average of 225.8 $\pm$ 2.2~mb obtained from a number of 
previous measurements \cite{pc_pion}. A total sample of 385.7k events has been
obtained after fiducial cuts on the beam emittance and on the longitudinal
vertex position.

%
% ****************************** Section 3.2 ****************************
%
\subsection{Acceptance coverage, binning and statistical errors}
\vspace{3mm}
\label{sec:accept}

\begin{figure}[h]
  \begin{center}
  	\includegraphics[width=15.5cm]{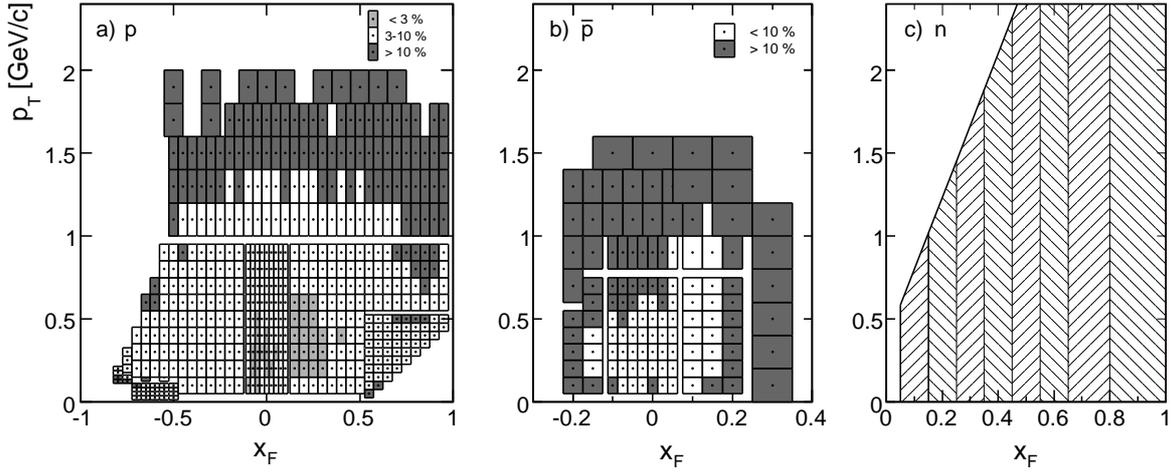}
 	\caption{Binning scheme in $x_F$ and $p_T$ together with information
             on the statistical errors for a) protons, b) anti-protons 
             and c) neutrons}
  	 \label{fig:accept}
  \end{center}
\end{figure}

The NA49 detector \cite{nim} covers a range of polar laboratory angles between
$\pm$45 degrees with a set of four Time Projection Chambers combining
tracking and particle identification, two of the TPC's being placed
inside superconducting magnets. While for anti-protons the accessible
range in $x_F$ and $p_T$ is essentially defined by the limited event statistics,
it has been possible to completely exploit the available range of
polar angle for protons. The corresponding binning schemes are shown
in Fig.~\ref{fig:accept} in the cms variables $x_F$ and $p_T$.

%    Fig.2

A rough indication of the effective statistical errors is given by the
shading of the bins. Neutrons have been detected in a forward hadronic
calorimeter \cite{pp_proton} in combination with proportional chambers vetoing
charged hadrons. Due to the limited resolution in transverse momentum
only $p_T$ integrated information in 8 bins in $x_F$ (Fig.~\ref{fig:accept}c) could be
obtained, after unfolding of the energy resolution. This coverage is
identical to the one in p+p interactions \cite{pp_proton} and allows for direct
yield comparison.

%
% ****************************** Section 4 ****************************
%
\section{Particle Identification}
\vspace{3mm}
\label{sec:pid}

\begin{figure}[h]
  \begin{center}
  	\includegraphics[width=8.5cm]{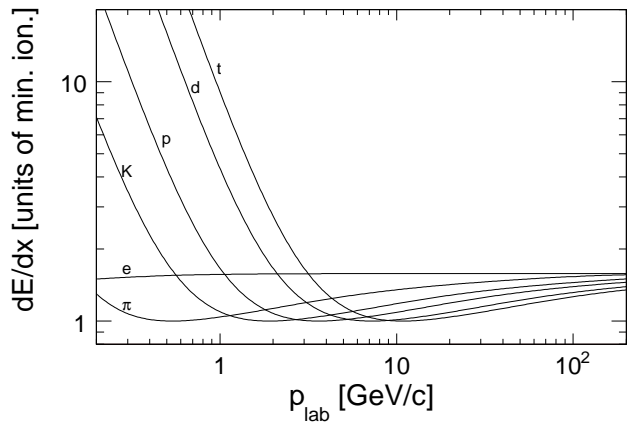}
 	  \caption{Parametrization of the mean truncated energy loss
             as a function of total lab momentum $p_{\textrm{lab}}$ for electrons, pions, kaons
             and protons. The situation for deuterons and tritons is also indicated}
  	 \label{fig:bb}
  \end{center}
\end{figure}

Due to the forward-backward asymmetry of p+A interactions, the
study of the backward hemisphere is of major interest for the
understanding of target fragmentation and intra-nuclear
cascading. Particle identification at negative $x_F$ is therefore
mandatory; it has to rely for the NA49 detector on the measurement 
of ionization energy loss in the TPC system. This method has been 
developed and described in detail for mesons and baryons in p+p 
collisions in \cite{pp_proton,pp_pion,pp_kaon} for $x_F >$~0. 
A substantial effort has been invested for the present study in its extension 
to the far backward direction down to the acceptance limit in $x_F$ imposed by 
the NA49 detector configuration. With decreasing $x_F$ the baryonic lab momentum
decreases below the region of minimum ionization where the
ionization energy loss increases like $1/\beta^2$ and
thereby successively crosses the deposits from kaons, pions and
electrons. This is shown in Fig.~\ref{fig:bb} for the the momentum dependence
of the parametrization of the mean truncated energy loss used in this
analysis.

%       Fig.3

\begin{figure}[h]
  \begin{center}
  	\includegraphics[width=5.cm]{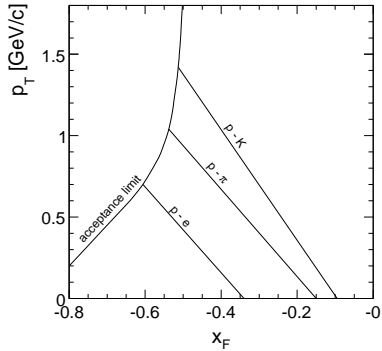}
 	\caption{Lines of equal energy loss for protons and kaons
            (p-K), protons and pions (p-$\pi$) and protons and electrons
            (p-e) as functions of $x_F$ and $p_T$, together with the acceptance
             limit of the NA49 detector}
  	 \label{fig:overlap}
  \end{center}
\end{figure}

In terms of $x_F$ and $p_T$, this cross-over pattern reflects into lines
of equal energy loss in the $x_F$--$p_T$ plane as shown in Fig.~\ref{fig:overlap}.

%    Fig.4     

\begin{figure}[h]
  \begin{center}
  	\includegraphics[width=9.5cm]{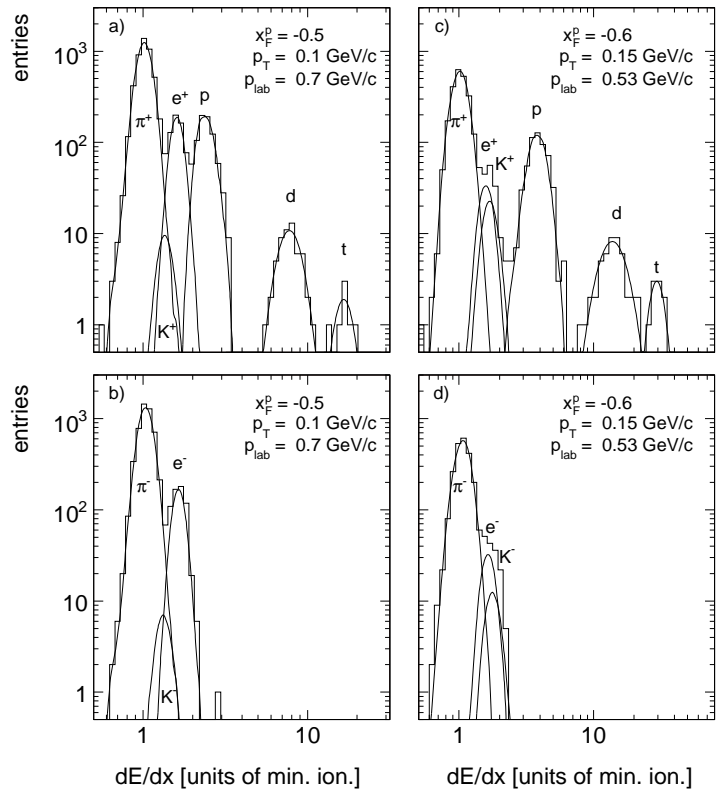}
 	\caption{Truncated energy loss distributions for a) positives and b) negatives at $x_F$~=~-0.5, $p_T$~=~0.1~GeV/c;
     and for c) positives and d) negatives at $x_F$~=~-0.6, $p_T$~=~0.15~GeV/c}
  	 \label{fig:backdedx}
  \end{center}
\end{figure}

The region above the line p-K allows for the standard multi-parameter
fits of the truncated energy loss distributions as described in the 
preceding publications \cite{pc_pion,pp_proton,pp_pion,pp_kaon}. The approximately triangular 
region below the line p-e permits the direct extraction of baryon yields partially
even without fitting. This is exemplified in Fig.~\ref{fig:backdedx} for two bins at
$x_F$~=~-0.5 and -0.6 and small $p_T$. 

%     Fig.5 

It is interesting to note that also the light ions deuteron and triton 
are here well separable practically without background. For anti-protons,
a direct measurement of the $\overline{\textrm{p}}$/p ratio becomes feasible down to 
values below 10$^{-3}$ in this region.

In order to extract proton yields from the energy loss distributions
in the intermediate region between the lines p-e and p-K, Fig~\ref{fig:overlap},
the particle ratios in each studied $x_F$/$p_T$ bin are of prime
importance. If the ratios p/$\pi$, K/$\pi$ and e/$\pi$ are known, proton
yields may be obtained from the total number of tracks even in
those bins where the proton energy loss equals the one
from electrons, pions or kaons. A two-dimensional interpolation of
the measured particle ratios over the full accessible phase space
has therefore been established. These ratios are obtained without
problem in the regions below the line p-e and above the line p-K
(Fig.~\ref{fig:overlap}) as well as in most intermediate bins where a sufficient
separation in $dE/dx$ of the different particle species is present.
Near the cross-over bins the measured ratios show a sharp increase
of the effective statistical fluctuations, an increase which has
been described in the discussion of the error matrix involved with
the multi-dimensional fitting procedure in \cite{pp_kaon}. As this effect
is of statistical and not of systematic origin, an interpolation
through the critical regions in $x_F$ and $p_T$ is applicable.

\begin{figure}[h]
  \begin{center}
  	\includegraphics[width=10cm]{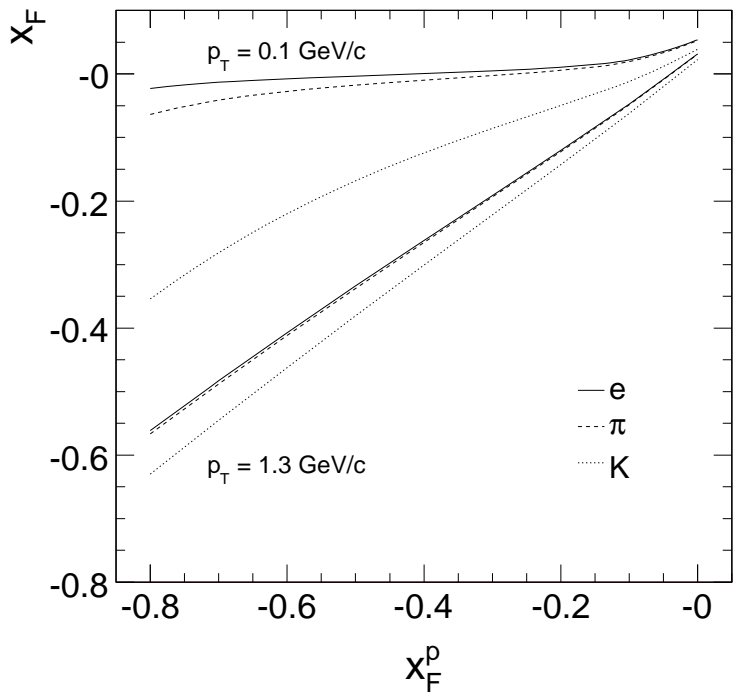}
 	\caption{$x_F$ for electrons, pions and kaons as a function of 
             proton $x_F^p$ for $p_T$~=~0.1~GeV/c (upper lines) and
             $p_T$~=~1.3~GeV/c (lower lines) }
  	 \label{fig:partxf}
  \end{center}
\end{figure}

It should be stressed here that the obtained particle ratios are 
non-physical in the sense that they use different phase space 
regions for each particle mass. For each $x_F$/$p_T$ bin the 
necessary transformation to total lab momentum is performed using 
the proton mass for each track. This means that electrons, pions 
and kaons from different effective $x_F$ values enter into the proton 
bin, with an asymmetry that increases with decreasing $x_F$ and $p_T$. 
This is quantified in Fig.~\ref{fig:partxf} where the effective mean $x_F$ 
for electrons, pions and kaons is shown as a function of proton $x_F^p$ for two
values of $p_T$.

%     Here Fig.6 

The lighter particles at small $p_T$ are thus effectively collected from the 
neighbourhood of $x_F$~=~0 with decreasing baryonic $x_F$.
For anti-protons this purely kinematic effect is unfavourable
for fitting as the effective $\overline{\textrm{p}}$/$\pi^-$ ratios quickly decrease 
below the percent level at low $p_T$, whereas the p/$\pi^+$ ratios stay
always above about 10\%, increasing rapidly with $p_T$ due to the rather
flat number distribution $dn/dx_F$. Proton and anti-proton extraction
are therefore regarded separately in the following Sects.~\ref{sec:prot_ext} and 
\ref{sec:aprot_ext}, respectively.

%
% ****************************** Section 4.1 ****************************
%
\subsection{Proton extraction}
\vspace{3mm}
\label{sec:prot_ext}

The problematics of the extraction of the yields of the four particle
species e, $\pi$, K and p from the measured overall truncated ionization energy loss distributions 
has been described in detail in the preceding publications 
\cite{pp_proton,pp_pion,pp_kaon,pc_pion,pc_discus}. 
In particular the estimation of the corresponding systematic and statistical errors 
has been treated in refs. \cite{pp_kaon,pp_proton}. The following sections describe the extraction 
of the different particle ratios as they are needed for the determination of 
the proton and anti-proton yields. The statistical errors of the particle ratios shown in the
following Figures are given by the number of extracted particles
per bin and do not contain the additional terms due to the
fitting process, see \cite{pp_kaon,pp_proton} for a detailed explanation.
The few percent of data points which exceed the quoted error
margins with respect to the two-dimensional interpolation
are due to these additional, purely statistical fluctuations.
They do not influence the quality of the ratio interpolation.

%
% ****************************** Section 4.1.1 ****************************
%
\subsubsection{e$^+$/$\mathbf \pi^+$ ratio}
\vspace{3mm}
\label{sec:e2pi}

\begin{figure}[h]
  \begin{center}
  	\includegraphics[width=11cm]{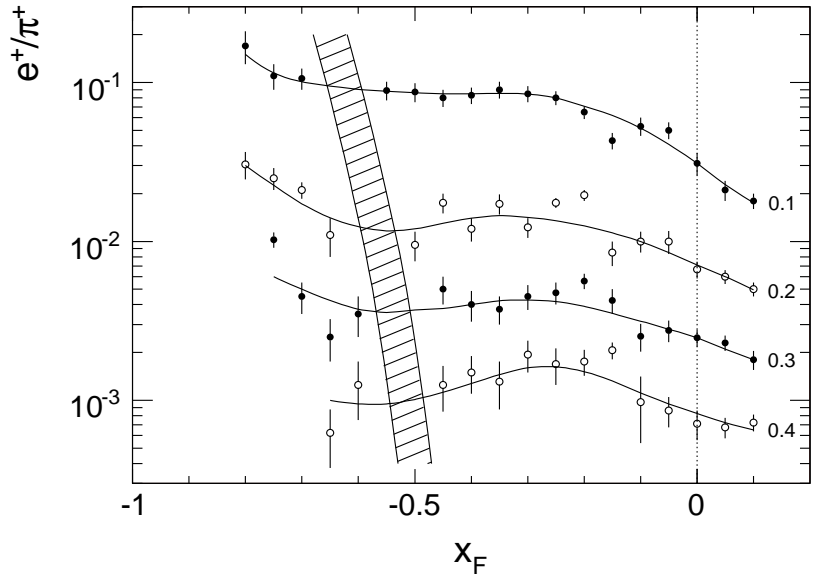}
 	\caption{e$^+$/$\pi^+$ ratio as a function of $x_F$ for four values of $p_T$
 	         from 0.1 to 0.4~GeV/c. For clarity of presentation, the ratios for
 	         subsequent $p_T$ values are divided by a factor of 2. The p-e$^+$
 	         ambiguity regions are indicated by the hatched area}
  	 \label{fig:e2pi}
  \end{center}
\end{figure}

The crossing of the proton $dE/dx$ through the practically constant 
electron energy loss at $p_{\textrm{lab}} \sim$~1~GeV/c is the least critical 
effect as the momentum dependence of the proton $dE/dx$ is a steep
function of lab momentum in this $p_{\textrm{lab}}$ range and as the e/$\pi$ ratio 
quickly decreases with increasing $p_T$, reaching the 1\% level 
already at $p_T >$~0.3~GeV/c. The e$^+$/$\pi^+$ ratio is shown in Fig.~\ref{fig:e2pi} as a 
function of $x_F$ for four values of $p_T$ together with the 
two-dimensional interpolation used. 

%    Fig.7

The hatched area indicates the position of the $dE/dx$ cross-over for each
$p_T$ value and evidently the ratios may be well interpolated through the
small affected $x_F$ regions.

%
% ****************************** Section 4.1.2 ****************************
%
\subsubsection{p/$\mathbf \pi^+$ ratio}
\vspace{3mm}
\label{sec:p2pi}

\begin{figure}[h]
  \begin{center}
  	\includegraphics[width=11cm]{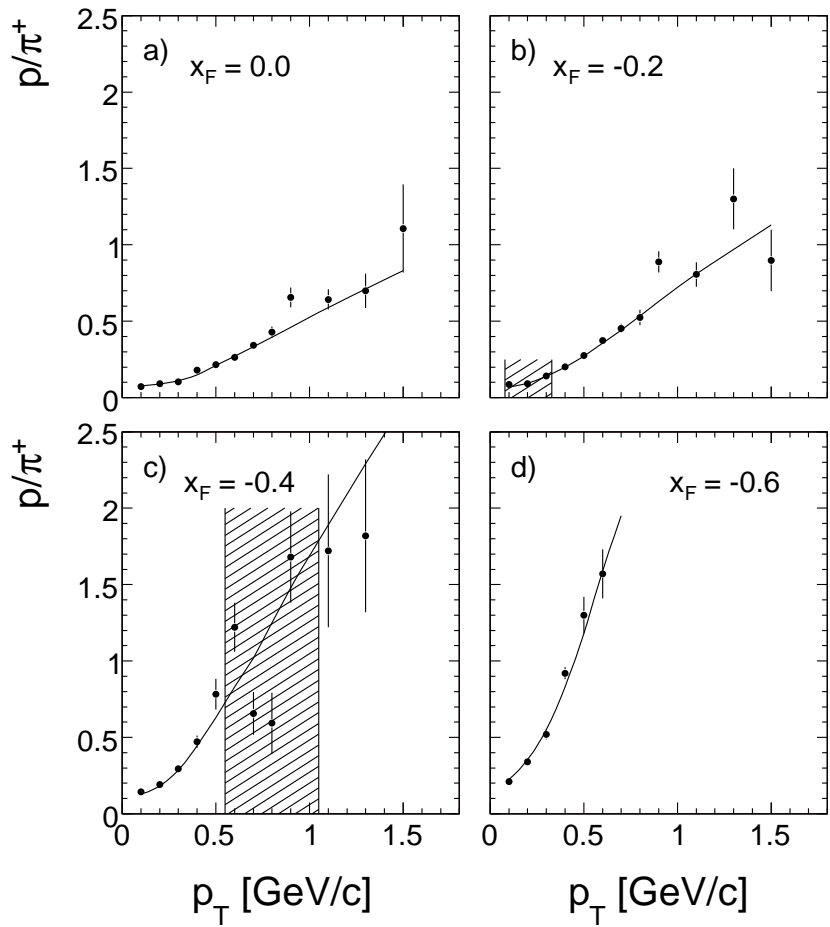}
 	\caption{p/$\pi^+$ ratios as a function of $p_T$ for a) $x_F$~=~0,
             b) $x_F$~=~-0.2, c) $x_F$~=~-0.4 and d) $x_F$~=~-0.6. The full lines present the
             two-dimensional data interpolation, the hatched areas between
             the vertical lines the regions affected by the p-$\pi$ and p-K
             ambiguities}
  	 \label{fig:pr2pi_pt}
  \end{center}
\end{figure}

Fitted p/$\pi^+$ ratios are presented in Fig.~\ref{fig:pr2pi_pt} for four $x_F$ values together 
with their two-dimensional interpolation as a function of $p_T$. While the
fit results yield stable $p_T$ dependences within their statistical errors
in the uncritical regions at $x_F$~=~0 and -0.6, the intermediate $x_F$
values at -0.2 and -0.4 show some additional fluctuation in the
cross-over regions indicated by the hatched areas which combine the
p-$\pi$ and p-K ambiguities.

 %    Fig.8 

Evidently the data interpolation describes the ratio properly through
the ambiguous $p_T$ areas.

\begin{figure}[h]
  \begin{center}
  	\includegraphics[width=12cm]{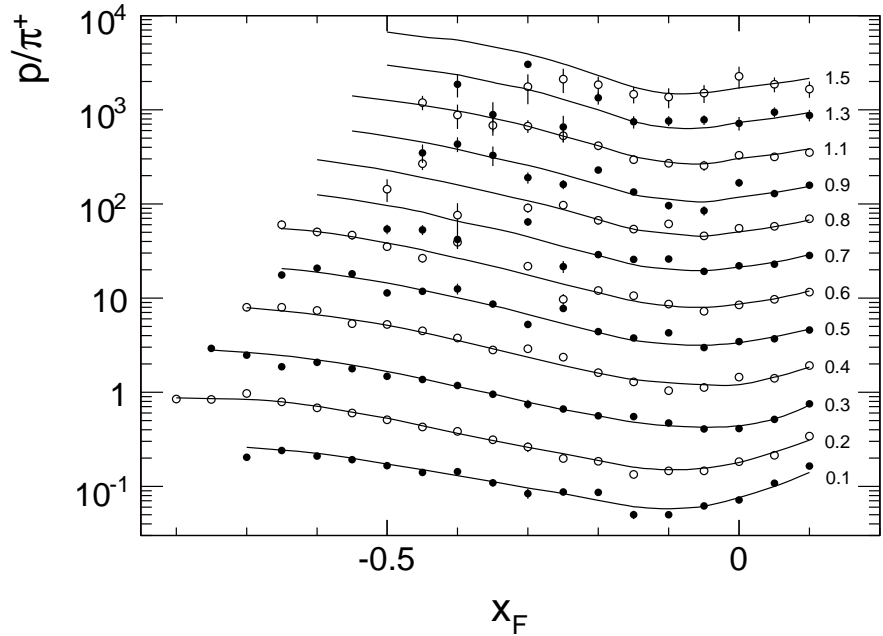}
 	\caption{p/$\pi^+$ ratios as functions of $x_F$ for fixed values of $p_T$ [GeV/c].
            Full lines: data interpolation. The ratios at successive $p_T$ 
            values are shifted by a factor of 2 for clarity of presentation}
  	 \label{fig:pr2pi_xfdist}
  \end{center}
\end{figure}

A complete picture over the full available backward phase space is 
given in Fig.~\ref{fig:pr2pi_xfdist} where the fitted p/$\pi^+$ ratios are shown as functions
of $x_F$ for different $p_T$ values together with the interpolations (full
lines). The ratios at successive $p_T$ values are shifted by a factor
of 2 for clarity of presentation.

%     Fig.9

\begin{figure}[b]
  \begin{center}
  	\includegraphics[width=12cm]{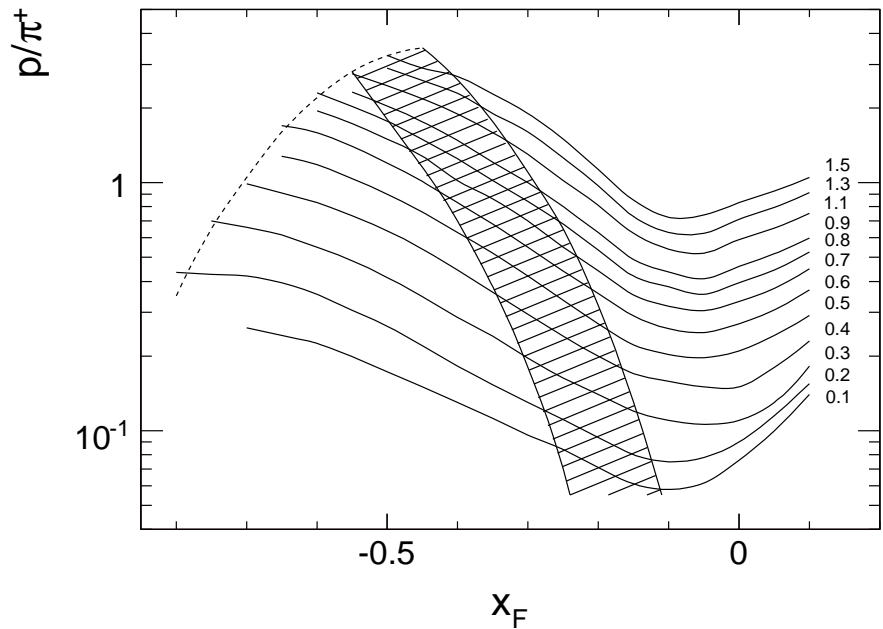}
 	\caption{Interpolated p/$\pi^+$ ratios as functions of $x_F$ for
            fixed values of $p_T$ [GeV/c]. Broken line: NA49 acceptance limit. Hatched
            area: region of p-$\pi$ and p-K ambiguity}
  	 \label{fig:pr2pi_xffit}
  \end{center}
\end{figure}

The complete situation for the data interpolation is finally presented
in Fig.~\ref{fig:pr2pi_xffit} with fixed vertical scale as a function of $x_F$ at different $p_T$ values.
Here the acceptance limit of the NA49 detector is given as the broken
line together with the region of p-$\pi$ and p-K ambiguity as hatched area.

%     Fig.10 

This plot again clarifies the way in which the critical cross-over
areas may be bridged by two-dimensional interpolation. 
 
%
% ****************************** Section 4.1.3 ****************************
%
\subsubsection{K$^+$/$\mathbf \pi^+$ ratio}
\vspace{3mm}
\label{sec:k2pi}

\begin{figure}[h]
  \begin{center}
  	\includegraphics[width=11cm]{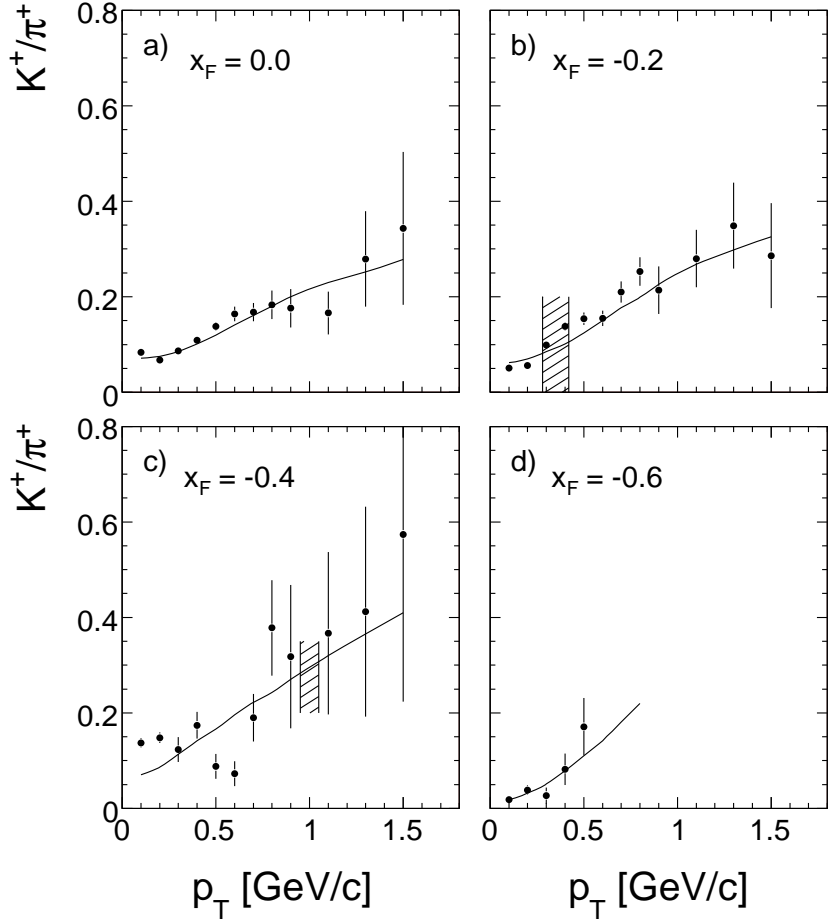}
 	\caption{K$^+$/$\pi^+$ ratios as a function of $p_T$ for four
             values of $x_F$, a) $x_F$~=~0, b) $x_F$~=~-0.2, c) $x_F$~=~-0.4 and d) $x_F$~=~-0.6.
             The regions of p-K and K-$\pi$ ambiguities are indicated as hatched
             areas in panels b) and c). The full lines represent the two-dimensional interpolation}
  	 \label{fig:k2pi_ptdist}
  \end{center}
\end{figure}

A situation quite similar to the p/$\pi^+$ ratio exists for the K$^+$/$\pi^+$
ratio. Again, there are regions of ambiguity against protons and
pions, but the influence of eventual systematic deviations on the
extraction of protons is small as the K$^+$/$\pi^+$ ratios are smaller than
the p/$\pi^+$ ratio by factors between 3 and 10. Fig.~\ref{fig:k2pi_ptdist} shows K$^+$/$\pi^+$
ratios as functions of $p_T$ for four $x_F$ values, where the lowest and
highest $x_F$ at -0.6 and 0 allow for unambiguous fits over the
full $x_F$ range, whereas the $x_F$ values at -0.2 and -0.4 suffer from
p-K and K-$\pi$ ambiguities in the hatched areas of $p_T$, with resulting
increased statistical fluctuations. The two-dimensional interpolation
is superimposed as full lines.

%     Fig.11

\begin{figure}[h]
  \begin{center}
  	\includegraphics[width=10.5cm]{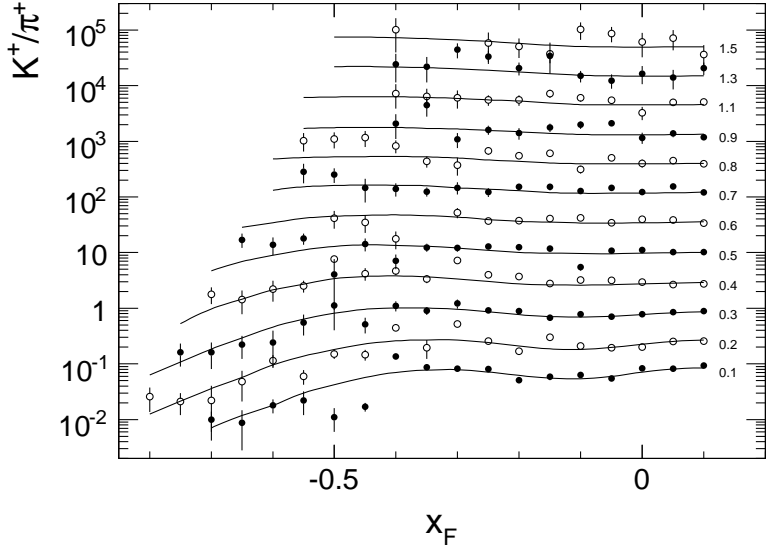}
 	\caption{K$^+$/$\pi^+$ ratios as functions of $x_F$ for fixed values of $p_T$ [GeV/c].
             Full lines: data interpolation. The ratios at successive $p_T$ 
             values are shifted by a factor of 3 for clarity of presentation}
  	 \label{fig:k2pi_xfdist}
  \end{center}
\end{figure}

All fitted values of K$^+$/$\pi^+$ are plotted in Fig.~\ref{fig:k2pi_xfdist} as a function of $x_F$
for fixed $p_T$. As in Fig.~\ref{fig:pr2pi_xfdist} the ratios at successive $p_T$ values are
shifted by 3 in order to sufficiently separate the measurements.
 
%     Fig.12  

\begin{figure}[h]
  \begin{center}
  	\includegraphics[width=10.5cm]{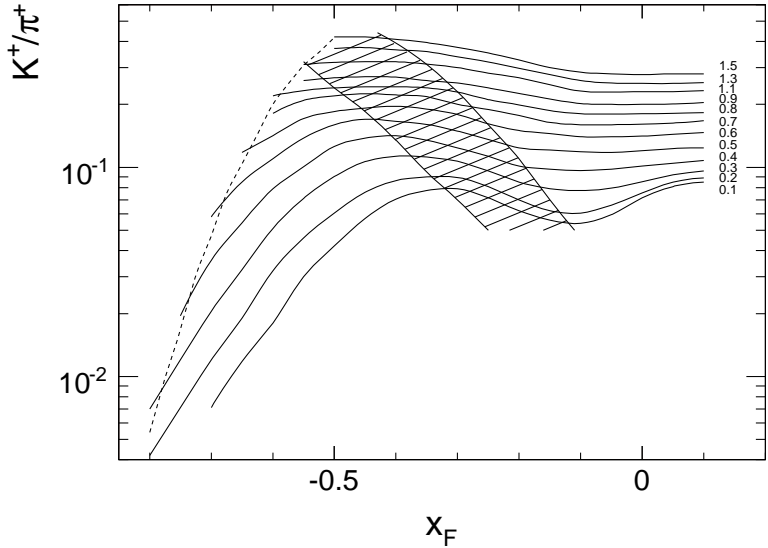}
 	\caption{Interpolated K$^+$/$\pi^+$ ratios as functions of $x_F$ for
             fixed values of $p_T$. Broken line: NA49 acceptance limit. Hatched
             area: region of K-$\pi$ and p-K ambiguity}
  	 \label{fig:k2pi_xffit}
  \end{center}
\end{figure}

Fig.~\ref{fig:k2pi_xffit} presents the overview of the interpolated K$^+$/$\pi^+$ ratios at fixed
vertical scale as a function of $x_F$ for fixed values of $p_T$ [GeV/c]. The broken
line represents the acceptance limits and the hatched area the region
of p-K and K-$\pi$ ambiguity.

%     Fig.13
%
% ****************************** Section 4.1.4 ****************************
%
\subsubsection{Proton extraction in the far forward region}
\vspace{3mm}
\label{sec:p_forw}

Due to the gap between the TPC detectors imposed by the operation with
heavy ion beams \cite{nim}, charged particles progressively leave the TPC
acceptance region at low $p_T$ for $x_{F} >$~0.55. Here, tracking is achieved
by the combination of a small "gap" TPC (GTPC) in conjunction with two
forward proportional chambers (VPC). The performance of this detector
combination is described in detail in \cite{pp_proton}. In the absence of particle
identification in this area one has to rely on external information
concerning the combined fraction of K$^+$ and $\pi^+$ in the total charged
particle yield. Several considerations help to establish reference
values for the (K$^+$ + $\pi^+$)/p ratios:

\begin{itemize}
 \item The (K$^+$  + $\pi^+$)/p ratios decrease very rapidly with increasing
       $x_F$ at all $p_T$, from about 10\% at $x_F$~=~0.6 to less than 1\% at $x_F$~=~0.9.
       Possible deviations from the used external reference data therefore
       introduce only small systematic effects in the extracted proton
       yield.
 \item Existing data may be used to come to a consistent estimation
       of the particle ratio. Direct measurements from Barton et al.\cite{barton}
       in p+C interactions cover the region from $x_F$~=~0.2 to 0.8 for
       $p_T$~=~0.3 and 0.5~GeV/c. Although the published invariant cross
       sections show sizeable deviations from the NA49 results, see
       Sect.~\ref{sec:comp}, the particle ratios of the two experiments compare well.
 \item The ratios also comply with measurements in p+p collisions, both
       from NA49 \cite{pp_proton,pp_pion,pp_kaon} and from Brenner et al. \cite{brenner} 
       at 100 and 175~GeV/c beam momentum.
\end{itemize}

\begin{figure}[h]
  \begin{center}
  	\includegraphics[width=15cm]{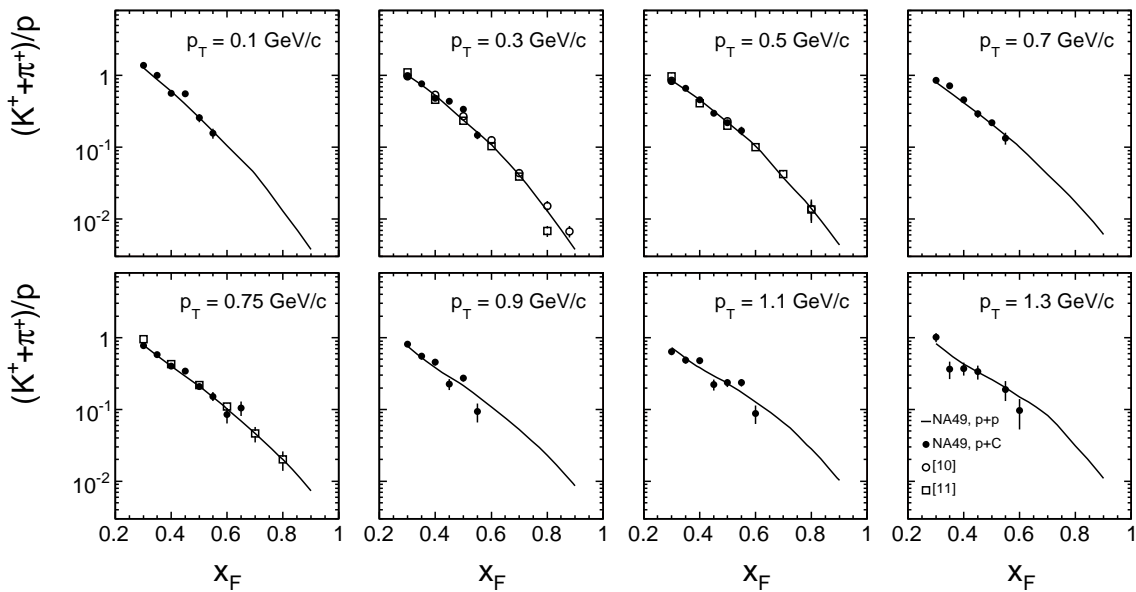}
 	\caption{Measured (K$^+$ + $\pi^+$)/p ratios as a function of $x_F$
             for eight values of $p_T$ between 0.1 and 1.3~GeV/c. The full
             lines give the results from NA49 in p+p collisions, the full
             and open circles the ones from NA49 and \cite{barton}, respectively.
             Open squares: results from \cite{brenner} in p+p interactions}
  	 \label{fig:kpi2p_xf}
  \end{center}
\end{figure}

An overview of the experimental situation is given in Fig.~\ref{fig:kpi2p_xf} which
shows the available measurements of the (K$^+$ + $\pi^+$)/p ratio as a function
of $x_F$ for eight values of $p_T$ between 0.1 and 1.3~GeV/c. 
The full lines give the combination of the NA49 data with Fermilab
and ISR data in p+p interactions as interpolated in \cite{pp_proton},
the open squares the p+p data of \cite{brenner}. An impressive
consistency on an about 10\% level between the experimental results from the 
two different reactions is apparent. This allows for a safe
extrapolation into the region above $x_F$~=~0.6 where particle
identification via $dE/dx$ is not available. 

%      Fig.14  

\begin{figure}[h]
  \begin{center}
  	\includegraphics[width=8.5cm]{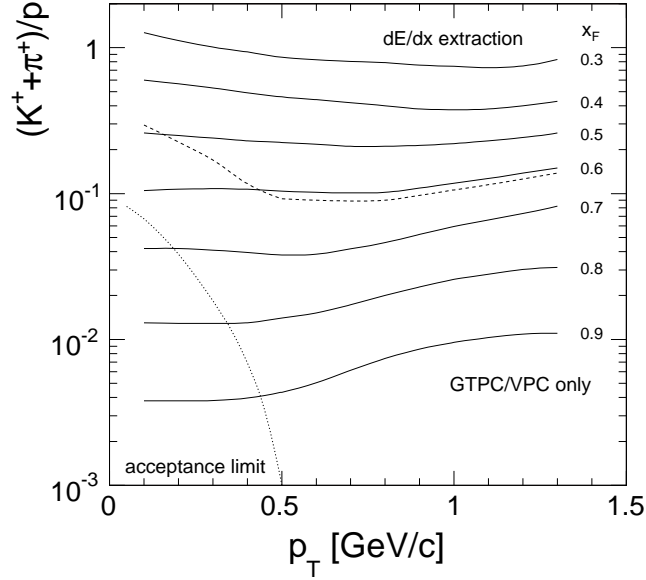}
 	\caption{Interpolated (K$^+$ + $\pi^+$)/p ratios as a function of
             $p_T$ for different values of $x_F$ between 0.3 and 0.9, full lines.
             Broken line: border between available TPC information and the
             GTPC/VPC combination (Tracking only). Dotted line: acceptance 
             limit of the NA49 detector}
  	 \label{fig:kpi2p_pt}
  \end{center}
\end{figure}

The interpolated (K$^+$ + $\pi^+$)/p ratios are again presented in Fig.~\ref{fig:kpi2p_pt}, 
here as a function of $p_T$ for several values of $x_F$ between
0.3 and 0.9. The limits of $dE/dx$ identification and NA49
acceptance are given by the broken and dotted lines, respectively.

%     Fig.15

This Figure demonstrates that the (K$^+$ + $\pi^+$)/p ratios show only a
small dependence on $p_T$. They are of order 10\% at the limit of
the $dE/dx$ identification and decrease rapidly to the 1\% level
at $x_F$~=~0.9. This implies that possible systematic differences
between p+p and the p+C interactions in the extrapolated region
should have effects on the percent level and below concerning
the extracted proton yields.

The equality of the meson/baryon ratio between p+p and the p+C
interactions may be taken as a first physics result of this
paper. It has two aspects: Firstly, 60\% of the minimum bias
p+C interactions correspond to single projectile collisions
inside the nucleus \cite{pc_discus}. These collisions should indeed produce
particle ratios equivalent to p+p interactions. Secondly, the
phenomenon of "stopping", that is of the transfer of particle
yields in multiple interactions towards the central region, 
is not limited to baryons but applies also to mesons \cite{pc_pion,pc_discus}.
Hence again an expected similarity in the meson/baryon ratios. 

%
% ****************************** Section 4.2 ****************************
%
\subsection{Anti-proton extraction}
\vspace{3mm}
\label{sec:aprot_ext}

\begin{figure}[h]
  \begin{center}
  	\includegraphics[width=12cm]{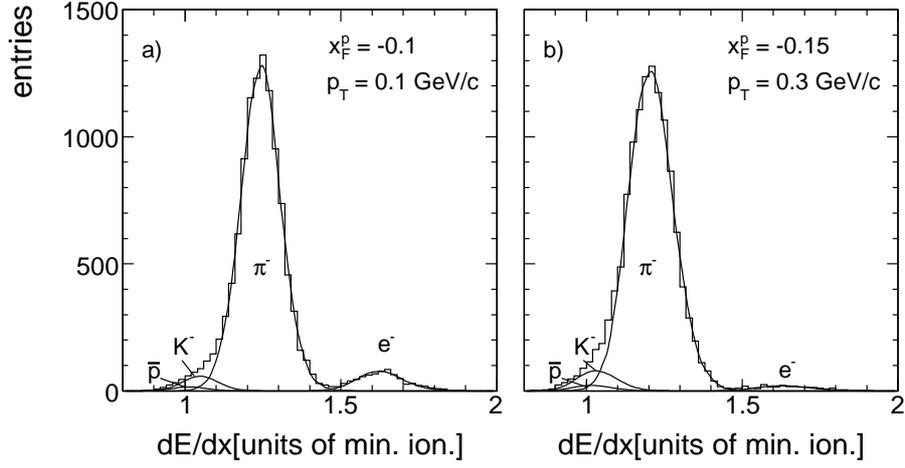}
 	\caption{$dE/dx$ distributions for negative particles
             a) $x_{F}$~=~-0.1, $p_T$~=~0.1 and b) $x_{F}$~=~-0.15, $p_T$~=~0.3}
  	 \label{fig:dedxneg}
  \end{center}
\end{figure}

As stated above the extension of the determination of anti-proton
yields into the backward hemisphere suffers from the fact that
the $\overline{\textrm{p}}$/$\pi^-$ and $\overline{\textrm{p}}$/K$^-$ ratios decrease
with decreasing $x_{F}$. This is shown by the energy loss distributions of two 
typical bins in $x_F$ and $p_T$ in Fig.~\ref{fig:dedxneg}.

%    Here Fig.16 

This effect is largely due to the asymmetry between the effective
$x_F$ for light particles and anti-protons due to the transformation
to the lab momentum using proton mass, see Fig.~\ref{fig:partxf}. 
Thus at $x_F^{\overline{\textrm{p}}}$~=~-0.2 pions are sampled close to 
maximum yield whereas the anti-proton cross section is steeply decreasing.

\begin{figure}[h]
  \begin{center}
  	\includegraphics[width=9.cm]{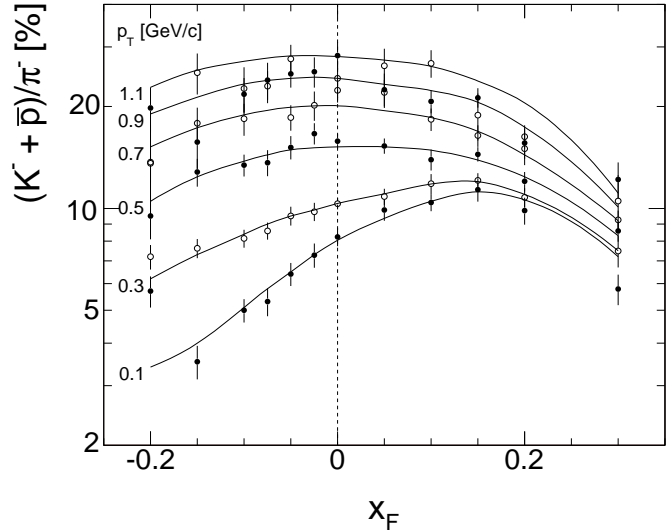}
 	\caption{(K$^-$ + $\overline{\textrm{p}}$)/$\pi^-$ ratios as a function of $x_{F}$ for fixed
              values of $p_T$. Full lines: two-dimensional interpolation of
              the fitted ratios}
  	 \label{fig:pk2pi}
  \end{center}
\end{figure}

While the extraction of pion yields therefore presents no problem in this
phase space region, the fits of kaon and anti-proton densities become
strongly correlated with sizeable uncertainties in their relative
position on the energy loss scale. The combined (K$^-$ + $\overline{\textrm{p}}$) 
yields however stay well defined with respect to the pions.
This is shown by the fitted (K$^-$ + $\overline{\textrm{p}}$)/$\pi^-$ ratios of Fig.~\ref{fig:pk2pi}.

%     Here Fig.17 

\begin{figure}[h]
  \begin{center}
  	\includegraphics[width=7.cm]{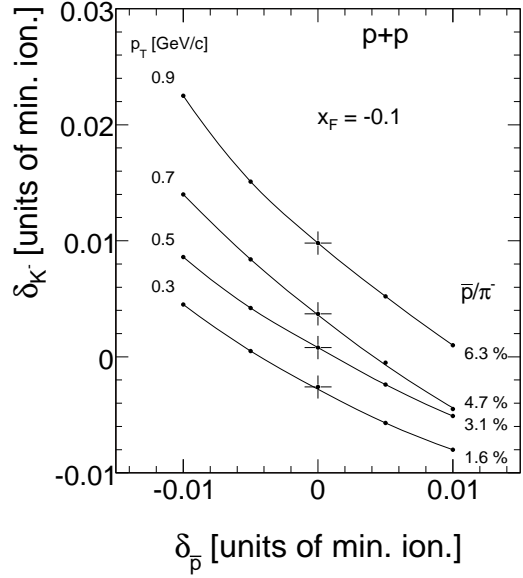}
 	\caption{Correlation between the relative displacements $\delta_{\overline{\textrm{p}}}$ 
 	         and $\delta_{\textrm{K}^-}$ in p+p collisions at \mbox{$x_{F}$~=~-0.1} for fixed
             values of $p_T$, imposing the forward-backward symmetry of cross 
             section in this interaction. The lines are given to guide the eye}
  	 \label{fig:shifts_pp}
  \end{center}
\end{figure}

In order to resolve the K$^-$-$\overline{\textrm{p}}$ ambiguity, the high statistics data
on $\overline{\textrm{p}}$ production in p+p interactions \cite{pp_proton} in order to fix 
the position of the K$^-$ and $\overline{\textrm{p}}$ peaks in the corresponding $dE/dx$ distributions.
In this symmetric configuration, the measured $\overline{\textrm{p}}$ cross
sections may be reflected into the backward hemisphere and thereby 
the correlation between the positions of the K$^-$ and $\overline{\textrm{p}}$ 
peaks extracted. The positions of these peaks is
expressed as their systematic deviations in the $dE/dx$ variable
from the Bethe-Bloch parametrization, $\delta_{\overline{\textrm{p}}}$ and $\delta_{\textrm{K}^-}$,
in units of minimum ionization. As shown in refs. \cite{pp_kaon,pp_proton}
these shifts are experimentally determined with an accuracy
of about 0.001 in the scale of minimum ionization. The correlation
is shown in Fig.~\ref{fig:shifts_pp} for an example at $x_{F}$~=~-0.1 for fixed values
of $p_T$ between 0.3 and 0.9~GeV/c, also indicating the corresponding
$\overline{\textrm{p}}$/$\pi^-$ ratios. 

%     Fig.18

Using the same correlation for the p+C data, effective $\overline{\textrm{p}}$/$\pi^-$ ratios
are obtained. The observed stability of these ratios over the full range
of the correlations is a strong test of the validity of the method. 

\begin{figure}[h]
  \begin{center}
  	\includegraphics[width=8cm]{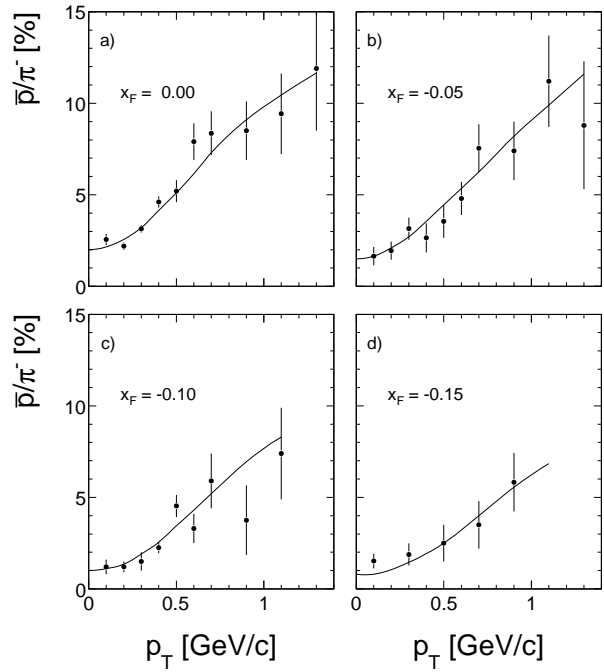}
 	\caption{$\overline{\textrm{p}}$/$\pi^-$ ratios as a function of $p_T$ for a) $x_{F}$~=~0, 
             b) $x_{F}$~=~-0.05, c) $x_{F}$~=~-0.1 and d) $x_{F}$~=~-0.2}
  	 \label{fig:p2pi_ptdist}
  \end{center}
\end{figure}

Fig.~\ref{fig:p2pi_ptdist} presents the obtained $\overline{\textrm{p}}$/$\pi^-$ 
ratios as a function of $p_T$ for
$x_{F}$~=~-0.05, -0.1 and -0.15, together with the directly fitted values
at $x_{F}$~=~0. Full lines: two-dimensional interpolation of the ratios.

%     Fig.19

\begin{figure}[h]
  \begin{center}
  	\includegraphics[width=9cm]{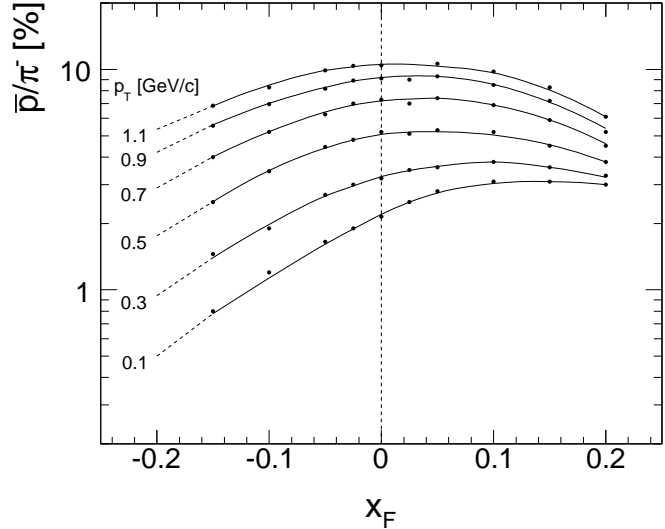}
 	\caption{$\overline{\textrm{p}}$/$\pi^-$ ratios as a function of $x_{F}$ for fixed values
             of $p_T$. The values at $x_{F}$~=~-0.2 are extrapolations using the broken lines}
  	 \label{fig:p2pi}
  \end{center}
\end{figure}

A complete picture of the $\overline{\textrm{p}}$/$\pi^-$ ratios used in this analysis is
given in Fig.~\ref{fig:p2pi} as a function of $x_{F}$ which combines the directly fitted
ratios in the forward hemisphere with the ones obtained using the
reflection method described above in the backward hemisphere.

%     Here Fig.20 .

Note that the values at $x_{F}$~=~-0.2 are obtained by extrapolation following
the broken lines. Note also that the applied method allows the
extraction of the ratios in the percent and sub-percent region.

%
% ****************************** Section 5 ****************************
%
\section{Evaluation of invariant cross sections and corrections}
\vspace{3mm}
\label{sec:corr}

The invariant cross section,

\begin{equation}
  f(x_F,p_T) = E(x_F,p_T) \cdot \frac{d^3\sigma}{dp^3} (x_F,p_T)
\end{equation}
is experimentally determined by the measured quantity \cite{pp_pion}

\begin{equation}
  f_{\textrm{meas}}(x_F,p_T,\Delta p^3) =
  E(x_F,p_T,\Delta p^3) \cdot \frac{\sigma_{\textrm{trig}}}
  {N_{\textrm{ev}}} \cdot
  \frac{\Delta n(x_F,p_T,\Delta p^3)}{\Delta p^3}~~~,
\end{equation}
where $\Delta p^3$ is the finite phase space element defined by the bin
width with $x_F$ and $p_T$ being defined in the bin center.

As described in \cite{pp_pion} several steps of normalization and correction
are necessary in order to make $f_{\textrm{meas}}(x_F,p_T,\Delta p^3)$ approach
$f(x_F,p_T)$. The determination of the trigger cross section and its
deviation from the total inelastic p+C cross section have been
discussed in \cite{pc_pion}. The following corrections for baryons have been
applied and will be discussed below:

\begin{itemize}
  \item treatment of the empty target contribution
  \item effect of the interaction trigger
  \item feed-down from weak decays of strange particles
  \item re-interaction in the target volume
  \item absorption in the detector material
  \item effects of final bin width
\end{itemize}

%
% ****************************** Section 5.1 ****************************
%
\subsection{Empty target contribution}
\vspace{3mm}
\label{sec:et_cor}

\begin{figure}[h]
  \begin{center}
  	\includegraphics[width=6.5cm]{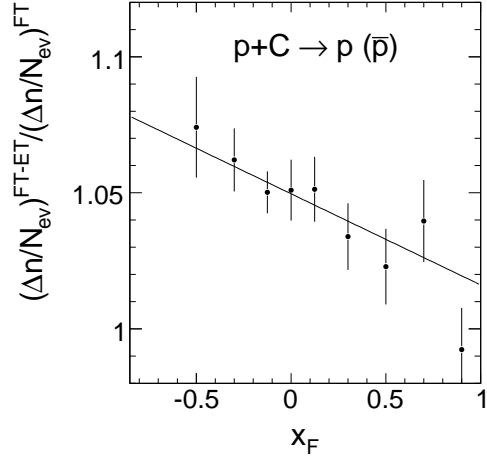}
 	\caption{Empty target correction as a function of $x_F$. The full
             line shows the chosen linear interpolation}
  	 \label{fig:empty}
  \end{center}
\end{figure}

This correction has been determined experimentally using the available
empty target data sample, as described in \cite{pc_pion}. The resulting correction
is essentially determined by the different amounts of empty events
in full and empty target condition. It is within errors $p_T$ independent
and equal for protons and anti-protons. It increases from about 2\% in 
in the far forward direction to about 7\% in the most backward region
as shown in Fig.~\ref{fig:empty}.

%     Here Fig.21 .

%
% ****************************** Section 5.2 ****************************
%
\subsection{Effect of the interaction trigger}
\vspace{3mm}
\label{sec:s4_cor}

\begin{figure}[h]
  \begin{center}
  	\includegraphics[width=9.5cm]{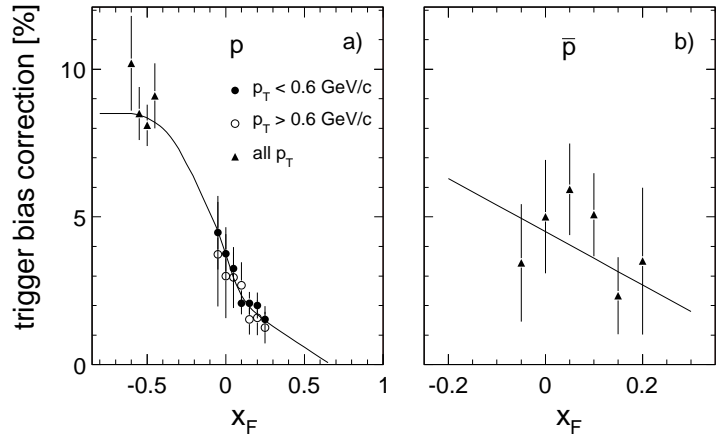}
 	\caption{Trigger bias correction as a function of $x_F$ for
             a) protons and b) anti-protons. The chosen interpolation is 
             given by the full lines}
  	 \label{fig:s4}
  \end{center}
\end{figure}

Due to the high trigger efficiency of 93\% \cite{pc_pion} this correction is small
compared to p+p interactions \cite{pp_pion}. It has been determined experimentally
by increasing the diameter of the trigger counter using the accumulated
data. Within its statistical uncertainty it is independent of $p_T$ and
similar for protons and anti-protons. Its $x_F$ dependence as shown in
Fig.~\ref{fig:s4} is following the expected trend \cite{pp_proton} where the fast decrease in
forward direction as compared to p+p collisions is due to the lower 
particle yields at high $x_F$. Note that the slope of
the $x_F$ dependence for anti-protons corresponds to the one for
protons. Both corrections have to increase in backward direction
due to the effects of hadronic factorisation, see also \cite{pp_pion}.

%     Here Fig.22 .

%
% ****************************** Section 5.3 ****************************
%
\subsection{Feed-down correction}
\vspace{3mm}
\label{sec:fd_cor}

\begin{figure}[h]
  \begin{center}
  	\includegraphics[width=15.5cm]{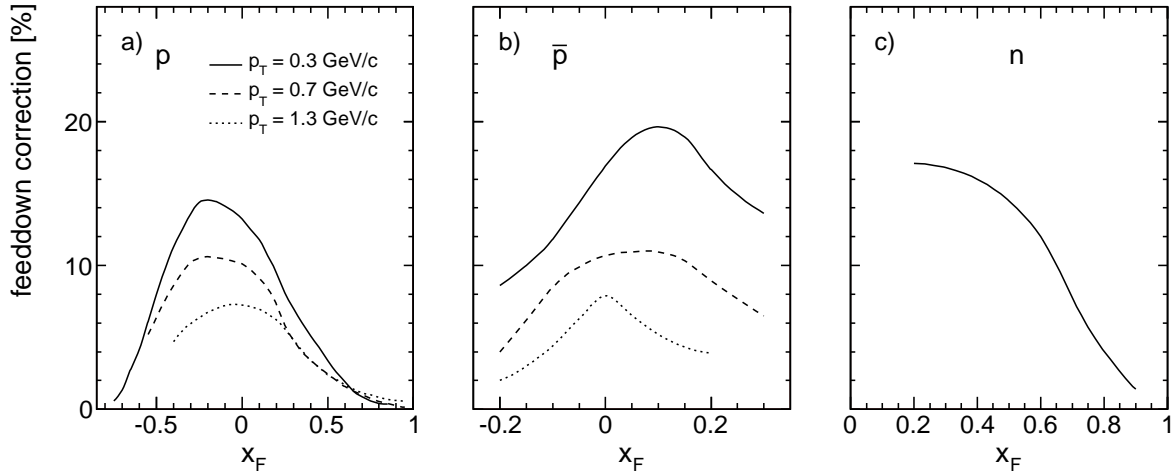}
 	\caption{Feed-down correction as a function of $x_F$ at
             different $p_T$ values for a) protons, b) anti-protons and
             c) neutrons, in the latter case integrated over $p_T$}
  	 \label{fig:fd}
  \end{center}
\end{figure}

The hyperon cross sections relative to p+p collisions established 
in \cite{pc_pion} for $x_F >$~0 have been used. These cross section ratios approach
at $x_F <$~0 the expected factor of 1.6 corresponding to the number of
intra-nuclear projectile collisions, in account of the fact that
for $\Lambda$ and $\overline{\Lambda}$ there is no isospin effect \cite{iso}. 
For the contribution from target fragmentation this ratio should
be constant into the backward hemisphere. For the determination
of the feed-down correction the corresponding yields have to be
folded with the on-vertex baryon reconstruction efficiency which  
reaches large values in the far backward hemisphere. The resulting
correction in percent of the total proton yield drops however
quickly below $x_F <$~-0.2 due to the decrease of the $\Lambda$ cross section
relative to protons and due to the fact that the important baryon
contribution from intra-nuclear cascading has no hyperon content.
The numerical values in percent of the baryon yields are shown in
Fig.~\ref{fig:fd} as a function of $x_F$.

%     Here Fig.23 .

In comparison to the pion data \cite{pc_pion} this correction reaches considerable
values of up to 20\% for the anti-protons and therefore constitutes,
together with the absolute normalization, the most important source 
of systematic uncertainty. 

%
% ****************************** Section 5.4 ****************************
%
\subsection{Re-interaction in the target}
\vspace{3mm}
\label{sec:tr_cor}

The carbon target has an interaction length of 1.5\%, which corresponds
to about half of the length of the hydrogen target used in p+p
collisions. The expected re-interaction correction is therefore
smaller than +0.5\% in the forward and -2\% in the backward hemisphere.
The values obtained in \cite{pp_proton} have therefore been downscaled accordingly. 

%
% ****************************** Section 5.5 ****************************
%
\subsection{Absorption in the detector material}
\vspace{3mm}
\label{sec:abs_cor}

\begin{figure}[h]
  \begin{center}
  	\includegraphics[width=11cm]{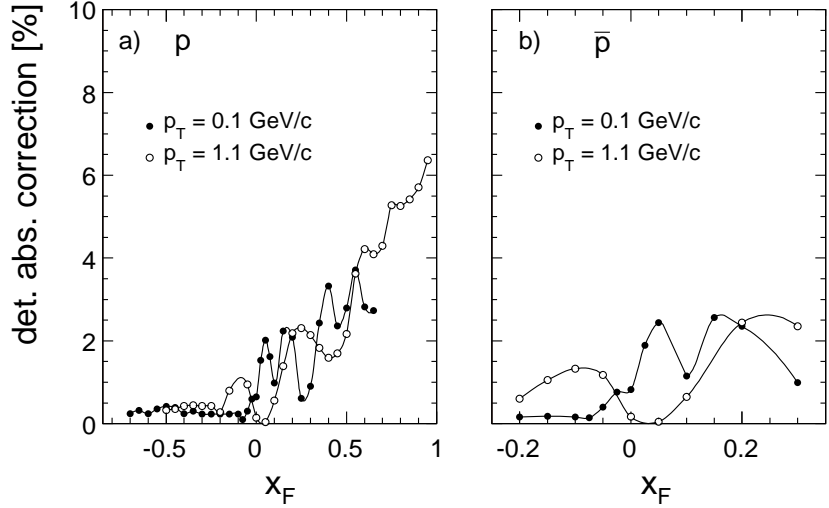}
 	\caption{Detector absorption correction as a function of
             $x_F$ at different $p_T$ values for a) protons and b) anti-protons}
  	 \label{fig:da}
  \end{center}
\end{figure}

The absorption losses in the detector material are equal to the ones
obtained in \cite{pp_proton}. Baryons in the newly exploited region in the far 
backward direction feature short track lengths in the first NA49 TPC
detector only and are not affected by any support structures; hence
the corresponding corrections are below the 1\% level as shown in
Fig.~\ref{fig:da}.

%     Here Fig.24

%
% ****************************** Section 5.6 ****************************
%
\subsection{Binning correction}
\vspace{3mm}
\label{sec:bin_cor}

\begin{figure}[h]
  \begin{center}
  	\includegraphics[width=11cm]{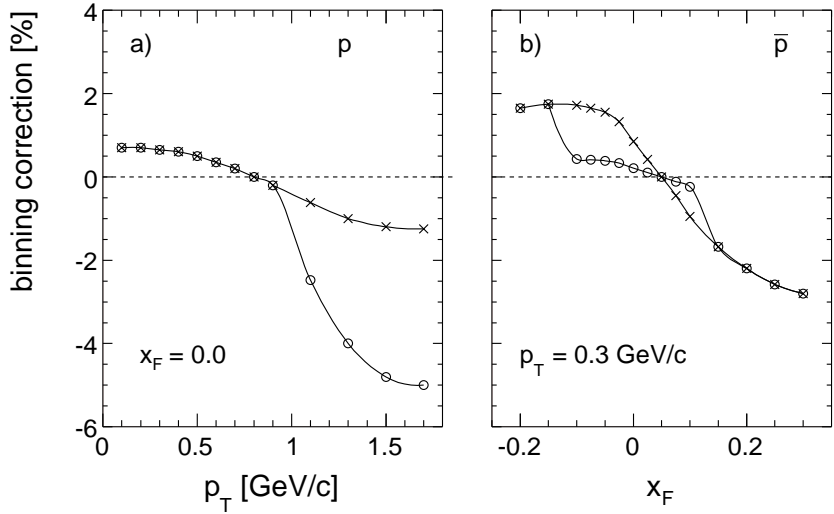}
 	\caption{Binning correction a) in $p_T$ for protons at $x_F$~=~0
             and b) in $x_F$ for anti-protons at $p_T$~=~0.3~GeV/c. The values 
             for the nominal bin widths of 0.1~GeV/c in $p_T$ and 0.05 in $x_F$ 
             are indicated by crosses, for the actually used bin widths by
             open circles}
  	 \label{fig:bincor}
  \end{center}
\end{figure}

The correction for finite bin width follows the scheme developed in 
\cite{pp_pion} using the local second derivative of the particle density
distribution. This correction stays, despite of the rather sizeable
bins used in some areas of the p+C data, generally below the $\pm$2\%
level, being negligible in the $x_F$ co-ordinate for protons due to
their rather flat $dn/dx_F$ distribution. Two typical distributions
of the $p_T$ correction for protons and for the $x_F$ correction for
anti-protons are shown in Fig.~\ref{fig:bincor} both for the nominal bin width
of 0.1~GeV/c in $p_T$ and 0.05 in $x_F$ and for the actually used bin
widths.

%     Here Fig.25 .

%
% ****************************** Section 5.7 ****************************
%
\subsection{Systematic errors}
\vspace{3mm}
\label{sec:syst}

\begin{figure}[h]
  \begin{center}
  	\includegraphics[width=12.5cm]{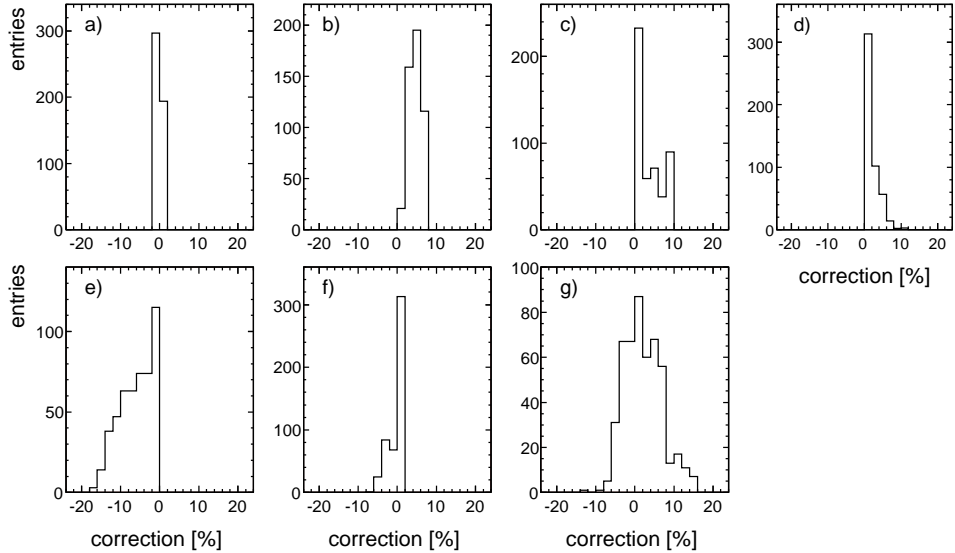}
 	\caption{Distribution of proton corrections for a) target re-interaction,
             b) empty target, c) trigger bias, d) absorption, e) feed-down,
             f) binning, g) total}
  	 \label{fig:cor_prot}
  \end{center}
\end{figure}

An estimation of the systematic errors induced by the absolute
normalization and the applied corrections is given in Table~\ref{tab:syst}.
For the proton and anti-proton extraction  in the newly exploited 
backward regions an additional systematic error due to particle 
identification is indicated. An upper limit of 7.0\% (8.5\%) for
protons (anti-protons) results from the linear addition of the
error sources, increasing to 10\% (14.5\%) in the backward region.
Quadratic summation results in the corresponding values of
3.7\% (4.2\%) and 4.7\% (7.3\%). Note that even the upper error limits given
by the extreme and improbable case of linear addition stay below
a ten percent margin for the charged baryons. The distribution of the numerical
values of the corrections in all measured bins is shown in Fig.~\ref{fig:cor_prot}
for protons.

%Here Table 1  Systematic errors
\begin{table}[h]
\renewcommand{\tabcolsep}{0.2pc} 
\renewcommand{\arraystretch}{0.9}
\small
%\footnotesize
%\scriptsize
\begin{center}
\begin{tabular}{|lccc|lcc|}
\hline
                                  &        &  p       & $\overline{\textrm{p}}$  &                       &  &   n         \\ \hline
Normalization                     &        & 2.5\%    & 2.5\%                    & Normalization         &  &  2.5\%   \\ 
Tracking efficiency               &        & 0.5\%    & 0.5\%                    &                       &   &    \\ 
Trigger bias                      &        & 0.5\%    & 2.0\%                    & Trigger bias          &  &  1\%     \\ 
Feed-down                         &        & 2.5\%    & 2.5\%                    & Feed-down             &  &  3\% \\ 
Detector absorption               & \multirow{2}{2mm}{\footnotesize $\biggr\}$}  
                                  & \hspace{-8mm} \multirow{2}{0.5mm}{0.5\%}            
                                  & \hspace{-8mm} \multirow{2}{0.5mm}{0.5\%}  & 
                                                        Detector absorption & \multirow{3}{2mm}{\footnotesize $\biggr\}$}&   \\           
Target re-interaction             &        &           &                         & Target re-interaction  &   &   0.5 -- 1.5\% \\ 
Binning                           &        & 0.5\%     & 0.5\%                   & Binning correction     &   &    \\ 
Particle ID backward              &        & 0 -- 6\%  & 2 -- 10\%          & Acceptance  &   &   0 -- 2\%  \\
              &        &     &                                             & Energy scale error     &   &   4 -- 8\%  \\ 
&        &            &                                                    & Energy resolution unfolding  &   &   3 -- 8\%   \\
\cline{1-4}
backward &&            &                                                 & Charged veto efficiency      &   &   2 -- 3\%  \\
Total(upper limit)                &        & 10.0\%    & 14.5\%             & Cluster overlap              &   &   2\%     \\
Total(quadratic sum)              &        & 4.7\%     & 7.3\%             & Hadron identification        &   &   2 -- 5\%   \\
\cline{1-4}
forward &&            &                                                    & K$_L^0$ contribution         &   &   0 -- 3\% \\
\cline{5-7}
Total(upper limit)                &        & 7.0\%     & 8.5\%             & Total (upper limit)          &   &   29\%  \\ 
Total(quadratic sum)              &        & 3.7\%     & 4.2\%             & Total (quadratic sum)        &  &    10\%   \\ \hline

\end{tabular}
\end{center}
\caption{Systematic errors}
\label{tab:syst}
\end{table}

%     Here Fig.26  

%
% ****************************** Section 6 ****************************
%
\section{Results on double-differential cross sections for p and $\overline{\textrm{p}}$}
\vspace{3mm}
\label{sec:results}

%
% ****************************** Section 6.1 ****************************
%
\subsection{Data tables}
\vspace{3mm}
\label{sec:tables} 

The binning scheme presented in Sect.~\ref{sec:setup} results in 491 and 121
data values for protons and anti-protons, respectively. These
are presented in Tables~\ref{tab:prot_cs} and \ref{tab:aprot_cs}.

%     Here Table 2 
\begin{table}
\renewcommand{\tabcolsep}{0.15pc} 
\renewcommand{\arraystretch}{0.9}
\scriptsize
\begin{center}
\begin{tabular}{|c|cr|cr|cr|cr|cr|cr|cr|cr|cr|cr|}
\hline
\multicolumn{21}{|c|}{$f(x_F,p_T), \Delta f$} \\ 
\hline
$p_T \backslash x_F$ & \multicolumn{2}{c|}{-0.8} & \multicolumn{2}{c|}{-0.75} & \multicolumn{2}{c|}{-0.7} & \multicolumn{2}{c|}{-0.65} & \multicolumn{2}{c|}{-0.6} & \multicolumn{2}{c|}{-0.55} & \multicolumn{2}{c|}{-0.5} & \multicolumn{2}{c|}{-0.45} & \multicolumn{2}{c|}{-0.4} & \multicolumn{2}{c|}{-0.35}\\ \hline
0.025 && && &316.4&6.00 &217.2&6.91 &213.3&6.65 &179.8&6.88 &173.0&6.61 && && &&\\
0.05 && && &308.1&8.17 &262.3&4.46 &254.7&5.50 &176.0&4.87 &166.8&6.09 && && &&\\
0.075 && && &313.1&8.49 &275.2&3.55 &229.3&3.70 &177.0&4.01 &163.1&3.93 && && &&\\
0.1 && && &368.4&13.0 &279.3&6.70 &224.9&5.12 &207.2&4.37 &165.3&5.32 &118.6&3.70 &102.0&3.76 &78.7&4.24\\
0.125 &465.1&11.1 &320.8&16.9 && && && && && && && &&\\
0.15 &539.9&10.7 &376.1&7.00 && &280.8&10.4 && &194.5&5.08 && && && &&\\
0.175 &437.5&9.88 && && && && && && && && &&\\
0.2 &363.8&6.34 &348.5&4.48 &306.4&4.67 &233.2&4.30 &221.7&5.22 &184.6&4.93 &143.9&5.41 &107.2&3.84 &86.4&3.59 &67.2&4.06\\
0.25 && &280.7&6.24 && && && && && && && &&\\
0.3 && &217.2&6.52 &222.0&6.87 &152.7&4.94 &150.0&6.18 &123.1&4.60 &94.4&5.45 &88.3&4.45 &69.5&3.55 &58.8&3.95\\
0.4 && && &141.3&7.45 &128.0&7.51 &110.2&7.74 &86.1&5.27 &72.7&5.46 &64.6&4.40 &51.0&4.81 &45.3&3.74\\
0.5 && && && &65.2&9.40 &72.6&8.54 &61.5&5.58 &53.9&5.37 &44.9&5.81 &36.7&5.03 &32.6&4.23\\
0.6 && && && &38.5&11.2 &37.9&10.8 &36.8&6.59 &32.8&7.74 &29.4&6.58 &28.4&5.20 &20.8&5.13\\
0.7 && && && && &21.2&13.3 &23.2&7.78 &23.0&7.37 &19.0&7.57 &17.6&7.48 &15.47&5.98\\
0.8 && && && && && &14.0&9.43 &11.65&7.96 &13.6&8.18 &11.10&8.85 & 9.60&6.00\\
0.9 && && && && && & 9.25&8.80 & 8.16&8.29 & 8.64&10.2 & 6.57&8.91 & 6.49&8.37\\
1.1 && && && && && && & 2.17&13.2 & 3.14&8.93 & 2.25&9.51 & 2.39&8.94\\
1.3 && && && && && && & 0.68&21.9 & 1.21&13.0 & 1.19&12.5 & 1.07&12.4\\
1.5 && && && && && && & 0.339&20.6 & 0.504&18.7 & 0.425&19.0 & 0.443&19.7\\
1.7 && && && && && && & 0.145&19.2 && && &&\\
1.9 && && && && && && & 0.056&28.9 && && &&\\
\hline
$p_T \backslash x_F$ & \multicolumn{2}{c|}{-0.3} & \multicolumn{2}{c|}{-0.25} & \multicolumn{2}{c|}{-0.2} & \multicolumn{2}{c|}{-0.15} & \multicolumn{2}{c|}{-0.1} & \multicolumn{2}{c|}{-0.075} & \multicolumn{2}{c|}{-0.05} & \multicolumn{2}{c|}{-0.025} & \multicolumn{2}{c|}{0.0} & \multicolumn{2}{c|}{0.025}\\ \hline
0.1 &61.6&6.65 &49.1&5.27 &38.1&4.51 &31.7&9.29 &28.36&2.98 &29.7&3.85 &24.15&2.88 &22.92&3.97 &21.12&4.08 &19.15&4.39\\
0.2 &53.6&6.66 &43.7&3.20 &32.6&4.15 &29.7&6.79 &25.5&4.18 &25.64&2.95 &23.08&2.07 &19.43&3.08 &20.45&2.97 &21.14&2.99\\
0.3 &44.5&4.61 &35.7&3.31 &30.3&3.56 &27.0&5.96 &22.49&3.67 &20.70&2.69 &19.72&1.87 &17.97&2.65 &17.85&2.66 &15.33&3.16\\
0.4 &36.4&4.00 &27.45&3.32 &23.07&3.59 &19.4&6.31 &17.38&3.65 &16.36&3.77 &15.54&2.60 &14.95&4.20 &14.84&2.58 &13.22&3.01\\
0.5 &26.5&4.12 &21.0&5.22 &17.51&3.68 &14.36&6.45 &13.32&3.86 &12.64&4.49 &12.59&2.67 &12.63&4.19 &11.53&4.31 &10.67&3.06\\
0.6 &17.9&5.65 &15.68&4.27 &12.48&4.39 &10.37&7.02 & 9.30&4.28 & 9.83&4.73 & 8.74&3.41 & 8.71&4.67 & 7.93&4.84 & 8.42&4.20\\
0.7 &13.43&5.97 &11.73&4.51 & 8.56&4.99 & 7.12&5.03 & 6.71&4.74 & 6.64&5.44 & 5.91&3.88 & 6.72&6.00 & 5.75&6.35 & 5.53&4.93\\
0.8 & 8.36&7.08 & 7.08&5.70 & 6.19&5.51 & 4.77&5.85 & 4.36&5.63 & 4.70&7.35 & 4.05&5.32 & 4.04&7.40 & 4.21&7.21 & 4.00&7.46\\
0.9 & 5.75&8.03 & 4.91&6.48 & 3.99&6.70 & 3.38&6.68 & 2.81&6.71 & 3.28&8.49 & 2.70&6.26 & 2.82&8.56 & 2.84&8.61 & 2.79&8.67\\
1.1 & 2.13&8.42 & 1.86&6.74 & 1.69&6.58 & 1.539&6.44 & 1.426&6.34 && & 1.269&6.26 && & 1.451&5.68 &&\\
1.3 & 0.84&12.6 & 0.859&11.0 & 0.703&9.78 & 0.608&10.0 & 0.568&9.42 && & 0.566&9.14 && & 0.543&8.88 &&\\
1.5 & 0.345&19.0 & 0.384&13.2 & 0.393&12.5 & 0.305&13.7 & 0.217&14.7 && & 0.241&13.2 && & 0.281&12.1 &&\\
1.7 & 0.166&20.4 && & 0.134&20.7 & 0.151&18.2 & 0.100&21.2 && & 0.116&19.1 && & 0.086&21.4 &&\\
1.9 & 0.080&20.2 && && && & 0.055&19.9 && && && & 0.056&18.6 &&\\
\hline
$p_T \backslash x_F$ & \multicolumn{2}{c|}{0.05} & \multicolumn{2}{c|}{0.075} & \multicolumn{2}{c|}{0.1} & \multicolumn{2}{c|}{0.15} & \multicolumn{2}{c|}{0.2} & \multicolumn{2}{c|}{0.25} & \multicolumn{2}{c|}{0.3} & \multicolumn{2}{c|}{0.35} & \multicolumn{2}{c|}{0.4} & \multicolumn{2}{c|}{0.45}\\ \hline
0.1 &22.69&4.19 &23.5&4.32 &24.0&4.93 &28.0&3.66 &31.2&3.51 &43.0&3.15 &44.5&3.52 &48.6&5.09 &58.2&3.57 &60.6&4.51\\
0.2 &19.38&3.21 &20.18&3.31 &23.45&3.58 &27.91&2.60 &29.52&2.69 &38.72&2.48 &45.3&2.48 &44.4&3.82 &51.1&3.30 &48.8&3.57\\
0.3 &15.74&2.93 &17.56&3.21 &17.20&3.41 &21.65&2.42 &24.82&2.29 &31.18&2.27 &37.0&2.75 &36.5&3.43 &40.0&3.03 &38.4&3.29\\
0.4 &14.06&3.02 &13.33&3.24 &15.04&3.21 &17.37&2.36 &20.98&2.21 &24.38&2.45 &27.38&2.79 &29.1&4.51 &30.86&2.98 &31.20&3.17\\
0.5 &11.34&3.07 &10.80&3.27 &11.41&3.35 &13.17&2.46 &15.92&2.24 &19.66&2.46 &20.22&2.92 &21.73&3.48 &22.84&3.12 &23.84&3.97\\
0.6 & 8.28&3.35 & 8.90&3.38 & 8.41&3.63 &10.09&2.60 &11.31&2.47 &12.67&2.80 &14.75&3.14 &14.74&3.85 &17.33&3.28 &16.61&4.34\\
0.7 & 5.58&3.88 & 5.46&4.03 & 6.42&3.91 & 6.46&3.04 & 6.34&4.25 & 8.23&3.23 & 9.58&3.62 &10.29&4.20 &11.00&3.83 &11.92&4.76\\
0.8 & 4.35&4.23 & 5.02&5.30 & 4.47&5.78 & 4.81&3.36 & 5.06&3.27 & 5.94&3.99 & 6.38&4.17 & 6.83&4.99 & 7.62&4.72 & 6.68&5.96\\
0.9 & 3.15&7.04 & 2.60&7.05 & 3.18&6.64 & 2.81&4.12 & 3.12&3.90 & 3.79&4.74 & 3.79&5.11 & 4.69&3.90 & 4.62&5.77 & 4.74&6.68\\
1.1 & 1.236&5.60 && & 1.230&5.66 & 1.311&5.64 & 1.212&5.81 & 1.556&5.85 & 1.557&6.27 & 1.536&6.33 & 1.582&6.25 & 1.71&7.10\\
1.3 & 0.532&9.13 && & 0.464&10.4 & 0.544&9.00 & 0.525&7.82 & 0.503&9.59 & 0.452&10.7 & 0.550&13.6 & 0.582&9.52 & 0.549&11.6\\
1.5 & 0.240&13.4 && & 0.207&14.7 & 0.217&13.5 & 0.203&12.2 & 0.245&13.1 & 0.237&14.1 & 0.218&15.4 & 0.216&16.4 & 0.182&18.8\\
1.7 & 0.146&16.8 && & 0.092&21.9 & 0.100&20.0 & 0.096&23.8 & 0.080&21.9 & 0.074&23.9 && & 0.089&24.2 & 0.048&37.1\\
1.9 && && & 0.0284&28.6 && && && & 0.0390&21.0 && & 0.0307&27.7 &&\\
\hline
$p_T \backslash x_F$ & \multicolumn{2}{c|}{0.5} & \multicolumn{2}{c|}{0.55} & \multicolumn{2}{c|}{0.6} & \multicolumn{2}{c|}{0.65} & \multicolumn{2}{c|}{0.7} & \multicolumn{2}{c|}{0.75} & \multicolumn{2}{c|}{0.8} & \multicolumn{2}{c|}{0.85} & \multicolumn{2}{c|}{0.9} & \multicolumn{2}{c|}{0.95}\\ \hline
0.05 && &80.3&12.5 && && && && && && && &&\\
0.1 &67.9&4.52 &94.8&8.11 &60.9&10.8 && && && && && && &&\\
0.15 && &65.6&7.88 &78.9&7.81 &71.7&8.64 && && && && && &&\\
0.2 &53.3&3.64 &60.2&7.12 &68.7&7.24 &62.0&8.03 &71.4&7.96 && && && && &&\\
0.25 && &51.8&6.92 &79.5&6.01 &55.8&7.60 &63.6&7.47 &70.3&7.50 && && && &&\\
0.3 &42.3&3.36 &43.7&6.88 &46.5&7.16 &59.9&6.74 &46.6&8.01 &46.4&8.45 &51.3&8.33 && && &&\\
0.35 && &34.6&7.16 &39.5&7.20 &48.4&6.93 &43.5&7.67 &43.4&8.06 &49.0&7.91 &41.8&8.91 && &&\\
0.4 &30.9&3.43 &34.7&6.68 &32.2&7.47 &40.2&7.09 &37.5&7.76 &32.3&8.70 &31.4&9.24 &35.7&9.09 &42.9&8.51 &&\\
0.45 && &27.3&7.12 &24.6&8.06 &28.5&7.96 &25.8&8.80 &31.3&8.42 &29.1&9.09 &29.2&9.41 &39.6&8.39 &47.1&7.76\\
0.5 &22.20&3.64 &22.2&7.50 &23.0&7.91 &25.7&7.98 &19.8&10.1 &17.9&10.5 &20.2&10.3 &22.3&10.2 &31.1&8.80 &45.0&7.54\\
0.6 &15.46&3.98 &15.48&4.36 &17.28&4.43 &16.41&4.85 &16.6&6.77 &16.0&7.20 &13.0&8.28 &14.9&8.01 &15.7&7.98 &25.0&4.82\\
0.7 &11.10&4.36 &11.38&4.70 &10.21&5.34 &11.28&5.40 &10.16&7.98 & 9.64&8.57 & 8.58&9.41 & 7.19&10.6 & 9.02&9.71 &14.10&5.95\\
0.8 & 8.07&4.77 & 6.82&5.68 & 6.66&6.17 & 5.62&7.14 & 6.37&9.41 & 7.11&9.33 & 5.92&10.5 & 5.52&11.3 & 3.72&12.6 & 7.18&7.83\\
0.9 & 4.04&6.34 & 4.50&6.58 & 4.08&7.39 & 4.05&7.91 & 3.12&12.6 & 3.64&12.2 & 3.01&14.0 & 3.07&14.3 & 2.89&15.2 & 3.94&10.0\\
1.1 & 1.53&6.54 & 1.73&6.70 & 1.64&7.41 & 1.26&8.91 & 1.19&9.67 & 1.27&13.1 & 1.13&14.4 & 1.11&15.1 & 1.01&16.4 & 1.04&12.5\\
1.3 & 0.590&9.45 & 0.510&11.1 & 0.657&10.6 & 0.368&14.9 & 0.465&14.0 & 0.531&18.6 & 0.372&22.9 & 0.273&27.7 & 0.53&20.9 & 0.451&17.4\\
1.5 & 0.191&14.8 & 0.145&18.6 & 0.107&23.6 & 0.125&23.6 & 0.132&24.3 & 0.202&27.7 & 0.136&35.4 & 0.072&50.0 & 0.216&30.1 & 0.186&25.0\\
1.7 & 0.068&21.3 & 0.056&26.7 & 0.045&33.4 & 0.023&50.0 & 0.034&44.7 & 0.054&50.0 & 0.045&57.7 && & 0.052&57.7 & 0.051&44.7\\
1.9 & 0.0078&35.3 && & 0.0117&40.9 && & 0.0084&57.7 && && && && &&\\
\hline
\end{tabular}
\end{center}
\caption{Invariant cross section, $f(x_F,p_T)$, in mb/(GeV$^2$/c$^3$) for protons in p+C collisions 
         at 158~GeV/c beam momentum. The relative statistical errors, $\Delta f$, are given in \%.
         The systematic errors are given in Table~\ref{tab:syst} }
\label{tab:prot_cs}
\end{table}

%     Here Table 3 
\begin{table}[t]
\renewcommand{\tabcolsep}{0.2pc} 
\renewcommand{\arraystretch}{1.05}
\small
\begin{center}
\begin{tabular}{|c|cr|cr|cr|cr|cr|cr|cr|}
\hline
\multicolumn{15}{|c|}{$f(x_F,p_T), \Delta f$} \\ 
\hline
$p_T \backslash x_F$ & \multicolumn{2}{c|}{-0.2} & \multicolumn{2}{c|}{-0.15} & \multicolumn{2}{c|}{-0.1} & \multicolumn{2}{c|}{-0.075} & \multicolumn{2}{c|}{-0.05} & \multicolumn{2}{c|}{-0.025} & \multicolumn{2}{c|}{0.0}\\ \hline
0.1 & 3.29&17.9 & 4.11&12.5 & 4.80&10.3 & 5.01&9.25 & 5.50&8.45 & 5.41&8.18 & 4.95&10.2\\
0.2 & 2.51&14.8 & 3.81&9.14 & 4.22&7.76 & 5.05&6.77 & 5.53&6.00 & 5.27&5.88 & 4.54&6.24\\
0.3 & 2.11&13.2 & 3.12&8.49 & 3.53&6.91 & 4.25&5.96 & 4.50&5.53 & 4.49&5.27 & 4.45&5.26\\
0.4 & 1.56&13.3 & 2.20&8.76 & 2.81&7.37 & 3.01&8.76 & 3.38&7.93 & 3.46&5.37 & 3.56&5.21\\
0.5 & 1.17&16.6 & 1.63&10.6 & 2.10&8.29 & 2.27&10.6 & 2.52&8.41 & 2.61&8.09 & 2.44&5.75\\
0.6 && & 1.19&13.6 & 1.40&10.1 & 1.48&12.1 & 1.75&10.8 & 1.80&9.15 & 1.89&8.70\\
0.7 & 0.617&14.0 & 0.82&13.8 & 1.04&12.2 & 1.14&13.3 & 1.13&12.8 & 1.24&11.9 & 1.35&11.3\\
0.9 & 0.250&19.6 & 0.359&14.9 & 0.440&11.8 & 0.537&12.9 & 0.522&12.5 & 0.531&12.0 & 0.532&12.8\\
1.1 & 0.103&27.4 & 0.146&21.2 & 0.203&16.2 && & 0.235&13.0 && & 0.207&17.9\\
1.3 & 0.041&40.6 & 0.069&28.9 & 0.078&26.0 && & 0.089&19.6 && & 0.091&25.7\\
1.5 && && && && && && & 0.037&40.8\\
\hline
\hline
$p_T \backslash x_F$ & \multicolumn{2}{c|}{0.025} & \multicolumn{2}{c|}{0.05} & \multicolumn{2}{c|}{0.1} & \multicolumn{2}{c|}{0.15} & \multicolumn{2}{c|}{0.2} & \multicolumn{2}{c|}{0.3} & \multicolumn{2}{c|}{ }\\ \hline
0.1 & 5.30&8.10 & 4.90&6.22 & 3.27&8.27 & 2.23&11.2 & 1.08&17.6 & 0.451&15.9 &&\\
0.2 & 4.88&6.06 & 4.66&4.55 & 3.05&6.16 & 1.86&8.84 & 1.28&11.8 && &&\\
0.3 & 4.33&5.37 & 3.52&4.31 & 2.52&5.63 & 1.60&7.78 & 1.08&10.6 & 0.317&11.1 &&\\
0.4 & 3.00&5.67 & 3.03&4.15 & 2.26&5.27 & 1.44&7.31 & 0.817&10.6 && &&\\
0.5 & 2.36&5.93 & 2.228&4.45 & 1.625&5.70 & 1.088&7.66 & 0.633&11.0 & 0.219&10.6 &&\\
0.6 & 1.44&10.1 & 1.56&7.10 & 1.227&6.15 & 0.913&7.87 & 0.566&10.9 && &&\\
0.7 & 0.98&13.1 & 1.091&8.05 & 0.970&6.72 & 0.616&9.02 & 0.407&12.1 & 0.126&15.0 &&\\
0.9 & 0.460&15.0 & 0.466&9.09 & 0.422&9.05 & 0.246&9.08 & 0.177&14.4 & 0.069&18.2 &&\\
1.1 && & 0.187&14.0 & 0.170&15.4 & 0.098&18.8 & 0.0563&16.1 & 0.0206&30.4 &&\\
1.3 && & 0.073&25.8 & 0.067&24.0 && & 0.0254&22.7 && &&\\
1.5 && && & 0.0209&33.6 && & 0.0121&31.5 && &&\\
\hline
\end{tabular}
\end{center}
\caption{Invariant cross section, $f(x_F,p_T)$, in mb/(GeV$^2$/c$^3$) for anti-protons in p+C collisions
         at 158~GeV/c beam momentum. The relative statistical errors, $\Delta f$, are given in \%.
         The systematic errors are given in Table~\ref{tab:syst}}
\label{tab:aprot_cs}
\end{table}

%
% ****************************** Section 6.2 ****************************
%
\subsection{Data interpolation}
\vspace{3mm}
\label{sec:interp}

As in the preceding publications concerning p+p \cite{pp_proton,pp_pion,pp_kaon} and
p+C \cite{pc_pion} interactions, a two-dimensional interpolation is
applied to the data. This interpolation has several aims:

\begin{itemize}
 \item It should reduce the local statistical data fluctuations
       by correlating several data points in the neighbourhood
       of each measured cross section, imposing the physics
       constraints of smoothness and continuity. 
 \item It should ensure stability 
       at the boundaries of the covered phase space regions.
 \item It should allow for an eventual slight extrapolation beyond
       the measured regions taking full account of the corresponding
       physics constraints. As an example the extrapolation to
       $p_T$~=~0 is strongly constrained by the fact that the invariant
       cross sections should approach this point with slope zero.
 \item It should allow the creation of a fine grid of interpolated
       values that may serve as a reference for the comparison
       of the yields of different particle species in the same
       hadronic collision or for the comparison of different
       types of hadronic interactions.
\end{itemize}

%
% ****************************** Section 6.2.1 ****************************
%
\subsubsection{The problem of analytic representation}
\vspace{3mm}
\label{sec:interp1}

A look at Figs.~\ref{fig:pt_dist_negxf}--\ref{fig:xf_dist_prot} 
shows that on the level of precision
and completeness achieved by the present experiment, the
measured cross sections show complex shapes both in their $p_T$ 
and in their $x_F$ dependences. These shapes exclude the description 
of the data with straight-forward analytic functions unless a complex array 
of rapidly varying fit parameters
would be established, with heavy constraints in order to fulfil 
the physics boundary conditions. There is a vast
literature using simple parametrizations like single or
double exponential fits or transverse mass ($m_T$, see Sect.~\ref{sec:rap}) fits 
for hadronic $p_T$ distributions, and power law distributions 
of the type $(1-x_F)^n$ for the $x_F$ dependences. 
Although these approaches have been claimed to carry 
physics relevance in terms of the concept of a fixed 
"hadronic temperature" in the case of transverse momentum or 
"counting rules" for the $x_F$ distributions in a purported connection 
to quark or even gluon structure functions, their impact 
on the understanding of soft hadronic physics has been doubtful 
to say the least. In the preceding publications 
\cite{pp_proton,pp_pion,pp_kaon,pc_pion,pc_discus} it has been shown 
that the shape of transverse momentum distributions against transverse mass
disproves the concept of a constant hadronic "temperature"
both in p+p and in p+C interactions, see also Sect.~\ref{sec:rap} below.
In particular in ref. \cite{pp_kaon} a detailed confrontation with the
effects of the decay of known resonances, in contrast to
the assumption of a "thermodynamic" behaviour, has been
discussed.

In view of this problematics, an alternative arithmetic
approach to the two-dimensional interpolation is eventually
given by cubic spline fits. These fits would of course, in
terms of an analytic description, at best offer a vast matrix
of parameters concerning the locally used third-order polynomials. 
In addition, the necessary introduction of the
physics constraints and boundary conditions would weigh
heavily against the practicability of such an approach. The
human eye, on the other hand, offers a welcome capability
to perform interpolation which is easily guided by the
physics constraints mentioned above. This approach is described below.

%
% ****************************** Section 6.2.2 ****************************
%
\subsubsection{Multi-step recursive eyeball fits}
\vspace{3mm}
\label{sec:interp2}

The data shown in Tables~\ref{tab:prot_cs} and \ref{tab:aprot_cs} have 
statistical errors between 2\% and about 30\% with distributions shown in
Fig.~\ref{fig:data_inter}a for p and in Fig.~\ref{fig:data_inter}b 
for $\overline{\textrm{p}}$. The corresponding mean values are 8\% and
12\%, respectively. With these error margins it is no problem
to produce interpolations via eyeball fits with a precision
of a fraction of the error bars, using an appropriate scale
for the invariant cross sections. These interpolations have to
describe the data simultaneously both in the $p_T$ and $x_F$ variables. 
An iterative, recursive method is therefore applied in both
dimensions. The quality of this eye fit procedure may be judged
by plotting the distribution of the differences between the
data points and the interpolation, normalized by the statistical
errors (normalized residuals). These distributions should be
Gaussians centred close to zero, as unlike in algebraic fits
the center at zero is not enforced, and with a variance close to
unity if the statistical errors are correctly estimated and
no additional systematics is introduced by the fit procedure.

%
% ****************************** Section 6.2.3 ****************************
%
\subsubsection{Statistical and correlated errors}
\vspace{3mm}
\label{sec:interp3}

\begin{figure}[h]
  \begin{center}
  	\includegraphics[width=14cm]{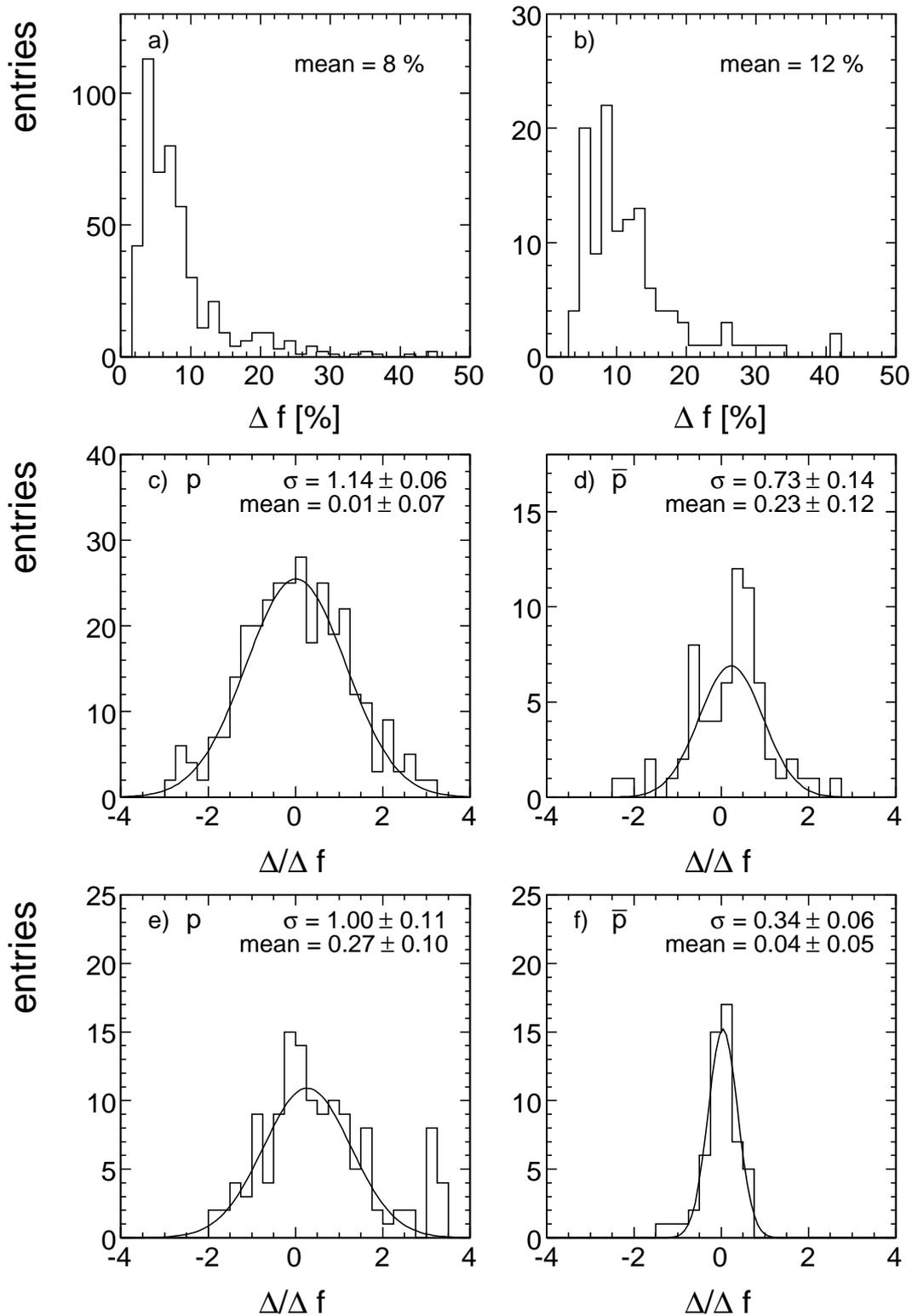}
 	\caption{Number distributions of the statistical
             errors given in Tables~\ref{tab:prot_cs} and \ref{tab:aprot_cs}, 
             a) for protons, b) for anti-protons. The normalized differences between data and
             interpolation (normalized residuals) are shown separately
             for the cross sections obtained by $dE/dx$ fitting, c) for
             protons, and d) for anti-protons, and for those bins where
             an interpolation of particle ratios has been used, e) for
             protons and f) for anti-protons}
  	 \label{fig:data_inter}
  \end{center}
\end{figure}

As described above in Sect.~\ref{sec:pid}, an interpolation of particle
ratios has been used in part of the backward phase space in
order to allow for the extraction of proton and anti-proton
yields in regions where particle identification via fits to
the ionization energy loss distributions are not reliable. 
In these regions the statistical errors become correlated and 
are in principle not any more directly related to the number 
of extracted particles in each bin. The regions of direct $dE/dx$ 
analysis and of the interpolation of particle ratios have therefore 
been separated and the normalized residuals are shown in Fig.~\ref{fig:data_inter} 
in corresponding separate number distributions.

%       Fig.27 

The uncorrelated residual distributions, Fig.~\ref{fig:data_inter}c and d, are
described by Gaussian fits with means of 0.01$\pm$0.07 for protons
and 0.23$\pm$0.12 for anti-protons, corresponding with respect to
the mean statistical errors to less than 1\% for protons and
2.7\% for anti-protons. Both values are within two standard
deviations with respect to the fit errors. The variances
of 1.14$\pm$0.06 and 0.73$\pm$0.14 are also within about two standard
deviations from the expected values of unity. The correlated
residual distributions, Fig.~\ref{fig:data_inter}e and f, are centred within 
2.7 standard deviations or 2\% for the protons and less than one
standard deviation or 0.5\% for the anti-protons. In both cases,
however, the variances are smaller than in the uncorrelated samples. 
While this difference is only on the one standard deviation
level for the protons, it is statistically significant for the
anti-protons. It should be mentioned here that in the multi-parameter
fits applied to the truncated energy loss distributions for the
extraction of particle yields, additional error sources of
statistical origin appear which depend on the relative particle
yields in each bin as well as on the position of each particle on 
the overall Bethe-Bloch parametrization. This effect has been
elaborated in detail in refs. \cite{pp_kaon,pp_proton} where it has been
demonstrated that an increase of the effective statistical errors
over the $1/\sqrt{N}$ estimator, where $N$ is the number of extracted
particles per bin, by several tens of percent is to be expected
in certain cases. The deviations observed in Figs.~\ref{fig:data_inter}c and d
are therefore within reasonable error margins.

Concerning the reduced variance of Fig.~\ref{fig:data_inter}f it has been
decided to keep the errors given in Table~\ref{tab:aprot_cs} at the level
of the estimator $1/\sqrt{N}$ in order to absorb eventual
systematic effects originating in the particle ratio interpolation.

%
% ****************************** Section 6.2.4 ****************************
%
\subsubsection{Tabulation of the interpolated cross sections}
\vspace{3mm}
\label{sec:interp4}

%   Table 4
\begin{table}
\renewcommand{\tabcolsep}{0.25pc} 
\renewcommand{\arraystretch}{0.85}
\scriptsize
\begin{center}
\begin{tabular}{|l|r@{.}l|r@{.}l|r@{.}l|r@{.}l|r@{.}l|r@{.}l|r@{.}l|r@{.}l|r@{.}l|r@{.}l|r@{.}l|}
\hline
$p_T \backslash x_F$ & \multicolumn{2}{c|}{0.0} & \multicolumn{2}{c|}{0.025} & \multicolumn{2}{c|}{0.05} & \multicolumn{2}{c|}{0.075} & \multicolumn{2}{c|}{0.1} & \multicolumn{2}{c|}{0.15} & \multicolumn{2}{c|}{0.2} & \multicolumn{2}{c|}{0.25} & \multicolumn{2}{c|}{0.3} & \multicolumn{2}{c|}{0.35} & \multicolumn{2}{c|}{0.4}\\ \hline
0.0 &21 & 75 &20 & 91 &21 & 30 &23 & 04 &24 & 68 &28 & 74 &34 & 31 &40 & 59 &46 & 28 &52 & 05 &57 & 20\\
0.05 &21 & 50 &20 & 82 &21 & 25 &22 & 83 &24 & 57 &28 & 54 &33 & 100 &40 & 31 &45 & 96 &51 & 57 &56 & 68\\
0.1 &21 & 11 &20 & 48 &21 & 06 &22 & 25 &23 & 96 &28 & 02 &33 & 30 &39 & 58 &45 & 23 &50 & 75 &55 & 90\\
0.15 &20 & 48 &19 & 93 &20 & 44 &21 & 50 &22 & 98 &27 & 07 &31 & 95 &38 & 32 &43 & 39 &48 & 91 &53 & 39\\
0.2 &19 & 61 &19 & 12 &19 & 47 &20 & 39 &21 & 80 &25 & 85 &30 & 30 &36 & 34 &40 & 78 &46 & 07 &50 & 40\\
0.25 &18 & 55 &17 & 96 &18 & 30 &19 & 07 &20 & 87 &24 & 12 &28 & 02 &33 & 68 &37 & 97 &42 & 60 &45 & 96\\
0.3 &17 & 32 &16 & 73 &16 & 88 &17 & 44 &18 & 73 &22 & 10 &25 & 55 &30 & 72 &34 & 71 &38 & 32 &40 & 87\\
0.35 &15 & 90 &15 & 29 &15 & 40 &15 & 68 &16 & 88 &19 & 83 &23 & 14 &27 & 63 &31 & 15 &33 & 76 &35 & 93\\
0.4 &14 & 30 &13 & 75 &13 & 85 &13 & 98 &14 & 94 &17 & 52 &20 & 63 &24 & 46 &27 & 25 &29 & 41 &30 & 93\\
0.45 &12 & 66 &12 & 26 &12 & 29 &12 & 52 &13 & 10 &15 & 22 &18 & 09 &21 & 40 &23 & 41 &25 & 14 &26 & 45\\
0.5 &11 & 03 &10 & 78 &10 & 87 &11 & 08 &11 & 44 &13 & 17 &15 & 50 &18 & 13 &19 & 83 &21 & 30 &22 & 31\\
0.55 &9 & 493 &9 & 320 &9 & 385 &9 & 499 &9 & 805 &11 & 23 &13 & 26 &15 & 26 &16 & 80 &18 & 01 &18 & 99\\
0.6 &8 & 080 &8 & 029 &7 & 988 &8 & 104 &8 & 364 &9 & 515 &11 & 13 &12 & 66 &14 & 07 &15 & 15 &16 & 05\\
0.65 &6 & 799 &6 & 709 &6 & 721 &6 & 893 &7 & 103 &8 & 043 &9 & 192 &10 & 43 &11 & 68 &12 & 63 &13 & 47\\
0.7 &5 & 655 &5 & 581 &5 & 642 &5 & 787 &6 & 004 &6 & 736 &7 & 541 &8 & 618 &9 & 670 &10 & 51 &11 & 15\\
0.75 &4 & 736 &4 & 706 &4 & 725 &4 & 802 &5 & 052 &5 & 513 &6 & 186 &6 & 957 &7 & 770 &8 & 559 &9 & 129\\
0.8 &3 & 958 &3 & 933 &3 & 948 &4 & 013 &4 & 182 &4 & 492 &4 & 994 &5 & 629 &6 & 229 &6 & 925 &7 & 352\\
0.85 &3 & 269 &3 & 271 &3 & 292 &3 & 322 &3 & 455 &3 & 651 &4 & 003 &4 & 523 &4 & 925 &5 & 463 &5 & 827\\
0.9 &2 & 707 &2 & 702 &2 & 725 &2 & 744 &2 & 821 &2 & 954 &3 & 217 &3 & 584 &3 & 903 &4 & 319 &4 & 575\\
0.95 &2 & 231 &2 & 253 &2 & 241 &2 & 251 &2 & 309 &2 & 390 &2 & 579 &2 & 816 &3 & 086 &3 & 353 &3 & 552\\
1.0 &1 & 843 &1 & 848 &1 & 843 &1 & 851 &1 & 890 &1 & 925 &2 & 063 &2 & 227 &2 & 429 &2 & 621 &2 & 738\\
1.05 &1 & 558 &1 & 519 &1 & 515 &1 & 519 &1 & 540 &1 & 543 &1 & 646 &1 & 777 &1 & 916 &2 & 053 &2 & 111\\
1.1 &1 & 273 &1 & 249 &1 & 240 &1 & 226 &1 & 249 &1 & 243 &1 & 304 &1 & 412 &1 & 508 &1 & 594 &1 & 642\\
1.15 &1 & 042 &1 & 030 &1 & 017 &0 & 9918 &0 & 9941 &0 & 9918 &1 & 029 &1 & 119 &1 & 187 &1 & 251 &1 & 287\\
1.2 &0 & 8604 &0 & 8422 &0 & 8249 &0 & 7988 &0 & 8025 &0 & 7896 &0 & 8193 &0 & 8886 &0 & 9406 &0 & 9782 &1 & 004\\
1.25 &0 & 7042 &0 & 6893 &0 & 6690 &0 & 6553 &0 & 6404 &0 & 6345 &0 & 6508 &0 & 7074 &0 & 7420 &0 & 7628 &0 & 7757\\
1.3 &0 & 5844 &0 & 5668 &0 & 5476 &0 & 5400 &0 & 5229 &0 & 5157 &0 & 5181 &0 & 5606 &0 & 5840 &0 & 5921 &0 & 5881\\
1.35 &0 & 4813 &0 & 4650 &0 & 4492 &0 & 4420 &0 & 4290 &0 & 4192 &0 & 4185 &0 & 4443 &0 & 4554 &0 & 4618 &0 & 4533\\
1.4 &0 & 3948 &0 & 3841 &0 & 3719 &0 & 3634 &0 & 3519 &0 & 3370 &0 & 3363 &0 & 3503 &0 & 3560 &0 & 3601 &0 & 3503\\
1.45 &0 & 3247 &0 & 3165 &0 & 3065 &0 & 2995 &0 & 2900 &0 & 2733 &0 & 2702 &0 & 2770 &0 & 2795 &0 & 2789 &0 & 2700\\
1.5 &0 & 2669 &0 & 2609 &0 & 2520 &0 & 2451 &0 & 2379 &0 & 2217 &0 & 2161 &0 & 2190 &0 & 2175 &0 & 2170 &0 & 2091\\
1.55 &0 & 2205 &0 & 2160 &0 & 2077 &0 & 2020 &0 & 1952 &0 & 1789 &0 & 1728 &0 & 1728 &0 & 1716 &0 & 1680 &0 & 1612\\
1.6 &0 & 1817 &0 & 1772 &0 & 1716 &0 & 1665 &0 & 1605 &0 & 1461 &0 & 1395 &0 & 1366 &0 & 1341 &0 & 1307 &0 & 1246\\
1.65 &0 & 1494 &0 & 1460 &0 & 1407 &0 & 1363 &0 & 1317 &0 & 1201 &0 & 1118 &0 & 1080 &0 & 1048 &0 & 1017 &0 & 09648\\
1.7 &0 & 1237 &0 & 1203 &0 & 1155 &0 & 1118 &0 & 1080 &0 & 09878 &0 & 09025 &0 & 08500 &0 & 08193 &0 & 07896 &0 & 07420\\
1.75 &0 & 1022 &0 & 09872 &0 & 09537 &0 & 09192 &0 & 08880 &0 & 08085 &0 & 07268 &0 & 06721 &0 & 06418 &0 & 06115 &0 & 05720\\
1.8 &0 & 08403 &0 & 08118 &0 & 07860 &0 & 07558 &0 & 07302 &0 & 06557 &0 & 05854 &0 & 05314 &0 & 05017 &0 & 04758 &0 & 04430\\
1.85 &0 & 06925 &0 & 06721 &0 & 06478 &0 & 06200 &0 & 06018 &0 & 05329 &0 & 04704 &0 & 04192 &0 & 03912 &0 & 03685 &0 & 03407\\
1.9 &0 & 05734 &0 & 05539 &0 & 05339 &0 & 05098 &0 & 04925 &0 & 04332 &0 & 03779 &0 & 03317 &0 & 03051 &0 & 02860 &0 & 02603\\
\hline
$p_T \backslash x_F$ & \multicolumn{2}{c|}{0.45} & \multicolumn{2}{c|}{0.5} & \multicolumn{2}{c|}{0.55} & \multicolumn{2}{c|}{0.6} & \multicolumn{2}{c|}{0.65} & \multicolumn{2}{c|}{0.7} & \multicolumn{2}{c|}{0.75} & \multicolumn{2}{c|}{0.8} & \multicolumn{2}{c|}{0.85} & \multicolumn{2}{c|}{0.9} & \multicolumn{2}{c|}{0.95}\\ \hline
0.0 &62 & 04 &65 & 72 &69 & 61 &72 & 06 &72 & 89 &73 & 57 &72 & 73 &73 & 74 &78 & 96 &93 & 85 &136 & 9\\
0.05 &61 & 47 &64 & 82 &68 & 82 &71 & 23 &72 & 06 &72 & 56 &72 & 06 &72 & 89 &77 & 88 &92 & 35 &135 & 0\\
0.1 &60 & 21 &62 & 76 &66 & 94 &69 & 29 &70 & 09 &71 & 23 &70 & 42 &70 & 74 &75 & 41 &89 & 62 &129 & 5\\
0.15 &57 & 11 &59 & 94 &63 & 49 &66 & 02 &66 & 94 &67 & 25 &66 & 94 &67 & 40 &72 & 02 &85 & 20 &122 & 9\\
0.2 &52 & 81 &55 & 29 &58 & 71 &61 & 33 &62 & 47 &61 & 47 &62 & 47 &62 & 76 &67 & 21 &79 & 88 &113 & 4\\
0.25 &47 & 39 &49 & 85 &52 & 81 &55 & 17 &56 & 06 &55 & 29 &55 & 81 &56 & 84 &61 & 15 &72 & 68 &102 & 9\\
0.3 &41 & 66 &43 & 72 &45 & 47 &47 & 72 &48 & 49 &48 & 16 &48 & 49 &49 & 28 &53 & 39 &64 & 18 &91 & 29\\
0.35 &36 & 20 &37 & 56 &38 & 61 &40 & 06 &40 & 80 &40 & 24 &40 & 06 &41 & 27 &44 & 61 &54 & 38 &77 & 88\\
0.4 &31 & 24 &31 & 97 &32 & 56 &33 & 40 &33 & 70 &33 & 09 &32 & 34 &33 & 09 &35 & 68 &44 & 40 &64 & 93\\
0.45 &27 & 02 &27 & 33 &27 & 52 &27 & 52 &27 & 46 &26 & 71 &25 & 86 &25 & 92 &28 & 67 &35 & 68 &52 & 77\\
0.5 &23 & 10 &23 & 05 &23 & 10 &22 & 89 &22 & 27 &21 & 41 &20 & 54 &20 & 40 &22 & 25 &27 & 89 &41 & 92\\
0.55 &19 & 66 &19 & 44 &19 & 22 &18 & 91 &18 & 10 &17 & 29 &16 & 43 &16 & 21 &17 & 48 &21 & 01 &32 & 54\\
0.6 &16 & 51 &16 & 32 &16 & 09 &15 & 58 &14 & 81 &14 & 05 &13 & 26 &13 & 02 &13 & 75 &15 & 94 &24 & 68\\
0.65 &13 & 89 &13 & 89 &13 & 26 &12 & 67 &12 & 10 &11 & 47 &10 & 71 &10 & 34 &10 & 80 &12 & 23 &18 & 38\\
0.7 &11 & 60 &11 & 47 &10 & 86 &10 & 27 &9 & 788 &9 & 177 &8 & 525 &8 & 312 &8 & 461 &9 & 320 &13 & 35\\
0.75 &9 & 493 &9 & 326 &8 & 723 &8 & 174 &7 & 811 &7 & 273 &6 & 740 &6 & 467 &6 & 568 &7 & 119 &9 & 941\\
0.8 &7 & 628 &7 & 408 &6 & 961 &6 & 493 &6 & 147 &5 & 658 &5 & 268 &5 & 043 &5 & 040 &5 & 438 &7 & 386\\
0.85 &6 & 059 &5 & 827 &5 & 504 &5 & 157 &4 & 816 &4 & 423 &4 & 099 &3 & 888 &3 & 912 &4 & 116 &5 & 400\\
0.9 &4 & 725 &4 & 586 &4 & 342 &4 & 078 &3 & 764 &3 & 417 &3 & 160 &2 & 983 &2 & 968 &3 & 136 &3 & 958\\
0.95 &3 & 668 &3 & 609 &3 & 370 &3 & 173 &2 & 922 &2 & 659 &2 & 370 &2 & 263 &2 & 215 &2 & 385 &2 & 934\\
1.0 &2 & 854 &2 & 802 &2 & 645 &2 & 474 &2 & 273 &2 & 026 &1 & 869 &1 & 729 &1 & 688 &1 & 809 &2 & 185\\
1.05 &2 & 190 &2 & 160 &2 & 067 &1 & 929 &1 & 769 &1 & 595 &1 & 428 &1 & 317 &1 & 290 &1 & 369 &1 & 631\\
1.1 &1 & 677 &1 & 646 &1 & 594 &1 & 491 &1 & 372 &1 & 229 &1 & 096 &1 & 006 &0 & 9827 &1 & 041 &1 & 217\\
1.15 &1 & 304 &1 & 281 &1 & 212 &1 & 141 &1 & 020 &0 & 9369 &0 & 8292 &0 & 7633 &0 & 7489 &0 & 7896 &0 & 9066\\
1.2 &1 & 006 &0 & 9698 &0 & 9171 &0 & 8500 &0 & 7681 &0 & 6929 &0 & 6262 &0 & 5857 &0 & 5747 &0 & 5990 &0 & 6767\\
1.25 &0 & 7541 &0 & 7340 &0 & 6877 &0 & 6418 &0 & 5787 &0 & 5184 &0 & 4674 &0 & 4464 &0 & 4369 &0 & 4554 &0 & 5028\\
1.3 &0 & 5787 &0 & 5491 &0 & 5157 &0 & 4780 &0 & 4339 &0 & 3843 &0 & 3587 &0 & 3386 &0 & 3345 &0 & 3455 &0 & 3753\\
1.35 &0 & 4420 &0 & 4173 &0 & 3885 &0 & 3593 &0 & 3254 &0 & 2900 &0 & 2733 &0 & 2586 &0 & 2555 &0 & 2627 &0 & 2802\\
1.4 &0 & 3384 &0 & 3165 &0 & 2920 &0 & 2688 &0 & 2451 &0 & 2210 &0 & 2073 &0 & 1971 &0 & 1952 &0 & 2006 &0 & 2091\\
1.45 &0 & 2585 &0 & 2385 &0 & 2200 &0 & 2006 &0 & 1838 &0 & 1677 &0 & 1580 &0 & 1512 &0 & 1487 &0 & 1529 &0 & 1576\\
1.5 &0 & 1979 &0 & 1809 &0 & 1657 &0 & 1505 &0 & 1385 &0 & 1275 &0 & 1201 &0 & 1163 &0 & 1141 &0 & 1160 &0 & 1181\\
1.55 &0 & 1512 &0 & 1369 &0 & 1246 &0 & 1128 &0 & 1041 &0 & 09692 &0 & 09156 &0 & 08860 &0 & 08718 &0 & 08799 &0 & 08880\\
1.6 &0 & 1155 &0 & 1029 &0 & 09299 &0 & 08461 &0 & 07842 &0 & 07386 &0 & 06993 &0 & 06736 &0 & 06613 &0 & 06644 &0 & 06675\\
1.65 &0 & 08799 &0 & 07860 &0 & 07038 &0 & 06345 &0 & 05921 &0 & 05577 &0 & 05281 &0 & 05137 &0 & 05017 &0 & 05017 &0 & 05005\\
1.7 &0 & 06736 &0 & 05921 &0 & 05314 &0 & 04758 &0 & 04461 &0 & 04192 &0 & 04024 &0 & 03933 &0 & 03859 &0 & 03806 &0 & 03753\\
1.75 &0 & 05157 &0 & 04492 &0 & 04003 &0 & 03609 &0 & 03353 &0 & 03180 &0 & 03067 &0 & 02997 &0 & 02920 &0 & 02887 &0 & 02854\\
1.8 &0 & 03939 &0 & 03415 &0 & 03037 &0 & 02738 &0 & 02526 &0 & 02423 &0 & 02343 &0 & 02263 &0 & 02236 &0 & 02200 &0 & 02150\\
1.85 &0 & 03002 &0 & 02585 &0 & 02288 &0 & 02072 &0 & 01912 &0 & 01838 &0 & 01765 &0 & 01709 &0 & 01708 &0 & 01669 &0 & 01638\\
1.9 &0 & 02293 &0 & 01961 &0 & 01748 &0 & 01568 &0 & 01440 &0 & 01398 &0 & 01333 &0 & 01308 &0 & 01295 &0 & 01272 &0 & 01249\\
\hline
\end{tabular}
\end{center}
\caption{Data interpolation for protons for positive $x_F$ in p+C collisions at 158 GeV/c beam momentum}
\label{tab:interp}
\end{table}

A complete set of interpolated cross sections can be found on the
web page "spshadrons" \cite{site} in steps of 0.05~GeV/c in $p_T$ and for
all $x_F$ values used in this publication. Comparable sets of
cross sections are given in \cite{site} for all other investigated
particle types and interactions. As an example Table~\ref{tab:interp} gives
a limited list of interpolated proton cross sections for the 22 $x_F$ values 
at $x_F \geq$~0 established in this paper. The lines through
the data points corresponding to the interpolation are obtained
using the smoothed interpolation procedure described in the
ROOT program package \cite{root}. It should be mentioned here 
that the region below $p_T$~=~0.05--0.1~GeV/c covered by the data has been
extended to $p_T$~=~0~GeV/c by extrapolation using the physics constraint
of zero slope at this point. A slight extrapolation in the high
$p_T$ region has also been applied using the basically exponential
behaviour of the cross sections in this area. As far as the
coverage of the high $x_F$ region at low $p_T$ which is outside the
NA49 acceptance is concerned, see the argumentation given in Sect.~\ref{sec:dist}.

%
% ****************************** Section 6.3 ****************************
%
\subsection{Dependence of the invariant cross sections on $\mathbf{p_T}$ and $\mathbf{x_F}$}
\vspace{3mm}
\label{sec:dist}

%     Here Fig.28 
\begin{figure}[b]
  \begin{center}
  	\includegraphics[width=15.cm]{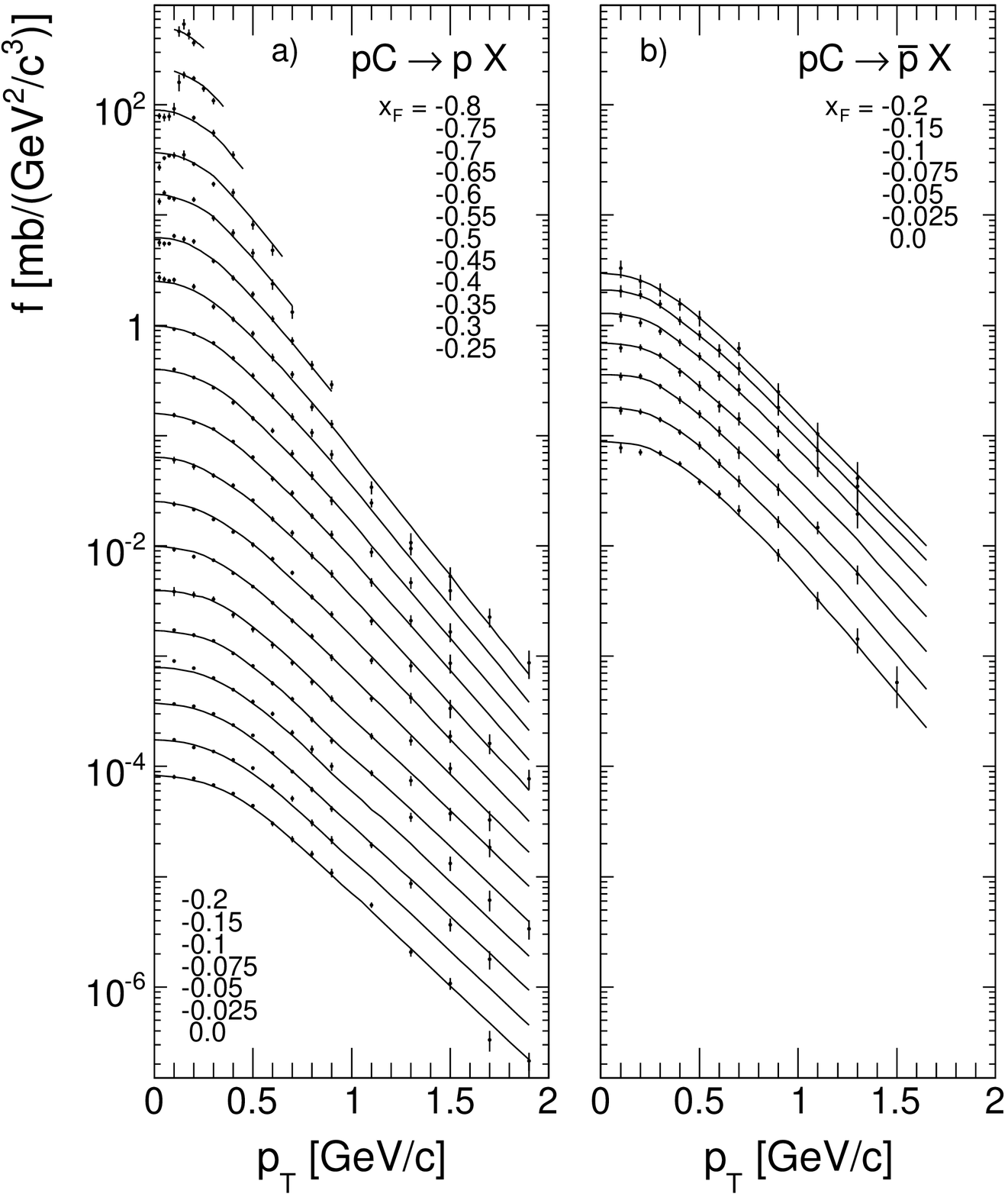}
 	\caption{Invariant cross sections and data interpolation
            (full lines) as a function of $p_T$ at fixed $x_F \leq$~0 for 
            a) protons and b) anti-protons produced in
            p+C collisions at 158~GeV/c. The displayed cross sections
            have been multiplied by a factor of 0.5 for successive values
            of $x_F$ for better visibility}
  	 \label{fig:pt_dist_negxf}
  \end{center}
\end{figure}

\begin{figure}[b]
  \begin{center}
  	\includegraphics[width=15.cm]{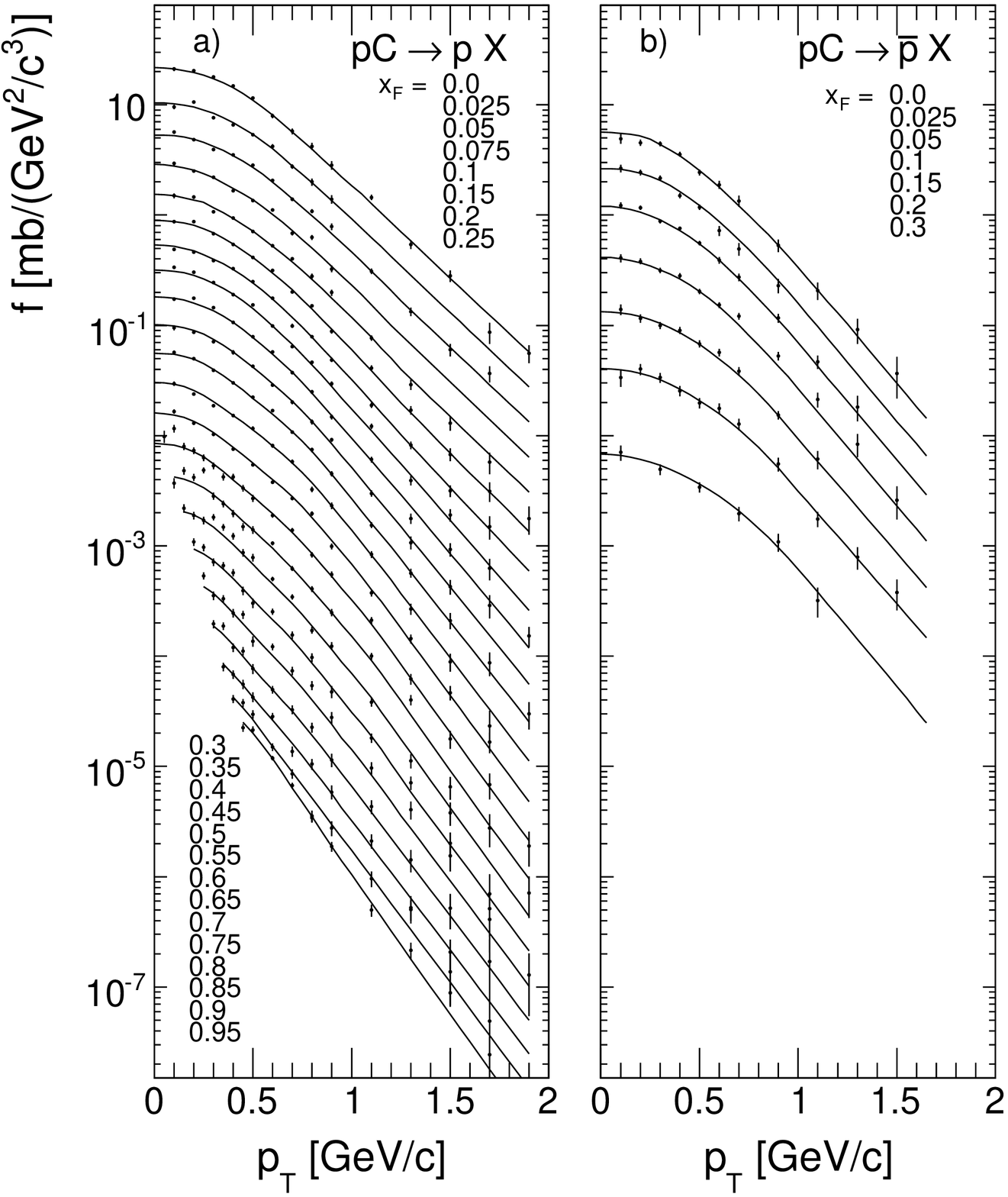}
 	\caption{Invariant cross sections and data interpolation
            (full lines) as a function of $p_T$ at fixed $x_F \geq$~0 for 
            a) protons and b) anti-protons produced in
            p+C collisions at 158~GeV/c. The displayed cross sections
            have been multiplied by a factor of 0.5 for successive values
            of $x_F$ for better visibility}
  	 \label{fig:pt_dist_posxf}
  \end{center}
\end{figure}

The distribution of the invariant cross section as a function of
$p_T$ is shown in Fig.~\ref{fig:pt_dist_negxf} for protons and anti-protons at negative $x_F$
and in Fig.~\ref{fig:pt_dist_posxf} at positive $x_F$, indicating the data interpolation
by full lines. For better visibility successive values in $x_F$ have
been multiplied by a factor 0.5.

%     Here Fig.29 

\begin{figure}[b]
  \begin{center}
  	\includegraphics[width=15.cm]{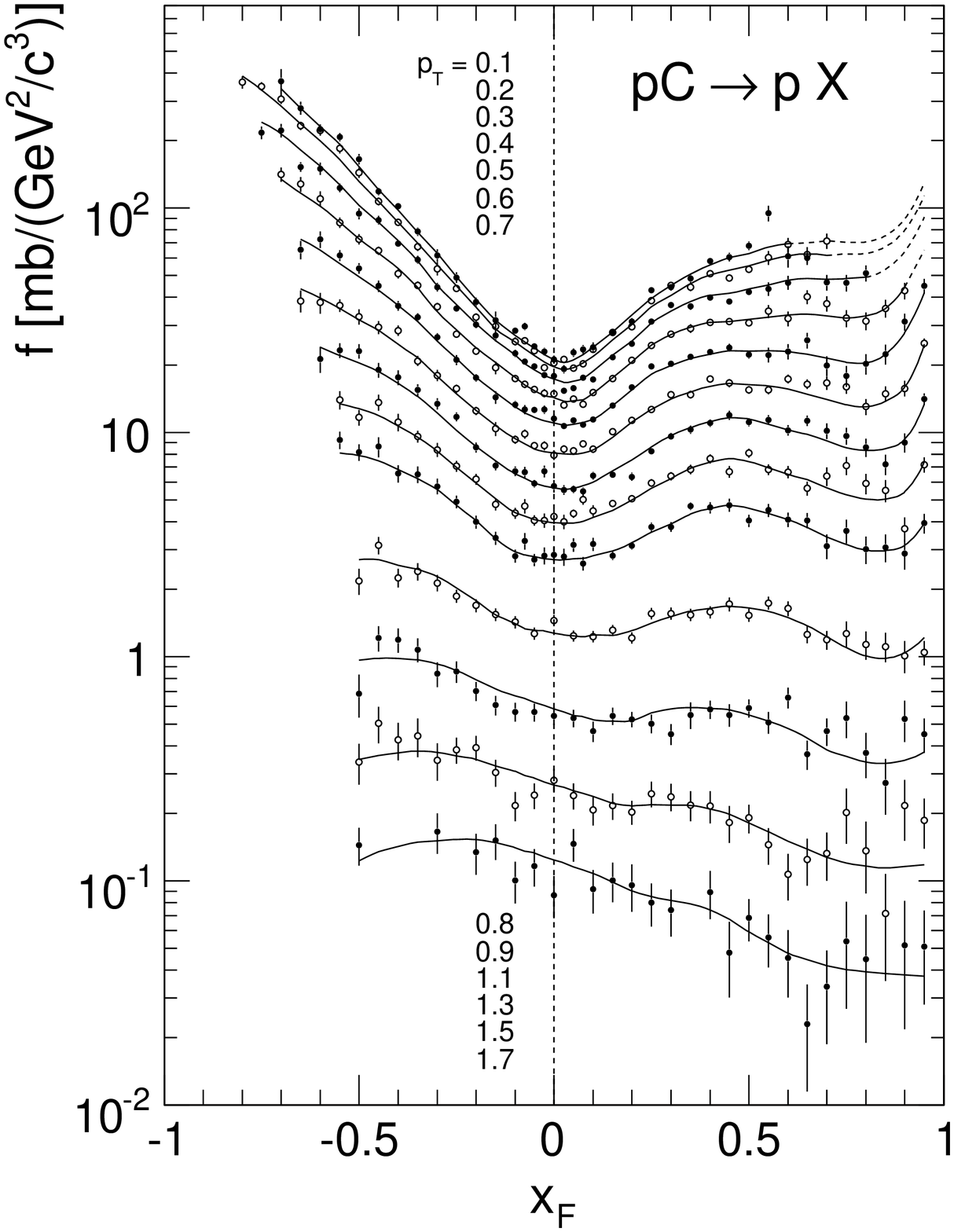}
 	\caption{Invariant cross sections as a function of $x_F$ at fixed
             $p_T$ for protons produced in p+C collisions at 158~GeV/c. The $p_T$
             values are to be correlated to the respective distributions in
             decreasing order of cross section. The broken lines at low $p_T$ and
             large $x_F$ indicate the extrapolation made by using p+p data}
  	 \label{fig:xf_dist_prot}
  \end{center}
\end{figure}

\begin{figure}[b]
  \begin{center}
  	\includegraphics[width=15.cm]{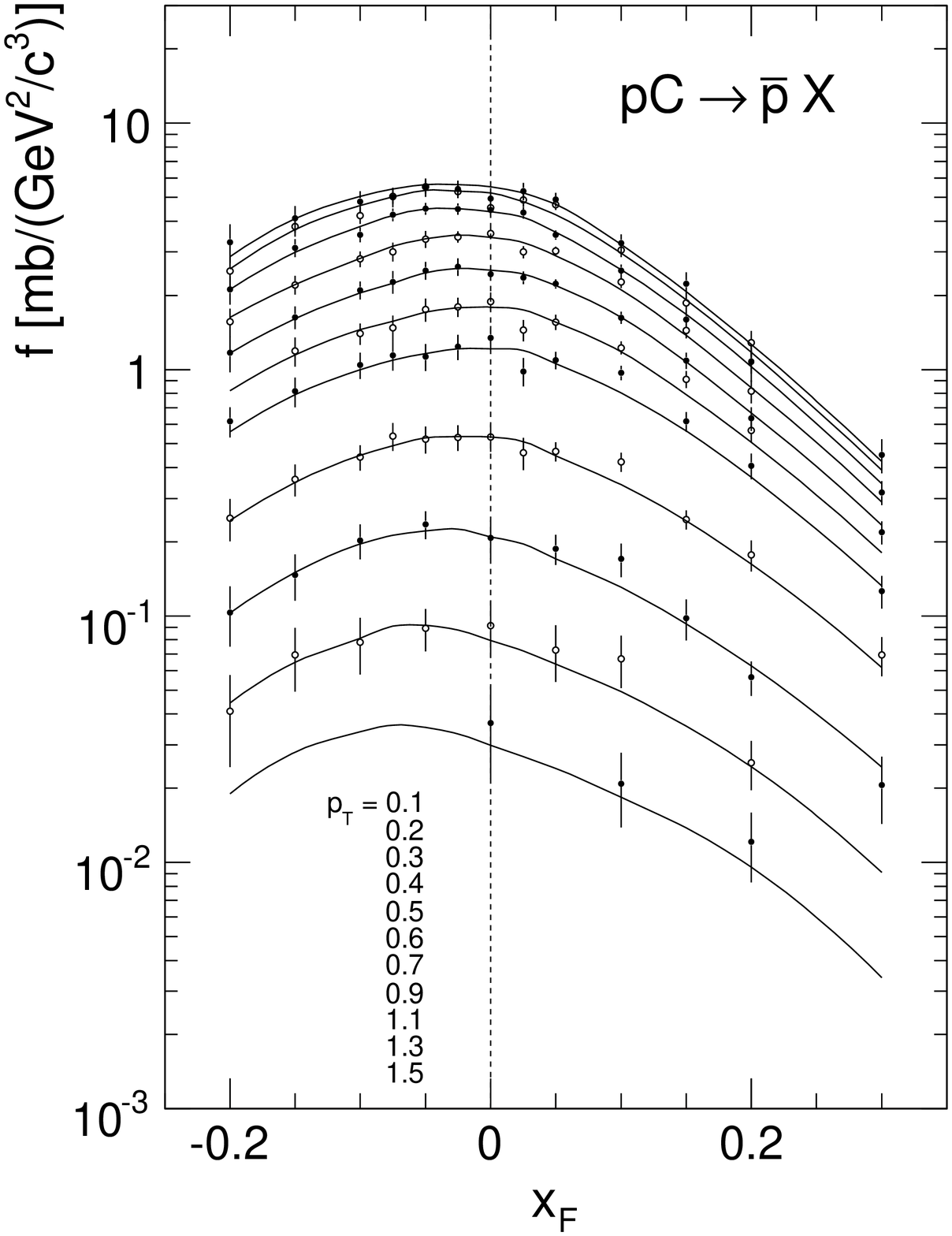}
 	\caption{Invariant cross sections as a function of $x_F$ at fixed
             $p_T$ for anti-protons produced in p+C collisions at 158~GeV/c. The $p_T$
             values are to be correlated to the respective distributions in
             decreasing order of cross section}
  	 \label{fig:xf_dist_aprot}
  \end{center}
\end{figure}

Corresponding $x_F$ distributions are presented in Fig.~\ref{fig:xf_dist_prot} for protons
and in Fig.~\ref{fig:xf_dist_aprot} for anti-protons.

%    Here Fig.30 

%     Here Fig.31 

The shape of the $p_T$ distributions resembles, for $x_F \geq$~0 (Fig.~\ref{fig:pt_dist_posxf}), the
one measured in p+p interactions \cite{pp_proton} including details of the deviation
from either exponential or Gaussian shape. In backward direction, 
Fig.~\ref{fig:pt_dist_negxf}, a more complex behaviour develops with a steepening up at
$x_F <$~-0.4. The basic asymmetry of the p+C interaction is more directly
visible in the $x_F$ distributions of Figs.~\ref{fig:xf_dist_prot} and \ref{fig:xf_dist_aprot}. 
While at high $p_T$ factors of 1.6--2 are typical between the backward and forward proton yields
at $x_F$~=~0.5--0.7, these factors grow to about 3--5 at $p_T$~=~0.1~GeV/c.
This allows a first view at the composition of p+A collisions from
projectile fragmentation in forward direction and target fragmentation
as well as intra-nuclear cascading in backward direction. A 
quantification and separation of these three basic ingredients will
be performed in Sects.~\ref{sec:hadr}, \ref{sec:hadr_net} and \ref{sec:pt_prot} below. For anti-protons the $x_F$ distributions
clearly peak at negative $x_F$. The forward-backward asymmetry is, at 
$|x_F|$~=~0.2, to first order $p_T$ independent and on the order of 1.6--1.9 
which is well above the asymmetry of protons at this $x_F$ indicating 
important effects from isospin and baryon number transfer. 

A remark concerning the
behaviour of the proton cross sections at $x_F >$~0.8 is in place here.
In this region a clear diffractive peak becomes visible which is
in shape equal to the one observed in p+p interactions \cite{pp_proton}.
This peak is to be expected from single projectile collisions
in the target nucleus which amount to about 60\% of all inelastic
p+C interactions \cite{pc_discus}. In the absence of comparable measurements
in the low $p_T$ region the relative shape of the extrapolated lines 
in the inaccessible area of phase space has been adjusted, after
re-normalization, to the one extracted in \cite{pp_proton} from existing p+p data. 
This extrapolation is indicated by the broken line segments in Fig.~\ref{fig:xf_dist_prot}.

%
% ****************************** Section 6.4 ****************************
%
\subsection{Rapidity and transverse mass distributions}
\vspace{3mm}
\label{sec:rap}

%     Here Fig.32
\begin{figure}[b]
  \begin{center}
  	\includegraphics[width=14.5cm]{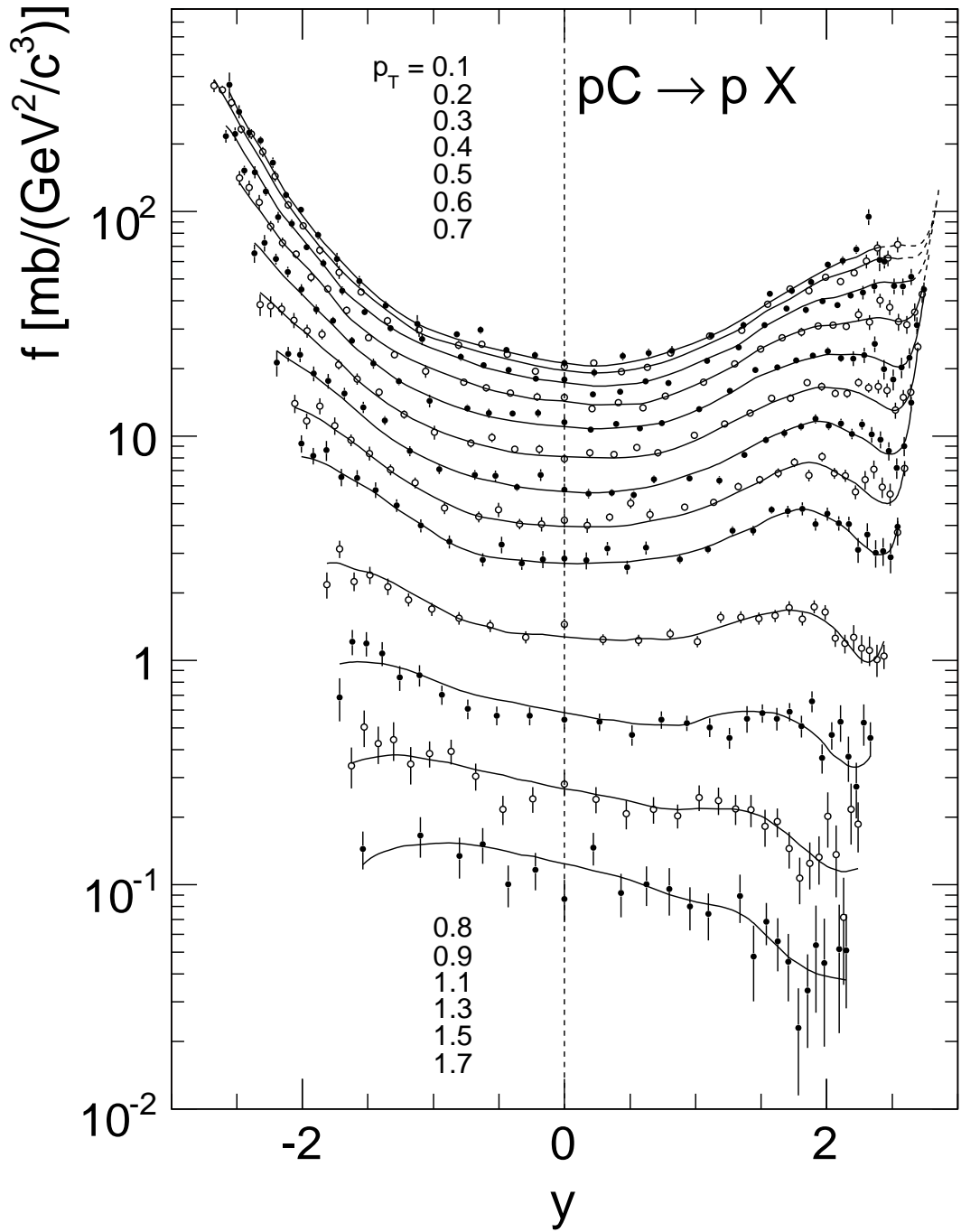}
 	\caption{Invariant cross sections as a function of $y$ at fixed
             $p_T$ for protons produced in p+C collisions at 158~GeV/c. The $p_T$
             values are to be correlated to the respective distributions in
             decreasing order of cross section. The broken lines at low $p_T$ and
             large $x_F$ indicate the extrapolation made by using p+p data}
  	 \label{fig:y_dist_prot}
  \end{center}
\end{figure}

The rapidity distribution for protons at fixed $p_T$ presented in Fig.~\ref{fig:y_dist_prot}
extends to the kinematic limit in forward direction and to -2.6 units 
in the target hemisphere. Again a clear view of the asymmetry of the
p+C interactions increasing with decreasing $p_T$ is evident.

\begin{figure}[b]
  \begin{center}
  	\includegraphics[width=15.cm]{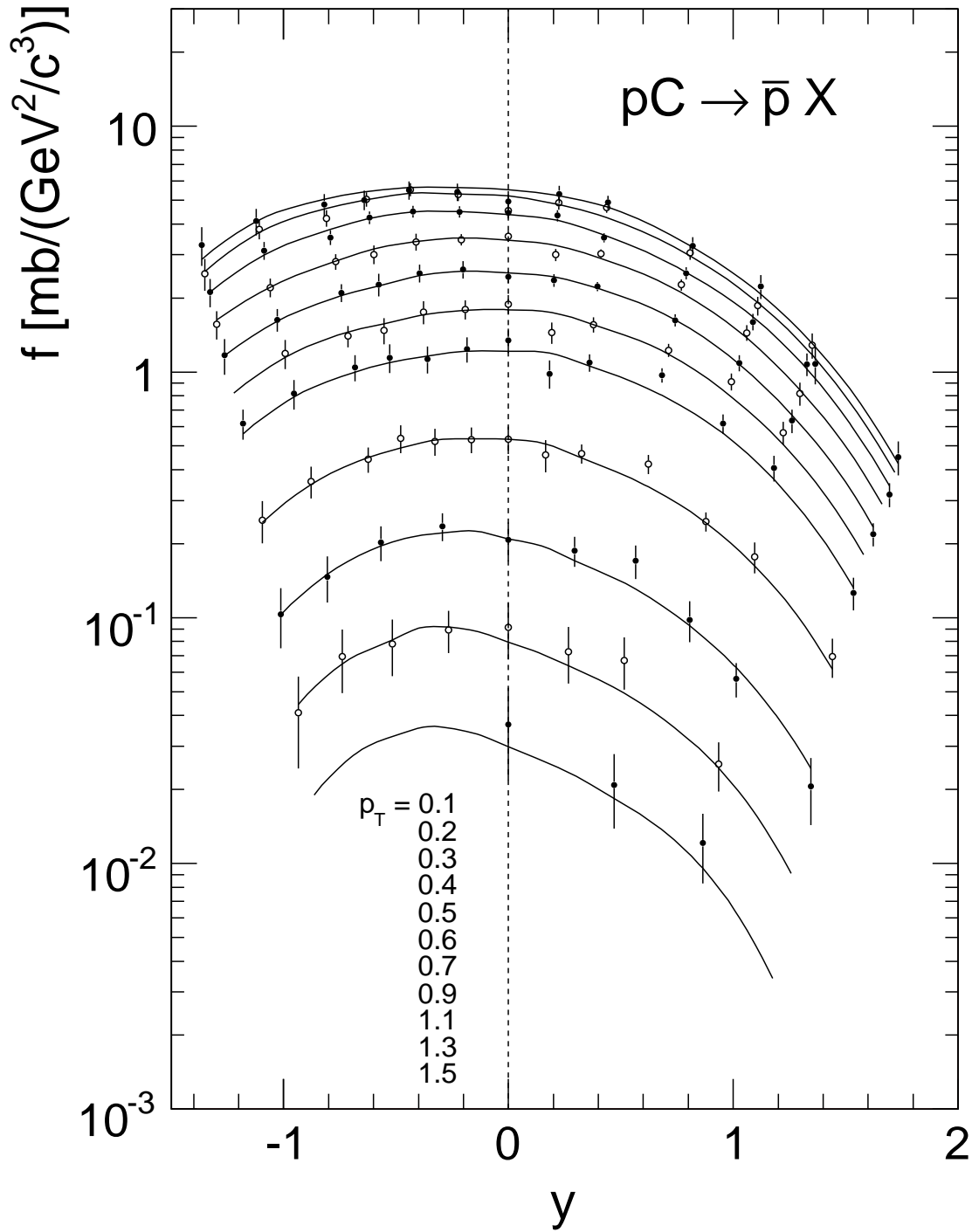}
 	\caption{Invariant cross sections as a function of $y$ at fixed
             $p_T$ for anti-protons produced in p+C collisions at 158~GeV/c. The $p_T$
             values are to be correlated to the respective distributions in
             decreasing order of cross section}
  	 \label{fig:y_dist_aprot}
  \end{center}
\end{figure}

The rapidity range of anti-protons is limited by statistics to
-1.4 to +1.8 units. The corresponding distribution as a function of
$y$ for fixed $p_T$ is shown in Fig.~\ref{fig:y_dist_aprot}.

%     Here Fig.33    

\begin{figure}[h]
  \begin{center}
  	\includegraphics[width=10.5cm]{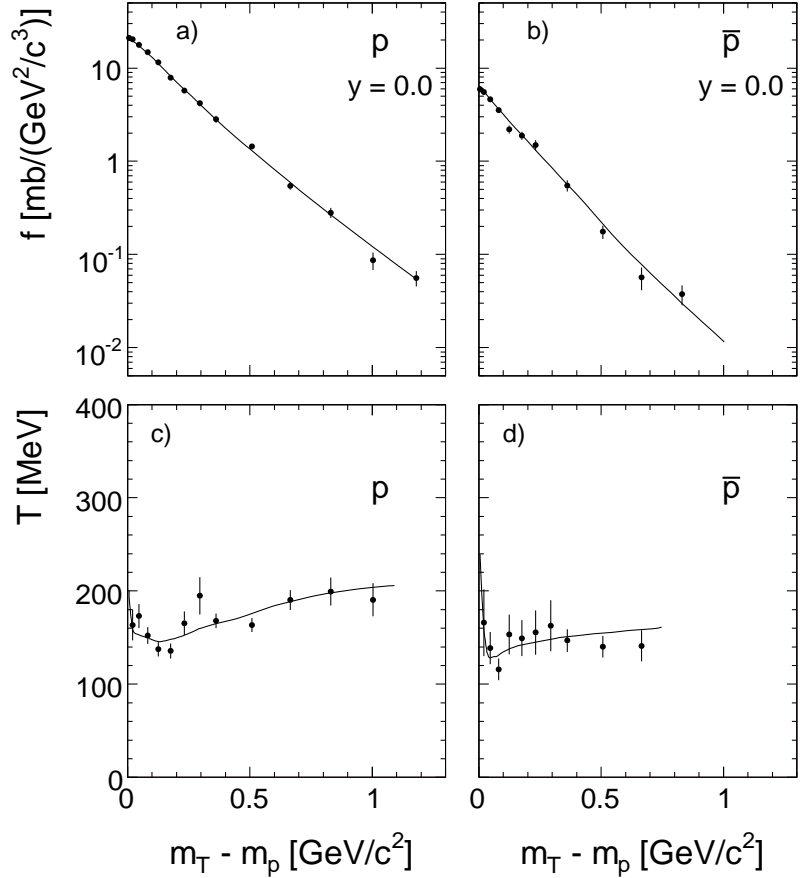}
 	\caption{Invariant cross section as a function of $m_T-m_p$ for
             a) protons and b) anti-protons. Panels c) and d) give the
             inverse slope parameters of the $m_T$ distributions as a function
             of $m_T-m_p$ for protons and anti-protons, respectively. The full
             lines represent the results of the data interpolation}
  	 \label{fig:mt_dist}
  \end{center}
\end{figure}

Transverse mass distributions at $y$~=~0, with $m_T = \sqrt{m_p^2 + p_T^2}$,
are presented in Fig.~\ref{fig:mt_dist} together with the local inverse slopes
obtained by exponential fits to three subsequent points. 
A situation very similar to p+p collisions emerges with a 
non-exponential behaviour and inverse slope parameters varying 
systematically by up to 60~MeV with the transverse mass.

%     Here Fig.34 

%
% ****************************** Section 6.5 ****************************
%
\subsection{Comparison to other experiments}
\vspace{3mm}
\label{sec:comp}

\begin{figure}[h]
  \begin{center}
  	\includegraphics[width=8cm]{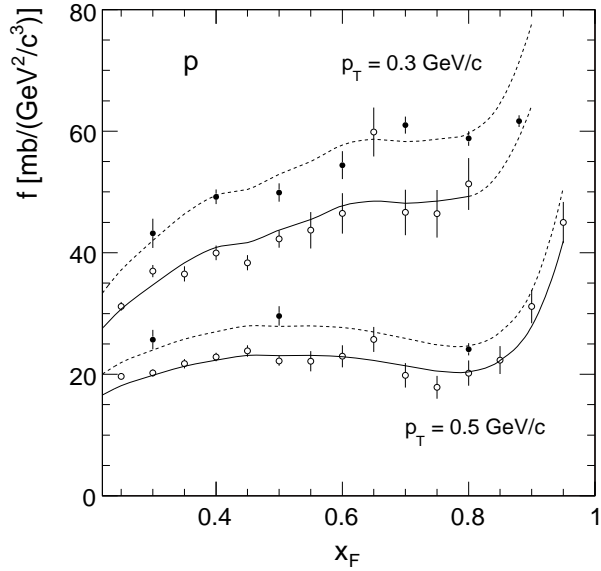}
 	\caption{Invariant proton cross sections from \cite{barton} (full circles)
             for $p_T$~=~0.3 and 0.5~GeV/c together with the measurements (open circles) and 
             data interpolation (full lines) of NA49, and the interpolation multiplied by
             1.21 (broken lines). A slight extrapolation of NA49 results at $p_T$~=~0.3~GeV/c is
             shown with broken line, see Sect.~\ref{sec:interp4}}
  	 \label{fig:bart_comp}
  \end{center}
\end{figure}

As stated in Sect.~\ref{sec:exp_sit} there are only two experiments providing
double differential data in the SPS energy range. The data set
\cite{bayukov} which is disjoint from the NA49 phase space coverage will
be discussed in detail in the next Sect.~\ref{sec:backext}. As far as the
Fermilab data of Barton et al.\cite{barton} are concerned there are 10
data points for protons and 4 data points for anti-protons 
available in overlapping phase space ranges. The situation for
protons is shown in Fig.~\ref{fig:bart_comp} where the invariant cross sections \cite{barton} (full circles)
at $p_T$~=~0.3 and 0.5~GeV/c are shown together with the measurements (open circles) and data 
interpolation (full lines) of the NA49 experiment as a function of $x_F$.

%    Here Fig.35

The comparison between the two experiments reveals very sizeable
systematic offsets with an average of +21\% or +6 standard deviations
averaged over all data points. This complies with the comparison
of pion yields \cite{pc_pion} with an average of +25\% or +3.6 standard 
deviations. The good agreement between the particle ratios
(K$^+$+$\pi^+$)/p demonstrated in Sect.~\ref{sec:p_forw} over the full comparable
range of $x_F$ and $p_T$ speaks indeed for a normalization problem as
the origin of the discrepancies which are not visible in the
results of p+p interactions \cite{pp_proton}.

%     Here Fig.36
\begin{figure}[b]
  \begin{center}
  	\includegraphics[width=7cm]{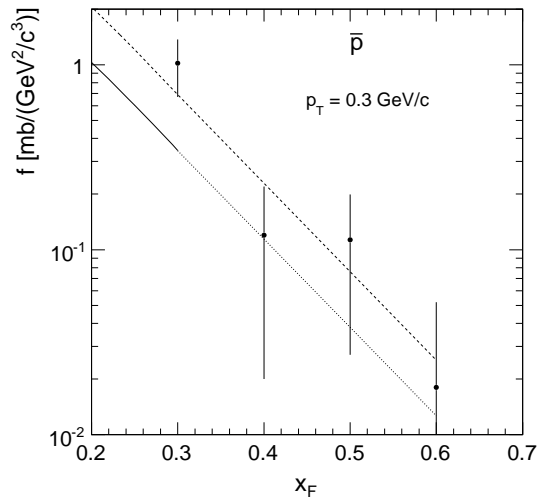}
 	\caption{Invariant anti-proton cross sections from \cite{barton}
             for $p_T$~=~0.3~GeV/c as a function of $x_F$ together with the
             data interpolation of NA49 (full line) which has been
             extrapolated to $x_F$~=~0.6 with an exponential function (dotted
             line). The NA49 reference multiplied by a factor of 2 is
             indicated by the dashed line}
  	 \label{fig:bart_aprot}
  \end{center}
\end{figure}

For anti-protons the situation is considerably less clear due to
the very large statistical errors of the data \cite{barton} indicating only
upper limits for some of the measurements. Fig.~\ref{fig:bart_aprot} gives the
$x_F$ dependence of these data for $p_T$~=~0.3~GeV/c together with the
NA49 interpolation (full line) which has been partially 
extrapolated using an exponential shape. The large upward shift
of the Fermilab data of about a factor of 2 (dashed line in
Fig.~\ref{fig:bart_aprot}) is compounded by the fact that due to the lower beam
momentum of 100~GeV/c these data should be expected to be about 
30--40\% below the NA49 cross sections \cite{pp_proton}. 

In this context it should be mentioned that already in p+p collisions
the Fermilab data \cite{brenner} were high by about +25\% for anti-protons.

%
% ****************************** Section 7 ****************************
%
\section{Data extension into the far backward direction}
\vspace{3mm}
\label{sec:backext}

As the backward acceptance of the NA49 detector is limited to the ranges of 
$x_F >$~-0.8 to $x_F >$~-0.5 at low and high $p_T$, respectively, it is desirable 
to extend this coverage into the far backward region to $x_F$ values down to 
and below -1. A detailed survey of existing experiments in the backward direction 
of p+C interactions has therefore been undertaken and is being published in an accompanying
paper \cite{pc_survey}. This survey establishes the detailed $s$-dependence
of measured cross sections for beam momenta between 1 and 400~GeV/c, 
for lab angles between 10 and 180~degrees, and
for lab momenta between 0.2 and 1.2~GeV/c. It shows in
particular that the baryon cross sections in the SPS energy
range from about 100 to 400~GeV/c beam momentum may be regarded as $s$-independent within 
tight systematic limits of less than two percent \cite{pc_survey}. This allows 
the combination of the extensive data set of the Fermilab experiment \cite{bayukov} 
at 400~GeV/c beam momentum and lab angles between 70 and 160~degrees with the NA49
data which span the angular range up to 40~degrees.

\begin{figure}[h]
  \begin{center}
  	\includegraphics[width=9.5cm]{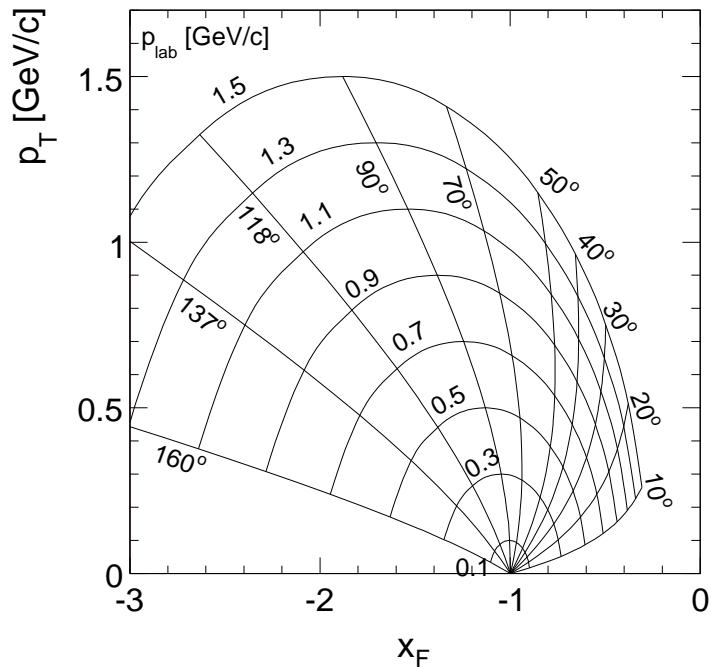}
 	\caption{Kinematics of fixed $p_{\textrm{lab}}$ and $\Theta_{\textrm{lab}}$ in the
             $x_F$/$p_T$ plane
             %. The NA49 acceptance is shown as hatched area
             }
  	 \label{fig:kin}
  \end{center}
\end{figure}

The relevant kinematic situation is presented in Fig.~\ref{fig:kin} where lines
of constant $p_{\textrm{lab}}$ and $\Theta_{\textrm{lab}}$ are shown in the $x_F$/$p_T$ plane. 

%     Here Fig.37 

\begin{figure}[h]
  \begin{center}
  	\includegraphics[width=6cm]{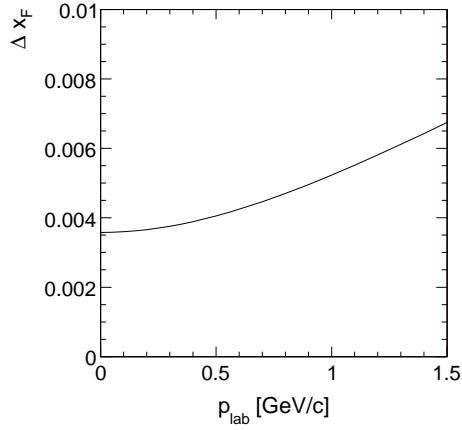}
 	\caption{Differences in $x_F$: $\Delta x_F = x_F^{158} - x_F^{400}$ resulting from the transformation
             into the cms system between beam momenta of 158 and 400~GeV/c
             as a function of $p_{\textrm{lab}}$. The differences are independent of
             $\Theta_{\textrm{lab}}$ in the range 70~$< \Theta_{\textrm{lab}} <$~160 degrees}
  	 \label{fig:kin_dif}
  \end{center}
\end{figure}

In the necessary transformation between the lab and cms frames
involved in Fig.~\ref{fig:kin} there is very little difference between the
beam momenta of 158 and 400~GeV/c. This is shown in Fig.~\ref{fig:kin_dif} which
gives the difference in $x_F$ as a function of $p_{\textrm{lab}}$ for the angular
range between 70 and 160~degrees.

%     Here Fig.38 

\begin{figure}[h]
  \begin{center}
  	\includegraphics[width=9cm]{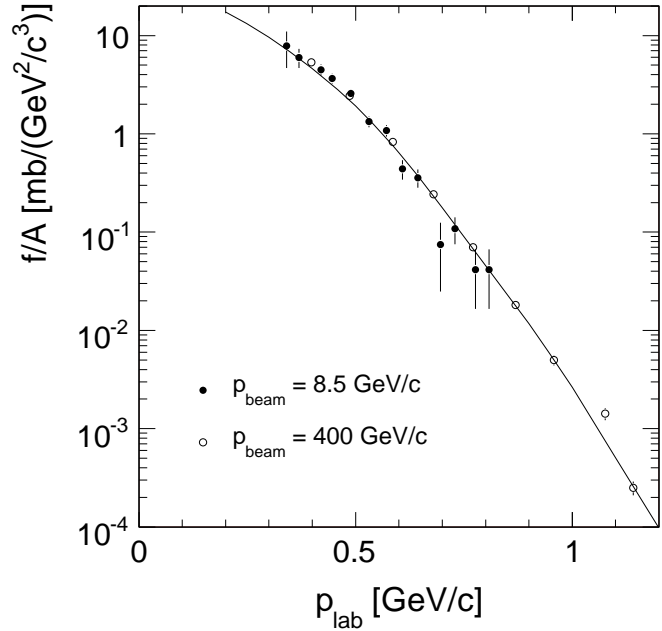}
 	\caption{Invariant proton cross sections \cite{burgov} at $\Theta_{\textrm{lab}}$~=~162
             degrees and 8.5~GeV/c beam momentum as a function of $p_{\textrm{lab}}$ in 
             comparison with the data from \cite{bayukov} at 160 degrees and 400~GeV/c
             beam momentum. The full line represents the interpolation of the Fermilab data}
  	 \label{fig:comp_burg}
  \end{center}
\end{figure}

One example of the apparent $s$-independence of the backward baryon
yields is shown in Fig.~\ref{fig:comp_burg} which compares data at 
$\Theta_{\textrm{lab}}$~=~162 and 160~degrees for 8.5 \cite{burgov} and 400~GeV/c beam momentum \cite{bayukov}.

%     Here Fig.39

Due to the flat angular distribution at $\Theta_{\textrm{lab}}$ around 160~degrees, see
Fig.~\ref{fig:pc_prot_theta}, the small angular difference in angle between the two measurements  
has negligible influence on the cross sections comparison. Also the SPS measurements
of "grey" protons by Braune et al. \cite{braune} show no dependence on beam momentum
in the range from 50 to 150~GeV/c over the complete $\Theta_{\textrm{lab}}$ range
from 10 to 159~degrees.

%
% ****************************** Section 7.1 ****************************
%
\subsection{NA49 results at fixed ${\mathbf \Theta_{\textrm{lab}}}$ and ${\mathbf p_{\textrm{lab}}}$ 
            combined with the data from \cite{bayukov}}
\vspace{3mm}
\label{sec:comb}

\begin{figure}[h]
  \begin{center}
  	\includegraphics[width=8.5cm]{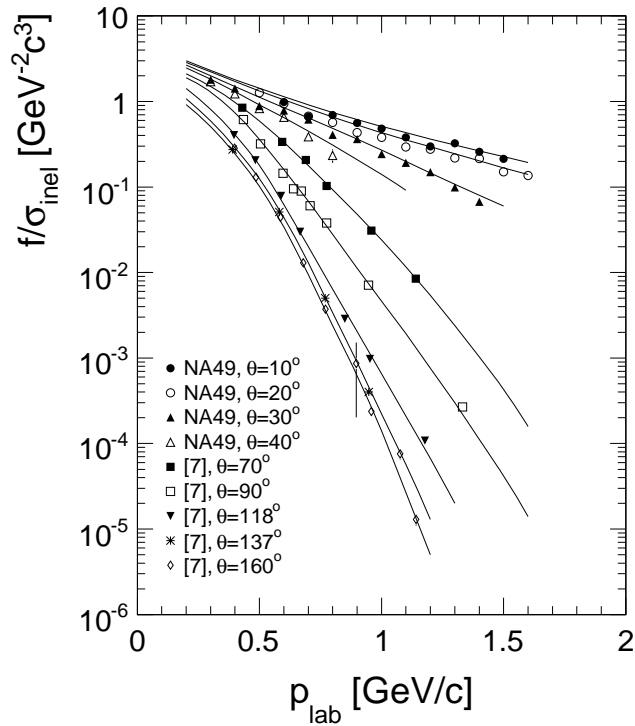}
 	\caption{Invariant proton cross sections from NA49 and \cite{bayukov}
             at lab angles between 10 and 160 degrees as a function of $p_{\textrm{lab}}$.
             The data interpolation at fixed angle is given by the full lines}
  	 \label{fig:pc_prot_plab}
  \end{center}
\end{figure}

The kinematic situation presented in Fig.~\ref{fig:kin} shows that the NA49
acceptance allows the measurement of proton yields as a function of
$p_{\textrm{lab}}$ up to $\Theta_{\textrm{lab}}$~=~40 degrees. The corresponding data values are
tabulated in Table~\ref{tab:angle}. These data are shown in Fig.~\ref{fig:pc_prot_plab} together 
with the Fermilab data \cite{bayukov} as a function of $p_{\textrm{lab}}$. 

%      Table 5   
\begin{table}[h]
\renewcommand{\tabcolsep}{0.6pc} 
\renewcommand{\arraystretch}{0.95}
\footnotesize
\begin{center}
\begin{tabular}{|c|cc|cc|cc|cc|}
\hline
\multicolumn{9}{|c|}{$f(\Theta_{\textrm{lab}},p_{\textrm{lab}}), \Delta f$} \\ 
\hline
$p_{\textrm{lab}} \backslash \Theta_{\textrm{lab}}$ & \multicolumn{2}{c|}{10$^o$} & \multicolumn{2}{c|}{20$^o$} & \multicolumn{2}{c|}{30$^o$} & \multicolumn{2}{c|}{40$^o$} \\ \hline    
0.3 &        &      &        &      & 1.7979 & 7.58 & 1.7087 & 6.20  \\ 
0.4 &        &      &        &      & 1.4154 & 7.07 & 1.2310 & 7.67  \\                      
0.5 &        &      & 1.2656 & 11.9 & 0.8836 & 7.67 & 0.8255 & 8.16  \\ 
0.6 & 0.9854 & 6.11 & 0.9423 & 6.94 & 0.7850 & 7.20 & 0.6522 & 8.22  \\
0.7 & 0.6654 & 4.81 & 0.6714 & 5.93 & 0.6167 & 7.30 & 0.3868 & 8.66  \\
0.8 & 0.6932 & 7.01 & 0.5700 & 6.93 & 0.4075 & 8.20 & 0.2339 & 17.8  \\
0.9 & 0.5627 & 5.12 & 0.4342 & 7.44 & 0.3631 & 7.99 &        &       \\
1.0 & 0.4805 & 7.89 & 0.3800 & 7.37 & 0.2436 & 9.11 &        &       \\ 
1.1 & 0.3810 & 6.96 & 0.2932 & 5.42 & 0.1924 & 9.72 &        &       \\
1.2 & 0.2975 & 7.23 & 0.2756 & 5.22 & 0.1493 & 10.4 &        &       \\  
1.3 & 0.3244 & 6.57 & 0.2179 & 5.42 & 0.0993 & 12.1 &        &       \\
1.4 & 0.2578 & 6.83 & 0.2147 & 7.82 & 0.0670 & 11.5 &        &       \\
1.5 & 0.2128 & 7.19 & 0.1500 & 8.87 &        &      &        &       \\
1.6 &        &      & 0.1358 & 8.98 &        &      &        &       \\ \hline
\end{tabular}
\end{center}
\caption{Invariant proton cross sections measured by NA49 at
         fixed values of $\Theta_{\textrm{lab}}$ between 10 and 40 degrees as a for 
         $p_{\textrm{lab}}$ values between from 0.3 to 1.6~GeV/c. The relative statistical errors, 
         $\Delta f$, are given in \%}
\label{tab:angle}
\end{table}

%     Here Fig.40 

Evidently the two data sets are complementary and offer for the first
time an almost complete angular coverage of the backward proton 
production with double differential cross sections.

\begin{figure}[h]
  \begin{center}
  	\includegraphics[width=11cm]{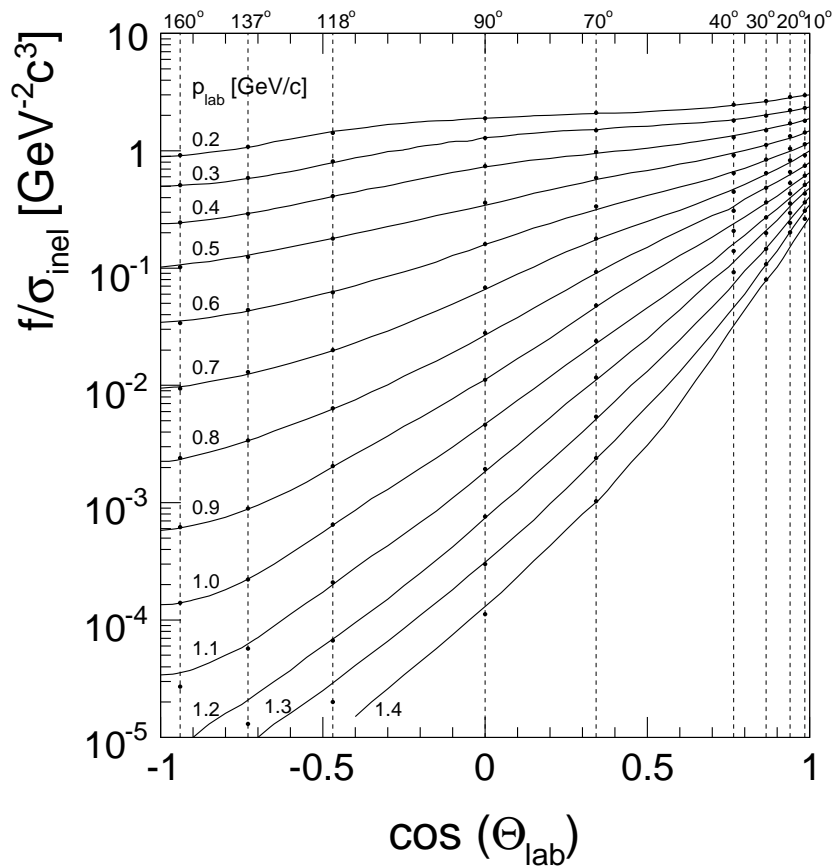}
 	\caption{Invariant proton cross sections of the combined NA49 
             and Fermilab data as a function of $\cos (\Theta_{\textrm{lab}})$ for fixed values 
             of $p_{\textrm{lab}}$ between 0.2 and 1.4~GeV/c. The data interpolation is shown 
             as full lines}
  	 \label{fig:pc_prot_theta}
  \end{center}
\end{figure}

A two-dimensional data interpolation has been performed as shown by
the full lines in Fig.~\ref{fig:pc_prot_plab}. This allows to produce the combined angular 
distribution as a function of $\cos (\Theta_{\textrm{lab}})$ for fixed values of
$p_{\textrm{lab}}$ presented in Fig.~\ref{fig:pc_prot_theta}.

%     Here Fig.41 .

\begin{figure}[h]
  \begin{center}
  	\includegraphics[width=7.5cm]{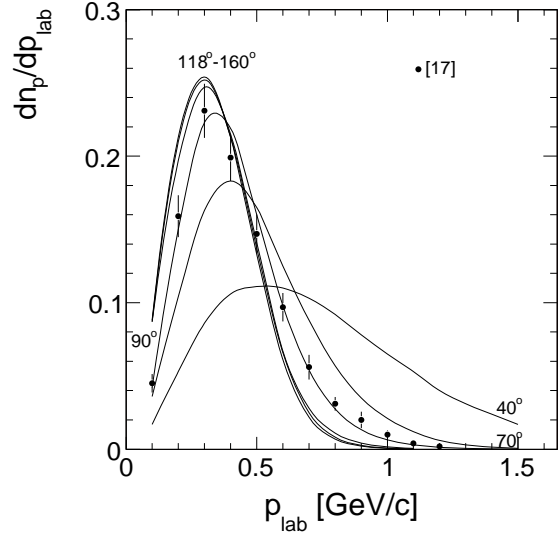}
 	\caption{Proton density distributions $dn_p/dp_{\textrm{lab}}$, normalized to unit area,
 	         as a function of $p_{\textrm{lab}}$ for fixed $\Theta_{\textrm{lab}}$. The bubble
             chamber data \cite{bailly} have been interpolated in steps of $p_{\textrm{lab}}$
             of 0.1~GeV/c and are shown as full circles}
  	 \label{fig:pc_norm_plab}
  \end{center}
\end{figure}

The proton density distributions $dn_p/dp_{\textrm{lab}}$ derived from the interpolated
invariant cross sections are shown as a function of $p_{\textrm{lab}}$ in Fig.~\ref{fig:pc_norm_plab},
normalized to unit area.

%     Here Fig.42 

These distributions are closely similar for 160~$> \Theta_{\textrm{lab}} >$~118 degrees
and develop a tail to large $p_{\textrm{lab}}$ values for angles smaller than
90 degrees indicating increasing  contributions from the fragmentation 
of the participant nucleons hit by the projectile. The "grey" proton 
momentum distribution measured in the EHS rapid cycling bubble
chamber \cite{bailly}, interpolated in steps of $p_{\textrm{lab}}$
of 0.1~GeV/c, are shown as full circles in Fig~\ref{fig:pc_norm_plab}. 
It is closely tracing the result at $\Theta_{\textrm{lab}}$~=~90~degrees.
In this case a strong momentum cut is introduced by requesting 
bubble densities at 1.3 minimum ionizing rejecting most of the
faster forward region. 

\begin{figure}[h]
  \begin{center}
  	\includegraphics[width=7cm]{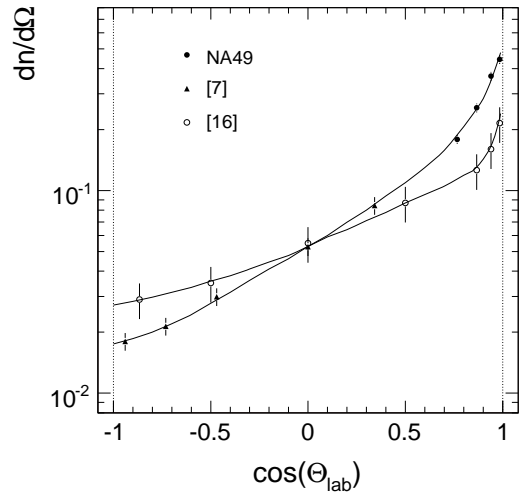}
 	\caption{Proton density $dn/d\Omega$ as a function of 
             $\cos (\Theta_{\textrm{lab}})$. Full circles: integrated combined data from 
             NA49 and \cite{bayukov}. Open circles: measurement by Braune et al. \cite{braune}.
             Lines are drawn to guide the eye. The four lab angles for the NA49 data 
             correspond to Fig.~\ref{fig:pc_prot_theta}}
  	 \label{fig:dndomega}
  \end{center}
\end{figure}

The $dn_p/dp_{\textrm{lab}}$ distributions shown in Fig.~\ref{fig:pc_norm_plab} may be integrated 
over $p_{\textrm{lab}}$ resulting in the proton densities $dn/d\Omega$ shown in Fig.~\ref{fig:dndomega} as a
function of $\cos (\Theta_{\textrm{lab}})$. Here the integration has been limited
to 0~$< p_{\textrm{lab}} <$~1.6~GeV/c as at low angles the target fragmentation
contribution will create a divergent behaviour and since the
comparison data do not contain this component. 

%     Here Fig.43   

The direct measurements from \cite{braune} also presented in Fig.~\ref{fig:dndomega} show
systematic deviations both at forward and at backward angles. As
this experiment uses an energy loss measurement with variable
threshold it is not clear to which extent it represents identified
proton yields. The authors in fact explain that a contribution from
"evaporation" particles ("black tracks" in emulsion work) cannot 
be excluded. Such a contribution would typically be characterized 
by a flatter angular distribution as compared to protons, see also 
the discussion on light ions in Sect.~\ref{sec:ions} showing very sizeable 
d/p ratios. As the d/p and t/p ratios decrease steeply with increasing $x_F$,
see Fig.~\ref{fig:dt2p}, the yield measured by \cite{braune} would
increase at $\Theta_{\textrm{lab}} >$~90$^\circ$ and decrease towards small
angles with respect to the one of identified protons. In addition, a $dE/dx$ 
cut-off in the detector of \cite{braune} at $p_{\textrm{lab}} \sim$~0.2~GeV/c will reduce 
the measured yields as a function of angle, see Fig.~\ref{fig:pc_norm_plab}.

%
% ****************************** Section 7.2 ****************************
%
\subsection{Double differential cross section $\mathbf {f(x_F,p_T)}$ as a function of $\mathbf {x_F}$}
\vspace{3mm}
\label{sec:cs}

%     Here Fig.44 
\begin{figure}[b]
  \begin{center}
  	\includegraphics[width=15cm]{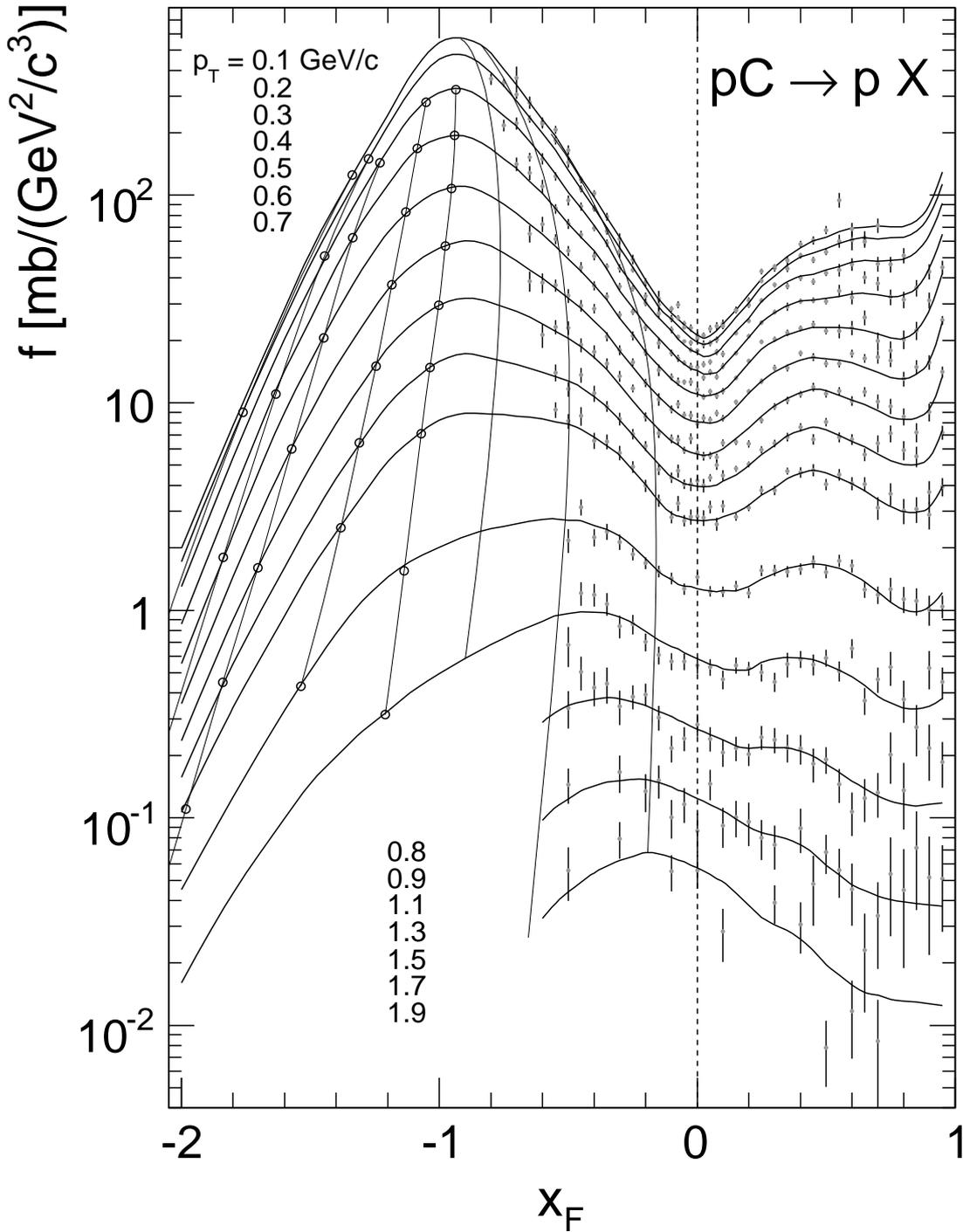}
 	\caption{Invariant cross sections at fixed $p_T$ as a function
             of $x_F$. Full circles: NA49 data, open circles: data from \cite{bayukov}. The thin lines
             show the cross section at fixed angles of 10$^o$, 30$^o$ and 50$^o$}
  	 \label{fig:xf_dist_all}
  \end{center}
\end{figure}

The invariant cross sections measured in $p_{\textrm{lab}}$ and $\Theta_{\textrm{lab}}$ may be
transformed into the $x_F$/$p_T$ variables following the kinematics shown 
in Fig.~\ref{fig:kin}. The corresponding $x_F$ distributions at fixed $p_T$ are 
presented in Fig.~\ref{fig:xf_dist_all} which shows the $x_F$ range down to -2.0. Both the
NA49 data (full circles) and the Fermilab results (open circles)
are plotted. The full lines in Fig.~\ref{fig:xf_dist_all} represent the
interpolation of the two data sets also covering the non-measured
angular region between 40 and 70~degrees, see Fig.~\ref{fig:pc_prot_theta}. The thin
lines indicate the position of the lab angles of 10, 30, and 50~degrees 
as well as the five angles measured by \cite{bayukov}. 

Fig.~\ref{fig:xf_dist_all} represents one of the main results of this paper.
It shows for the first time a complete coverage of the baryonic phase space in p+A collisions, from $x_F$~=~-2 up to the
kinematic limit for the projectile fragmentation at $x_F \sim$~+1. Several features merit comment:

\begin{itemize}
  \item The invariant proton cross sections extend far below the
        kinematic limit for target fragmentation at $x_F$~=~-1
  \item There is no indication of a diffractive structure with
        a peak at $x_F \sim$~-1 as it would be expected from the prompt
        fragmentation of the hit target nucleons -- on the other hand, at $x_F >$~+0.9 there is a diffractive
        peak in the projectile fragmentation region, see \cite{pp_proton} for
        comparison with p+p interactions
  \item The backward cross sections peak at $x_F \sim$~-0.9, not at $x_F \sim$~-1
        indicating a sizeable longitudinal momentum transfer in the
        nuclear fragmentation region
  \item At low $p_T$ or low transverse momentum transfer however, the
        lines of constant lab angle are compatible as expected with a
        convergence towards $x_F$~=~-1
\end{itemize}

A detailed discussion of these features including the de-composition
of the measured proton yields into the basic components of projectile,
target and nuclear fragmentation, is presented in Sects.~\ref{sec:hadr_net} and \ref{sec:pt_prot}. 

%
% ****************************** Section 8 ****************************
%
\section{Light ions: deuterons and tritons}
\vspace{3mm}
\label{sec:ions}

As shown in Sect.~\ref{sec:pid}, Fig.~\ref{fig:backdedx}, the particle 
identification via energy loss measurement in the NA49 TPC system also 
allows the extraction of deuteron and triton yields. The accessible
kinematic region covers lab momenta from the detector acceptance limit 
at about 0.25~GeV/c up to the crossing of the energy loss distributions 
with the ones for electrons, Fig.~\ref{fig:bb}, at about 
$p_{\textrm{lab}}$~=~2~GeV/c for deuterons and 3~GeV/c for tritons. 

As for the proton cross sections, this range is complementary
to the Fermilab experiment \cite{frankel} which gives light ion cross
sections in the lab angular range from 70 to 160~degrees,
at $p_{\textrm{lab}}$ from 0.5 to 1.3~GeV/c for deuterons and from 0.7 to
1.3~GeV/c for tritons. The NA49 data offer the advantage of
reaching low $p_T$ down to 0.1~GeV/c and of covering the forward
region from $\Theta_{\textrm{lab}}$~=~40~degrees down to about 3~degrees. 
This allows for the first time to trace the extension of 
nuclear fragmentation into light ions towards the central
region of particle production.

%
% ****************************** Section 8.1 ****************************
%
\subsection{Ion to proton ratios}
\vspace{3mm}
\label{sec:dt2p}

In order to clearly bring out the last aspect mentioned above the deuteron 
and triton yields are given here as ratios to the proton yields in each bin 
of $p_{\textrm{lab}}$ and $\Theta_{\textrm{lab}}$,

\begin{align}
  R_d(x_F,p_T) &= \frac{(dn/dp_{\textrm{lab}}d\Omega)_d}{(dn/dp_{\textrm{lab}}d\Omega)_p}  \\
  R_t(x_F,p_T) &= \frac{(dn/dp_{\textrm{lab}}d\Omega)_t}{(dn/dp_{\textrm{lab}}d\Omega)_p}
\end{align}

These density ratios are given as functions of $x_F$ and $p_T$
using the proton mass in the transformation from lab to cms 
variables. They thus give directly the relative contribution
of the light ions with respect to protons in the $x_F$/$p_T$ bins
shown in Fig.~\ref{fig:accept} and Table~\ref{tab:prot_cs}. In forming the ion/proton density
ratios, most of the data corrections, Sect.~\ref{sec:corr}, drop out with
the exception of absorption which is small but increased for
ions, and of hyperon feed-down which is of course only applicable
to protons. The resulting ratios are shown as functions of $x_F$ and
$p_T$ in Table~\ref{tab:d2p} for deuterons and in Table~\ref{tab:t2p} for tritons.
Most of the systematic errors cancel in the ratios of 
Tables~\ref{tab:d2p} and \ref{tab:t2p}, with the
exception of the feed-down (only for protons) and the
detector and target absorption. The resulting systematic
uncertainty has been estimated to less than 3\% for deuterons and
less than 5\% for tritons which is small compared to the given
statistical errors. 

%   Table 6 
\begin{table}[h]
\renewcommand{\tabcolsep}{0.2pc} 
\renewcommand{\arraystretch}{1.1}
\scriptsize
\begin{center}
\begin{tabular}{|c|cr|cr|cr|cr|cr|cr|cr|cr|cr|cr|cr|cr|}
\hline
%\multicolumn{13}{|c|}{$R_d(x_F,p_T), \Delta R_d$} \\ 
\multicolumn{25}{|c|}{$R_d(x_F,p_T) [\%], \Delta R_d [\%]$} \\ 
\hline
$p_T \backslash x_F$ & \multicolumn{2}{c|}{-0.8} & \multicolumn{2}{c|}{-0.75} & \multicolumn{2}{c|}{-0.7} & \multicolumn{2}{c|}{-0.65} & \multicolumn{2}{c|}{-0.6} & \multicolumn{2}{c|}{-0.55} & \multicolumn{2}{c|}{-0.5} & \multicolumn{2}{c|}{-0.45} & \multicolumn{2}{c|}{-0.4} & \multicolumn{2}{c|}{-0.35} & \multicolumn{2}{c|}{-0.3} & \multicolumn{2}{c|}{-0.25}\\ \hline
0.1 &  11.5&2.1 &  10.8&2.2 &   9.08&1.9 &  14.46&2.6 &  5.04&1.4 &  6.20&1.6 &  8.10&0.9 &  5.76&0.9 &  4.80&0.9 &  5.85&1.0 &  4.85&1.1 &      &  \\
0.2 &  12.0&1.3 &  14.0&1.5 &  12.4 &1.2 &   9.95&1.2 &  8.59&1.1 &  8.02&1.1 &  6.88&0.8 &  8.23&1.1 &  5.09&0.9 &  7.39&1.7 &  4.18&1.5 &      &  \\
0.3 &  10.3&2.0 &  11.7&1.4 &   8.8 &1.1 &   9.72&1.1 &  7.52&1.1 &  8.41&1.1 &  8.00&1.1 &  4.42&1.0 &  4.95&0.8 &  4.18&1.0 &  3.35&1.2 &  2.90&0.6  \\
0.4 &      &&       8.5&2.1 &   6.7 &1.1 &  12.32&1.5 &  8.12&1.3 &  8.01&1.2 &  8.29&1.7 &  6.39&1.2 &  5.58&1.2 &  4.95&1.0 &  1.72&0.5 &      &  \\
0.5 &      &&          &&       7.7 &1.9 &   5.97&1.2 &  7.07&1.2 &  5.91&1.4 &  6.98&1.7 &  3.43&1.1 &  8.10&1.5 &  4.02&0.9 &  1.90&0.6 &      &  \\
0.6 &      &&          &&           &&      11.43&2.5 &  6.46&2.9 &  7.04&1.8 &  4.56&1.4 &  4.22&1.1 &  2.65&0.9 &  5.03&1.3 &  1.79&0.8 &      &   \\
0.7 &      &&          &&           &&           &&      9.26&4.0 &  9.28&2.9 &  7.38&2.5 &  7.65&2.4 &  3.72&1.5 &      &&          &&          &  \\
0.8 &      &&          &&           &&           &&          &&      8.84&3.4 &  2.67&1.4 &  3.68&1.5 &  5.20&2.1 &      &&          &&          &   \\
0.9 &      &&          &&           &&           &&          &&      4.88&2.0 &  3.73&1.5 &      &&          &&          &&          &&          &  \\
1.1 &      &&          &&           &&           &&          &&          &&      3.84&2.6 &      &&          &&          &&          &&          &          \\
\hline
\end{tabular}
\end{center}
\caption{Deuteron to proton density ratios $R_d(x_F,p_T)$ and their statistical errors $\Delta R_d$ in percent 
         as a function of $x_F$ and $p_T$ using proton mass in the transformation from lab to
         cms system }
\label{tab:d2p}
\end{table}

%   Table 7 
\begin{table}[h]
\renewcommand{\tabcolsep}{0.2pc} 
\renewcommand{\arraystretch}{1.1}
\scriptsize
\begin{center}
\begin{tabular}{|c|cr|cr|cr|cr|cr|cr|cr|cr|cr|cr|cr|}
\hline
%\multicolumn{13}{|c|}{$R_t(x_F,p_T), \Delta R_t$} \\ 
\multicolumn{23}{|c|}{$R_t(x_F,p_T) [\%], \Delta R_t [\%]$} \\ 
\hline
$p_T \backslash x_F$ & \multicolumn{2}{c|}{-0.8} & \multicolumn{2}{c|}{-0.75} & \multicolumn{2}{c|}{-0.7} & \multicolumn{2}{c|}{-0.65} & \multicolumn{2}{c|}{-0.6} & \multicolumn{2}{c|}{-0.55} & \multicolumn{2}{c|}{-0.5} & \multicolumn{2}{c|}{-0.45} & \multicolumn{2}{c|}{-0.4} & \multicolumn{2}{c|}{-0.35} & \multicolumn{2}{c|}{-0.3} \\ \hline
0.1 &  1.22&0.1 &      &&      1.74&0.9 &  0.79&0.5 &      &&      1.18&0.7 &  1.00&0.3 &  0.44&0.2 &  0.31&0.2 &  0.45&0.3 &      &         \\
0.2 &  1.12&0.4 &  2.24&0.6 &  0.96&0.3 &  2.39&0.5 &  1.80&0.5 &  0.96&0.4 &  0.76&0.2 &  1.22&0.4 &  0.28&0.2 &  0.34&0.3 &      &         \\
0.3 &  1.73&0.8 &  2.14&0.6 &  2.46&0.6 &  2.50&0.6 &  1.27&0.4 &  1.61&0.5 &  0.44&0.2 &  1.06&0.4 &  0.15&0.1 &  0.28&0.2 &  0.41&0.3  \\
0.4 &      &&      1.02&0.7 &  0.82&0.4 &  1.35&0.5 &  1.27&0.5 &  0.69&0.3 &  0.99&0.5 &  0.43&0.3 &  0.73&0.4 &  0.40&0.3 &  0.47&0.2  \\
0.5 &      &&          &&          &&      2.08&0.8 &  1.16&0.5 &  0.33&0.3 &  0.67&0.4 &  0.38&0.3 &      &&      0.20&0.2 &  0.36&0.2  \\
0.6 &      &&          &&          &&          &&          &&      0.45&0.4 &      &&      0.48&0.4 &      &&          &&          &         \\
\hline
\end{tabular}
\end{center}
\caption{Triton to proton density ratios $R_t(x_F,p_T)$ and their statistical errors $\Delta R_t$ in percent 
         as a function of $x_F$ and $p_T$ using proton mass in the transformation from lab to cms system}
\label{tab:t2p}
\end{table}

In these data tables the cut-offs imposed by the NA49 acceptance
at $\Theta_{\textrm{lab}} \sim$~40~degrees below $x_F \sim$~-0.5 and by the upper limit
on $p_{\textrm{lab}}$ imposed by the energy loss measurement at $x_F \gtrsim$~-0.5 are
discernible. In addition the fast decrease of the ratios towards
higher $x_F$ limits the extraction of tritons due to the low overall
event statistics. 

%
% ****************************** Section 8.2 ****************************
%
\subsection{Comparison to the Fermilab data \cite{frankel}}
\vspace{3mm}
\label{sec:ions_comp}

\begin{figure}[h]
  \begin{center}
  	\includegraphics[width=12.5cm]{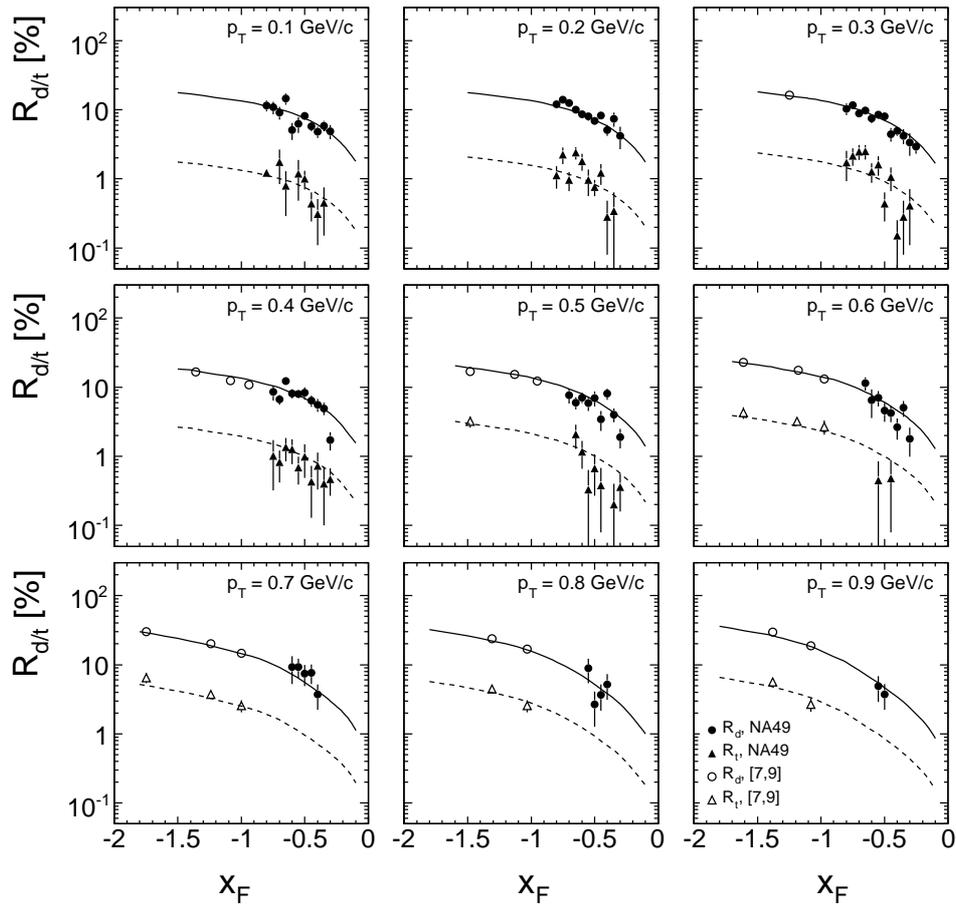}
 	\caption{Deuteron and triton to proton density ratios $R_d$ and
             $R_t$ as a function of $x_F$ for fixed values of $p_T$ between 0.1 and 
             0.9~GeV/c. The full lines give the two-dimensional data interpolation
             established for $R_d$, the broken lines are the same multiplied by the
             suppression factors, assumed $x_F$ independent, shown in Fig.~\ref{fig:trit_supp}}
  	 \label{fig:dt2p}
  \end{center}
\end{figure}

The density ratios $R_d(x_F,p_T$) and $R_t(x_F,p_T)$ are presented in Fig.~\ref{fig:dt2p}
as a function of $x_F$ in comparison to the Fermilab data which are
available above $p_T \sim$~0.3 GeV/c.

%  Fig.45

\begin{figure}[h]
  \begin{center}
  	\includegraphics[width=7cm]{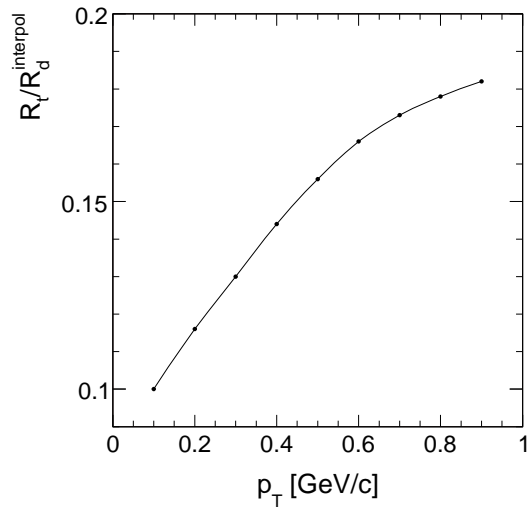}
 	\caption{Triton over deuteron ratio $R_t$/$R_d^{\textrm{interpol}}$
             as a function of $p_T$, assumed $x_F$ independent.
             The full line is drawn to guide the eye}
  	 \label{fig:trit_supp}
  \end{center}
\end{figure}

As for the proton data, the good consistency with the Fermilab data
in the $p_T$ range above 0.3~GeV/c is confirmed for $R_d$ by the 
two-dimensional data interpolation performed by eyeball fits and shown as the full lines in Fig.~\ref{fig:dt2p}.
The density ratio $R_t$ shows a similar consistency with \cite{frankel} over the smaller available 
region of comparison due to the higher $p_{\textrm{lab}}$ cut-off in the Fermilab 
data. With respect to $R_d$, $R_t$ is suppressed by a factor of 0.1 to 0.2
depending on $p_T$ but within statistics independent of $x_F$. This is
demonstrated by the broken lines in Fig.~\ref{fig:dt2p} which represent the 
interpolation of $R_d$ with the ratio $R_t$/$R_d^{\textrm{interpol}}$, assumed $x_F$ independent,
given as a function of $p_T$ in Fig.~\ref{fig:trit_supp}.

%      Fig.46

%
% ****************************** Section 8.3 ****************************
%
\subsection{$\mathbf {x_F}$ and $\mathbf {p_T}$ dependences}
\vspace{3mm}
\label{sec:ions_dep}

\begin{figure}[h]
  \begin{center}
  	\includegraphics[width=10cm]{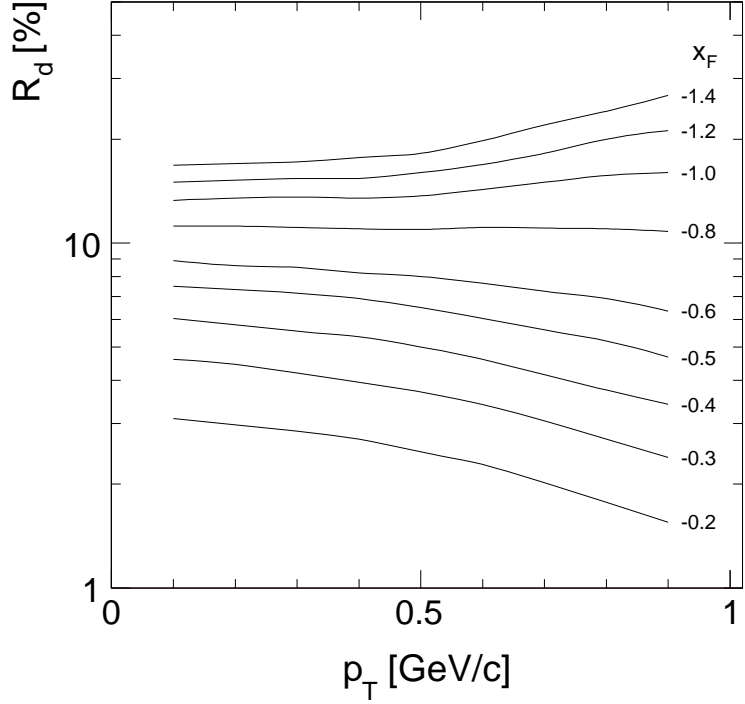}
 	\caption{The density ratio $R_d$ as a function of $p_T$ for fixed values of $x_F$
             between -1.4 and -0.2}
  	 \label{fig:d2p_pt}
  \end{center}
\end{figure}

The $x_F$ dependences of the density ratios presented in Fig.~\ref{fig:dt2p} show
a rather complex behaviour. Evidently deuterons and tritons, if
seen as nuclear fragments as opposed to coalescing from produced
baryons, reach far towards central production. In consequence the 
separation of nuclear fragmentation and coalescence is not an
easy especially as fragmentation products, as shown by the
tentative extrapolation of the data interpolation towards $x_F$~=~-0.1,
might well represent a contribution in the sub-percent range even
at $x_F$~=~0 and already for the light Carbon nuclei at SPS energy. 
The $p_T$ dependence of $R_d$ at fixed $x_F$ is as well non-trivial, as 
shown in Fig.~\ref{fig:d2p_pt}.

%     Here Fig.47 

Evidently $R_d$ tends to decrease with increasing $p_T$ for $x_F >$~-0.8 and
with an inverted tendency for $x_F <$~-0.8. The $R_t$/$R_d$ ratio
on the other hand clearly increases with $p_T$ at all available
$x_F$ values, see Fig.~\ref{fig:trit_supp}.  

%
% ****************************** Section 9 ****************************
%
\section{Particle ratios}
\vspace{3mm}
\label{sec:ratios}

The already published data on p+C \cite{pc_pion,pc_discus} and 
p+p \cite{pp_proton,pp_pion,pp_kaon} interactions
allow for a very detailed study of particle ratios. In a first
overview baryonic ratios will be investigated in this section,
both concerning anti-proton/proton ratios in elementary and nuclear
collisions,

\begin{align}
  R_{\overline{\textrm{p}}\textrm{p}}^{\textrm{pC}} &= f_{\overline{\textrm{p}}}^{\textrm{pC}}(x_F,p_T)/f_\textrm{p}^{\textrm{pC}}(x_F,p_T)  \\
  R_{\overline{\textrm{p}}\textrm{p}}^{\textrm{pp}} &= f_{\overline{\textrm{p}}}^{\textrm{pp}}(x_F,p_T)/f_\textrm{p}^{\textrm{pp}}(x_F,p_T) 
\end{align}
and baryon density ratios directly comparing p+C and p+p reactions,

\begin{align}
  R_{\overline{\textrm{p}}} &= f_{\overline{\textrm{p}}}^{\textrm{pC}}(x_F,p_T)/f_{\overline{\textrm{p}}}^{\textrm{pp}}(x_F,p_T) \\
  R_\textrm{p}              &= f_\textrm{p}^{\textrm{pC}}(x_F,p_T)/f_\textrm{p}^{\textrm{pp}}(x_F,p_T).
\end{align}

\begin{figure}[h]
  \begin{center}
  	\includegraphics[width=14cm]{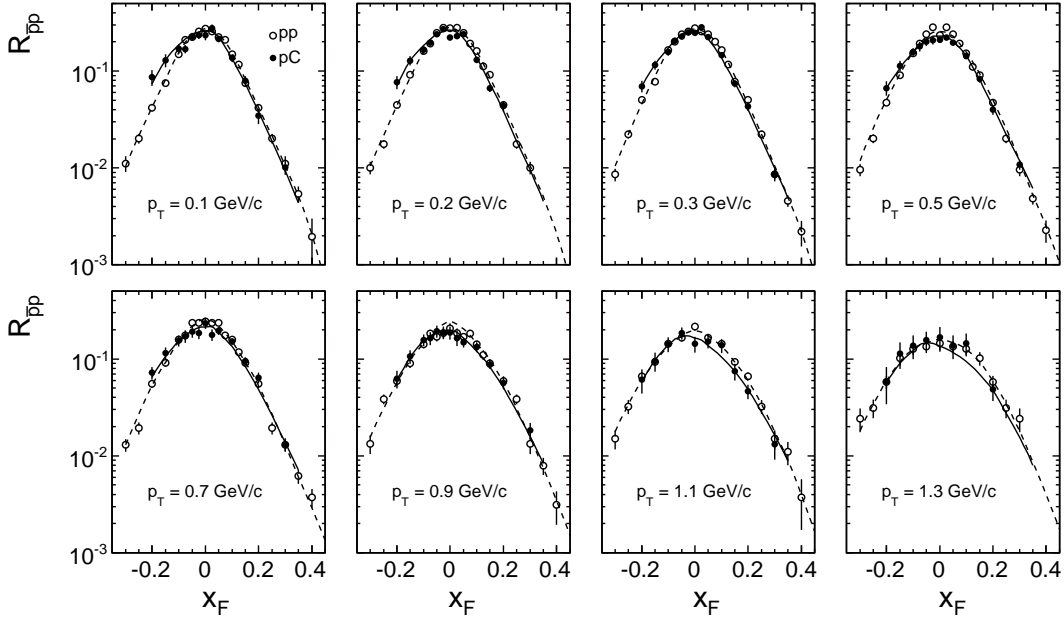}
 	\caption{The anti-proton to proton ratio $R_{\overline{\textrm{p}}\textrm{p}}$ as a function of $x_F$ for 8 values of $p_T$
             comparing p+C (closed circles) and p+p (open circles) interactions.
             The corresponding ratios of the data interpolations are shown
             as full lines (p+C) and broken lines (p+p)}
   	 \label{fig:a2p_all}
  \end{center}
\end{figure}

Fig.~\ref{fig:a2p_all} shows $R_{\overline{\textrm{p}}\textrm{p}}^{\textrm{pC}}$ (closed circles) and 
$R_{\overline{\textrm{p}}\textrm{p}}^{\textrm{pp}}$ (open circles)
together with the corresponding ratios of the data interpolation
(full and broken lines) as a function of $x_F$ for several bins of $p_T$.

%     Here Fig.48

\begin{figure}[h]
  \begin{center}
  	\includegraphics[width=6.cm]{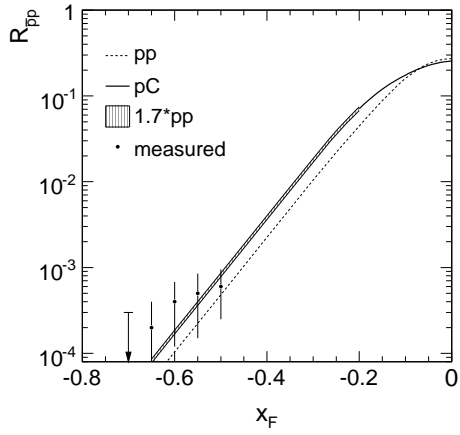}
 	\caption{The anti-proton to proton ratio $R_{\overline{\textrm{p}}\textrm{p}}^{\textrm{pC}}$ in the intervals -0.7~$< x_F <$~-0.5 
             and 0.1~$< p_T <$~0.4~GeV/c. As illustration,
             the measured ratios $R_{\overline{\textrm{p}}\textrm{p}}$ from p+p interactions (broken
             line, \cite{pp_proton}) are given together with the data interpolation (full line down to $x_F$~=~-0.2).
             The hatched area corresponds to the expected increase by a factor of 1.7
             from isospin effect, with an assumed uncertainty of 5\%}
  	 \label{fig:a2p_xf}
  \end{center}
\end{figure}

This comparison reveals that the ratios are within errors equal in
the projectile hemisphere $x_F \gtrsim$~0.1 with the exception of the two
highest $p_T$ bins. This means that the transfer of projectile baryon number
("stopping") towards the central region is equal for protons and
for anti-protons. In the backward hemisphere there are distinct
differences between p+C and p+p interactions at low to medium $p_T$.
This is a result of isospin effects between the isoscalar C and
the p targets. In fact it is known that the anti-proton yields increase
and the proton yields decrease in neutron fragmentation \cite{iso} with
both effects increasing $R_{\overline{\textrm{p}}\textrm{p}}^{\textrm{pC}}$ compared 
to $R_{\overline{\textrm{p}}\textrm{p}}^{\textrm{pp}}$. The clean
extraction of $\overline{\textrm{p}}$ and p yields in the far backward direction, 
$x_F <$~-0.4, see Sect.~\ref{sec:pid} and Fig.~\ref{fig:backdedx}, allows the extension of 
$R_{\overline{\textrm{p}}\textrm{p}}^{\textrm{pC}}$ to the $x_F$ range -0.7~$< x_F <$~-0.5. 
This is presented in Fig.~\ref{fig:a2p_xf} in the $p_T$ interval 0.1~$< p_T <$~0.4~GeV/c.

%     Here Fig.49

The broken line represents $R_{\overline{\textrm{p}}\textrm{p}}^{\textrm{pp}}$ \cite{pp_proton}, 
the full line the measured $R_{\overline{\textrm{p}}\textrm{p}}^{\textrm{pC}}$. The expected increase 
of the $\overline{\textrm{p}}$ yield by a factor of 1.7 ($\overline{\textrm{p}}$ increase by a factor 
of 1.33 \cite{iso} and p decrease by a factor of 1.3 \cite{pp_proton}) is indicated by a hatched area
assuming a 5\% uncertainty in the respective factors.
The fact that the measured backward ratio follows closely
the expectation from target fragmentation shows that there is no
$\overline{\textrm{p}}$ production from nuclear cascading. This is in 
agreement with the upper limit of $R_{\overline{\textrm{p}}\textrm{p}}^{\textrm{pC}}$ 
of 10$^{-4}$ to 10$^{-5}$ given in \cite{niki} for p+$^{181}$Ta at 90 degrees laboratory angle.

\begin{figure}[h]
  \begin{center}
  	\includegraphics[width=12.cm]{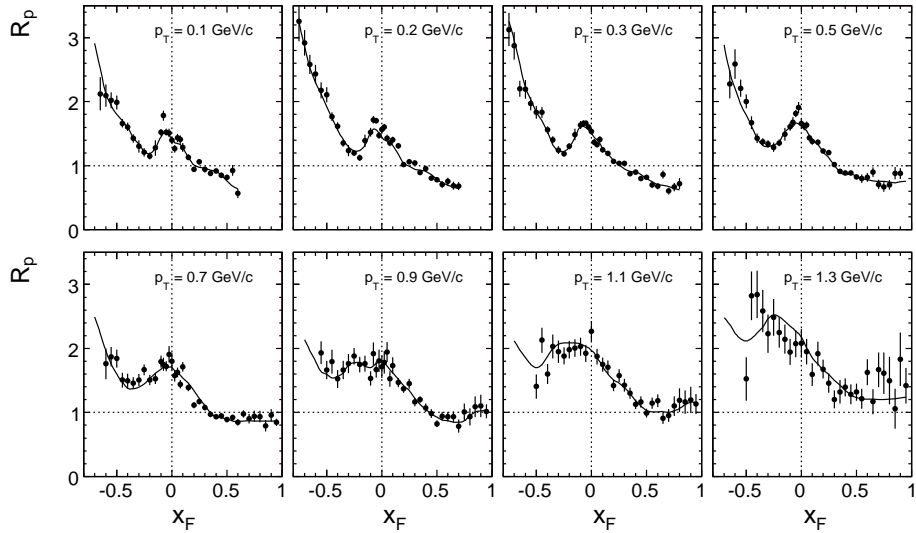}
 	\caption{The proton ratio $R_p$ as a function of
            $x_F$ for fixed $p_T$ values between 0.1 and 1.3~GeV/c. The full
            lines give the ratios for the corresponding data interpolations}
  	 \label{fig:c2p_prot}
  \end{center}
\end{figure}

\begin{figure}[h]
  \begin{center}
  	\includegraphics[width=12.cm]{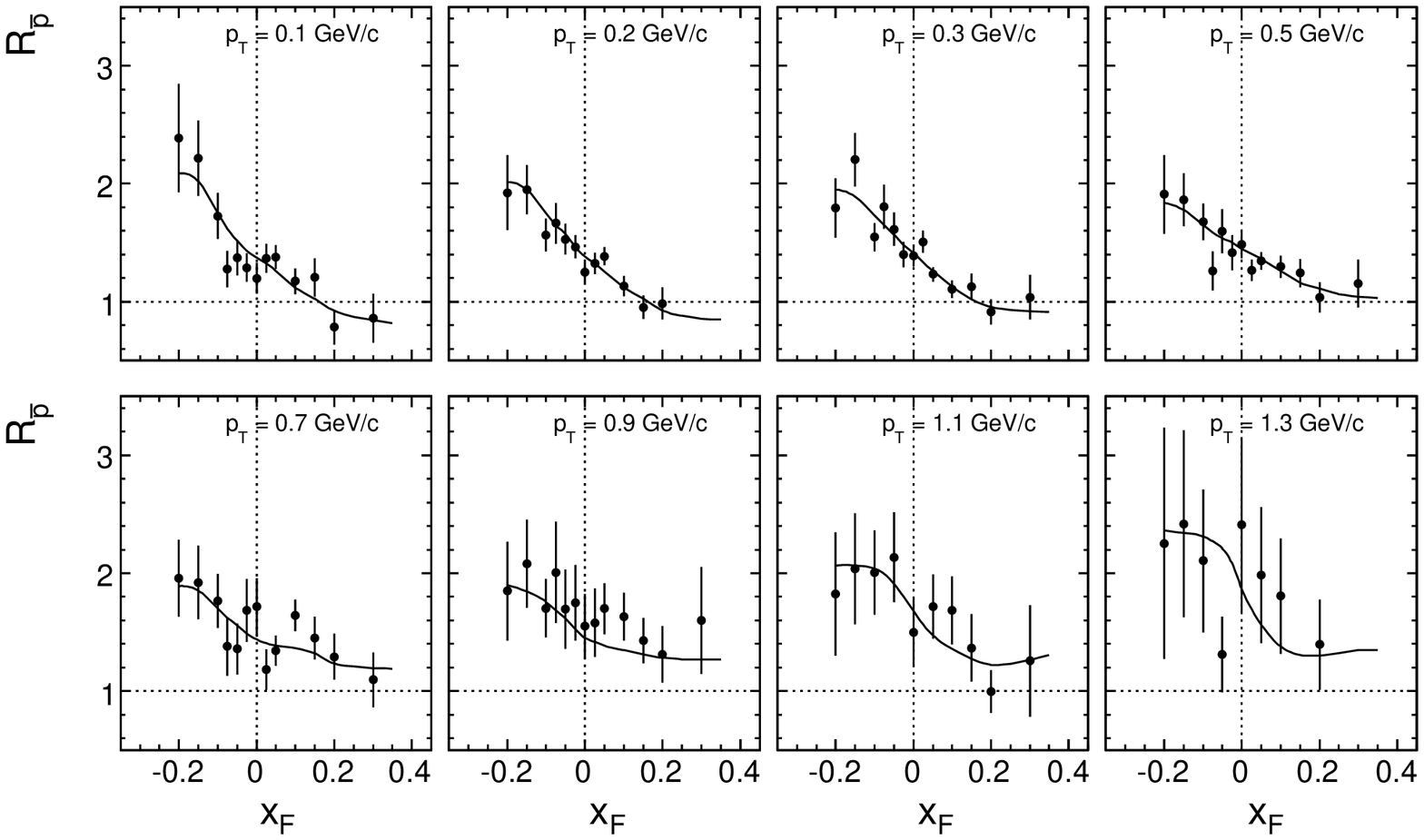}
 	\caption{The anti-proton ratio $R_{\overline{\textrm{p}}}$ as a function of
             $x_F$ for fixed $p_T$ values between 0.1 and 1.3~GeV/c. The full
             lines give the ratios for the corresponding data interpolations}
  	 \label{fig:c2p_aprot}
  \end{center}
\end{figure}

In contrast to the $\overline{\textrm{p}}$/p ratios discussed above, the baryon yields
proper exhibit an important evolution when passing from p+p to
p+C interactions. This is presented as functions of $x_F$ for fixed values of $p_T$
in Fig.~\ref{fig:c2p_prot} for $R_\textrm{p}$ and in 
Fig.~\ref{fig:c2p_aprot} for $R_{\overline{\textrm{p}}}$.

%     Fig.50

%     Fig.51

Looking first at $R_\textrm{p}$, Fig.~\ref{fig:c2p_prot}, where the data give access to a very
wide range of $x_F$, three distinct zones may be distinguished.

A first area is governed by the projectile fragmentation, $x_F \gtrsim$~0.1. 
Here the transfer of baryon number from the forward to the central region is 
clearly evident, with a suppression of the far forward yield by a factor of 0.6 at 
low to medium $p_T$ followed by a steady increase of proton density with increasing 
$p_T$ until the densities exceed the p+p values at all $x_F$ for $p_T >$~1~GeV/c.

A second area is characterized by the fragmentation of the target nucleons 
hit by the projectile, -0.5~$< x_F <$~0. Here the pile-up of
produced hadrons in the central area due to the 1.6 mean projectile
collisions in the C nucleus is visible, followed by a steady decrease
of proton density in the backward direction due to the contribution of neutron fragmentation.

A third component is characterized by the steep increase of proton
density for $x_F <$~-0.2, at low to medium $p_T$. This region is governed
by the intra-nuclear cascading of baryons ("grey" protons) in the
lab momentum range up to about 2~GeV/c. This component dies out
as expected with increasing transverse momentum.

Due to the smaller $x_F$ range experimentally available for the
anti-protons, the corresponding ratio $R_{\overline{\textrm{p}}}$ is more restricted.
It has however been demonstrated (Fig.~\ref{fig:a2p_xf}) that here no cascading component exists. 
As argued above, the evolution of the projectile fragmentation ($x_F \gtrsim$~0.1) 
follows closely the one for protons including
the increase with transverse momentum. As far as the target
fragmentation is concerned, the increase of anti-proton density
beyond the pile-up from the 1.6 average projectile collisions
to about a factor of 2 is expected from isospin symmetry.

A further, more detailed argumentation concerning this phenomenology
can be found in the discussion Sect.~\ref{sec:hadr} below. 

%
% ****************************** Section 10 ****************************
%
\section{Integrated data}
\vspace{3mm}
\label{sec:ptint}

%
% ****************************** Section 10.1 ****************************
%
\subsection{$\mathbf {p_T}$ integrated distributions}
\vspace{3mm}
\label{sec:ptint_data}

%       Table 8, proton table
\begin{table}[b]
\footnotesize
\renewcommand{\tabcolsep}{0.4pc} % enlarge column spacing
\renewcommand{\arraystretch}{0.95} % enlarge line spacing
\begin{center}
\begin{tabular}{|r@{}l|cc|cc|cc|cc||r@{}l|c||r@{}l|c|}
\hline
\multicolumn{16}{|c|}{p} \\
\hline
\multicolumn{2}{|c|}{$x_F$}& $F$ [mb$\cdot$c]& $\Delta$& $dn/dx_F$& $\Delta$& $\langle p_T \rangle$& $\Delta$
      & $\langle p_T^2 \rangle$ & $\Delta$ & \multicolumn{2}{c|}{$y$}  &  $dn/dy$ & \multicolumn{2}{c|}{$y$}  & $dn/dy$\\ \hline
-0.&8   &  81.6  & 4.8 &  1.400 & 4.8 &  0.391 & 2.1 &  0.208 & 1.2 & -3&.2 &  0.21949 &   0&.8 &  0.12875  \\
-0.&75  &  73.2  & 4.5 &  1.337 & 4.5 &  0.402 & 1.8 &  0.220 & 2.9 & -3&.1 &  0.43696 &   0&.9 &  0.13405 \\
-0.&7   &  65.4  & 3.8 &  1.278 & 3.8 &  0.414 & 1.6 &  0.232 & 2.5 & -3&.0 &  0.76998 &   1&.0 &  0.14070  \\
-0.&65  &  57.0  & 2.9 &  1.198 & 2.9 &  0.428 & 1.3 &  0.249 & 2.3 & -2&.9 &  1.09283 &   1&.1 &  0.14866 \\
-0.&6   &  50.4  & 2.9 &  1.142 & 2.9 &  0.442 & 1.2 &  0.266 & 2.1 & -2&.8 &  1.23706 &   1&.2 &  0.15804 \\
-0.&55  &  43.8  & 2.0 &  1.079 & 2.0 &  0.458 & 0.82 &  0.285 & 1.5 & -2&.7 &  1.20270 &   1&.3 &  0.16829 \\
-0.&5   &  37.7  & 2.1 &  1.018 & 2.1 &  0.477 & 0.91 &  0.308 & 1.8 & -2&.6 &  1.06937 &   1&.4 &  0.18043 \\
-0.&45  &  32.8  & 1.8 &  0.975 & 1.8 &  0.494 & 0.93 &  0.329 & 1.9 & -2&.5 &  0.91240 &   1&.5 &  0.19426 \\
-0.&4   &  28.22 & 1.7 &  0.935 & 1.7 &  0.508 & 0.80 &  0.348 & 1.7 & -2&.4 &  0.77367 &   1&.6 &  0.20672 \\
-0.&35  &  23.92 & 1.6 &  0.892 & 1.6 &  0.523 & 0.85 &  0.369 & 1.7 & -2&.3 &  0.65943 &   1&.7 &  0.21800 \\
-0.&3   &  19.83 & 1.8 &  0.844 & 1.8 &  0.538 & 0.94 &  0.391 & 1.9 & -2&.2 &  0.56603 &   1&.8 &  0.22892 \\
-0.&25  &  16.57 & 1.4 &  0.818 & 1.4 &  0.547 & 0.76 &  0.403 & 1.7 & -2&.1 &  0.49330 &   1&.9 &  0.23773 \\
-0.&2   &  13.88 & 1.4 &  0.808 & 1.4 &  0.554 & 0.68 &  0.414 & 1.5 & -2&.0 &  0.43602 &   2&.0 &  0.24016 \\
-0.&15  &  11.80 & 2.1 &  0.825 & 2.1 &  0.554 & 0.92 &  0.415 & 1.7 & -1&.9 &  0.38870 &   2&.1 &  0.23634 \\
-0.&1   &  10.31 & 1.4 &  0.873 & 1.4 &  0.551 & 0.60 &  0.411 & 1.3 & -1&.8 &  0.34565 &   2&.2 &  0.23645 \\
-0.&075 &   9.71 & 1.6 &  0.900 & 1.5 &  0.551 & 0.59 &  0.410 & 1.1 & -1&.7 &  0.30284 &   2&.3 &  0.23584 \\
-0.&05  &   9.26 & 1.1 &  0.927 & 1.0 &  0.549 & 0.63 &  0.407 & 1.4 & -1&.6 &  0.26883 &   2&.4 &  0.23145 \\
-0.&025 &   8.90 & 1.6 &  0.938 & 1.6 &  0.550 & 0.61 &  0.407 & 1.2 & -1&.5 &  0.24220 &   2&.5 &  0.22097 \\
 0.&0   &   8.67 & 1.5 &  0.931 & 1.4 &  0.551 & 0.59 &  0.407 & 1.1 & -1&.4 &  0.22093 &   && \\
 0.&025 &   8.49 & 1.5 &  0.893 & 1.4 &  0.555 & 0.68 &  0.412 & 1.3 & -1&.3 &  0.20292 &   && \\
 0.&05  &   8.53 & 1.2 &  0.852 & 1.2 &  0.555 & 0.63 &  0.412 & 1.3 & -1&.2 &  0.18852 &   && \\
 0.&075 &   8.68 & 1.3 &  0.804 & 1.3 &  0.555 & 0.64 &  0.412 & 1.2 & -1&.1 &  0.17704 &&&\\
 0.&1   &   9.07 & 1.3 &  0.768 & 1.3 &  0.553 & 0.75 &  0.409 & 1.6 & -1&.0 &  0.16671 &&&\\
 0.&15  &  10.16 & 1.0 &  0.712 & 0.9 &  0.545 & 0.54 &  0.396 & 1.2 & -0&.9 &  0.15776 &&&\\
 0.&2   &  11.60 & 0.9 &  0.677 & 0.9 &  0.538 & 0.54 &  0.385 & 1.2 & -0&.8 &  0.14975 &&&\\
 0.&25  &  13.42 & 1.0 &  0.663 & 0.9 &  0.531 & 0.45 &  0.374 & 1.0 & -0&.7 &  0.14307 &&&\\
 0.&3   &  14.87 & 1.1 &  0.633 & 1.1 &  0.528 & 0.53 &  0.370 & 1.2 & -0&.6 &  0.13782 &&&\\
 0.&35  &  16.20 & 1.4 &  0.604 & 1.4 &  0.525 & 0.68 &  0.365 & 1.4 & -0&.5 &  0.13390 &&&\\
 0.&4   &  17.20 & 1.1 &  0.569 & 1.1 &  0.521 & 0.60 &  0.360 & 1.3 & -0&.4 &  0.13026 &&&\\
 0.&45  &  17.74 & 1.4 &  0.527 & 1.4 &  0.519 & 0.65 &  0.357 & 1.3 & -0&.3 &  0.12706 &&&\\
 0.&5   &  17.93 & 1.3 &  0.483 & 1.3 &  0.511 & 0.63 &  0.348 & 1.2 & -0&.2 &  0.12485 &&&\\
 0.&55  &  17.96 & 2.2 &  0.442 & 2.2 &  0.499 & 1.1 &  0.334 & 1.9 & -0&.1 &  0.12318 &&&\\
 0.&6   &  17.90 & 2.4 &  0.406 & 2.4 &  0.488 & 0.98 &  0.321 & 1.7 &  0&.0 &  0.12157 &&&\\
 0.&65  &  17.59 & 2.6 &  0.369 & 2.6 &  0.479 & 0.87 &  0.310 & 1.7 &  0&.1 &  0.12065 &&&\\
 0.&7   &  17.00 & 3.0 &  0.332 & 3.0 &  0.470 & 1.3 &  0.300 & 2.3 &  0&.2 &  0.11939 &&&\\
 0.&75  &  16.48 & 3.4 &  0.301 & 3.4 &  0.461 & 1.3 &  0.290 & 2.7 &  0&.3 &  0.11861 &&&\\
 0.&8   &  16.40 & 3.7 &  0.282 & 3.7 &  0.455 & 1.2 &  0.282 & 2.6 &  0&.4 &  0.11907 &&&\\
 0.&85  &  17.42 & 4.2 &  0.282 & 4.2 &  0.449 & 1.0 &  0.274 & 2.2 &  0&.5 &  0.11975 &&&\\
 0.&9   &  20.6  & 4.0 &  0.315 & 4.0 &  0.440 & 1.3 &  0.261 & 3.0 &  0&.6 &  0.12154 &&&\\
 0.&95  &  29.5  & 3.4 &  0.428 & 3.4 &  0.432 & 0.75 &  0.248 & 1.7 &  0&.7 &  0.12474 &&&\\ \hline
\end{tabular}
\end{center}
\caption{$p_T$ integrated invariant cross section $F$ [mb$\cdot$c], density 
         distribution $dn/dx_F$, mean transverse momentum $\langle p_T \rangle$ [GeV/c], 
         mean transverse momentum squared $\langle p_T^2 \rangle$ [(GeV/c)$^2$] 
         as a function of $x_F$, as well as density distribution 
         $dn/dy$ as a function of $y$ for p. The statistical uncertainty $\Delta$ 
         for each quantity is given in \% as an upper limit considering 
         the full statistical error of each measured $p_T$/$x_F$ bin.
         The systematic errors are 3.7 \%, Table~\ref{tab:syst}}
\label{tab:int_prot}
\end{table}

The $p_T$ integrated non-invariant and invariant yields
are defined by:

\begin{align}
  \label{eq:int}
  dn/dx_F &= \pi/\sigma_{\textrm{inel}} \cdot \sqrt{s}/2 \cdot \int{f/E \cdot dp_T^2} \nonumber \\
  F &= \int{f \cdot dp_T^2}  \\
  dn/dy &= \pi/\sigma_{\textrm{inel}} \cdot \int{f \cdot dp_T^2} \nonumber  
\end{align} 
with $f= E \cdot d^3\sigma/dp^3$, the invariant double differential cross
section. The integrations are performed numerically using the
two-dimensional data interpolation (Sect.~\ref{sec:interp}) in the range
0~$< p_T <$~1.9~GeV/c. Tables~\ref{tab:int_prot} 
and \ref{tab:int_aprot} give the numerical values and the first and second moments of the 
$p_T$ distributions as functions of $x_F$ and rapidity for protons
and anti-protons, respectively. The relative errors quoted are 
given in percent using the full statistical errors of the measured 
points. They thus present upper limits as the data interpolation 
in $x_F$ will reduce the statistical fluctuations for the $p_T$ 
integrated quantities.

%       Table 9 anti-proton table
\begin{table}[h]
\footnotesize
\renewcommand{\tabcolsep}{0.4pc} % enlarge column spacing
\renewcommand{\arraystretch}{1.0} % enlarge line spacing
\begin{center}
\begin{tabular}{|r@{}l|cc|cc|cc|cc||r@{}l|c||r@{}l|c|}
\hline
\multicolumn{16}{|c|}{$\overline{\textrm{p}}$} \\
\hline
\multicolumn{2}{|c|}{$x_F$}& $F$ [mb$\cdot$c]& $\Delta$& $dn/dx_F$& $\Delta$& $\langle p_T \rangle$& $\Delta$
      & $\langle p_T^2 \rangle$ & $\Delta$ & \multicolumn{2}{c|}{$y$}  &  $dn/dy$ & \multicolumn{2}{c|}{$y$}  &  $dn/dy$ \\ \hline
-0&.2   &  0.910 & 5.7 &  0.0533 & 5.7 &  0.528 & 3.3 &  0.375 & 6.4 & -1&.1 &  0.01399 &   0&.4 &  0.02215 \\
-0&.15  &  1.282 & 3.9 &  0.0905 & 3.8 &  0.524 & 2.1 &  0.371 & 4.2 & -1&.0 &  0.01692 &   0&.5 &  0.02054 \\
-0&.1   &  1.627 & 3.2 &  0.1395 & 3.1 &  0.518 & 1.7 &  0.361 & 3.4 & -0&.9 &  0.01915 &   0&.6 &  0.01876  \\
-0&.075 &  1.780 & 3.6 &  0.1676 & 3.7 &  0.513 & 1.6 &  0.353 & 3.2 & -0&.8 &  0.02118 &   0&.7 &  0.01689 \\
-0&.05  &  1.902 & 3.1 &  0.1940 & 3.0 &  0.507 & 1.4 &  0.344 & 2.7 & -0&.7 &  0.02284 &   0&.8 &  0.01492 \\
-0&.025 &  1.940 & 2.9 &  0.2094 & 2.8 &  0.505 & 1.4 &  0.338 & 2.7 & -0&.6 &  0.02419 &   0&.9 &  0.01295 \\
 0&.0   &  1.903 & 2.9 &  0.2095 & 2.7 &  0.502 & 1.6 &  0.333 & 3.4 & -0&.5 &  0.02546 &   1&.0 &  0.01106 \\
 0&.025 &  1.815 & 3.0 &  0.1959 & 2.8 &  0.505 & 1.6 &  0.335 & 3.2 & -0&.4 &  0.02638 &   1&.1 &  0.00924 \\
 0&.05  &  1.607 & 2.2 &  0.1642 & 2.0 &  0.507 & 1.4 &  0.338 & 2.9 & -0&.3 &  0.02690 &   1&.2 &  0.00748 \\
 0&.1   &  1.183 & 2.4 &  0.1014 & 2.3 &  0.524 & 1.2 &  0.359 & 2.5 & -0&.2 &  0.02711 &   1&.3 &  0.00595 \\
 0&.15  &  0.798 & 3.0 &  0.0562 & 3.0 &  0.543 & 1.6 &  0.382 & 3.3 & -0&.1 &  0.02702 &   1&.4 &  0.00457 \\
 0&.2   &  0.506 & 4.1 &  0.0295 & 4.0 &  0.558 & 2.1 &  0.403 & 3.7 &  0&.0 &  0.02656 &   1&.5 &  0.00332 \\
 0&.3   &  0.179 & 6.0 &  0.0076 & 6.0 &  0.570 & 2.8 &  0.428 & 5.0 &  0&.1 &  0.02609 &   1&.6 &  0.00225 \\
  &&&&&&&&&&  0&.2 &  0.02524 &  1&.7 &  0.00132 \\
  &&&&&&&&&&  0&.3 &  0.02381 &&&\\ \hline
\end{tabular}
\end{center}
\caption{$p_T$ integrated invariant cross section $F$ [mb$\cdot$c], density 
         distribution $dn/dx_F$, mean transverse momentum $\langle p_T \rangle$ [GeV/c], 
         mean transverse momentum squared $\langle p_T^2 \rangle $ [(GeV/c)$^2$] 
         as a function of $x_F$, as well as density distribution 
         $dn/dy$ as a function of $y$ for $\overline{\textrm{p}}$. The statistical uncertainty 
         $\Delta$ for each quantity is given in \% as an upper limit considering 
         the full statistical error of each measured $p_T$/$x_F$ bin.
         The systematic errors are 4.2 \%, Table~\ref{tab:syst}}
\label{tab:int_aprot}
\end{table}

Concerning the extension of the kinematic coverage into the far
backward direction for protons, $x_F <$~-0.8, see Sect.~\ref{sec:backext}, 
additional integrated quantities are given in Table~\ref{tab:int_prot_back} for 
-2.0~$\leq x_F \leq$~-0.9. Again, the data interpolation is used to obtain 
the numerical values.

%       Table 10 
\begin{table}[h]
\footnotesize
\renewcommand{\tabcolsep}{0.4pc} % enlarge column spacing
\renewcommand{\arraystretch}{1.0} % enlarge line spacing
\begin{center}
\begin{tabular}{|l|cc|cc|cc|cc|}
\hline
\multicolumn{9}{|c|}{p} \\
\hline
$x_F$& $F$& $\Delta$& $dn/dx_F$& $\Delta$& $\langle p_T \rangle$& $\Delta$
      & $\langle p_T^2 \rangle$ & $\Delta$ \\ \hline
-2.0 &   0.466 &   3.0 &   0.00323 &   3.0 &   0.482 &   2.0 &   0.320 &   3.0 \\  
-1.8 &   1.816 &   3.0 &   0.01398 &   3.0 &   0.457 &   2.0 &   0.283 &   3.0 \\  
-1.6 &   6.51  &   3.0 &   0.0563  &   3.0 &   0.435 &   2.0 &   0.251 &   3.0 \\  
-1.5 &  11.46  &   3.0 &   0.1057  &   3.0 &   0.424 &   2.0 &   0.240 &   3.0 \\  
-1.4 &  19.49  &   3.0 &   0.1926  &   3.0 &   0.417 &   2.0 &   0.230 &   3.0 \\  
-1.3 &  31.6   &   3.0 &   0.336   &   3.0 &   0.409 &   2.0 &   0.220 &   3.0 \\  
-1.2 &  48.9   &   3.0 &   0.564   &   3.0 &   0.400 &   2.0 &   0.211 &   3.0 \\  
-1.1 &  71.0   &   3.0 &   0.890   &   3.0 &   0.384 &   2.0 &   0.197 &   3.0 \\  
-1.0 &  88.8   &   3.0 &   1.224   &   3.0 &   0.374 &   2.0 &   0.189 &   3.0 \\  
-0.9 &  93.0   &   3.0 &   1.422   &   3.0 &   0.378 &   2.0 &   0.194 &   3.0 \\ \hline
\end{tabular}
\end{center}
\caption{$p_T$ integrated invariant cross section $F$ [mb$\cdot$c], density 
         distribution $dn/dx_F$, mean transverse momentum $\langle p_T \rangle$ [GeV/c], 
         mean transverse momentum squared $\langle p_T^2 \rangle $ [(GeV/c)$^2$] 
         as a function of $x_F$ for protons in the far backward region. The estimated 
         relative statistical uncertainties $\Delta$ are given in \%.
         The systematic errors are estimated to be 5 \% }
\label{tab:int_prot_back}
\end{table}

In this region the estimation of the statistical uncertainties
is non-trivial as the measured data have been transformed from
the laboratory co-ordinates ($p_{\textrm{lab}}$,$\Theta_{\textrm{lab}}$) 
to the cms quantities ($x_F$,$p_T$) using a two-dimensional interpolation. 
The given errors have been obtained from the total number of measured protons at 
each lab angle \cite{frankel} which varies from about 1100 to 3600. The
values given in Table~\ref{tab:int_prot_back} are therefore to be regarded as upper
limits. 

\begin{figure}[h]
  \begin{center}
  	\includegraphics[width=15cm]{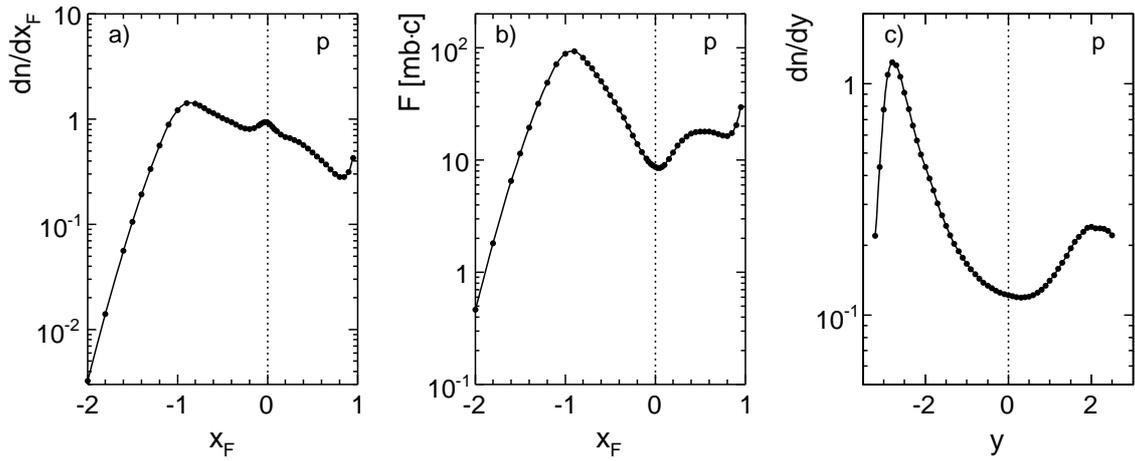}
 	\caption{$p_T$ integrated distributions a) $dn/dx_F$, b) $F$ and c) $dn/dy$
             for protons as a function of $x_F$ and $y$, respectively. Full lines: data interpolation}
  	 \label{fig:ptint_prot}
  \end{center}
\end{figure}

\begin{figure}[h]
  \begin{center}
  	\includegraphics[width=15cm]{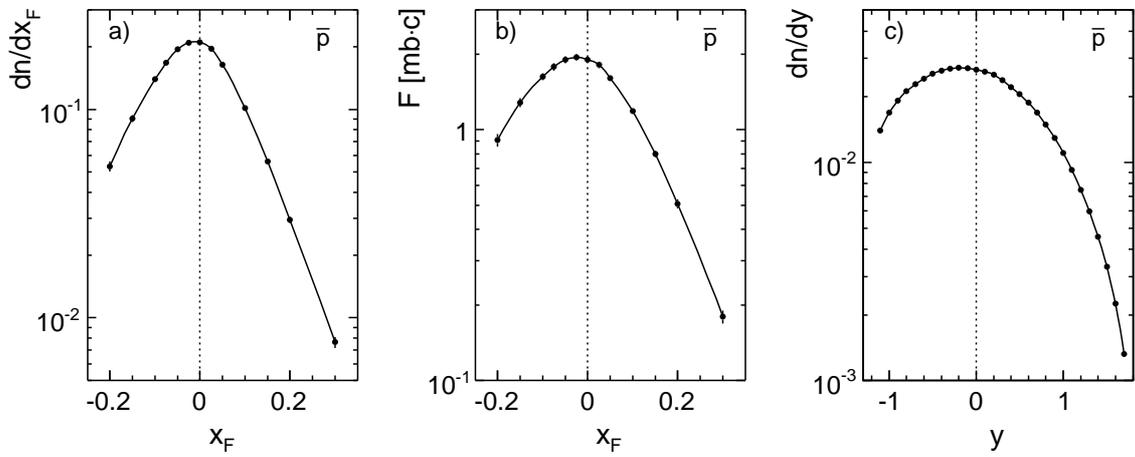}
 	\caption{$p_T$ integrated distributions a) $dn/dx_F$, b) $F$ and c) $dn/dy$
             for anti-protons as a function of $x_F$ and $y$, respectively. Full lines: data interpolation}
  	 \label{fig:ptint_aprot}
  \end{center}
\end{figure}

The corresponding distributions for $dn/dx_F$, $F$ and $dn/dy$ are shown in 
Fig.~\ref{fig:ptint_prot} for protons and in Fig.~\ref{fig:ptint_aprot} for anti-protons.

%     Fig.52

%      Fig.53

\begin{figure}[h]
  \begin{center}
  	\includegraphics[width=6cm]{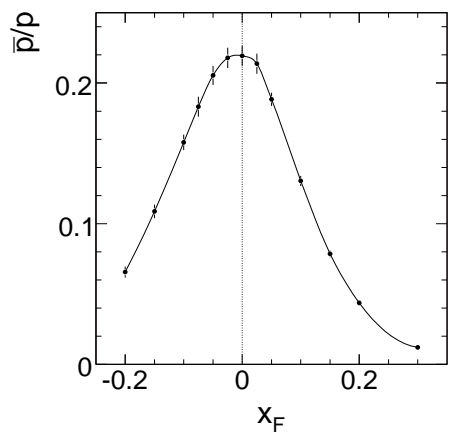}
 	\caption{The ratio of $p_T$ integrated $\overline{\textrm{p}}$ and p yields as a function of $x_F$. 
 	         Full lines: data interpolation}
  	 \label{fig:ratint}
  \end{center}
\end{figure}

\begin{figure}[h]
  \begin{center}
  	\includegraphics[width=11.cm]{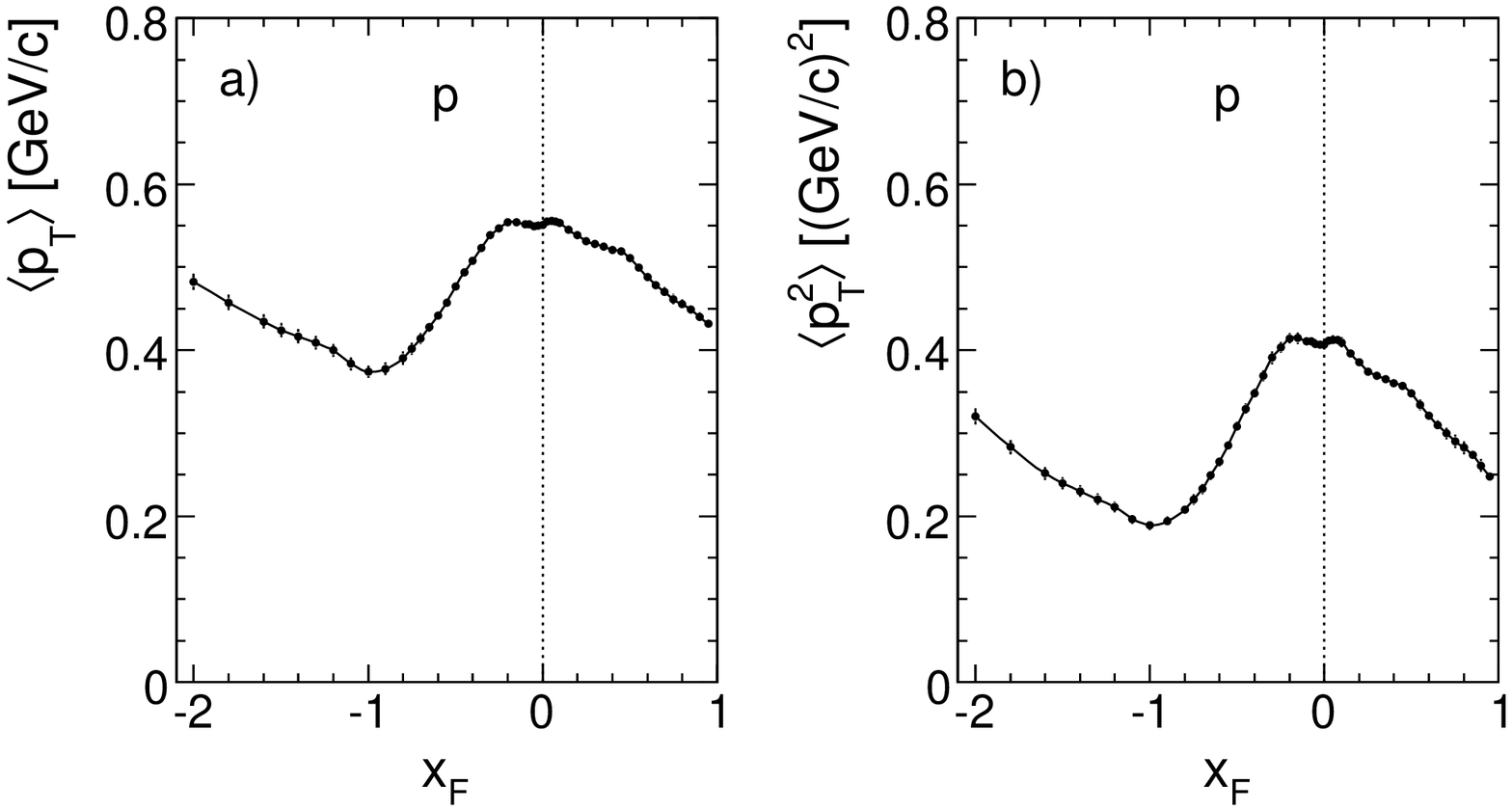}
 	\caption{a) mean $p_T$ and b) mean $p_T^2$ for protons as a function of
             $x_F$. Full lines: data interpolation}
  	 \label{fig:meanpt_prot}
  \end{center}
\end{figure}

\begin{figure}[h]
  \begin{center}
  	\includegraphics[width=11.cm]{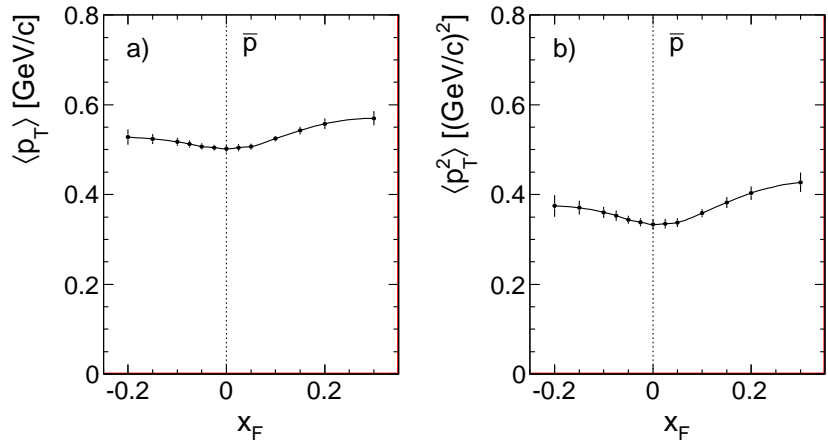}
 	\caption{a) mean $p_T$ and b) mean $p_T^2$ for anti-protons as a function of
             $x_F$. Full lines: data interpolation}
  	 \label{fig:meanpt_aprot}
  \end{center}
\end{figure}

The ratio of $p_T$ integrated $\overline{\textrm{p}}$ and p yields as a function of $x_F$ are shown in Fig.~\ref{fig:ratint}. The first and second moments $\langle p_T \rangle$ and $\langle p_T^2 \rangle$ are presented
as a function of $x_F$ in Fig.~\ref{fig:meanpt_prot} for protons and in Fig.~\ref{fig:meanpt_aprot}
for anti-protons.

%     Fig.54

%     Fig.55

%     Fig.56

%
% ****************************** Section 10.2 ****************************
%
\subsection{Centrality dependence}
\vspace{3mm}
\label{sec:ngrey}

The detection of "grey" protons in the centrality detector of NA49 \cite{nim}
allows the study of particle densities as a function of the
number $n_{\textrm{grey}}$ of protons in the lab momentum range of approximately
0.15 to 1.2~GeV/c in the backward hemisphere. The number distribution $dN/dn_{\textrm{grey}}$ has been shown
in \cite{pc_pion} to be a steep function of $n_{\textrm{grey}}$ with only 20\% and 5\% of
all events at $n_{\textrm{grey}}$~=~1 and 2, respectively. The determination of
double differential cross sections is therefore not feasible in
this experiment. As already shown in \cite{pc_pion} for pions, the extraction
of $p_T$ integrated yields as a function of $x_F$ by fitting the $dE/dx$ 
distributions over the complete range of $p_T$ is however feasible
in a limited range of $x_F$. This range is determined by the variation
of total momentum with $p_T$ and extends from $x_F$~=~-0.2 to +0.65 for 
protons, the upper limit being imposed by the progressive loss of 
acceptance at low $p_T$. Two samples with $n_{\textrm{grey}} \geq$~1 and 
$n_{\textrm{grey}} \geq$~2 have been selected. The resulting proton density 
distributions $dn/dx_F$ are given after correction in Table~\ref{tab:xf_ncd}, with the additional use of
the complete data sample (see also \cite{pc_pion}) which allows for a
precise control of eventual systematic effects. It should be realized that the
measurement of "grey" protons is confined to the region $x_F <$~-0.2 \cite{pc_pion}
and does therefore not interfere with the results presented in Table~\ref{tab:xf_ncd}.

%      Table 11 
\begin{table}[h]
\footnotesize
\renewcommand{\tabcolsep}{0.4pc} % enlarge column spacing
\renewcommand{\arraystretch}{1.0} % enlarge line spacing
\begin{center}
\begin{tabular}{|r@{}l|cc|cc|cc|}
\hline
\multicolumn{2}{|c|}{$x_F$}& $(dn/dx_F)^{\textrm{all}}$& $\Delta$& $(dn/dx_F)^{n_{\textrm{grey}}\geq 1}$& $\Delta$& 
                              $(dn/dx_F)^{n_{\textrm{grey}}\geq 2}$& $\Delta$ \\ \hline
 -0.&2  &  0.797 &  3.8 &  1.210 &  5.3 &  1.388 &  8.2 \\
 -0.&15 &  0.858 &  2.4 &  1.225 &  5.4 &  1.461 &  8.8 \\
 -0.&1  &  0.827 &  3.3 &  1.230 &  5.2 &  1.401 & 10.1 \\
 -0.&05 &  0.852 &  3.2 &  1.247 &  4.6 &  1.373 &  8.9 \\
  0.&0  &  0.923 &  3.0 &  1.167 &  2.7 &  1.285 &  6.0 \\
  0.&05 &  0.867 &  1.5 &  1.143 &  2.4 &  1.201 &  4.6 \\
  0.&1  &  0.774 &  1.4 &  0.975 &  2.2 &  1.055 &  4.0 \\
  0.&15 &  0.725 &  1.4 &  0.824 &  2.4 &  0.903 &  4.7 \\
  0.&2  &  0.678 &  1.5 &  0.800 &  2.4 &  0.838 &  4.8 \\
  0.&25 &  0.682 &  1.5 &  0.758 &  2.5 &  0.738 &  5.0 \\
  0.&3  &  0.636 &  1.6 &  0.649 &  2.7 &  0.615 &  5.5 \\
  0.&35 &  0.594 &  1.5 &  0.597 &  2.9 &  0.541 &  6.1 \\
  0.&4  &  0.579 &  1.6 &  0.532 &  3.0 &  0.519 &  6.2 \\
  0.&45 &  0.510 &  1.6 &  0.462 &  3.2 &  0.426 &  6.8 \\
  0.&5  &  0.478 &  1.7 &  0.451 &  3.3 &  0.424 &  7.2 \\
  0.&55 &  0.441 &  1.8 &  0.384 &  3.7 &  0.360 &  7.5 \\
  0.&6  &  0.423 &  2.8 &  0.388 &  5.7 &  0.261 & 13.7 \\
  0.&65 &  0.383 &  3.4 &  0.312 &  6.8 &  0.298 & 13.4 \\ \hline
\end{tabular}
\end{center}
\caption{$dn/dx_F$ distributions for protons as a function of $x_F$ for
         the complete data sample and the selections $n_{\textrm{grey}} \geq$~1 and
         $n_{\textrm{grey}} \geq$~2. The relative statistical errors are given in \%.
         The systematic errors correspond to Table~\ref{tab:int_prot}}
\label{tab:xf_ncd}
\end{table}

\begin{figure}[h]
  \begin{center}
  	\includegraphics[width=12cm]{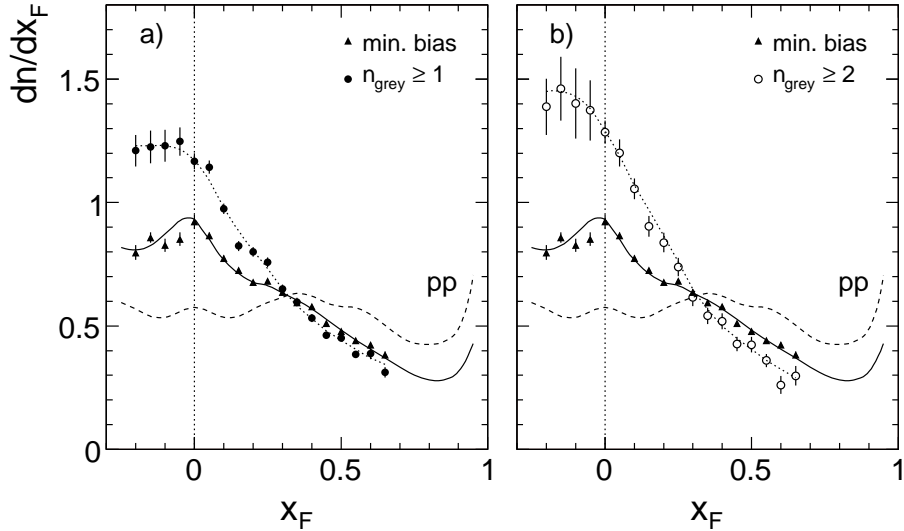}
 	\caption{$dn/dx_F$ for protons as a function of $x_F$ for the 
             complete data sample and for the conditions a) $n_{\textrm{grey}} \geq$~1 and
             b) $n_{\textrm{grey}} \geq$~2. The broken line gives the density in p+p collisions
             \cite{pp_proton}, the full line the integration of the interpolated minimum bias p+C data.
             The full and open circles give the measured yield with the indicated $n_{\textrm{grey}}$
             selection, the dotted lines are plotted to guide the eye}
  	 \label{fig:xf_ncd}
  \end{center}
\end{figure}

As shown in Fig.~\ref{fig:xf_ncd}, the results for the total data sample are
compatible within errors with the integration of the data
interpolation, Table~\ref{tab:int_prot} and Fig.~\ref{fig:ptint_prot}.

%      Fig.57

As seen from Fig.~\ref{fig:xf_ncd}, a systematic and smooth variation of the
proton density distributions is evident when passing from
p+p to minimum bias p+C and to centrality enhanced p+C 
interactions. A well defined cross-over with equal
number density is visible at $x_F$~=~0.25--0.3. Above this value
the densities decrease progressively with increasing number
of projectile collisions by up to a factor of 0.5 at the
experimentally accessible limit of $x_F$~=~0.6. Below $x_F$~=~0.3
this trend is inverted with a density increase of up to
factors of 2.5 at negative $x_F$. 

\begin{figure}[h]
  \begin{center}
  	\includegraphics[width=12cm]{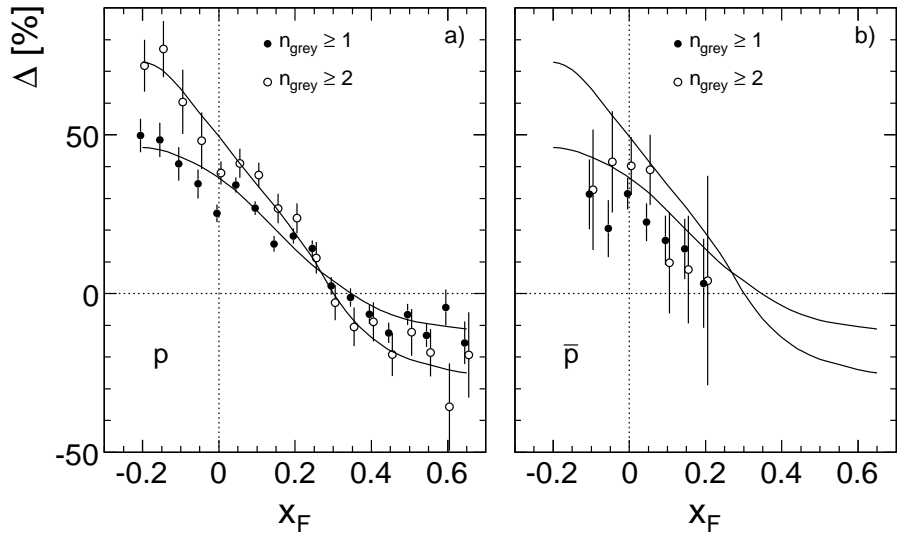}
 	\caption{Relative difference $\Delta$ in \% between $n_{\textrm{grey}}$ selected
             and minimum bias samples as a function of $x_F$ for a) proton and 
             b) anti-protons. The lines are shown to guide the eye. The lines in panel b)
             repeat those from panel a)}
  	 \label{fig:xf_ncd_dif}
  \end{center}
\end{figure}

A similar behaviour is also seen for anti-protons, although here
only a reduced range in $x_F$ is accessible due to the limiting
statistical uncertainties. This is shown in Fig.~\ref{fig:xf_ncd_dif} where the
relative difference in baryon density between the $n_{\textrm{grey}}$ selected
and minimum bias samples is given in \% as a function of $x_F$
for protons (panel a) and anti-protons (panel b). 

%      Fig.58

The interplay between baryon number transfer from the projectile
hemisphere and pile-up of baryons from the target fragmentation,
including isospin effects, will be discussed in more detail in Sect.12
of this paper.

%
% ****************************** Section 10.3 ****************************
%
\subsection{Neutron data}
\vspace{3mm}
\label{sec:neut}

The detection of forward neutrons in the Ring Calorimeter of
NA49 \cite{nim} has been introduced and described in detail in \cite{pp_proton}.
The neutron analysis concerning the separation of electromagnetic
and hadronic deposits, the veto against charged particles, the
calorimeter calibration and the energy resolution unfolding
may be directly applied to the p+C interactions. The $x_F/p_T$ acceptance of the calorimeter
is shown in Fig.~\ref{fig:accept}. For the total neutron yields at $x_F \leq$~0.3
an extrapolation beyond the available $p_T$ window has been performed using corresponding
proton $p_T$ distributions, see also \cite{pp_proton} for this procedure.
This extrapolation concerns $p_T$ values in the ranges 0.8~$< p_T <$~2~GeV/c 
at $x_F$~=~0.1 up to 1.6~$< p_T <$~2~GeV/c at $x_F$~=~0.3.
The corrections for feed-down from weak decays are shown
in Fig.~\ref{fig:fd}. The empty target, trigger
bias and re-interaction corrections are equal to the ones for
protons \cite{pp_proton}. The contributions from $K^0_L$ decay and anti-neutron
production are obtained from the charged kaon data which are 
available from NA49 \cite{pc_kaon} and from the isospin argumentation
using the anti-proton data explained in \cite{pp_proton}.

The resulting neutron densities $dn/dx_F$ are listed in Table~\ref{tab:neut}
together with their ratio to the results from p+p collisions \cite{pp_proton}.

%       Table 12
\begin{table}[h]
\footnotesize
\renewcommand{\tabcolsep}{0.4pc} % enlarge column spacing
\renewcommand{\arraystretch}{1.0} % enlarge line spacing
\begin{center}
\begin{tabular}{|r@{}l|cc|cc|}
\hline
\multicolumn{2}{|c|}{$x_F$}& $dn/dx_F$& $\Delta$& $(dn/dx_F)^{\textrm{pC}}/(dn/dx_F)^{\textrm{pp}}$& $\Delta$ \\ \hline
 0.&1  &  0.621 &  21.4 &   1.29 &  5.0 \\
 0.&2  &  0.482 &  15.1 &   1.18 &  3.2 \\
 0.&3  &  0.389 &  13.6 &   1.03 &  3.1 \\
 0.&4  &  0.303 &  12.7 &   0.93 &  3.2 \\
 0.&5  &  0.268 &  12.8 &   0.83 &  3.6 \\
 0.&6  &  0.221 &  11.0 &   0.75 &  4.1 \\
 0.&75 &  0.194 &  11.7 &   0.68 &  5.2 \\
 0.&9  &  0.128 &  29.1 &   0.59 &  8.4 \\ \hline
\end{tabular}
\end{center}
\caption{$p_T$ integrated neutron density distribution
         $dn/dx_F$ as a function of $x_F$ and the ratio of neutron
         densities in p+C and p+p interactions. The relative errors
         are given in \%. They are governed by the systematic 
         uncertainties quoted in Table~\ref{tab:syst}}
\label{tab:neut}
\end{table}

\begin{figure}[h]
  \begin{center}
  	\includegraphics[width=12cm]{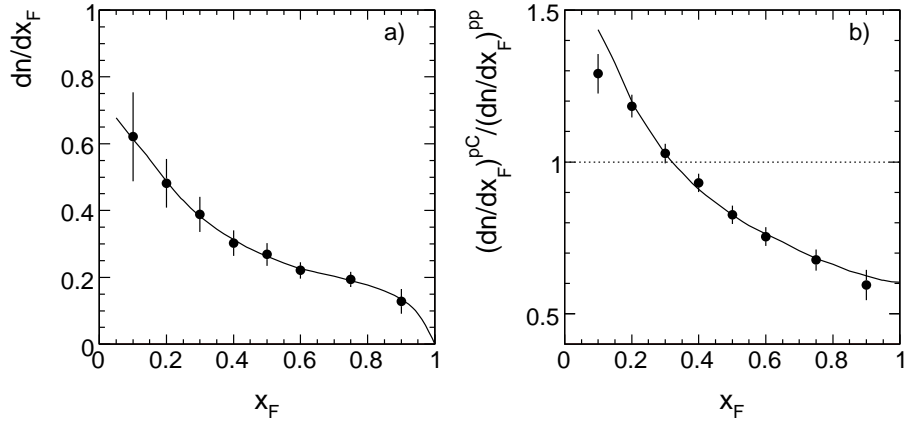}
 	\caption{a) $p_T$ integrated neutron density distribution
             $dn/dx_F$. Full line: data interpolation; b) neutron density ratio between p+C and p+p
             interactions as a function of $x_F$. The full line shows
             the corresponding density ratio for protons}
  	 \label{fig:ptint_neut}
  \end{center}
\end{figure}

The $p_T$ integrated neutron density distribution $dn/dx_F$ as a 
function of $x_F$ is presented in Fig.~\ref{fig:ptint_neut}a together with a data interpolation (full line).
Due to the absence of charge exchange processes in hadronic interactions at SPS energy \cite{pc_survey}
this interpolation is constrained to density zero at $x_F$~=~1.
Of particular interest in this context is the ratio between
neutron densities in p+C and p+p interactions in comparison
to the same ratio for protons as shown in Fig.~\ref{fig:ptint_neut}b.

%       Fig.59

Evidently protons and neutrons show
within errors the same behaviour when passing from p+p
to p+C collisions. The density ratio is equal to 1 at
$x_F$~=~0.3, increasing to 1.3 at $x_F$~=~0.1 and decreasing to
0.6 at $x_F$ towards +1. The latter value corresponds to
the expected fraction of single projectile collisions
in p+C interactions derived from the nuclear density
distribution \cite{pc_discus}. 
%
% ****************************** Section 10.4 ****************************
%
\subsection{Comparison to other data}
\vspace{3mm}
\label{sec:neut_comp}

Two data sets are available for comparison with the NA49 integrated
data in the SPS energy range. The first set from the ACCMOR
collaboration \cite{accmor} gives the density distribution $dn/dx_F$ of protons 
for p+A collisions from Be to U nuclei at 120~GeV/c beam momentum. 
A second, very recent publication from the MIPP collaboration \cite{mipp}
provides neutron densities $dn/dx_F$ for p+C interactions again at
120~GeV/c beam momentum at the FERMILAB Main Injector.

%
% ****************************** Section 10.4.1 ****************************
%
\subsubsection{ACCMOR data \cite{accmor}}
\vspace{3mm}
\label{sec:accmor}

\begin{figure}[h]
  \begin{center}
  	\includegraphics[width=10cm]{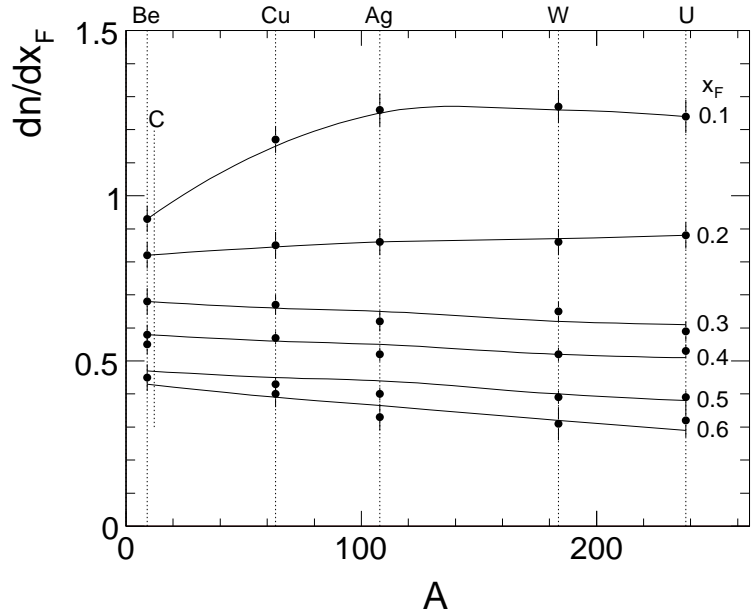}
 	\caption{Proton density $dn/dx_F$ as a function of mass number $A$
             for p+Be, p+Cu, p+Ag, p+W and p+U nuclei for different values of
             $x_F$. The lines describe the data interpolation used
             for the determination of the corresponding p+C cross
             sections}
  	 \label{fig:ptint_accmor}
  \end{center}
\end{figure}

The ACCMOR collaboration \cite{accmor} measured proton densities $dn/dx_F$ with
a proton beam at 120~GeV/c momentum at the CERN SPS for five
different nuclei (Be, Cu, Ag, W, U) in the $x_F$ range from 0.07 to 0.6.
These data may be interpolated from Be to C. This is demonstrated 
in Fig.~\ref{fig:ptint_accmor} which gives the proton density as a function 
of mass number $A$ for five $x_F$ values from 0.1 to 0.6.

%       Fig.60 

\begin{figure}[h]
  \begin{center}
  	\includegraphics[width=8.cm]{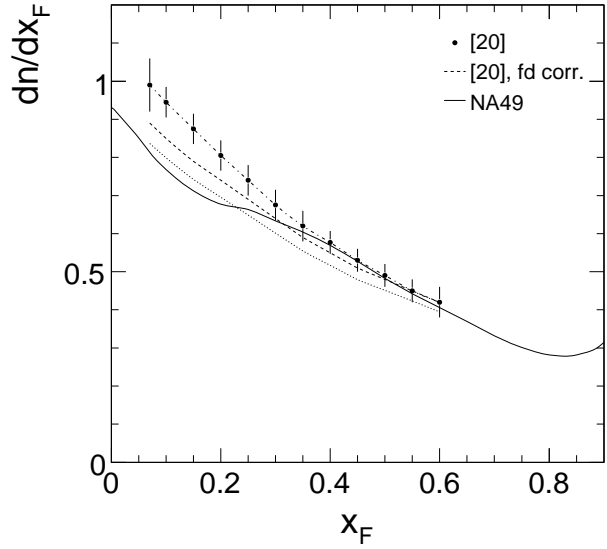}
 	\caption{Proton density $dn/dx_F$ as a function of $x_F$
             interpolated to p+C from the ACCMOR data. The dash-dotted
             line through the ACCMOR results is drawn to guide the
             eye, the full line shows the NA49 result. The broken
             line gives the ACCMOR result after a tentative subtraction 
             of proton feed-down from weak decays. The dotted line corresponds
             to the modification of the total inelastic cross section for
             p+Be collisions with 6\%, see Fig.~\ref{fig:inel_cs}. The dash-dotted line 
             through the ACCMOR results is drawn to guide the eye}
  	 \label{fig:ptint_comp}
  \end{center}
\end{figure}

The data interpolation of $dn/dx_F$ to p+C interactions is shown 
in Fig.~\ref{fig:ptint_comp} as a function of $x_F$ in comparison to the NA49 results,
Table~\ref{tab:int_prot} and Fig.~\ref{fig:ptint_prot} above.

%       Fig.61

\begin{figure}[h]
  \begin{center}
  	\includegraphics[width=7.5cm]{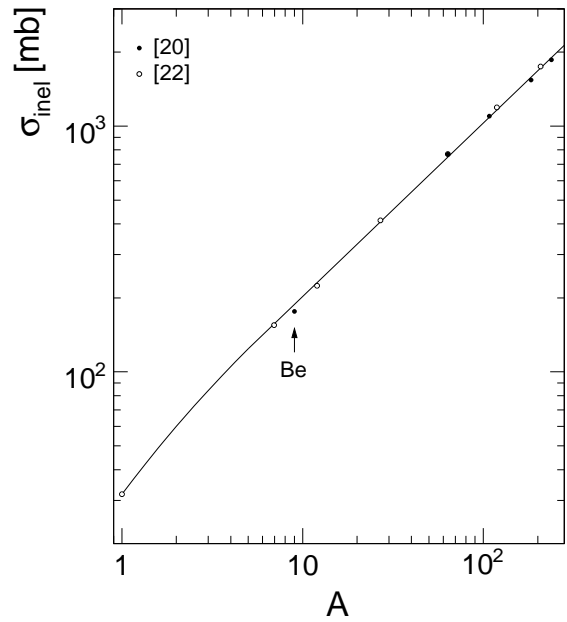}
 	\caption{Data summary of measured total inelastic cross sections of p+A 
 	         interactions as a function of $A$, \cite{accmor,carroll}}
  	 \label{fig:inel_cs}
  \end{center}
\end{figure}

The ACCMOR data show an upward deviation which increases from
about 0 at $x_F$~=~0.6 to about 20\% at the lower data limit of $x_F$~=~0.07.
This deviation looks similar to the feed-down correction for
protons from weak hyperon decays used for the NA49 data, Sect.~\ref{sec:fd_cor},
which reaches 15\% at small $x_F$ and decreases rapidly to 0
at $x_F \sim$~0.7. As such a correction is not mentioned in \cite{accmor}
and as the geometrical layout of the experiment is similar to 
the one of NA49 as far as the position of the tracking elements
is concerned ($\sim$~3~m distance between target and first tracking
station) it has been tentatively assumed that the on-vertex
reconstruction efficiency of decay products might be similar
in both experiments. Subtracting the feed-down correction from
NA49 results in the broken line in Fig.~\ref{fig:ptint_prot}. This 
line is about 8\% above the NA49 results for $x_F <$~0.4. At this point a look at the 
total inelastic cross sections as a function of $A$ from different 
references \cite{accmor,carroll}, Fig.~\ref{fig:inel_cs}, shows that 
the cross section for p+Be given in \cite{accmor} falls low by about 
6\% compared to the interpolation of other available data \cite{carroll}. 
A corresponding correction of the cross section is shown as dotted line in 
Fig.~\ref{fig:ptint_comp}

%       Fig.62

In conclusion it may be stated that the ACCMOR results are
compatible with the NA49 results within the quoted systematic
uncertainties.

%
% ****************************** Section 10.4.2 ****************************
%
\subsubsection{MIPP data \cite{mipp}}
\vspace{3mm}
\label{sec:mipp}

\begin{figure}[h]
  \begin{center}
  	\includegraphics[width=11.5cm]{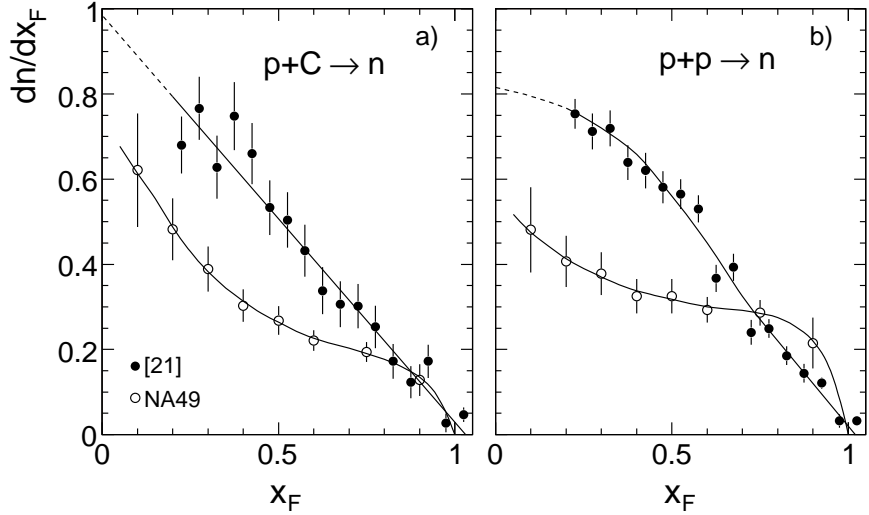}
 	\caption{Comparison of MIPP neutron densities $dn/dx_F$
             to the NA49 results as a function of $x_F$, a) for p+C
             interactions and b) for p+p interactions. Hand interpolations
             through the respective data points are shown as full lines}
  	 \label{fig:mipp_comp}
  \end{center}
\end{figure}

Very recently new data on neutron densities $dn/dx_F$ have become
available from the MIPP collaboration at the FERMILAB Main 
Injector for p+p and p+A interactions \cite{mipp}. The MIPP neutron 
densities $dn/dx_F$ for p+C interactions at 120~GeV/c beam momentum 
are compared to the NA49 results in Fig.~\ref{fig:mipp_comp}a, the densities
for p+p collisions at 84~GeV/c beam momentum in Fig.~\ref{fig:mipp_comp}b.

%       Fig.63

\begin{figure}[h]
  \begin{center}
  	\includegraphics[width=11.5cm]{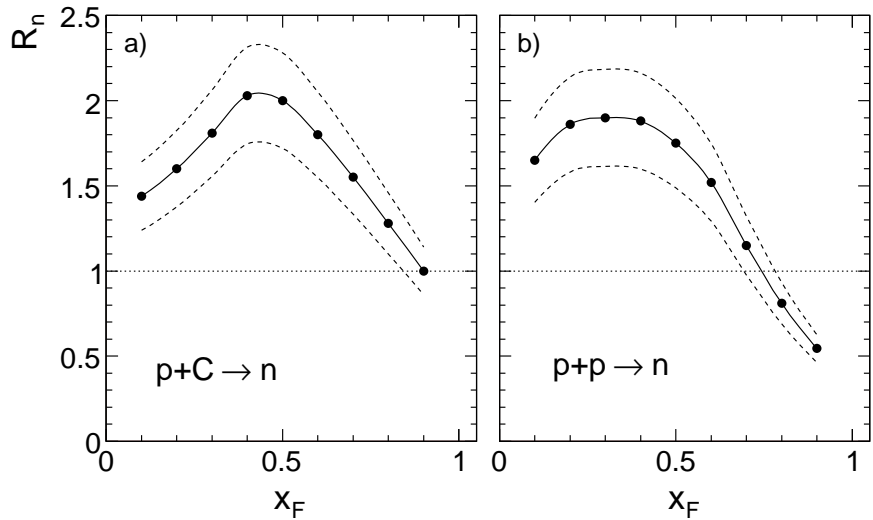}
 	\caption{Density ratio $R_n = (dn/dx_F)^{\textrm{MIPP}}/(dn/dx_F)^{\textrm{NA49}}$
             as a function of $x_F$, a) for p+C, b) for p+p interactions.
             The error bands shown are based on the statistical 
             uncertainties of the data points}
  	 \label{fig:mipp_rat}
  \end{center}
\end{figure}

Important deviations are visible between the two experiments both
for p+C and p+p collisions. This is quantified in Fig.~\ref{fig:mipp_rat} where
the ratio

\begin{equation}
   R_n = (dn/dx_F)^{\textrm{MIPP}}/(dn/dx_F)^{\textrm{NA49}}
\end{equation}
between the two respective interpolations is shown as a function of $x_F$.

%       Fig.64  

The density ratios show a remarkable similarity between p+C
and p+p interactions, varying systematically from 1.4--1.6
at the lower $x_F$ limit to a maximum of 1.9--2.0 at $x_F \sim$~0.4
and decreasing to 0.5--1.0 at the upper kinematic limit.
The total neutron yields in the forward hemisphere obtained
by an extrapolation of the measured densities towards $x_F$~=~0
as shown in Fig.~\ref{fig:mipp_comp} deviate by about 60\% for both reactions,
with the integrated MIPP neutron yield resulting in 1 neutron
per event for p+p collisions at 84~GeV/c beam momentum.

\begin{figure}[h]
  \begin{center}
  	\includegraphics[width=8cm]{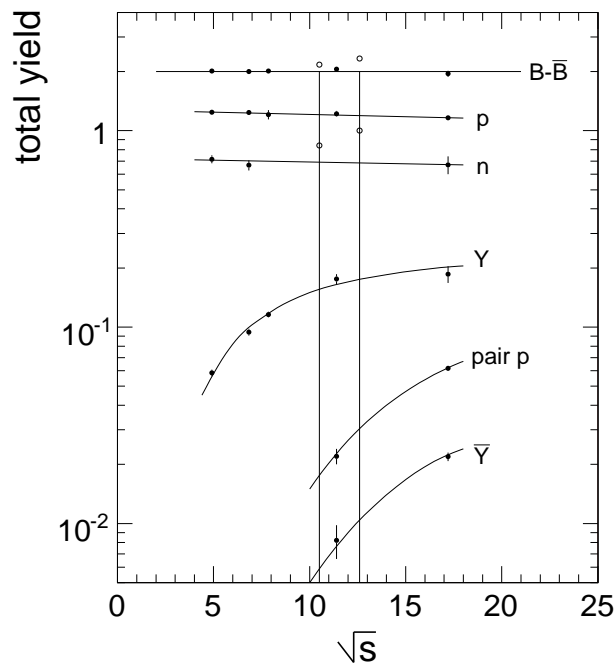}
 	\caption{Total yields of protons, neutrons, hyperons,
             pair produced protons and anti-hyperons, \cite{blobel,blobel1,zabrodin,bogo,ammosov,ammosov1},
             as well as the total number of net baryons (B-$\overline{\textrm{B}}$) as a function
             of $\sqrt{s}$ between 4.9 and 17.2~GeV (full circles). The integrated
             neutron yields from MIPP \cite{mipp} at 58 and 84~GeV/c beam 
             momentum are indicated as open circles}
  	 \label{fig:bn}
  \end{center}
\end{figure}

These large deviations pose a problem for baryon number 
conservation which may be verified for p+p interactions using
published total yields for protons, neutrons, hyperons, pair 
produced protons and pair produced hyperons. The available 
data are plotted in Fig.~\ref{fig:bn} as a function of $\sqrt{s}$ from
\cite{blobel,blobel1} at 12 and 24~GeV/c beam momentum, \cite{zabrodin,bogo}
at 32~GeV/c beam momentum, \cite{ammosov,ammosov1,abramov} at 69~GeV/c 
beam momentum and from NA49. 

%       Fig.65

The large neutron yields published by MIPP are clearly out of 
proportion with respect to the available data and are incompatible
with baryon number conservation.

%
% ****************************** Section 11 ****************************
%
\section{A phenomenological study of baryon and anti-baryon production in p+C interactions 
         over the complete phase space}
\vspace{3mm}
\label{sec:two_intro}

In the following sections a detailed phenomenological
analysis of the experimental results presented above
will be carried out. In this context it may be useful
to recall some basic ingredients of p+A interactions and their relation to the elementary nucleon-nucleon
collisions. The following short subsections should
be regarded as a recollection of known facts as well
as an introduction to the more detailed and quantitative
studies described in the subsequent sections.

%
% ****************************** Section 11.1 ****************************
%
\subsection{The three components contributing to p+A collisions}
\vspace{3mm}
\label{sec:two_intro1}

There is today no doubt that the final state of p+A
collisions is built up from three basic components:

\begin{itemize}
 \item The fragmentation of the projectile particle
 \item The fragmentation of those target nucleons which are
       hit by the projectile
 \item The nuclear component which is created by the momentum
       transfer from the projectile via the hit target nucleons
       into the nucleus.
\end{itemize}

It is important to realize that different time scales
are acting in the hadronization of the mentioned components:
the fragmentation time scales are short with respect to
the nuclear cascading processes. In consequence, as will
be shown in detail below, only those secondaries from the
target fragmentation which have small momenta in the target rest frame are involved in the cascading process.

\begin{figure}[h]
  \begin{center}
        \includegraphics[width=10cm]{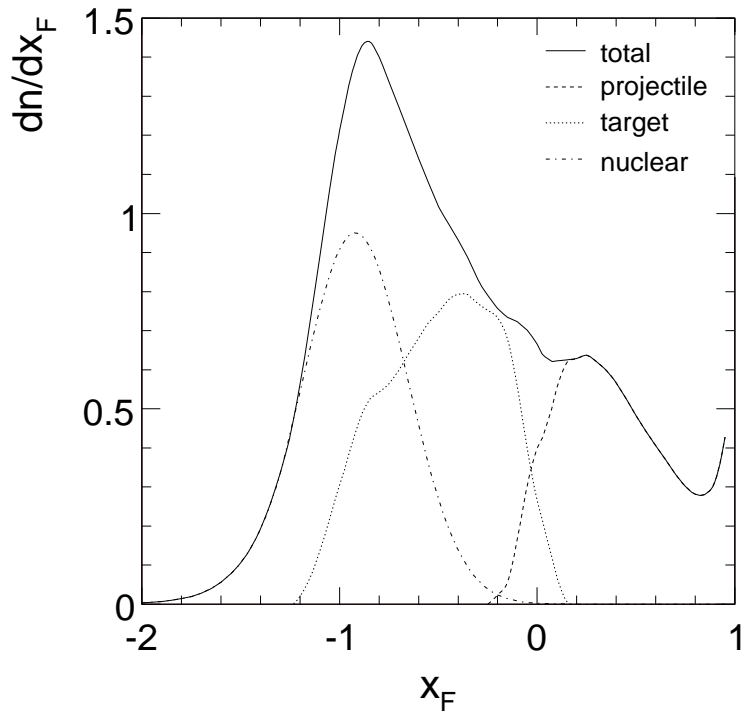}
        \caption{Nuclear, target and projectile components for protons in p+C collisions}
        \label{fig:components}
  \end{center}
\end{figure}

A schematic view of this situation is shown in Fig.~\ref{fig:components} for
final state protons which gives, in addition to the total
measured proton yield (full line), its separation into the three basic components of p+A interactions: 
the projectile fragmentation (broken line), target fragmentation (dotted
line), and nuclear component (dash-dotted line).

%    Fig.66

The separation and extraction of the three components will be described below in a largely model-independent way that essentially only relies on experimental input. The term "model-independent" will be explained and become clear in the course of the following argumentation. The asymmetric character of the p+A interactions is in this context to be seen as an asset and not as a complication, as it permits -  in contrast to the symmetric A+A collisions - a
straight-forward, parameter-free experimental separation of the components with only one single assumption concerning the target contribution.

%
% ****************************** Section 11.2 ****************************
%
\subsection{Intra-nuclear collisions}
\vspace{3mm}
\label{sec:two_intro2}

The projectile undergoes subsequent collisions on its way through
the target nucleus. The number of these collisions is given
by the variable $\nu$ with mean value $\langle \nu \rangle$. In the past, quite a
number of assumptions have been made about this multiple
collision process, ranging from the splitting-up of the
projectile into quarks upon the first collision to the constancy 
of the total interaction cross section through
the subsequent collisions. In the latter hypothesis, the
variable $\nu$ is calculable on an event-by-event basis using
the measured density distribution of nucleons inside the
nucleus \cite{pc_discus}. Its mean value is then also accessible from 
the total inelastic p+A interaction cross section in its relation 
to the p+p cross section \cite{pc_discus},

\begin{equation}
  \langle \nu \rangle = A \frac{\sigma(\textrm{pp})}{\sigma(\textrm{pA})} 
\end{equation}

In this case, it stands to reason to assume the hadronization
of the hit target nucleons to be equivalent to the hadronization
of the elementary nucleon-nucleon interaction, of course up to
intra-nuclear cascading of the low momentum secondaries evoked
above, and taking full account of isospin symmetry. This hypothesis allows 
the straight-forward prediction of particle densities in 
the target fragmentation region (Fig.~\ref{fig:components}) which is directly amenable 
to experimental verification. It is the only assumption made in 
the following analysis and it will be tested quantitatively using
the available experimental results both from p+p \cite{pp_proton} and p+C
interactions.

%
% ****************************** Section 11.3 ****************************
%
\subsection{Net baryons and baryon/anti-baryon pairs}
\vspace{3mm}
\label{sec:two_intro3}

Final state baryon production appears in two categories: "net"
baryons which are linked to the presence of a projectile baryon
and have to obey baryon number conservation, and 
baryon/anti-baryon pairs. Experimentally the "net" baryon yield
may be defined by the difference between the total yield of
a baryon species and the corresponding yield of pair-produced
baryons of the same species. Whereas the sum of all net baryon
yields is constrained by baryon number conservation, there
is no limit to the production of baryon/anti-baryon pairs
which rises sharply from the threshold at about $\sqrt{s}$~=~6~GeV 
through the SPS energy range and flattens towards
collider energies in a way characteristic of the production
of heavy hadronic systems.

In the case of protons, it has to be realized that
the baryon/anti-baryon yield is not only given by 
proton/anti-proton pair production, but has also additional
components containing neutrons and anti-neutrons. In the
following, protons and neutrons which are produced as
baryon/anti-baryon pairs are therefore denoted by:

\begin{equation}
\begin{split}              
   \textrm{pair produced protons}  &= \widetilde{\textrm{p}}\\
   \textrm{pair produced neutrons} &= \widetilde{\textrm{n}}
\end{split}              
\end{equation}

In general these pairs form isospin triplets with the
isospin-3 components shown in Table~\ref{tab:iso}.

%   Table 13
\begin{table}[h]
  \begin{center}
    \begin{tabular}{cccc}
    \hline
  $I_3$   &        -1         &             0              &        1                \\  \hline
 \multirow{2}{25mm}{baryon pairs}  
    &   $\overline{\textrm{p}}\,\widetilde{\textrm{n}}$  &  $\overline{\textrm{p}}\,\widetilde{\textrm{p}}$  & 
        $\overline{\textrm{n}}\,\widetilde{\textrm{p}}$    \\
    &                                                  &  $\overline{\textrm{n}}\,\widetilde{\textrm{n}}$  &         \\ \hline 
	\end{tabular}
  \end{center}
  \caption{Isospin structure of baryon/anti-baryon pair production}
  \label{tab:iso}
\end{table}

This phenomenology has been studied by NA49 using the comparison
of p+p and n+p interactions \cite{iso}. Here it has been shown that the
anti-proton yield increases by about a factor of 1.6 when
exchanging the $I_3$~=~+1/2 against an $I_3$~=~-1/2 projectile, thereby
enhancing the $I_3$~=~-1 against the $I_3$~=~+1 combination in Table~\ref{tab:iso}.

As the central $\overline{\textrm{p}}$/p ratio reaches about 23\% for minimum bias
p+C interactions, the definition of net protons needs a careful
analysis of the corresponding isospin factors, as shown below.

%
% ****************************** Section 11.4 ****************************
%
\subsection{Overlap between the components of fragmentation}
\vspace{3mm}
\label{sec:two_intro4}

In the separation of the different components of final state
hadronization characterized for example by the target and
the projectile hemispheres, the eventual overlap between these
hemispheres, Fig.~\ref{fig:components}, plays an important role. For any quantitative work the exact extent and shape of the respective overlap functions has to be known and it will be shown below that this is indeed possible in a completely model-independent way, using basic conservation rules. In the case of final state pions this has been achieved in ref. \cite{pc_discus} using charge conservation and isospin symmetry, using also $\pi$+p interactions. For net protons, baryon number conservation may be invoked, and for anti-protons the internal isospin structure of the fragmentation process has been used. In both cases this is discussed in detail
below.

Several well-known experimental findings present sharp constraints
to the phenomenon of component overlap, in particular at SPS energy:

\begin{itemize}
 \item The absence, at SPS energy, of charge and flavour exchange
 \item The absence of long-range correlations between the target and
       projectile hemispheres, for $|x_F| >$~0.2
 \item The independence of the target fragmentation on the nature of
       the projectile particle and vice versa (hadronic factorization)
 \item The presence of short-range forward-backward multiplicity
       correlations at $|x_F| <$~0.2
\end{itemize}

Within the experimental uncertainties, a common overlap function
which is limited to 0.2 units of $x_F$ has been measured for net protons
and anti-protons.

%
% ****************************** Section 11.5 ****************************
%
\subsection{Extraction of the projectile and nuclear components}
\vspace{3mm}
\label{sec:two_intro5}

Given the hypothesis for the target fragmentation described above,
and given the fact that this hypothesis may be tested against
the experimental results, the projectile and nuclear components
follow, without additional assumptions, from the subtraction of
the target fragmentation from the measured total proton or
anti-proton yields. For the protons it is useful to extract
the net proton projectile component as in this case a direct cross-check of baryon number conservation becomes possible. The extraction will be performed both for the $p_T$ integrated and for the double-differential yields. With respect to the
interplay of the target fragmentation and nuclear components
in the far backward direction, the modification and extent in $x_F$
of the diffractive target component will be determined.

%
% ****************************** Section 11.6 ****************************
%
\subsection{Error estimation}
\vspace{3mm}
\label{sec:two_intro6}

The present analysis makes use of experimental results both from
p+p and from p+C interactions. In both cases the results from
the two-dimensional data interpolation and of its integration
over transverse momentum are exploited. With mean statistical errors 
of 8\% (12\%) in p+C and 5\% (11\%) in p+p for protons and
anti-protons, respectively, the local fluctuations of the data
interpolation are about a factor of 3 lower. For the $p_T$
integration, the statistical errors have an upper limit of
order 1\%.The systematic uncertainties are estimated to
3.7\% (4.5\%) in p+C and 2.5\% (3.3\%) in p+p, where a part of
these quantities is of common origin. In the data comparison
between the two reactions, a systematic error of about 5\% is therefore estimated.

%
% ****************************** Section 12 ****************************
%
\section{Two component mechanism of baryon and baryon pair production}
\vspace{3mm}
\label{sec:two_comp}

In preparation of the separation of the different components
contributing to the overall baryon and anti-baryon yields in p+C
interactions, a model-independent study of the target and projectile components in p+p collisions will be performed in this section. This study allows the extraction of the baryonic overlap functions introduced in Sect.~\ref{sec:two_intro4} above. The section
will also contain a detailed comparison to the results from
a microscopic simulation code.

A large sample of 4.8 million inelastic events is available from
NA49 \cite{pp_proton,pp_pion,pp_kaon} with both proton/anti-proton 
and neutron identification in the final state. This allows for the selection of sub-samples
of events with defined net baryon number either in the projectile
or in the target hemisphere by tagging baryons at sufficiently large $|x_F|$. Sufficiently large means 
in this context $x_F$ values where the yield of pair produced baryons is low enough to ensure
negligible background. For protons this condition is fulfilled
for $|x_F| >$~0.35 where the $\overline{\textrm{p}}$/p ratio is less than 0.5\% \cite{pp_proton}. The following
ranges of $x_F$ have been used for net proton and neutron selection:

\begin{align}
  \textrm{Projectile hemisphere:}  & \quad \textrm{protons}   & 0.35 <  &x_F < 0.5  \label{eq:fp}\\
                                   & \quad \textrm{neutrons}  & 0.5 <   &x_F < 0.7  \label{eq:fn}  \\
  \textrm{Target hemisphere:}      & \quad \textrm{protons}   & -0.75 < &x_F < -0.6 \label{eq:bp} 
\end{align}

These ranges are given by the constraints of acceptance and proton identification via 
$dE/dx$, see Sect.~\ref{sec:pid}, and by the limited energy resolution of the hadron calorimetry 
for neutrons. In the following the measured baryonic double and single differential densities
obtained with net baryon constraint,

\begin{equation}
\begin{split}
  \rho^c(x_F,p_T) &= \frac{d^2n}{dx_Fdp_T} \\
  \rho^c_{\textrm{int}}(x_F) &= \frac{dn}{dx_F}\quad,
\end{split}
\end{equation}
will be described by their ratio to the corresponding inclusive yields,

\begin{equation}
\begin{split}
  &\rho^i(x_F,p_T)\\
  &\rho^i_{\textrm{int}}(x_F)
\end{split}
\end{equation}
by

\begin{equation}              
\begin{split}              
   R^c(x_F,p_T) &= \frac{\rho^c(x_F,p_T)}{\rho^i(x_F,p_T)} \\
   R^c_{\textrm{int}}(x_F)  &= \frac{\rho^c_{\textrm{int}}(x_F)}{\rho^i_{\textrm{int}}(x_F)}
\end{split} 
\label{eq:rat_corr}             
\end{equation}

The ratio $R^c_{\textrm{int}}(x_F)$ is obtained by direct yield extraction over the complete $p_T$ range 
from 0 to 1.9~GeV/c, while $R^c(x_F,p_T)$ extracts the ratios in distinct
$x_F/p_T$ bins. This allows for the study of eventual $p_T$
dependences. The double differential cross sections are obtained, due to limited statistics, 
in the range 0.1~$< p_T <$~0.7~GeV/c, and the resulting ratios are averaged over 
this $p_T$ range resulting in the ratios:

\begin{equation}              
   R^c_{\textrm{av}}(x_F) = \langle R^c(x_F,p_T)\rangle \quad.
   \label{eq:rat_av}
\end{equation}

In a first sub-section the production of anti-protons in these
samples will be studied, as the determination of the net proton
yields has to rely on the estimation of the cross sections of
pair-produced protons.

%
% ****************************** Section 12.1 ****************************
%
\subsection{Anti-protons with final-state net baryon constraint}
\vspace{3mm}
\label{sec:two_comp_aprot}

\begin{figure}[!h]
  \begin{center}
  	\includegraphics[width=7.5cm]{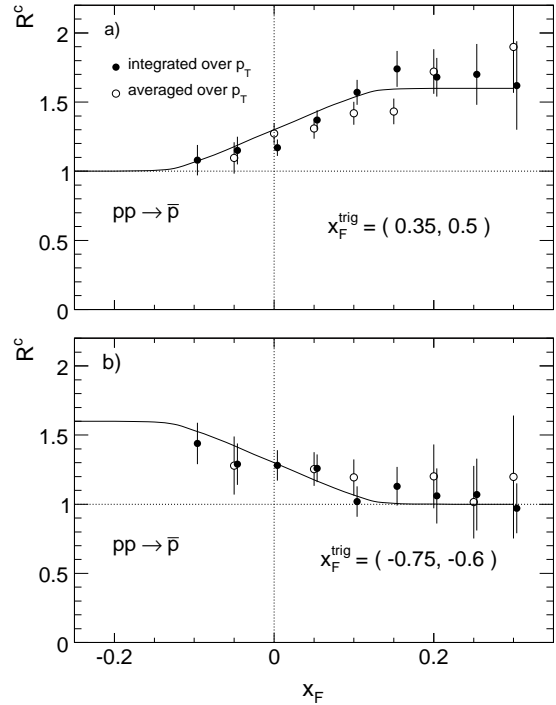}
 	\caption{Yield ratios $R^c_{\textrm{int}}(x_F)$ and $R^c_{\textrm{av}}(x_F)$
             for anti-protons for a) forward and b) backward
             proton selection, as a function of $x_F$. The full lines are
             forward-backward mirror symmetric and represent the re-normalized
             overlap function shown in Fig.~\ref{fig:overlapf} and Table~\ref{tab:overlap_proj}. 
             For definition of $R^c_{\textrm{int}}(x_F)$ and $R^c_{\textrm{av}}(x_F)$ see equations
             (\ref{eq:rat_corr}) and (\ref{eq:rat_av})}
  	 \label{fig:arat}
  \end{center}
\end{figure}

The ratios $R^c_{\textrm{av}}(x_F)$ and $R^c_{\textrm{int}}(x_F)$ for anti-protons
are presented in Fig.~\ref{fig:arat} for forward (\ref{eq:fp}) and backward 
(\ref{eq:bp}) proton selection.

%      Fig.67 

\begin{figure}[h]
  \begin{center}
  	\includegraphics[width=7.cm]{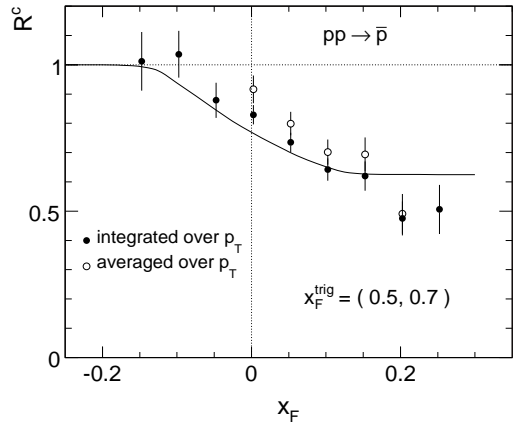}
 	\caption{Yield ratios $R^c_{\textrm{int}}(x_F)$ and $R^c_{\textrm{av}}(x_F)$
             for anti-protons for forward neutron tagging, as a function
             of $x_F$. The full line represents the prediction using the
             extracted overlap function, Fig.~\ref{fig:arat}. For definition
             of $R^c_{\textrm{int}}(x_F)$ and $R^c_{\textrm{av}}(x_F)$ see equations
             (\ref{eq:rat_corr}) and (\ref{eq:rat_av})}
  	 \label{fig:arat_neut}
  \end{center}
\end{figure}

For both net proton selections, a distinct correlation between the
anti-proton yields and the presence of a tagged proton in the
respective hemisphere is evident. The yield ratios reach values
of 1.6 in the far backward and forward directions, respectively.
The excess over 1 is halved at $x_F$~=~0 and the ratio goes to 1 in the respective opposite $x_F$ regions.
It has been verified that this correlation
is not induced by asymmetries in the azimuthal acceptance of the
detector. It is therefore due to an isospin effect as expected
from the iso-triplet nature of baryon pair production, Table~\ref{tab:iso}. Indeed
the presence of a net proton will unbalance the isospin structure
present in the inclusive event sample towards the $I_3$~=~-1 component
of the corresponding heavy mesonic state and thereby enhance the
$\overline{\textrm{p}}\,\widetilde{\textrm{n}}$ yield in the selected hemisphere. 
From this argument follows a strict prediction for the tagging with net neutrons:
in this case the $I_3$~=~+1 component, hence the $\overline{\textrm{n}}\,\widetilde{\textrm{p}}$ 
combination, should be favoured, and the anti-proton yield should be reduced
accordingly with respect to the inclusive sample. This is indeed
the case as shown in Fig.~\ref{fig:arat_neut} for forward neutron tagging, see (\ref{eq:fn}).

%      Fig.68 

This finding has two interesting consequences: it allows for the first
time the establishment of the range and shape of the anti-proton overlap 
function in a completely model-independent way. Secondly it has direct 
consequences for the possible production mechanism of baryon pairs in hadronic interactions.

\begin{figure}[h]
  \begin{center}
  	\includegraphics[width=7.cm]{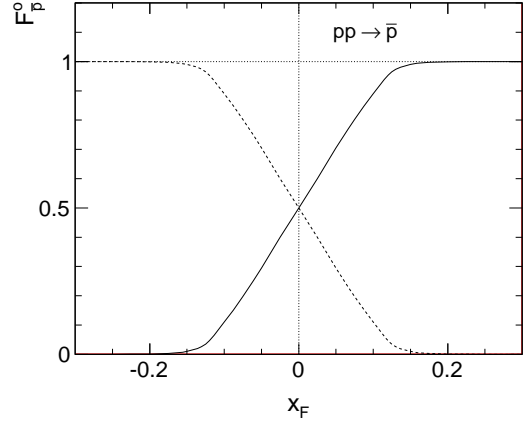}
 	\caption{Anti-proton overlap function $F^o_{\overline{\textrm{p}}}$ as a
             function of $x_F$ for projectile fragmentation (full line) and
             target fragmentation (broken line)}
  	 \label{fig:overlapf}
  \end{center}
\end{figure}

The hand interpolated lines given in Figs.~\ref{fig:arat} and \ref{fig:arat_neut} may be
redefined as the anti-proton overlap function $F^o_{\overline{\textrm{p}}}$ which is normalized in the
ordinate to the range 0~$\leq F^o_{\overline{\textrm{p}}} \leq$~1, as shown in Fig.~\ref{fig:overlapf}
and tabulated in Table~\ref{tab:overlap_proj}. This function is strongly constrained by the forward-backward
symmetry in p+p interactions. It has to pass through 0.5 at $x_F$~=~0 and it should reach
the values of 0 or 1 at $|x_F| \sim$~0.2 as imposed by the absence of long range hadronic correlations
beyond this limit \cite{pc_discus}. In addition, the
constraint $F^o_{\overline{\textrm{p}}}(+x_F) = 1 - F^o_{\overline{\textrm{p}}}(-x_F)$ has to be fulfilled.
This leaves, regarding the experimental error bars, in particular
those for the net proton yields (Fig.~\ref{fig:net_rat}) resulting in the same
$x_F$ dependence, an error margin of about 5\% for the intermediate
$x_F$ values.

A remark concerning the choice of the longitudinal variable might
be in place here. Throughout this paper the Feynman $x_F$ variable
has been chosen to describe the longitudinal momentum dependences.
This choice has several reasons:

\begin{itemize}
 \item $x_F$ describes, at least to first order, the observed scaling of the
       hadronic cross section with cms energy.
 \item It is orthogonal with respect to the transverse momentum.
 \item The range of the forward-backward correlations, as for example
       the forward-backward multiplicity correlations \cite{pc_discus} has been
       shown to be $s$-independent in the $x_F$ variable.
 \item If changing from $x_F$ to the rapidity variable $y$, the energy
       scaling of the results as well as the orthogonality to $p_T$ is
       lost. As shown in detail in ref. \cite{pc_discus} different rapidity limits
       for each interaction energy have to be introduced to describe
       the multiplicity correlations. In addition and of course,
       seeing rapidity as basically an angular variable, strong
       correlations with $p_T$ of all physics phenomena depending
       on longitudinal momentum are introduced.
\end{itemize}

%      Fig.69  

%  Table 14
\begin{table}[h]
  \begin{center}
    \begin{tabular}{r@{}lr@{}l||r@{}lr@{}l}
    \hline
    \multicolumn{2}{c}{$x_F$}  &  \multicolumn{2}{c||}{$F^o_{\overline{\textrm{p}}}$} &  
    \multicolumn{2}{c}{$x_F$}  &  \multicolumn{2}{c}{$F^o_{\overline{\textrm{p}}}$}  \\ \hline
    -0.&25       &           0.&0                  &   0.&025      &         0.&6                     \\
    -0.&225      &           0.&0003               &   0.&05       &         0.&705                   \\
    -0.&2        &           0.&0008               &   0.&075      &         0.&803                   \\
    -0.&175      &           0.&0027               &   0.&1        &         0.&890                   \\
    -0.&15       &           0.&010                &   0.&125      &         0.&965                   \\
    -0.&125      &           0.&035                &   0.&15       &         0.&990                   \\
    -0.&1        &           0.&110                &   0.&175      &         0.&9973                  \\
    -0.&075      &           0.&197                &   0.&2        &         0.&9992                  \\
    -0.&05       &           0.&295                &   0.&225      &         0.&9997                  \\
    -0.&025      &           0.&4                  &   0.&25       &         1.&0                     \\
    -0.&0        &           0.&5                  &     &         &           &                      \\
    \hline
	\end{tabular}
  \end{center}
  \caption{Anti-proton overlap function for projectile fragmentation}
  \label{tab:overlap_proj}
\end{table}

Concerning the production mechanism of baryon pairs, the strong isospin correlation 
both with the final state net baryons and with the hadronic projectile in the initial 
state speaks against the central
production from the quark-gluon sea, as for instance gluon fusion.
Concerning current hadronization models using string fragmentation,
the ad-hoc introduction of diquark systems in the colliding baryons
is necessary in order to describe the net baryon production. In
addition quark/anti-quark and diquark/anti-diquark pickup processes
have to be introduced for baryon pair production, in close resemblance to 
the description of hadronization in e$^+$+e$^-$ annihilation, with a multitude of 
adjustable parameters and doubtful predictive power, see Sect.~\ref{sec:jam}.

On the other hand the presence of heavy, high spin mesonic states in the early 
stage of hadronization of the highly excited baryonic systems in the p+p interaction 
presents a natural explanation for the observed correlations. In fact most observed 
heavy meson resonances have a $\overline{\textrm{p}}\,\widetilde{\textrm{p}}$
decay branching fraction, or have been discovered in the inverse $\overline{\textrm{p}}$+p 
annihilation process. It is interesting to note that baryon pair production, 
via the high effective mass involved, probes a rather primordial phase of hadronization as compared 
to the lighter final state hadrons. This again favours the observed strong isospin correlations. 
In this context it should be recalled that a high mass mesonic origin of baryon pairs was indeed 
proposed as early as 35 years ago by Bourquin and Gaillard \cite{bourquin} in order to
describe the observed inclusive yields.

%
% ****************************** Section 12.2 ****************************
%
\subsection{Pair-produced protons}
\vspace{3mm}
\label{sec:two_comp_pair_prot}

\begin{figure}[h]
  \begin{center}
  	\includegraphics[width=8cm]{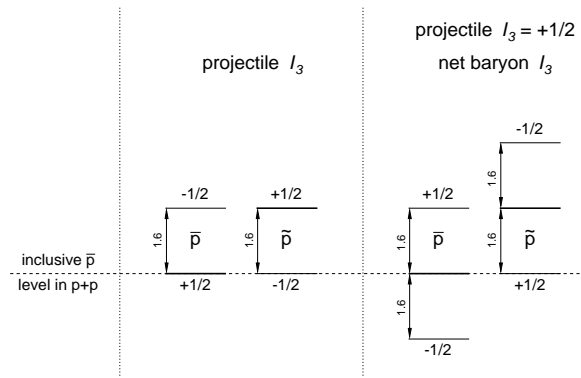}
 	\caption{Yield levels of anti-protons ($\overline{\textrm{p}}$) and pair produced
             protons ($\widetilde{\textrm{p}}$) with respect to the inclusive 
             $\overline{\textrm{p}}$ yield in p+p interactions, left: for a change 
             of projectile $I_3$ from +1/2 to -1/2, right: for a change of the trigger 
             net baryon $I_3$ from +1/2 to -1/2}
  	 \label{fig:iso_scheme}
  \end{center}
\end{figure}

Following the isospin structure of baryon pair production, see Table~\ref{tab:iso},
anti-protons ($\overline{\textrm{p}}$) and pair produced protons ($\widetilde{\textrm{p}}$) 
are coupled in their yields by isospin symmetry with the $I_3$ component of either 
the projectile or the trigger net baryon in the respective hemisphere. This is shown 
schematically in Fig.~\ref{fig:iso_scheme} where the $\overline{\textrm{p}}$ and $\widetilde{\textrm{p}}$ 
yields are presented with respect to the inclusive anti-proton level in p+p interactions.

%      Fig.70 

Evidently the definition of "net" protons, which needs the knowledge
of pair-produced proton yields, is linked to the measured $\overline{\textrm{p}}$ 
yields in a non-trivial fashion. As far as the overlap function between the
forward and backward hemispheres is concerned, it is of course the same
for $\overline{\textrm{p}}$ and $\widetilde{\textrm{p}}$ and has to be referred 
to the yield levels indicated in Fig.~\ref{fig:iso_scheme}.

%
% ****************************** Section 12.3 ****************************
%
\subsection{Net proton feed-over}
\vspace{3mm}
\label{sec:two_comp_feedover}

With the above clarification of the yields of pair produced baryons,
the inclusive net proton yield may now be defined as:

\begin{equation}
 \rho^{\textrm{net}}_{\textrm{p}}(x_F,p_T) = \rho^{\textrm{incl}}_{\textrm{p}}(x_F,p_T) - \rho^{\textrm{incl}}_{\widetilde{\textrm{p}}}(x_F,p_T)
 \label{eq:netp}
\end{equation}

As $\rho^{\textrm{incl}}_{\widetilde{\textrm{p}}}$ is 1.6 times higher than 
$\rho^{\textrm{incl}}_{\overline{\textrm{p}}}$, Fig.~\ref{fig:iso_scheme}, this means a substantial 
decrease of the central net proton density with respect to the simple subtraction 
of the $\overline{\textrm{p}}$ yield. The consequence of this for 
the evolution of $\rho^{\textrm{net}}_{\textrm{p}}$ with $\sqrt{s}$ has been
demonstrated in \cite{iso} where it has been shown that the central net invariant
proton cross section approaches zero in the ISR energy range.

The net proton feed-over and the corresponding overlap function may be
determined, as for the anti-proton case Sect.~\ref{sec:two_comp_aprot} above, by fixing
a net proton in the projectile or target hemisphere following the
selection criteria, (\ref{eq:fp}) and (\ref{eq:bp}) above, respectively. This ensures the
absence of net protons in the corresponding hemisphere and results in the
constrained net proton density:

\begin{equation}
 \rho^{c,\textrm{net}}_{\textrm{p}}(x_F,p_T) = \rho^c_{\textrm{p}}(x_F,p_T) - \rho^c_{\widetilde{\textrm{p}}}(x_F,p_T)
\end{equation}

These density distributions may be normalized by dividing by the inclusive
yield, (\ref{eq:netp}), resulting in the ratio

\begin{equation}
  R^{c,\textrm{net}}_{\textrm{p}}(x_F,p_T) = \frac{\rho^{c,\textrm{net}}_{\textrm{p}}(x_F,p_T)}{\rho^{\textrm{net}}_{\textrm{p}}(x_F,p_T)}
  \label{eq:ratp}
\end{equation}

\begin{figure}[h]
  \begin{center}
  	\includegraphics[width=9.8cm]{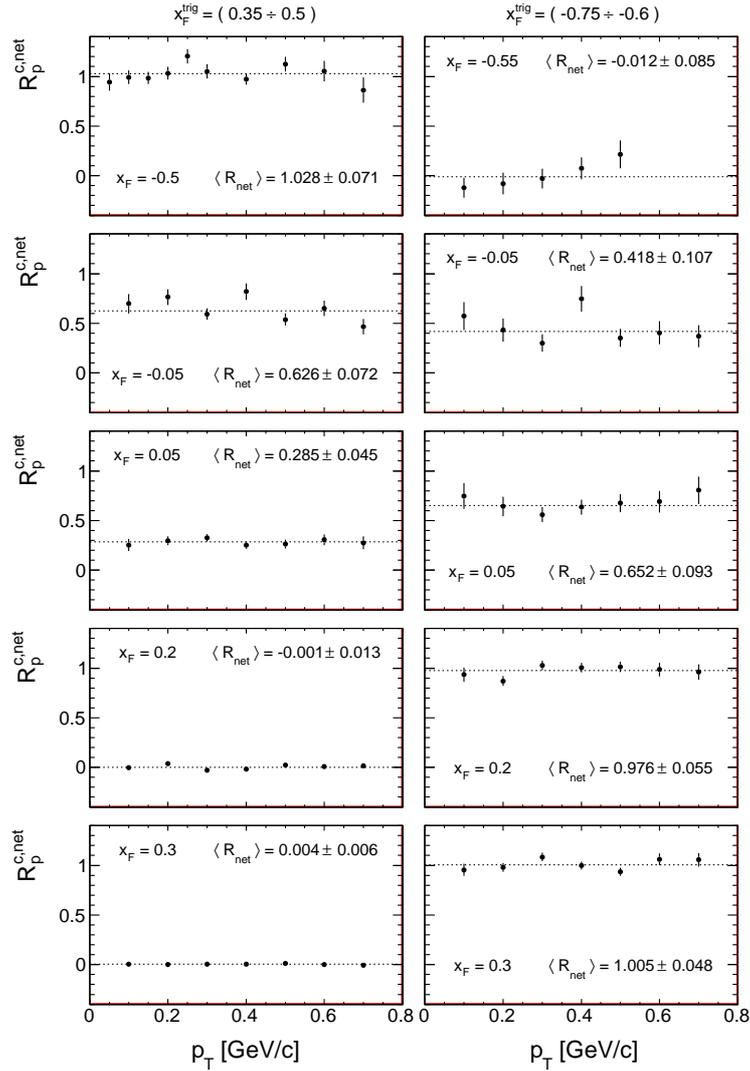}
 	\caption{Ratio $R^{c,\textrm{net}}_{\textrm{p}}(x_F,p_T)$ between constrained and 
 	         inclusive net proton densities as a function of $p_T$ at different values of
             $x_F$ for net proton trigger in the projectile hemisphere (left side panels) and 
             for trigger in the target hemisphere (right side panels)}
  	 \label{fig:net_rat_pt}
  \end{center}
\end{figure}

In a first instance, it can be shown that $R^{c,\textrm{net}}_{\textrm{p}}(x_F,p_T)$ is
independent on $p_T$ in the range 0.1~$< p_T <$~0.7 GeV/c, see Fig.~\ref{fig:net_rat_pt}

%      Fig.71 

\begin{figure}[h]
  \begin{center}
  	\includegraphics[width=6.5cm]{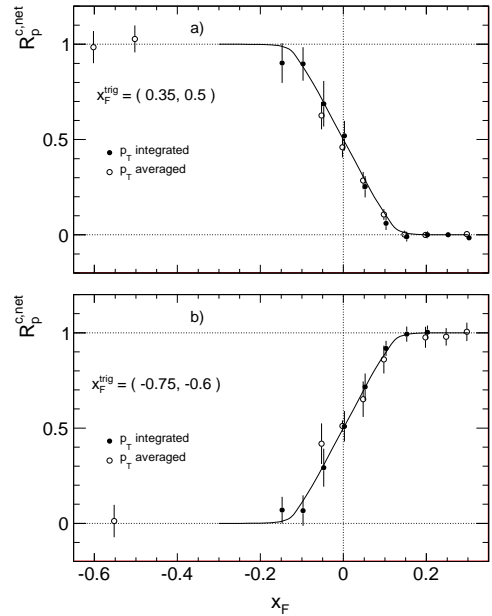}
 	\caption{$p_T$ integrated and $p_T$ averaged constrained net proton
             density ratios $R^{c,\textrm{net}}_{\textrm{p}}$ as a function of $x_F$, 
             a) for forward net proton selection and
             b) for backward net proton selection. The full lines shown represent
             the overlap functions presented in Fig.~\ref{fig:overlapf}}
  	 \label{fig:net_rat}
  \end{center}
\end{figure}

The $x_F$ dependence for the $p_T$ integrated ratio $R^{c,\textrm{net}}_{\textrm{p}}(x_F)$ and
the $p_T$ averaged ratio $\langle R^{c,\textrm{net}}_{\textrm{p}}(x_F,p_T) \rangle$ are shown 
in Fig.~\ref{fig:net_rat}. The full lines in this Figure represent the same overlap function
as extracted for anti-protons, Fig.~\ref{fig:overlapf}. In addition and due
to the higher statistics available for protons, the independence of
the overlap function on transverse momentum has been demonstrated, Fig.~\ref{fig:net_rat_pt}. 
This gives an interesting connection to the origin
of proton yields from the decay of heavy resonances, which
predicts $p_T$ independence in contrast to pions as decay products
of the same resonances, see ref. \cite{pc_discus}. 

%    Fig.72

%
% ****************************** Section 12.4 ****************************
%
\subsection{A remark concerning resonance decay}
\vspace{3mm}
\label{sec:res_decay}

The model-independent extraction of the baryonic feed-over and the
experimental determination of the corresponding overlap functions, as
described in the preceding sections, may be extended to resonance
decay. As in fact most if not all final state baryons stem from resonance
decay it is of interest to investigate the consequences of the hadronic
two-component mechanism observed for the final state baryons also for the
parent generation of resonances. This will be demonstrated below using
as examples the well measured $\Delta^{++}$ resonance for the proton and 
a mesonic resonance approximated by measured tensor meson characteristics
for the anti-proton feed-over.

%
% ****************************** Section 12.4.1 ****************************
%
\subsubsection{Protons from $\mathbf {\Delta^{++}}$ decay}
\vspace{3mm}
\label{sec:res_decay_prot}

\begin{figure}[h]
  \begin{center}
  	\includegraphics[width=11.5cm]{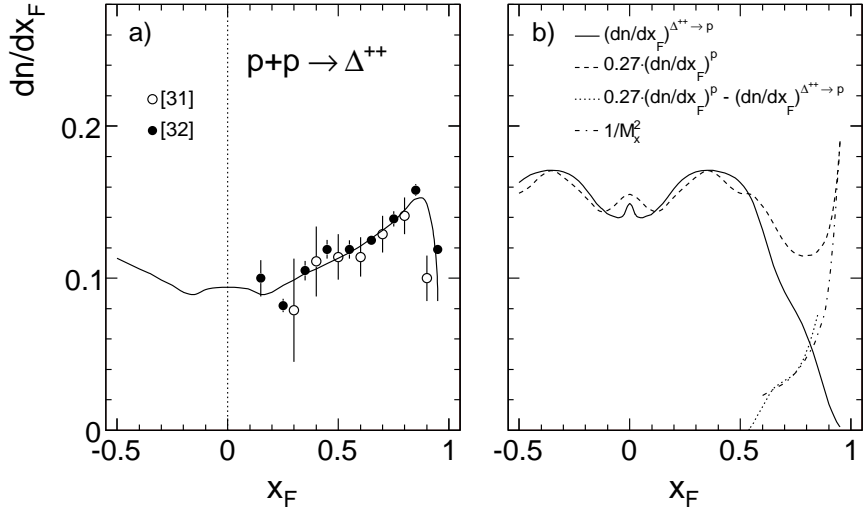}
 	\caption{Hand-interpolated, $p_T$ integrated density distributions $dn/dx_F$ as a function
             of $x_F$ a) for $\Delta^{++}$(1232), b) for the decay protons from 
             $\Delta^{++} \rightarrow \textrm{p} + \pi^+$ (full lines). The broken line in panel b gives the total
             inclusive proton yield multiplied by a factor 0.27, the dotted line
             the difference between the two distributions. Also indicated in panel
             b is the $1/M_x^2$ distribution (see text) as dash-dotted line}
  	 \label{fig:delta_xfdist}
  \end{center}
\end{figure}

The $p_T$ integrated density distribution $dn/dx_F(x_F)$ of the $\Delta^{++}$(1232)
resonance has been rather precisely measured by a number of experiments
in the SPS and ISR energy ranges as shown by two examples \cite{breakstone,aguilar} in
Fig.~\ref{fig:delta_xfdist}a. In both cases the full $p_T$ range from 0 to 2~GeV/c 
was available. The integration of the interpolated full line yields a total
inclusive cross section of 7~mb. The corresponding decay proton distribution from 
$\Delta^{++} \rightarrow \textrm{p} + \pi^+$ (100\% branching fraction) is presented
in Fig.~\ref{fig:delta_xfdist}b together with the measured total inclusive proton yield as
measured by NA49 \cite{pp_proton}, multiplied by the factor 0.27.

%       Fig.73 

This plot demonstrates several important features:

\begin{itemize}
 \item The shape of the decay proton distribution from $\Delta^{++}$ follows, for
       $|x_F| <$~0.6, very closely (to within about 5\%) the total inclusive proton
       distribution re-normalized by a factor 0.27. This means that 27\% of
       all protons in this $x_F$ range stem from $\Delta^{++}$ decay alone.
 \item The difference between the decay and inclusive distributions at $|x_F| >$~0 
       is well described by a $1/M_x^2$ form typical of single diffraction
       with $M_x^2 \sim s(1-x_F)$, dash-dotted line in panel b).
 \item Taking account of the other $\Delta$ states $\Delta^+$(1232) and $\Delta^0$(1232)
       with branching fractions of 66\% and 33\% into protons, respectively,
       this means that more than 40\% of all non-diffractive protons cascade
       down from $\Delta$ resonances
 \item The shape difference between the $dn/dx_F$ distributions of the mother
       resonance and the daughter protons exemplifies the very effective
       baryon number transfer towards the central $x_F$ region in resonance
       decay even for low-$Q$ resonances like the $\Delta$.
\end{itemize}

A number of further comments are due in this context. The absence of
charge and flavour exchange in inelastic p+p interactions at SPS
energy implies that the observed $\Delta$ resonances are not directly
produced at least in the primordial phase of target and projectile
excitation. They rather turn up as decay products of N$^*$ resonances.
This fact has been experimentally proven in a number of high precision
studies of single and double diffraction into p$\pi^+\pi^-$ states from PS to
ISR energies \cite{idschok,denegri,baksay,conta}. In these final states which cover an $|x_F|$
range from 1 to about 0.6 \cite{conta}, the initial $I$, $I_3$ state is 1/2, +1/2.
Although the sub-channels p$\pi^+$ and p$\pi^-$ show clear $\Delta$(1232) signals,
those are completely contained in the decay mass spectra of a series
of N$^*$ resonances (N$^*$(1440), N$^*$(1520), N$^*$(1680)). In the more central
area of hadronization, this clear isospin signature will become diluted
and contain 1/2, -1/2 states, see also the isospin correlations
discussed above. Nevertheless the production and decay of N$^*$ resonances
without intermediate $\Delta$ states will provide another important source
of final state nucleons.

\begin{figure}[h]
  \begin{center}
  	\includegraphics[width=11cm]{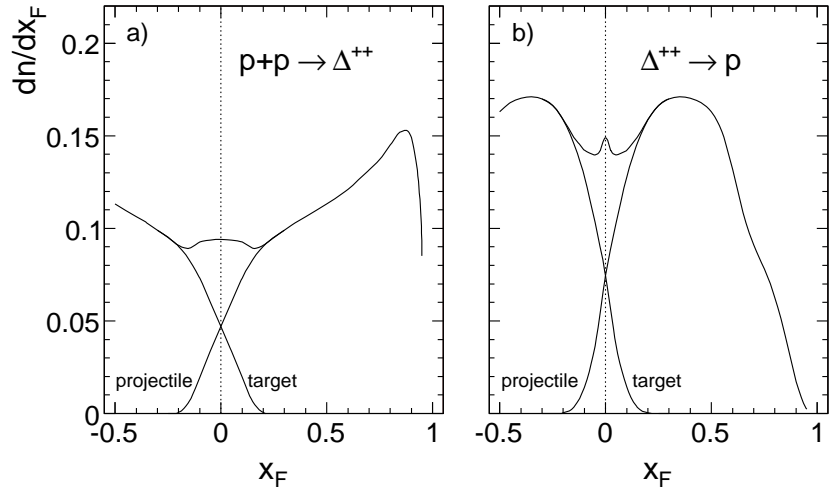}
 	\caption{Density distributions $dn/dx_F$  as a function of $x_F$
             a) for the target and projectile components of $\Delta^{++}$ and 
             b) for the decay protons}
  	 \label{fig:delta_prot_twocomp}
  \end{center}
\end{figure}

The two-component mechanism of hadronization should therefore also be
considered for baryon resonances with the constraint to reproduce
the measured overlap function for final state protons. This is
fulfilled for $\Delta^{++}$ by the separation of the production cross section
into a target and a projectile component as indicated in Fig.~\ref{fig:delta_prot_twocomp}a.
The corresponding $x_F$ distributions for the decay protons are given in
Fig.~\ref{fig:delta_prot_twocomp}b.

%       Fig.74 

\begin{figure}[h]
  \begin{center}
  	\includegraphics[width=11cm]{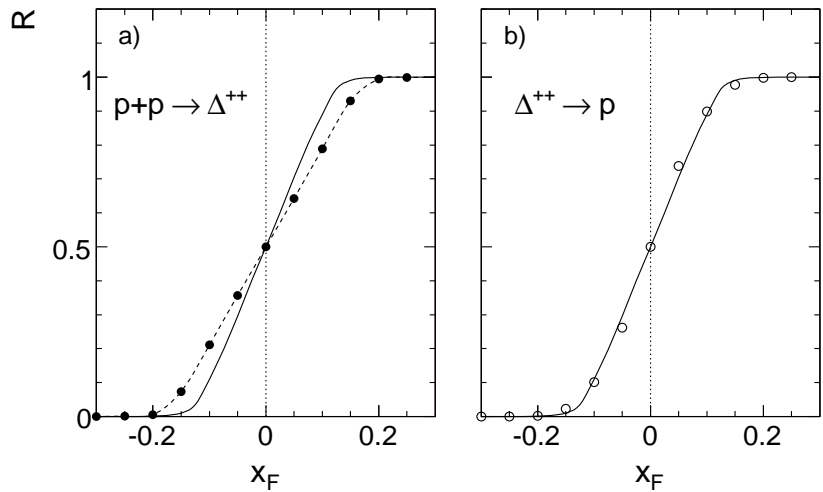}
 	\caption{Density ratios $R = (dn/dx_F)_{\textrm{proj}}/(dn/dx_F)^{\textrm{incl}}$
             a) for $\Delta^{++}$, full circles and dashed line, and b) for the
             decay protons from $\Delta^{++}$ decay, open circles, as a function of $x_F$.
             The measured overlap function, see Fig.~\ref{fig:net_rat}, is indicated as the full
             line in both panels}
  	 \label{fig:delta_prot_rat}
  \end{center}
\end{figure}

This choice results in the ratios $R = (dn/dx_F)_{\textrm{proj}}/(dn/dx_F)^{\textrm{incl}}$
between projectile component and inclusive distribution shown in Fig.~\ref{fig:delta_prot_rat}.

%       Fig.75 

Evidently the measured proton feed-over is precisely reproduced, Fig.~\ref{fig:delta_prot_rat}b,
if the overlap function of the mother resonance is chosen slightly
wider in $x_F$, Fig.~\ref{fig:delta_prot_rat}a, indicating a certain mass dependence in the $x_F$
scale. In this context the dependence on decay particle mass exhibited
by the much reduced feed-over for the decay pions from $\Delta^{++}$, as
elaborated in \cite{pc_discus}, should be recalled here.

%
% ****************************** Section 12.4.2 ****************************
%
\subsubsection{Anti-protons from heavy meson decay}
\vspace{3mm}
\label{sec:res_decay_aprot}

\begin{figure}[h]
  \begin{center}
  	\includegraphics[width=11cm]{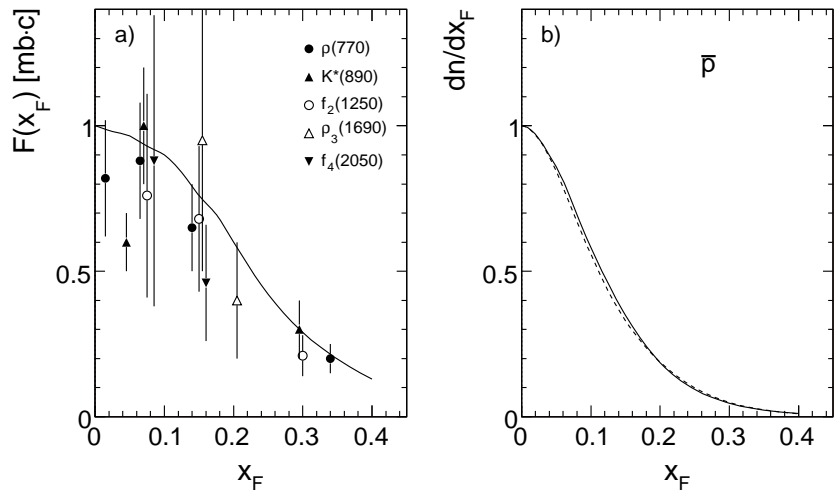}
 	\caption{a) Invariant $x_F$ distribution $F(x_F)$ of a mesonic state
             with 2.5~GeV mass (full line) compared to various vector and
             tensor mesons \cite{suzuki} normalized to unity at $x_F$~=~0, b) resulting
             inclusive density distribution $dn/dx_F$ for the decay anti-protons (full line)
             compared to the measured yields in p+p interactions \cite{pp_proton} (broken line)}
  	 \label{fig:tens_aprot}
  \end{center}
\end{figure}

A complementary approach may be used for the production of anti-protons
from heavy mesonic states \cite{bourquin}. In fact a sizeable number of states above
the p$\overline{\textrm{p}}$ mass threshold at 1.88~GeV/c$^2$ have been observed \cite{amsler}
both in hadronic interactions and in the inverse p$\overline{\textrm{p}}$ annihilation
into final state hadrons. While due to the large width and density of
these states their direct experimental detection in mass spectra
is difficult, the application of Partial Wave Analysis permits their
localization and determination of quantum numbers. In the following a
hypothetical state at 2.5~GeV mass with a Breit-Wigner width of
0.25~GeV and a two-body decay into p$\overline{\textrm{p}}$ is used in order to study
the corresponding daughter $x_F$ distributions and overlap functions.
An invariant $x_F$ distribution consistent with the one for vector and
tensor mesons published by Suzuki et al. \cite{suzuki} yields the inclusive
density distribution $dn/dx_F$ for the decay anti-protons shown in
Fig.~\ref{fig:tens_aprot}.

%      Here Fig.76

\begin{figure}[h]
  \begin{center}
  	\includegraphics[width=11cm]{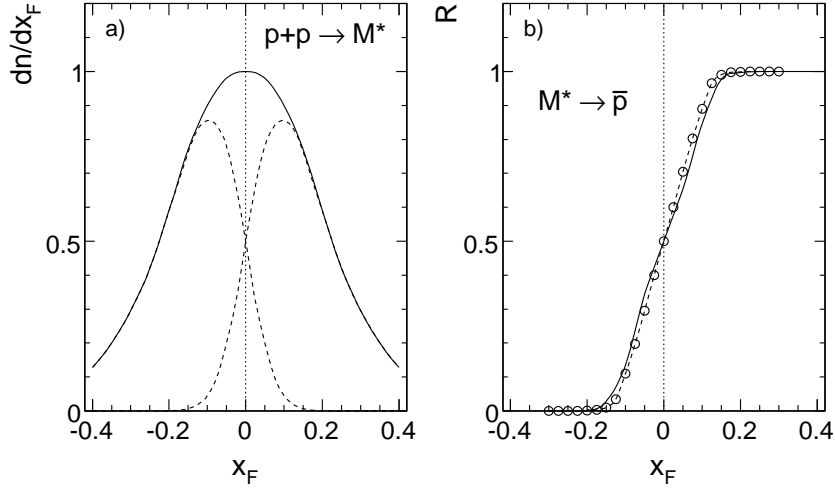}
 	\caption{a) Density distribution $dn/dx_F$ for the heavy mesonic
             state (full line) normalized to 1 at $x_F$~=~0 decomposed into a target
             and a projectile component (broken lines), b) the corresponding
             overlap function $R(x_F)$ (full line) superimposed with the measurement
             (open circles and broken line)}
  	 \label{fig:tens_aprot_rat}
  \end{center}
\end{figure}

Evidently this choice reproduces perfectly the measured inclusive
anti-proton yields measured by NA49 \cite{pp_proton}. Imposing the target-projectile
decomposition for the yield distribution of the heavy meson as shown in
Fig.~\ref{fig:tens_aprot_rat}a, the overlap function for the decay baryons reproduces
closely the measured feed-over for anti-protons, Fig.~\ref{fig:tens_aprot_rat}b.

%       Fig.77 

As far as the normalization of the resulting inclusive yields is
concerned, it may be stated that -- compared to the $\overline{\textrm{p}}$ cross section
of 1.2~mb \cite{pp_proton} -- the measurement of \cite{suzuki} indicates 1.6~mb for the f$_4$(2050)
state alone, and the mass dependence of the tensor meson production given
in \cite{drijard} an f$_4$(2050) cross section in excess of 5~mb at ISR energy.
There are, however, almost no measurements of the branching fraction
into baryon pairs, and if so, they vary by large factors, for the
f$_4$(2050) for instance from 50\% \cite{lamsa} to 0.2\% \cite{rozanska}. The amount to which
anti-protons cascade from heavy meson decay has therefore to stay
an open question.

%
% ****************************** Section 12.5 ****************************
%
\subsection{Comparison to a microscopic simulation code}
\vspace{3mm}
\label{sec:jam}

In the preceding NA49 publications concerning p+p and p+C interactions 
\cite{pc_pion,pc_discus,pp_proton,pp_pion,pp_kaon} a strictly model-independent 
approach to the interpretation of these extensive and precise data sets 
has been followed. Comparison to the multitude of existing
microscopic simulation codes has therefore been avoided in 
clear appreciation of the fact that the actual understanding
(or, better, lack of understanding) of the theoretical
foundations of the soft sector of QCD calls for improved 
experimental information rather than ad hoc parametrizations.
Given the detailed data concerning baryons both on the inclusive
level and on the level of baryonic correlations contained
in \cite{pp_proton} and in the above discussion, it might however be useful to 
confront the obtained results with one specific microscopic
simulation in order to obtain some idea of the precision and
predictive power reached in such approaches.   
 
The relatively recent code named JAM \cite{jam} has been chosen for 
this comparison. This code is supposed to describe hadronization
in elementary and nuclear interactions over the complete cms energy
range from 1~GeV up to collider energies. It uses, in the SPS energy
range, soft string excitation following the HIJING \cite{hijing} approach which in
turn relies on the string fragmentation mechanism developed in the 
PYTHIA/LUND \cite{pythia} environment.

%
% ****************************** Section 12.5.1 ****************************
%
\subsubsection{Inclusive baryon and anti-baryon density distributions $dn/dx_F$}
\vspace{3mm}
\label{sec:jam_incl}

\begin{figure}[h]
  \begin{center}
  	\includegraphics[width=10cm]{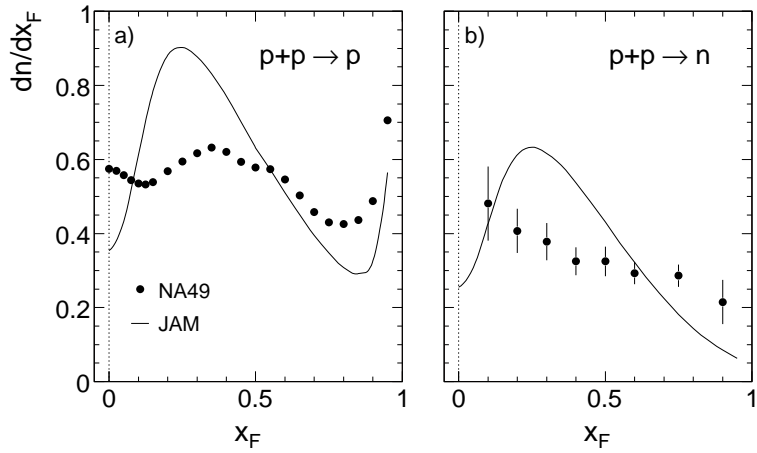}
 	\caption{Density distributions $dn/dx_F$ as functions of $x_F$
             a) for protons and b) for neutrons (full circles) in p+p
             collisions compared to the prediction from JAM (full lines)}
  	 \label{fig:pn_in_pp}
  \end{center}
\end{figure}

The measured proton and neutron density distributions $dn/dx_F$ in
p+p interactions are confronted in Fig.~\ref{fig:pn_in_pp} with the predictions
from JAM.

%      Fig.78 

Large systematic deviations between prediction and data are visible 
at low $x_F$ (-40\%), at medium $x_F\sim$~0.25 (+50\%) and at high $x_F$ (-30\%) for
the protons. Systematic deviations of similar or bigger size are also 
seen for the HSD and UrQMD codes \cite{anticic}.

\begin{figure}[h]
  \begin{center}
  	\includegraphics[width=10cm]{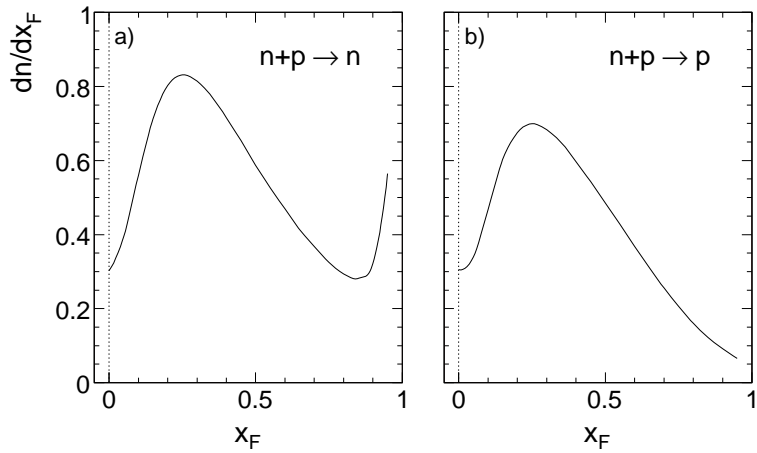}
 	\caption{Density distributions $dn/dx_F$ as functions of $x_F$
             a) for neutrons and b) for protons in n+p collisions from JAM}
  	 \label{fig:pn_in_np}
  \end{center}
\end{figure}

The neutron yields are evidently obtained from the proton densities 
by a constant multiplicative factor of 0.69 with the exception of the 
large $x_F$ region where a diffractive component with a $1/M_x^2$ behaviour, 
see Sect.~\ref{sec:res_decay_prot}, is added to the protons. The deviations are +60\% 
at medium $x_F$ and more than -100\% at large $x_F$ for the neutrons. A look 
at the density distributions for n+p interactions, Fig.~\ref{fig:pn_in_np}, shows only
approximate isospin symmetry which would impose that neutrons from
proton fragmentation should be equal to protons from neutrons in the
$x_F$ regions beyond the target-projectile overlap. The same should of 
course be true for neutrons from neutron beam and protons from proton 
beam.

%      Fig.79 

\begin{figure}[h]
  \begin{center}
  	\includegraphics[width=9.8cm]{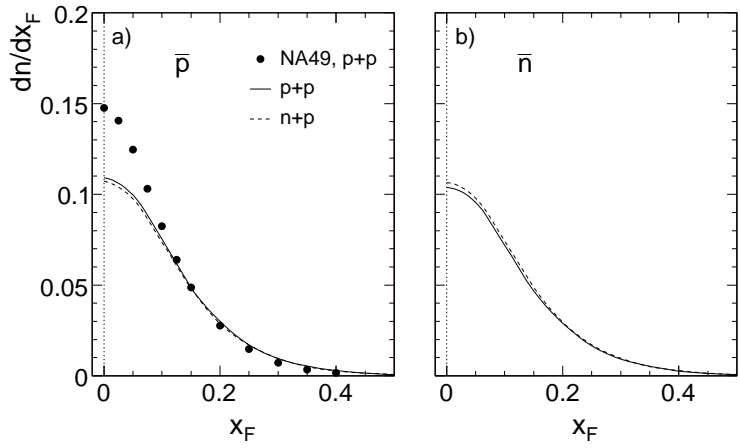}
 	\caption{a) anti-proton and b) anti-neutron distributions $dn/dx_F$
             as functions of $x_F$ from p+p and n+p interactions. The anti-proton
             distribution measured by NA49 is shown in panel a) with full circles. The full and 
             broken lines give the results of the JAM model for p+p and n+p interactions, respectively}
  	 \label{fig:antib}
  \end{center}
\end{figure}

The anti-baryon densities shown in Fig.~\ref{fig:antib} show an interesting pattern,
the anti-proton and anti-neutron yields being identical for both projectile-target combinations. 

%      Fig.80 

This means that only the $I_3$~=~0 combinations p$\overline{\textrm{p}}$ and 
n$\overline{\textrm{n}}$ are allowed which are produced with equal yields. 
The comparison with the measured anti-proton density in p+p collisions shows
a sizeable underestimation of the yield by 35\% at $x_F$~=~0 and an
equally large overestimation at $x_F \sim$~0.3. Invoking the isospin
effect measured in n+p interactions \cite{iso} this difference will
increase to 50\% in this reaction. 

%
% ****************************** Section 12.5.2 ****************************
%
\subsubsection{Baryonic correlations}
\vspace{3mm}
\label{sec:jam_corr}

While it might be a straightforward possibility to remedy the observed
discrepancies between the prediction and the inclusive data by 
modifying some of the many parameters involved in the simulation 
codes, the correlation data will probe the "physics" input on a
deeper level. This applies especially to the measured isospin
effects.

\begin{figure}[h]
  \begin{center}
  	\includegraphics[width=10.5cm]{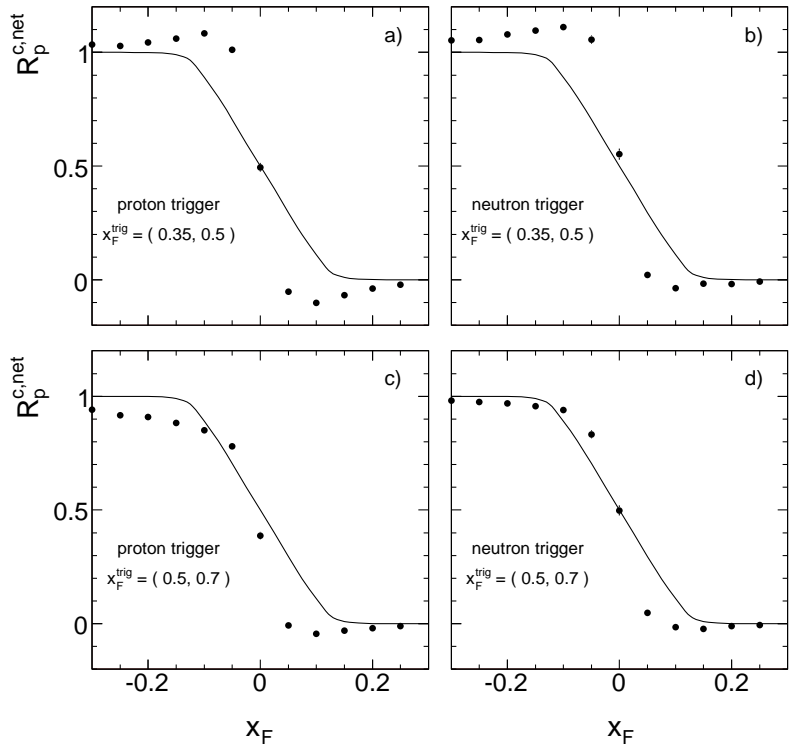}
 	\caption{Net proton overlap function $R^{c,\textrm{net}}_{\textrm{p}}(x_F)$
             as a function of $x_F$ for a) proton trigger at $x_F$~=~0.35 to 0.5, 
             b) neutron trigger at $x_F$~=~0.35 to 0.5, c) proton trigger at $x_F$~=~0.5 to 0.7 
             and d) neutron trigger at $x_F$~=~0.5 to 0.7. The full lines
             represent the measured function, Sect.~\ref{sec:feedover}, the 
             points come from the JAM simulation code}
  	 \label{fig:net_proton}
  \end{center}
\end{figure}

A first comparison concerns the net proton density correlated with
a trigger baryon in the projectile hemisphere resulting in the overlap
function $R^{c,\textrm{net}}_{\textrm{p}}$, (\ref{eq:ratp}), as shown in Fig.~\ref{fig:net_proton} for the two
trigger $x_F$ bins defined in Sect.~\ref{sec:two_comp}, Eqs.~\ref{eq:fp} and \ref{eq:fn}.

%      Fig.81 

The microscopic simulation results in a feed-over behaviour
which only reaches to about $\pm$0.06 in $x_F$. This is considerably
sharper than the measured behaviour (full lines in Fig.~\ref{fig:net_proton}) and
corresponds to the pionic feed-over extracted in \cite{pc_discus}.
In addition there is an asymmetric long range tail that extends
up to and beyond $|x_F| \sim$~0.3 and is different both for proton
and neutron trigger and for the two trigger $x_F$ bins in contrast
to the data.

\begin{figure}[h]
  \begin{center}
  	\includegraphics[width=10.5cm]{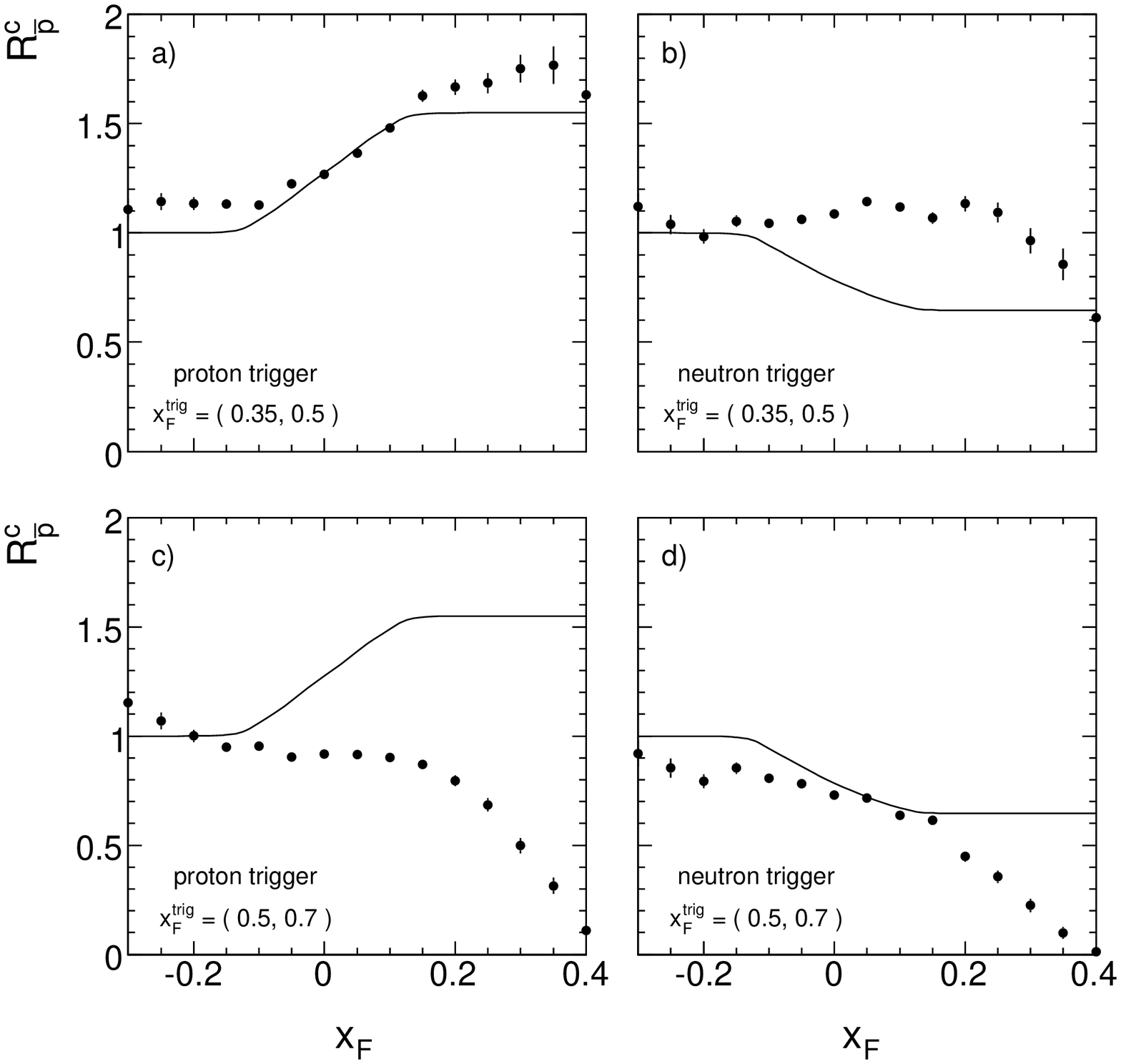}
 	\caption{Anti-proton density ratio $R^c_{\overline{\textrm{p}}}(x_F)$ as a function
             of $x_F$ for a) proton trigger at $x_F$~=~0.35 to 0.5, 
             b) neutron trigger at $x_F$~=~0.35 to 0.5, c) proton trigger at $x_F$~=~0.5 to 0.7 
             and d) neutron trigger at $x_F$~=~0.5 to 0.7. The full lines represent the measured
             function, Sect.~\ref{sec:two_comp_aprot}, the points come from the JAM 
             simulation code}
  	 \label{fig:net_aprot}
  \end{center}
\end{figure}

The anti-proton feed-over behaviour characterized by the ratio
$R^c_{\overline{\textrm{p}}}(x_F)$ of correlated to inclusive densities, Eq~\ref{eq:rat_corr}, is
presented in Fig.~\ref{fig:net_aprot} again for the two available trigger $x_F$ bins
and for proton and neutron trigger.

%     Here Fig.82 

A rather complicated pattern concerning the simulation emerges.
Evidently the symmetric isospin effect observed experimentally between 
proton and neutron triggers is not reproduced although there is a 
general reduction of the density ratio in the trigger hemisphere
with neutron triggers. As only isospin singlet baryon pairs
are generated, see Sect.~\ref{sec:jam_incl}, any isospin effect is not really
expected. The strong suppression of the density ratio in the high-$x_F$
trigger bin already starting at $x_F \sim$~0.2 indicates an effect of
energy-momentum conservation in the baryon pair simulation which
is probably the result of the details of string fragmentation.

In conclusion the microscopic simulation results in major deviations
from the data both on the level of the inclusive and of the correlated 
yields. The absence of a proper treatment of isospin effects both
concerning the $I_3$ component of the projectile and of the final state
baryons is flagrant. This puts into doubt the application of this
approach to p+A and especially A+A interactions where the neutron
component is preponderant.

%
% ****************************** Section 13 ****************************
%
\section{Proton and anti-proton feed-over in p+C interactions}
\vspace{3mm}
\label{sec:feedover}

\begin{figure}[h]
  \begin{center}
  	\includegraphics[width=8.cm]{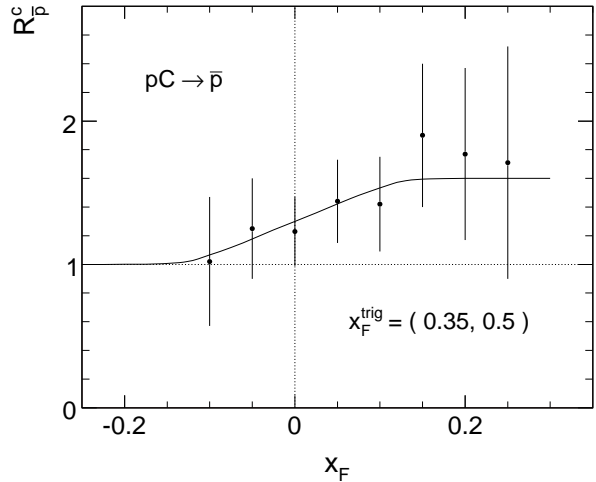}
 	\caption{Anti-proton density ratio $R^c_{\overline{\textrm{p}}}(x_F)$ as a
            function of $x_F$. The full line represents the feed-over in p+p events, see Fig.~\ref{fig:arat}}
  	 \label{fig:arat_pc}
  \end{center}
\end{figure}

The study presented above for p+p collisions may be repeated for p+C
interactions, albeit with reduced statistical significance due to the
smaller available data sample. The density ratio for anti-protons,
$R^c_{\overline{\textrm{p}}}(x_F)$ with forward proton tagging is presented in Fig.~\ref{fig:arat_pc}
where the full line represents the feed-over function measured in p+p
interactions, see Fig.~\ref{fig:arat}.

%      Fig.83 

Evidently the measurement reproduces the shape and extent of the $\overline{\textrm{p}}$
overlap in p+p collisions.

Net proton tagging in the backward hemisphere is not possible for p+A interactions 
as protons from intra-nuclear cascading ("grey" protons) which prevail in this area dilute the correlation.

\begin{figure}[h]
  \begin{center}
  	\includegraphics[width=8.cm]{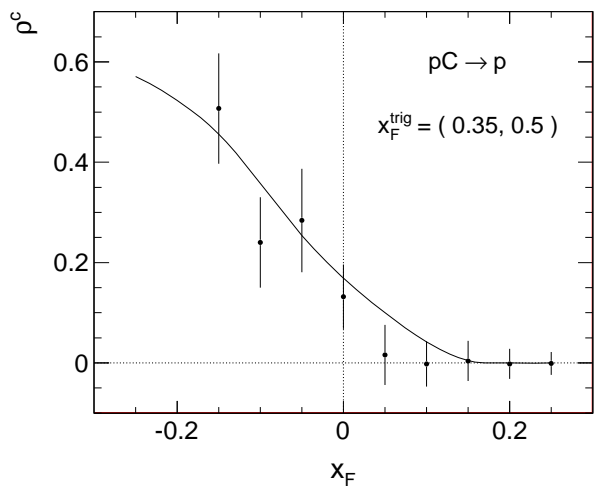}
 	\caption{Net proton density in p+C collisions with forward net proton
             constraint as a function of $x_F$. The full line corresponds to the net
             density in p+p interactions}
  	 \label{fig:prot_pc}
  \end{center}
\end{figure}

For the net proton density $\rho^{c,\textrm{net}}_{\textrm{p}}(x_F)$ with tagging in the
projectile hemisphere, a result which might look surprising at first
view is found as shown in Fig.~\ref{fig:prot_pc}.

%      Fig.84 

\begin{figure}[h]
  \begin{center}
  	\includegraphics[width=8cm]{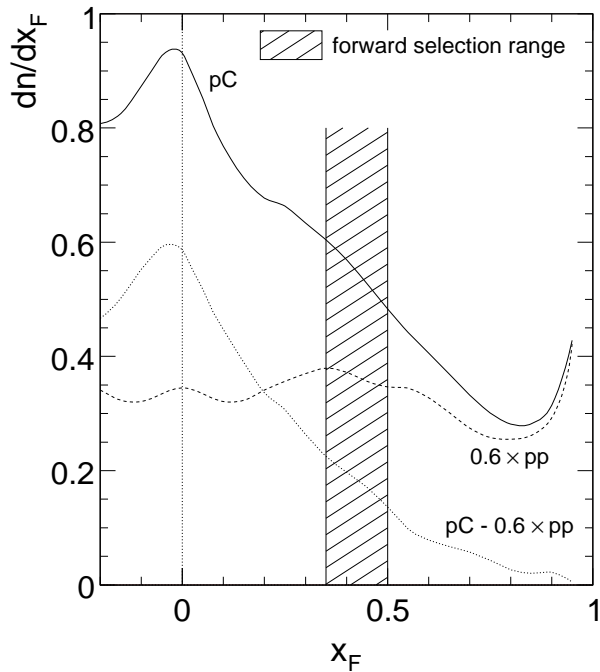}
 	\caption{The inclusive proton yield $dn/dx_F(x_F)$ in p+C collisions (full line),
             $0.6 \times dn/dx_F(x_F)$ from p+p (broken line), and their difference (dotted line). 
             The inclusive proton yield $dn/dx_F(x_F)$ in p+p collisions is shown as well 
             with full line. The hatched area represent the tagging region}
  	 \label{fig:xf_dist_sel}
  \end{center}
\end{figure}

Indeed the constrained net proton density reproduces the one found in
p+p interactions down to $x_F$~=~-0.2. This reveals an additional internal correlation 
effect generally present in minimum bias p+A interactions. By selecting a rather 
forward proton, see (\ref{eq:fp}), single collisions corresponding
to peripheral interactions are enhanced. In fact 60\% of the minimum bias p+C 
events correspond to single interactions of the projectile proton with
nuclear participants \cite{pc_discus}. The situation is clarified by Fig.~\ref{fig:xf_dist_sel}
which presents the inclusive proton yield $dn/dx_F(x_F)$ in p+C collisions (full line),
$0.6 \times dn/dx_F(x_F)$ from p+p (broken line), and their difference (dotted line). 
It is evident that the bulk of protons in the tagging region (hatched area) 
comes from single collisions, with the fraction of multiple collisions in addition 
biased against very inelastic interactions which would favour low-$x_F$ protons 
by enhanced baryon number transfer. It may however be stated that also for 
p+C interactions with forward proton tagging the constrained net proton density 
dies out at $x_F >$~0.2.

%      Fig.85 

%
% ****************************** Section 14 ****************************
%
\section{Hadronization in p+C collisions: anti-protons}
\vspace{3mm}
\label{sec:hadr}

The two-component hadronization mechanism studied experimentally
in the preceding section allows, in conjunction with the precision
data on anti-proton production in p+p interactions published by 
NA49 \cite{pp_proton}, the confrontation of the measured $\overline{\textrm{p}}$ yields in p+C
collisions with a straight-forward prediction based on elementary
reactions. This is simplified in the case of anti-protons by the
fact that there is, in contrast to proton and pion production, no 
contribution from nuclear cascading (see Sect.~\ref{sec:ratios} above). In
consequence the superposition of target and projectile fragmentation
should suffice to completely describe the observed cross sections.
In a first step it may be assumed that the projectile contribution 
corresponds exactly to the one in p+p interactions. For the target
contribution the same basic assumption may be made with two
important additional constraints taking account, firstly, of the
multiple intra-nuclear collisions of the projectile, and secondly,
of the isospin factor involved in the fragmentation of the neutrons
contained in the nucleus.

The mean number of projectile collisions $\langle \nu \rangle$ in minimum bias p+C
collisions has been investigated in \cite{pc_discus} using three different
methods:

\begin{itemize}
 \item A  Monte Carlo simulation using the measured nuclear density
       profile as input with the result $\langle \nu \rangle$~=~1.6
 \item The measured p+C inelastic cross section which gives, via the relation
       
       \begin{equation}
          \langle\nu\rangle = \frac{A\cdot\sigma(pp)}{\sigma(pA)}                   
       \end{equation}
       an estimation of $\langle \nu \rangle$ under the assumption that the intra-nuclear
       inelastic interaction cross section stays constant for all subsequent
       projectile collisions, resulting in $\langle \nu \rangle$~=~1.68
 \item The measured increase of pion yields in the backward hemisphere
       using the fact that the contribution from intra-nuclear cascading
       as well as the one from projectile fragmentation die out at 
       $x_F \sim$~-0.1, yielding $\langle \nu \rangle$~=~1.6.
\end{itemize}

The isospin factor for $\overline{\textrm{p}}$ production from the isoscalar C nucleus  
may be calculated, using the measured increase of $\overline{\textrm{p}}$ yields in
n+p collisions $g^n_{\overline{\textrm{p}}}$ \cite{iso} as 

\begin{equation}
   g^C_{\overline{\textrm{p}}} = 0.5( 1 + g^n_{\overline{\textrm{p}}} ) = 1.3 ,
\end{equation}
with $g^n_{\overline{\textrm{p}}}$~=~1.6. The combined overall factor to be 
applied to the target component is 1.6$\times$1.3 = 2.08.

%
% ****************************** Section 14.1 ****************************
%
\subsection{${\mathbf p_T}$ integrated density $\mathbf {dn/dx_F(x_F})$}
\vspace{3mm}
\label{sec:hadr_ptint}

\begin{figure}[h]
  \begin{center}
  	\includegraphics[width=9.cm]{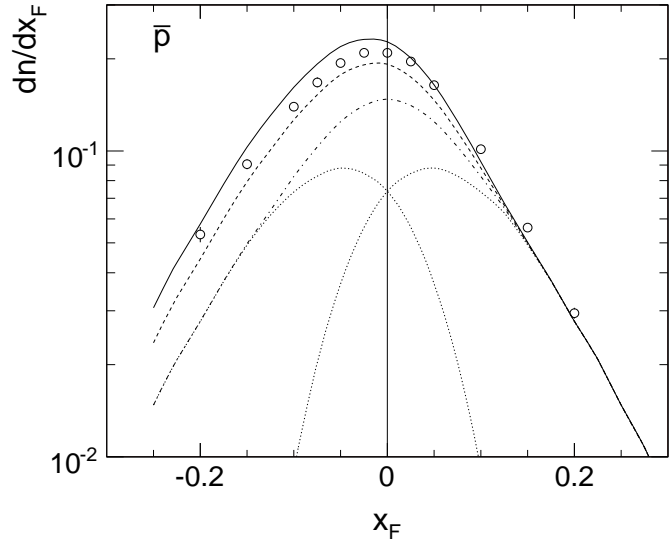}
 	\caption{Measured $p_T$ integrated anti-proton density $dn/dx_F(x_F)$
             in p+C interactions (open circles) confronted with the
             superposition of target and projectile components in p+p
             collisions (dotted lines) and total yield (dash-dotted line), with multiplication factors 
             $\langle \nu \rangle$~=~1.6 and 1 for target and projectile, respectively,
             broken line, and with the additional isospin factor 1.3
             for the target component (full line)}
  	 \label{fig:aprot_twocomp}
  \end{center}
\end{figure}

The evolution of the $p_T$ integrated anti-proton density $dn/dx_F(x_F)$
from the elementary p+p to the p+C interactions using the 
superposition of target and projectile components discussed above 
is presented in Fig.~\ref{fig:aprot_twocomp}. The $p_T$ integration was
performed both for the p+p and p+C data in the range 0~$< p_T <$~1.7~GeV/c.

%      Fig.86 

\begin{figure}[h]
  \begin{center}
  	\includegraphics[width=9cm]{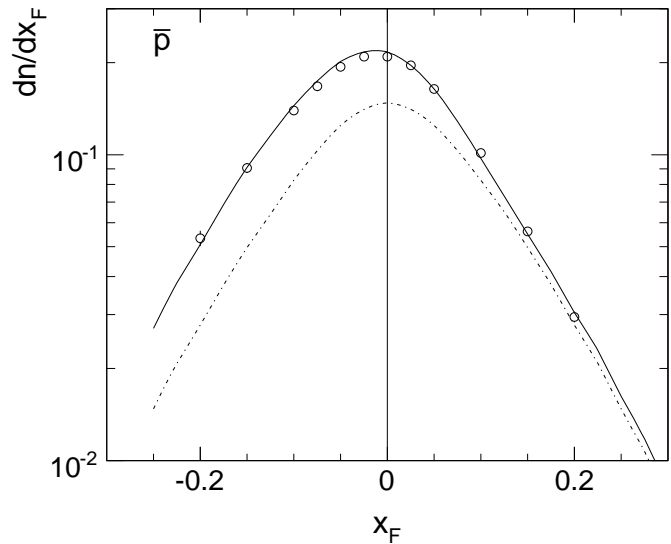}
 	\caption{Measured $p_T$ integrated anti-proton density $dn/dx_F(x_F)$
             in p+C interactions compared with a target-projectile 
             superposition scheme with factors 1.84 and 1.1, respectively,
             for the target and projectile components. The yield from p+p 
             is shown with dash-dotted line}
  	 \label{fig:aprot_twocomp_mod}
  \end{center}
\end{figure}

This most straight-forward superposition picture evidently
reproduces the measured $\overline{\textrm{p}}$ densities quite closely, overestimating
them by about 10\% in the target area and underestimating them
by the same amount in the projectile region. This can be remedied
by decreasing the target multiplication factor by 12\% from 2.08 
to 1.84 and by increasing the projectile contribution by 10\%
resulting in a reasonable fit of the experimental data compatible
with their statistical errors as shown in Fig.~\ref{fig:aprot_twocomp_mod}.

%     Fig.87  

The increase of the $\overline{\textrm{p}}$ yield by 10\% in the projectile hemisphere
with respect to the direct estimation from p+p collisions is a first 
interesting consequence of this study. It is equal to the increase of 
the pion yields deduced in \cite{pc_discus}.

The reduction of the target contribution by about 12\% with respect 
to the simple superposition of elementary hadronization processes as 
characterized by the mean number of collisions $\langle \nu \rangle$ and isospin symmetry,
is a second important result. The magnitude of this reduction being
relatively small it is nevertheless on the limit allowed by the
experimental determination both of $\langle \nu \rangle$ \cite{pc_discus} and of 
the isospin effect on anti-proton production from neutrons \cite{iso}.

Regarding the excitation mechanism of colliding hadrons by the
exchange of gluons or gluonic (charge and flavour-less) objects,
the observed effects are however to be expected. The projectile
interacts in its multiple collisions subsequently with "fresh"
nucleons which did not undergo previous exchanges. Hence its effective
excitation level will increase with $\langle \nu \rangle$. The projectile on the
other hand suffers in each collision a loss of its gluonic component
such that less excitation energy with the target nucleons can be exchanged in subsequent
interactions. In this sense the nucleus may be regarded as a gluon
filter, the description of this phenomenology by the term of "energy
loss" giving only a very general and somewhat misleading impression.

The anti-proton yields regarded here are especially sensitive to
this notion as at SPS energy the $s$-dependence of the production
cross section is still rather steep with about 10\% per GeV in $\sqrt{s}$
\cite{bourquin}. This is in contrast to the production of mesons, with 5\% per GeV 
for mean kaons \cite{pp_kaon} and only 3\% per GeV for mean pions \cite{rossi}. 
In this sense the study of pion yields in the target hemisphere approaches 
a precise measure of $\langle \nu \rangle$ \cite{pc_discus} whereas the observed anti-proton cross 
section indicates an effective loss in $\sqrt{s}$ of about 1~GeV in 
target fragmentation for p+C collisions. Evidently the extension of 
this study to heavier nuclei is of considerable interest in this 
respect. The data on p+Pb collisions with controlled centrality 
available from NA49 will illuminate this point, as is already 
visible in the preliminary results shown in \cite{gabor}.       

%
% ****************************** Section 14.2 ****************************
%
\subsection{Double differential invariant densities $\mathbf {f(x_F,p_T)/\sigma_{\textrm{inel}}}$}
\vspace{3mm}
\label{sec:hadr_ptdist}

\begin{figure}[h]
  \begin{center}
  	\includegraphics[width=12.5cm]{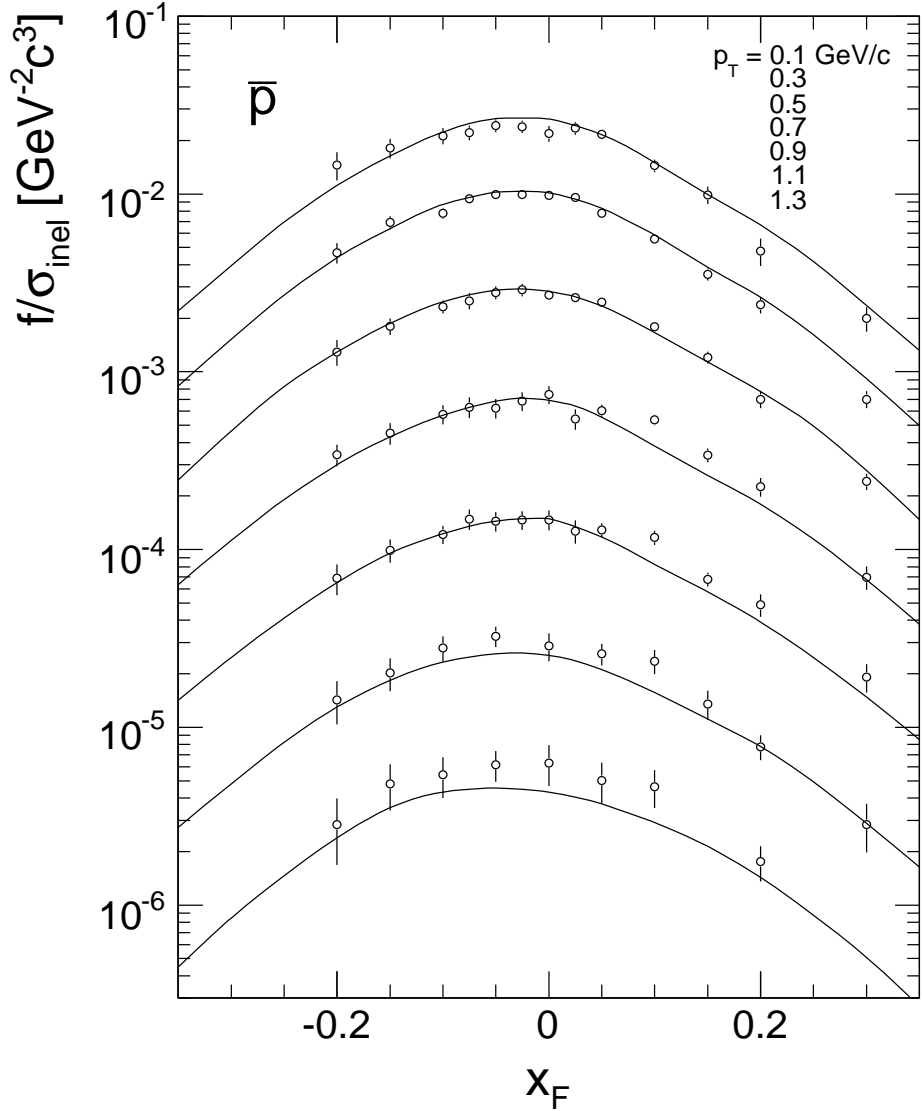}
 	\caption{Double differential invariant anti-proton densities
             per inelastic event (open circles) compared to the prediction
             from the two component fragmentation mechanism (full lines)
             as a function of $x_F$, for different values of $p_T$ between 0.1
             and 1.3~GeV/c. The distributions for different $p_T$ values are
  	         successively scaled down by 2 for better separation}
  	 \label{fig:aprot_xfdist}
  \end{center}
\end{figure}

Further details of the two-component hadronization mechanism detailed
above become visible if comparing the double differential invariant
densities to the optimized prediction from target and projectile
fragmentation obtained from the $p_T$ integrated yields, see Fig.~\ref{fig:aprot_twocomp_mod}. 
This is shown in Fig.~\ref{fig:aprot_xfdist} where the invariant cross sections $f(x_F,p_T)$ 
per inelastic event (open circles) are presented together with the 
predicted densities (full line).

%      Fig.88 

Three main features may be extracted from these plots:

\begin{itemize}
 \item The target component reproduces, within the experimental errors,
       the densities predicted from elementary interactions for $p_T <$~1~GeV/c.
 \item The projectile component reveals a definite $p_T$ dependence. The
       measurements fall below the prediction for $p_T \lesssim$~0.5~GeV/c and
       increase smoothly above the prediction above this value.
 \item In the range of $p_T >$~1~GeV/c the excess of the projectile component
       starts to extend well into the target hemisphere down to $x_F \sim$~-0.2.
\end{itemize}

\begin{figure}[h]
  \begin{center}
  	\includegraphics[width=15cm]{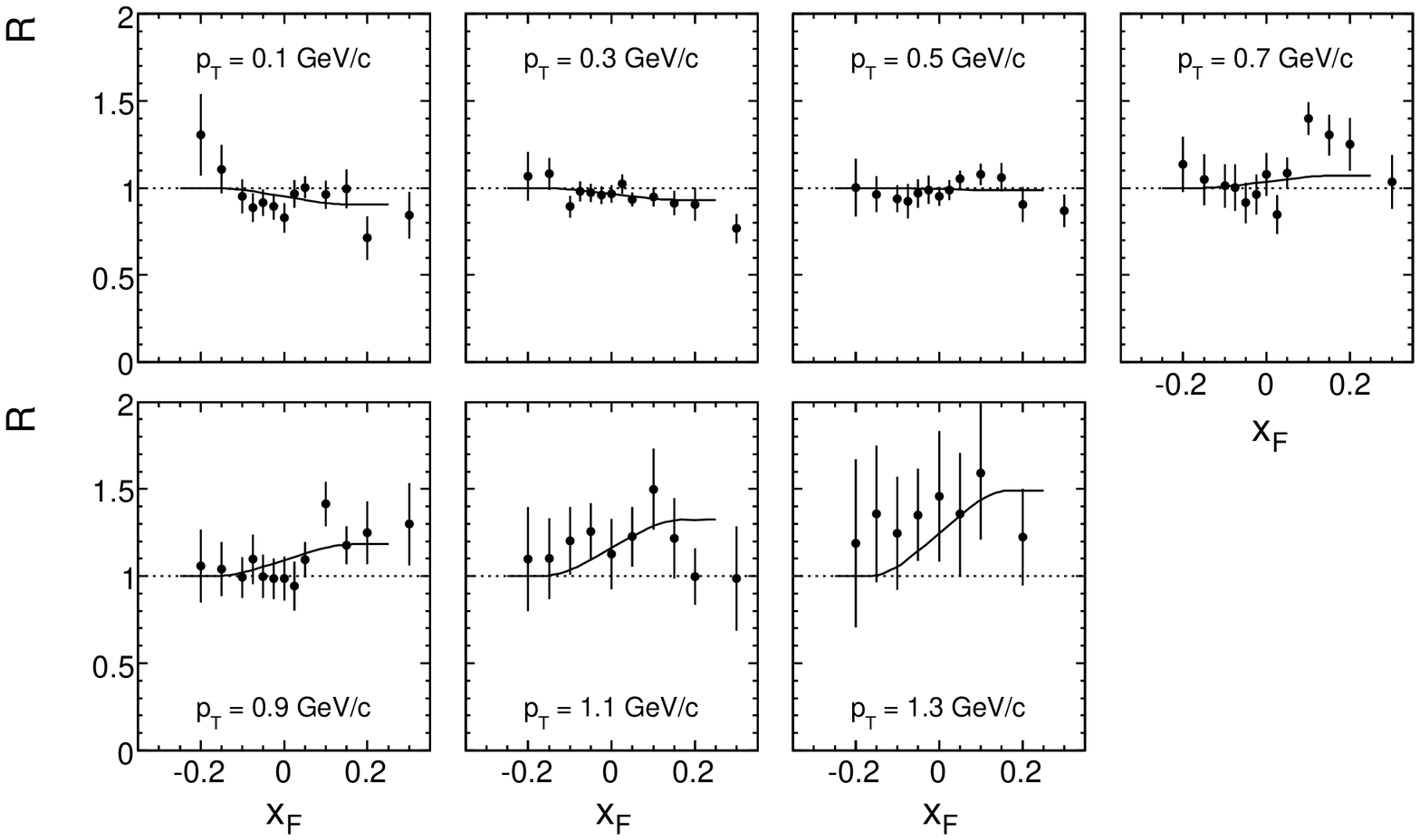}
 	\caption{Ratio $R$ between measured and predicted anti-proton
             densities for different $p_T$ values between 0.1 and 1.3~GeV/c
             as a function of $x_F$. The full lines represent the parametrization
             $R = ( 1 - F^o_{\overline{\textrm{p}}} ) + R^{\textrm{proj}}F^o_{\overline{\textrm{p}}}$ }
  	 \label{fig:aprot_xfrat}
  \end{center}
\end{figure}

This behaviour is detailed in Fig.~\ref{fig:aprot_xfrat} which presents the ratio $R$ between
measurement and prediction for the $p_T$ values shown in Fig.~\ref{fig:aprot_xfdist}.

%      Fig.89 

\begin{figure}[h]
  \begin{center}
  	\includegraphics[width=12cm]{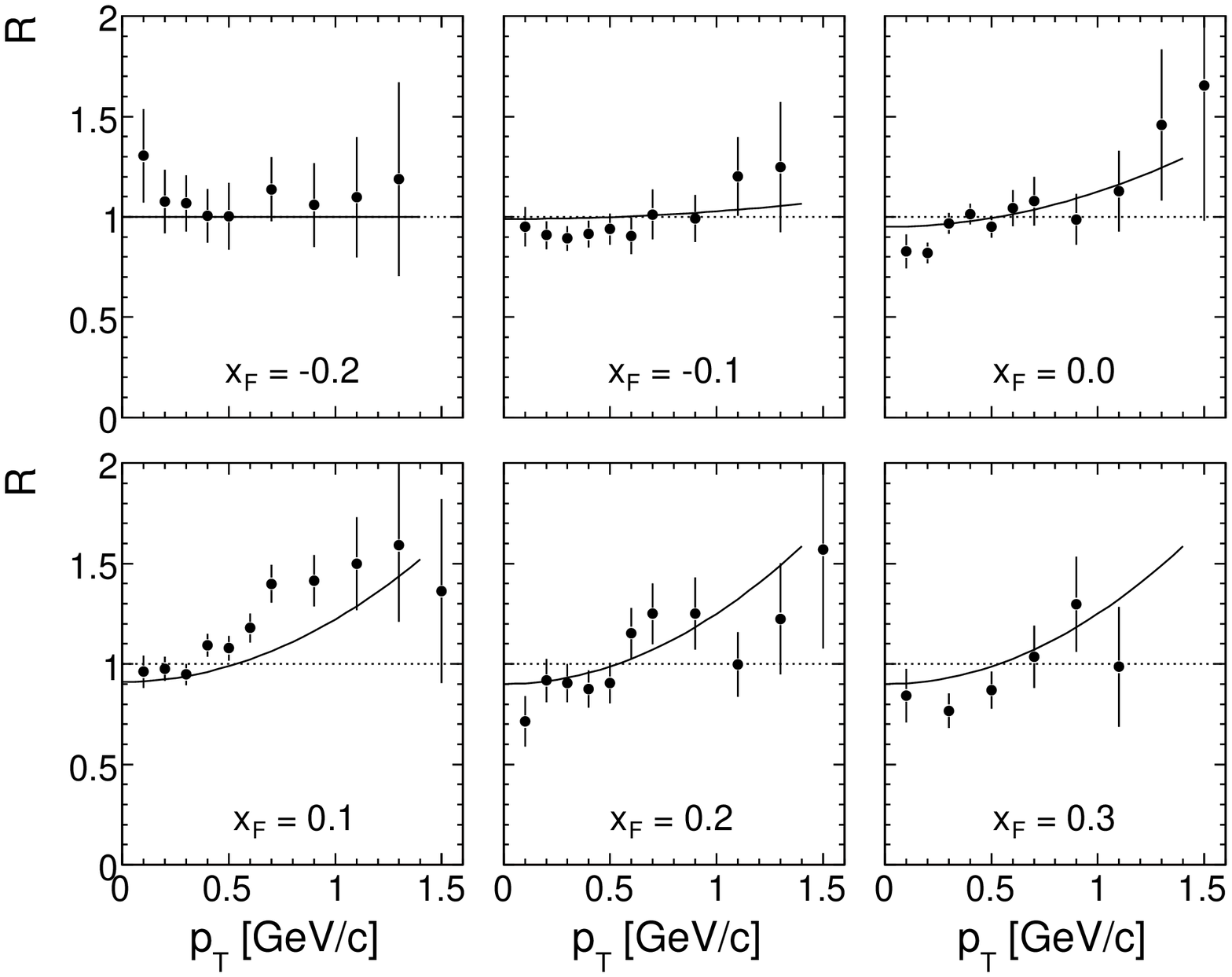}
 	\caption{Ratio $R$ between measured and predicted anti-proton
             densities for different $x_F$ values as a function of $p_T$.
             The full lines represent the parametrization
             $R = ( 1 - F^o_{\overline{\textrm{p}}} ) + R^{\textrm{proj}}F^o_{\overline{\textrm{p}}}$}
  	 \label{fig:aprot_ptrat}
  \end{center}
\end{figure}

The same ratio is plotted against $p_T$ for different values of $x_F$ in
Fig.~\ref{fig:aprot_ptrat}.

%      Fig.90 

It should be stressed here that these experimental results are in
strong support of the independent target-projectile fragmentation
in p+C interactions. The modification of the $p_T$ distribution of
the projectile component which superimposes itself to the 10\%
increase in total yield, indicates that the Cronin effect, whose
onset is visible here, is limited to the projectile hadronization.
The increase of yields at higher $p_T$ which extends well into the
target hemisphere might be due to an extension of the feed-over
range with $p_T$ (see also \cite{pc_discus} for pions), although the limited 
statistics does not allow for quantitative statements. The increase of $R$ in the projectile hemisphere may be
parametrized as

\begin{equation}
   \label{eq:pt_param}
   R^{\textrm{proj}} = 0.9 + 0.35p_T^2 
\end{equation}
Its modification in the transition to the target hemisphere
is then predicted by the projectile overlap function $F^o_{\overline{\textrm{p}}}$
(Fig.~\ref{fig:overlapf}) resulting in the full lines indicated in Figs.~\ref{fig:aprot_xfrat}
and \ref{fig:aprot_ptrat}. 

\begin{figure}[h]
  \begin{center}
  	\includegraphics[width=12cm]{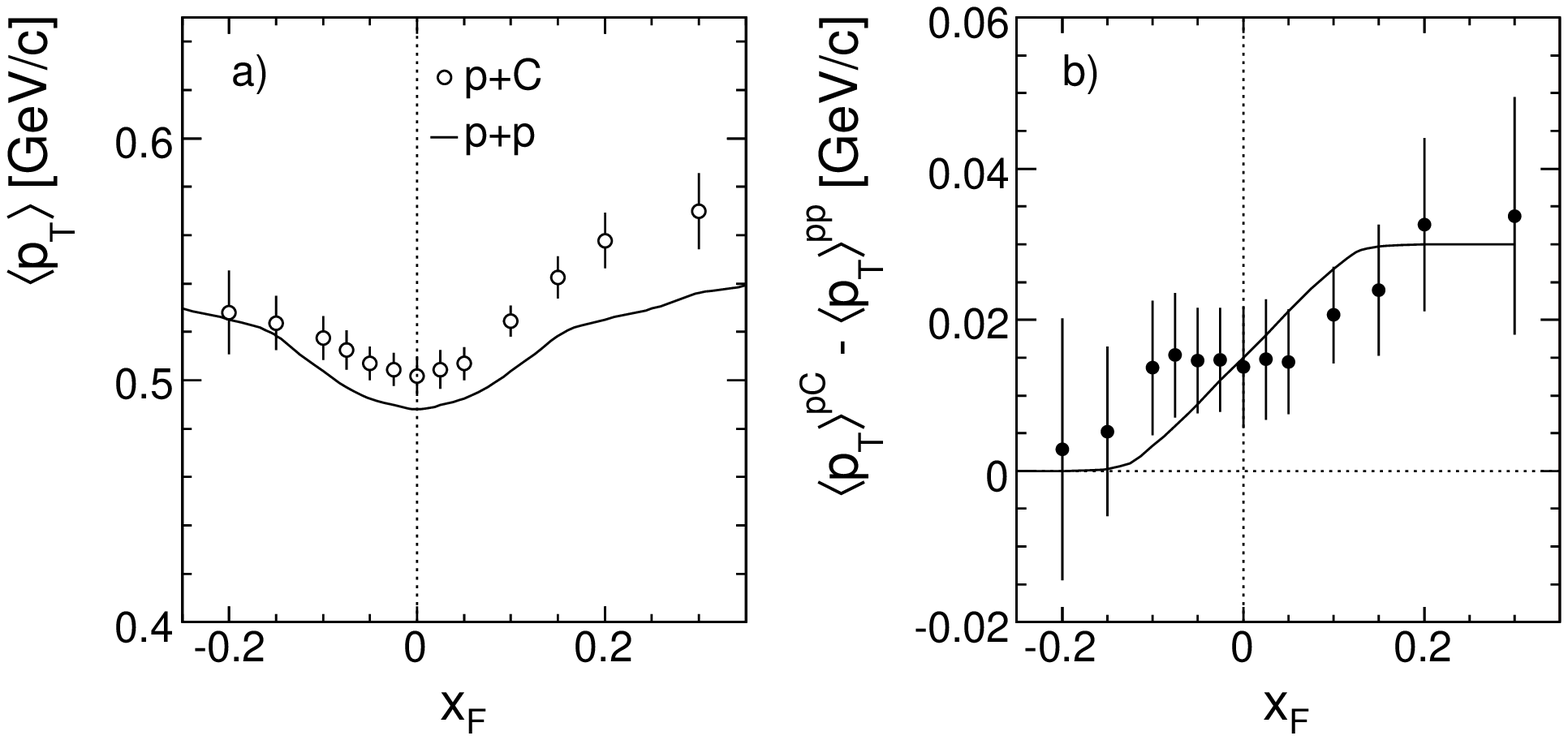}
 	\caption{a) comparison of average $p_T$ between p+p 
             collisions (full line) and p+C interactions (data points),
             b) difference in $\langle p_T \rangle$ between p+C and p+p interactions.
             The full line in panel b) represent the anti-proton overlap function $F^o_{\overline{\textrm{p}}}$
             from Sect.~\ref{sec:two_comp_aprot}}
  	 \label{fig:aprot_meanpt}
  \end{center}
\end{figure}

Further information is contained in the detailed comparison of
the average $p_T$ values to p+p interactions given in Fig.~\ref{fig:aprot_meanpt}.

%      Fig.91 

The convergence of the difference in $\langle p_T \rangle$ towards zero in the
backward hemisphere, Fig.~\ref{fig:aprot_meanpt}b, and its
description by the overlap function $F^o_{\overline{\textrm{p}}}$, Fig.~\ref{fig:overlapf},
is to be regarded as yet another
manifestation of the two-component
mechanism of hadronization as discussed in Sect.~\ref{sec:two_comp}.

%
% ****************************** Section 15 ****************************
%
\section{Proton production in p+C collisions: ${\mathbf p_T}$ integrated yields}
\vspace{3mm}
\label{sec:hadr_net}

As in the preceding section on anti-protons, proton production
will be first discussed using the $p_T$ integrated yields $dn/dx_F$
in order to clearly visualize and separate the three basic 
contributions to the overall proton cross section. After
establishing the net proton density by subtracting the yield
of pair produced protons, the fragmentation of the hit target
nucleons, the projectile fragmentation and the contribution
from intra-nuclear cascading will be treated in turn. The $p_T$ integrations used in this section were
performed both for the p+p and p+C data in the range 0~$< p_T <$~1.7~GeV/c.
%
% ****************************** Section 15.1 ****************************
%
\subsection{Pair produced protons and net proton density}
\vspace{3mm}
\label{sec:pair_prot}

\begin{figure}[h]
  \begin{center}
  	\includegraphics[width=7.5cm]{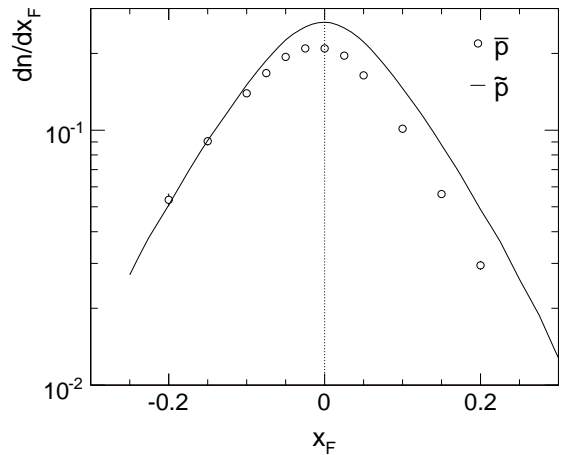}
 	\caption{Density distribution $dn/dx_F(x_F)$ for pair
             produced protons ($\,\widetilde{\textrm{p}}\,$) as a function of $x_F$ (full line).
             The data points are the measured anti-proton densities. The systematic error
              margin for the full line (predicted $\,\widetilde{\textrm{p}}\,$) corresponds to 7.2 \%}
  	 \label{fig:prot_pair_xfdist}
  \end{center}
\end{figure}

It has been shown above that the iso-triplet nature of baryon
pair production imposes a detailed follow-up of isospin effects
all through the hadronization process. This means of course
that the yield of anti-protons is in general not identical
to the yield of pair-produced protons. It is therefore not
sufficient to subtract the anti-proton density as established
in the preceding section from the proton yield in order to 
obtain the net proton density. In the case of the asymmetric
p+C interaction, the equality of anti-proton and pair-produced
proton yields is only valid for the target part of the overall
cross section due to isoscalar carbon nucleus, invoking isospin 
symmetry. For the projectile component on the other hand this
does not apply. In the absence of precision data for anti-proton
production in n+C collisions it may be assumed that the same 
enhancement of pair-produced protons over anti-protons as in 
p+p interactions applies where a factor of 1.6 has been 
established \cite{iso}. This modifies the superposition scheme for
anti-protons described in Sect.~\ref{sec:hadr} by increasing the projectile
contribution from 1.1 to 1.76 with respect to the input
p+p densities. The resulting $p_T$ integrated density distribution   
for pair produced protons (hereafter denoted as $\,\widetilde{\textrm{p}}\,$) is
shown in Fig.~\ref{fig:prot_pair_xfdist}.

%      Fig.92 

This distribution is in contrast to Fig.~\ref{fig:pn_in_pp} almost
symmetric and the comparison to the measured anti-proton 
densities demonstrates that this isospin effect is by no means 
negligible.

\begin{figure}[h]
  \begin{center}
  	\includegraphics[width=7.cm]{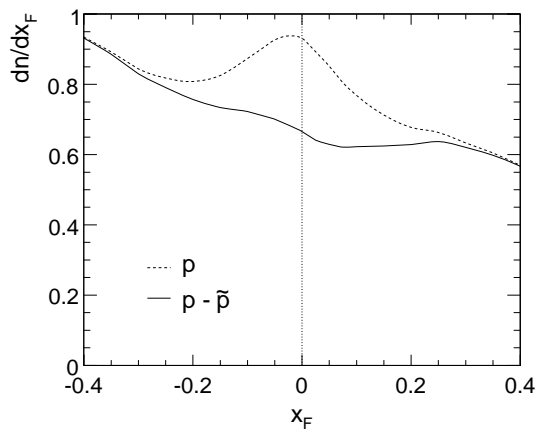}
 	\caption{Density distribution $dn/dx_F(x_F)$ for protons
             (broken line) and for net protons (full line) as a function
              of $x_F$ in the range -0.4~$< x_F <$~+0.4. The systematic error
              margin corresponds to 4.3 \%}
  	 \label{fig:incl_pair}
  \end{center}
\end{figure}

The corresponding net proton density distribution results from the
subtraction of the pair produced protons from the total proton
density distribution, Table~\ref{tab:int_prot} and Fig.~\ref{fig:ptint_prot}. 
It is presented in Fig.~\ref{fig:incl_pair} in the $x_F$ range influenced by 
pair production, -0.4~$< x_F <$~+0.4.

%      Fig.93 

The net proton density shows, in contrast to the total proton
yield, a smooth behaviour around $x_F$~=~0.

%
% ****************************** Section 15.2 ****************************
%
\subsection{Target and projectile components in net proton production}
\vspace{3mm}
\label{sec:targ_proj}

The fragmentation of those nucleons in the carbon nucleus which
are hit by the through-going projectile, here called the "target
component", is a quantity which should be closely related to
the proper superposition of net proton production in the
elementary p+p and p+n interactions, multiplied by the number
of projectile interactions. Such a superposition has been shown
to describe the target component of the anti-proton yields,
Sect.~\ref{sec:hadr}, up to a loss of about 12\% specific to multiple
collisions and related to the strong $s$-dependence of baryon pair
production. For baryons this loss should be negligible as the
baryon density is to first order $s$-independent at SPS energy.

For the prediction of the target component, knowledge about the two
basic contributions from

\begin{equation}
   \textrm{p+p} \rightarrow \,\widetilde{\textrm{p}}\,\quad \textrm{and} \quad \textrm{p+n} \rightarrow \,\widetilde{\textrm{p}}\,  
\end{equation}
is needed. The latter process may be related, via isospin symmetry,
to the reaction

\begin{equation}
   \textrm{p+p} \rightarrow \,\widetilde{\textrm{n}}\,                     
\end{equation}
which is measured by the NA49 experiment \cite{pp_proton}. The resulting prediction
for the isospin averaged net proton density from p+p collisions
is described in the following section.

%
% ****************************** Section 15.2.1 ****************************
%
\subsubsection{Isospin averaged net proton density from p+p collisions}
\vspace{3mm}
\label{sec:net_targ}

\begin{figure}[h]
  \begin{center}
  	\includegraphics[width=7.cm]{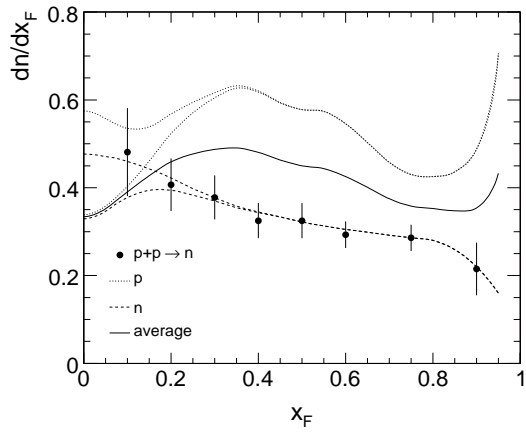}
 	\caption{Total and net proton density $dn/dx_F(x_F)$,
             dotted lines, total and net neutron densities (broken
             lines) together with the measured neutron cross sections
             (data points from \cite{pp_proton}) and the isospin averaged net 
              proton density(full line), as a function of $x_F$. The systematic error
              margin is 2.5 \% for protons and 4 \% for the predicted target component}
  	 \label{fig:prot_neut_net}
  \end{center}
\end{figure}

The $p_T$ integrated proton and neutron densities $dn/dx_F$ as measured
by the NA49 experiment are shown in Fig.~\ref{fig:prot_neut_net} as dotted and broken 
lines. The corresponding net proton yield is obtained by subtracting
1.6 times the measured anti-proton yield. For the net neutron yield,
the subtraction of the measured anti-proton yield is indicated by
the isospin symmetry of baryon pair production. In the case of the
isoscalar carbon nucleus, a simple average of the two distributions
is to be performed as indicated by the full line in Fig.~\ref{fig:prot_neut_net}.

%      Fig.94 
%
% ****************************** Section 15.2.2 ****************************
%
\subsubsection{Predicted target component of net proton production}
\vspace{3mm}
\label{sec:net_targ_pred}

\begin{figure}[h]
  \begin{center}
  	\includegraphics[width=6.cm]{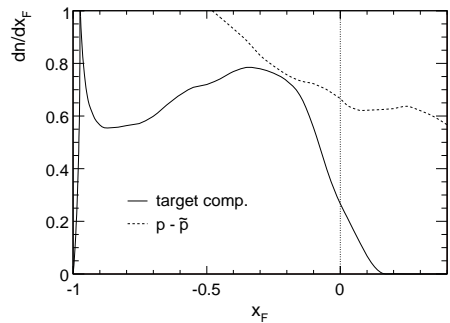}
 	\caption{Total net proton density $dn/dx_F(x_F)$, broken line,
             and predicted target contribution (full line) as a function
             of $x_F$}
  	 \label{fig:targ_net}
  \end{center}
\end{figure}

In order to obtain the predicted target component of net proton
production in p+C interaction, the predicted yield from p+p
collisions has to be multiplied by the number of projectile
collisions $\langle \nu \rangle$~=~1.6 and by the target overlap function, Sect.~\ref{sec:feedover}.
The resulting distribution is shown in Fig.~\ref{fig:targ_net} together with
the total net proton yield, Sect.~\ref{sec:pair_prot}.

%      Fig.95 

As a first result it may be seen that the target component nearly
saturates the total yield at $x_F \sim$~-0.2.

%
% ****************************** Section 15.3 ****************************
%
\subsection{The projectile component of net proton production}
\vspace{3mm}
\label{sec:net_proj}

\begin{figure}[h]
  \begin{center}
  	\includegraphics[width=7.cm]{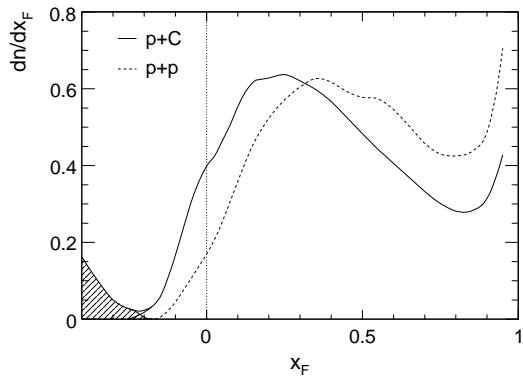}
 	\caption{Net proton density $dn/dx_F(x_F)$ for the projectile
             component in p+C interactions (full line) and in p+p
             collisions (broken line) as a function of $x_F$. The shaded
             area indicates the onset of the contribution from the
              nuclear component in p+C interactions. The systematic error
              margin is 2.5 \% in p+p and 5 \% in p+C collisions}
  	 \label{fig:proj_net}
  \end{center}
\end{figure}

The subtraction of the predicted target component from the overall
net proton density, Fig.~\ref{fig:targ_net}, allows now for the extraction of the
projectile component of the p+C interaction. This is demonstrated
in Fig.~\ref{fig:proj_net} which shows the net projectile component (full line)
in comparison with the net projectile component in p+p collisions
(broken line, Sect.~\ref{sec:feedover}).

%      Fig.96 

Several features of these distributions are noteworthy:

\begin{itemize}
 \item This study allows for the first time the isolation of the net
       proton projectile component in p+A collisions over the full 
       phase space.
 \item In principle this net proton component is strongly constrained
       by baryon number conservation in comparison to p+p collisions,
       to the extent that the surface under the two distributions
       should be equal up to second order effects like a relative 
       increase of neutron or hyperon production.  
 \item In fact the integrated yields are 0.547 and 0.52 net protons
       per inelastic event for p+p and p+C interactions, respectively.
       In view of the multi-step methodology involved in extracting
       this experimental result, the agreement to within about 5\%
       is certainly compatible with the systematic uncertainties.   
 \item A relative increase of the net neutron yield, which is in
       principle not excluded in multiple hadronic interactions, is
       improbable as it has been shown, see Sect.~\ref{sec:neut}, that the
       baryon number transfer is identical for neutrons and protons.
 \item A relative increase of hyperon production can as well only
       have limited influence as an increase of $\Lambda$ and $\Sigma$
       production by 50\% would only reduce the observed difference
       from 5\% to 4\%.
 \item The shapes of the two distributions are rather similar, with
       a downward shift of about 0.15 units in $x_F$ in p+C except for 
       the diffractive region which is governed by single projectile
       collisions
 \item At $x_F <$~-0.2 there is a steep increase of the target-subtracted
       density (shaded region in Fig.~\ref{fig:proj_net}) which is due to the tail of 
       protons from intra-nuclear cascading. This contribution will
       be discussed in detail below (Sect.~\ref{sec:net_nucl}). It is however
       to be noted that a clean separation of the projectile component
       from both the target and the nuclear cascading contributions
       is being achieved at SPS energy. At lower cms energies, the
       nuclear component will extend into higher $x_F$ ranges, covering
       $x_F$~=~0 at AGS energy, as it scales in $p_{\textrm{lab}}$ rather than $x_F$.
\end{itemize}

\begin{figure}[h]
  \begin{center}
  	\includegraphics[width=7cm]{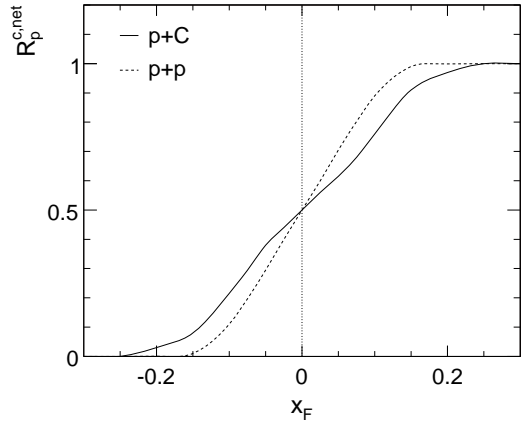}
 	\caption{Projectile overlap function $R^{c,\textrm{net}}_{\textrm{p}}(x_F)$ as a function of $x_F$
             from p+C interactions (full line) compared to the same
             function from p+p interactions (broken line)}
  	 \label{fig:rat_net}
  \end{center}
\end{figure}

It is interesting to extract the projectile overlap function
from the ratio $R^{c,\textrm{net}}_{\textrm{p}}$, see (\ref{eq:ratp}), between the projectile
component and the symmetrized total density. This is shown
in Fig.~\ref{fig:rat_net} in comparison with the corresponding function in 
p+p collisions.

%      Fig.97 

As a further important result of this study it may be stated
that in multiple collisions the width of the proton feed-over
from the projectile to the target hemisphere widens. In view
of the discussion of resonance decay in Sect.~\ref{sec:res_decay} above,
this would be compatible with an increase of the effective
mass in the process of projectile excitation.  

%
% ****************************** Section 15.4 ****************************
%
\subsection{The nuclear component of net proton production}
\vspace{3mm}
\label{sec:net_nucl}

If the subtraction of the target component from the total net proton
density distribution reveals the projectile component at $x_F >$~-0.2,
it also should allow for the extraction of the nuclear component
at $x_F \lesssim$~-0.2. Here the nuclear component is defined as the retarded
proton density produced by the intra-nuclear cascading of nucleons 
following the momentum transfer from the projectile to the nucleus, 
as opposed to the prompt fragmentation of those nucleons which are 
hit by the projectile.

\begin{figure}[h]
  \begin{center}
  	\includegraphics[width=7cm]{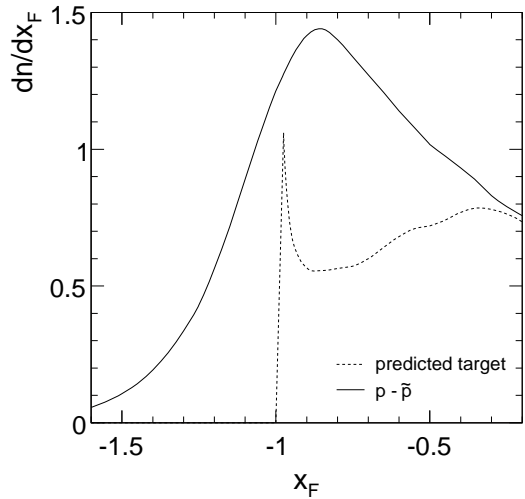}
 	\caption{Net proton density $dn/dx_F$ as a function of $x_F$
             in the range -1.6~$< x_F <$~-0.2. Full line: total measured yield,
             broken line: predicted target component}
  	 \label{fig:net_prot}
  \end{center}
\end{figure}

The situation is clarified in Fig.~\ref{fig:net_prot} which shows the total net proton
density $dn/dx_F$ for $x_F <$~-0.2 (full line) together with the target
component discussed in the preceding section (broken line), Fig.~\ref{fig:targ_net}.

%       Fig.98 

\begin{figure}[h]
  \begin{center}
  	\includegraphics[width=7cm]{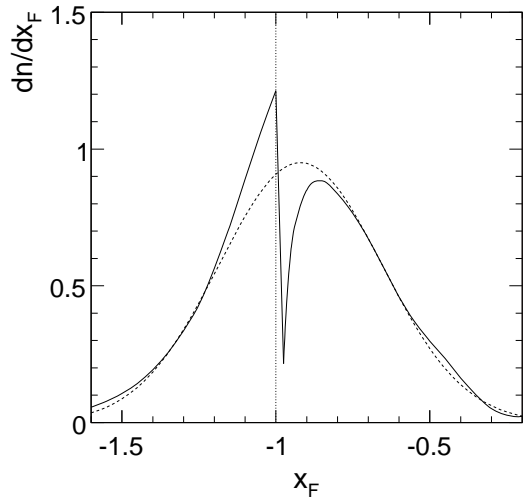}
 	\caption{Net proton density $dn/dx_F$ as a function of $x_F$ 
             in the region -1.6~$< x_F <$~-0.2 resulting from the subtraction of
             the predicted target density from the total measured proton
             yield (full line) and a Gaussian fit to the difference in
             the two regions -1.6~$< x_F <$~-1.2 and -0.8~$< x_F <$~-0.2, broken line. The systematic error
              margin corresponds to 7 \%}
  	 \label{fig:targ_subtr}
  \end{center}
\end{figure}

The subtraction of  the target component from the total yield, Fig.~\ref{fig:targ_subtr} (full line), 
results in a rather particular pattern.

%       Fig.99 

This pattern features two distinct regions in $x_F$. For the region
of $\pm$0.2 units around $x_F$~=~-1, an abrupt bipolar instability
arises from the presence of a diffractive peak in the predicted
target component, whereas for the regions -1.6~$< x_F <$~-1.2 and 
-0.8~$< x_F <$~-0.2 a smooth $x_F$ distribution results which is well
fitted by a Gaussian of the form

\begin{equation}
   \label{eq:nucl_gaus}
   \frac{dn}{dx_F} = 0.95 e^{-\frac{(x_F+0.92)^2}{2 \cdot 0.265^2}}
\end{equation}
centered at $x_F$~=~-0.92 with an rms of 0.265 as shown by the broken 
line in Fig.~\ref{fig:targ_subtr}. 

\begin{figure}[h]
  \begin{center}
  	\includegraphics[width=7.cm]{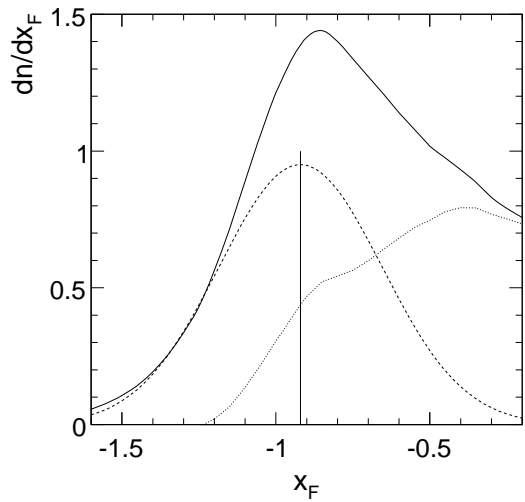}
 	\caption{Net proton densities $dn/dx_F$ in the region 
             -1.6~$< x_F <$~-0.2 as a function of $x_F$, full line total measured
             yield, broken line Gaussian fit to the target density in
             the regions -1.6~$< x_F <$~-1.2 and -0.8~$< x_F <$~-0.2, dotted line
             difference of the two distributions}
  	 \label{fig:targ_subtr_gaus}
  \end{center}
\end{figure}

The diffractive component of the predicted target density should 
show up in any measurement of the total proton density in the region
around $x_F$~=~-1 if the nuclear component would have as expected a
smooth behaviour across this region. Assuming the Gaussian fit to
describe this smooth behaviour one may tentatively subtract the
Gaussian shape from the total proton density as shown in Fig.~\ref{fig:targ_subtr_gaus}.

%       Fig.100

\begin{figure}[h]
  \begin{center}
  	\includegraphics[width=7.5cm]{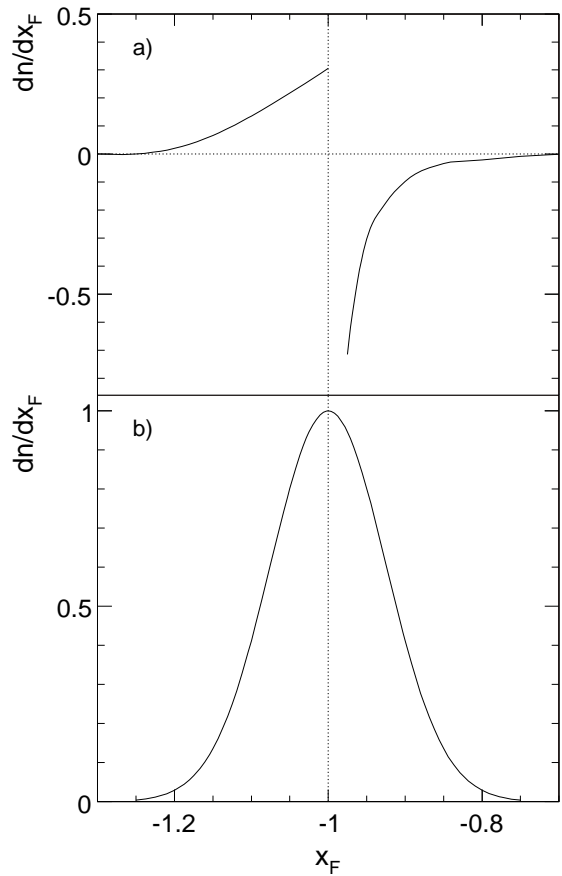}
 	\caption{a) Difference between the predicted target component
             (Fig.~\ref{fig:net_prot}, broken line) and the modified component obtained by
             subtraction of the Gaussian nuclear component (Fig.~\ref{fig:targ_subtr_gaus}, dotted
             line). b) Gaussian longitudinal proton density distribution
             as obtained in quasi-elastic scattering of alpha-alpha
             collisions at $\sqrt{s}$~=~30~GeV/nucleon at the CERN ISR \cite{bell}}
  	 \label{fig:targ_mod}
  \end{center}
\end{figure}

This results in a modified target component (dotted line in Fig.~\ref{fig:targ_subtr_gaus}) 
which reaches down to $x_F$ values at $\sim$~-1.2 and goes smoothly through
$x_F$~=~-1. The difference between the modified and predicted target
component is bipolar around $x_F$~=~-1 and conserves, to first order,
the total integrated density, as shown in Fig.~\ref{fig:targ_mod}.

%        Fig.101  

In this context the action of the Fermi motion of the hit nucleons on 
the momentum distribution of protons in the diffractive peak
should be recalled. Fermi motion may be approximated by a Gaussian
momentum distribution of about 75~MeV/c rms width in the nuclear rest
system. This translates at SPS energy to an rms width of about 0.085 in $x_F$. 
Low momentum transfer processes like quasi-elastic
scattering or diffraction convolute with this momentum distribution.
This leads to a smearing of the peak structure characterized
by a base width of $\pm$2-3 times the Fermi rms around $x_F$~=~-1.
This is quantified by the Gaussian $x_F$ distribution with rms 0.085 
in Fig.~\ref{fig:targ_mod}b which covers exactly the range of the observed deviation 
from the predicted diffractive peak. This convolution has been studied
in quasi-elastic alpha-alpha scattering at the CERN ISR \cite{bell}
where it was shown that low momentum transfer re-scattering processes 
have only small influence on the longitudinal width in contrast
to a substantial broadening of the transverse momentum distribution,
see Sect.14 below.

The extracted width of the nuclear component in longitudinal momentum 
with an rms of 0.265 units in $x_F$ is much larger than the spread
expected from Fermi motion. This feature is known from quite a
number of experiments since some decades \cite{pc_survey} but is shown
here for the first time to be symmetric, however not around $x_F$~=~-1 
but around $x_F$~=~-0.92. This substantial transfer of baryon number
needs comment. One effective mechanism contributing to the
transfer of nucleons in longitudinal momentum is certainly
the production of pions in the intra-nuclear cascade. In the
excitation of $I$~=~3/2 states at pion threshold, the decay products
are located at $x_F$~=~-0.85 and $x_F$~=~-0.15, respectively, for protons
and pions, and for isobars at $x_F \sim$~-1. For isobar masses above threshold 
the $x_F$ values of the decay baryons will shift further up.

The production of pions in the nuclear cascade has indeed been
studied in the accompanying publication \cite{pc_survey}. The backward pion
yields have been separated into the two components of target
fragmentation which may be predicted from elementary hadronic
collisions, and of nuclear cascading, in close similarity to
the argumentation carried out here for baryons. As a result
the total number of pions in the nuclear component has been
determined to 0.3 per inelastic event.

The Gaussian distribution of the nuclear component, Fig.~\ref{fig:targ_subtr_gaus},
integrates to 0.6 protons per inelastic event, see 
Sect.~\ref{sec:total_yields} below. Isospin symmetry and the absence of charge exchange
in the primary p+C interaction at SPS energy ask for an equal number 
of neutrons to be produced. This means that every fourth proton in 
the nuclear cascade produces a pion. In the decay of the isospin 
3/2 isobars involved with pion production, the decay protons are 
transferred to $x_F$~=~-0.85 at pion threshold with the decay pions 
being centred at $x_F$~=~-0.15 for low lab momenta. Integrating over 
the pion transverse momentum distribution this center shifts down 
to about -0.2 \cite{pc_survey} with a rather long tail towards lower $x_F$. The 
corresponding nucleons will be placed at $x_F$~=~-0.8 and above. The 
center of the nuclear component at $x_F$~=~-0.92 is therefore compatible 
with pion production in the nuclear cascade via isobar excitation 
near threshold and the subsequent decay into baryons and pions. The 
symmetry of the nuclear component would however ask for a bigger 
pion yield at about one pion for every second proton. Here
de-excitation processes without pion emission like final state 
isobar-nucleon re-scattering could be involved to reduce the
pion yield.    

In conclusion of the preceding sections on $p_T$ integrated proton
distributions it may be stated that a three-component mechanism
of convincing internal consistency has been established in a
quantitative and precise way. These three components are defined by

\begin{itemize}
  \item The fragmentation of the projectile which obeys baryon number 
        conservation and shows a transfer in longitudinal momentum
        corresponding to 0.15 units in Feynman $x_F$.
  \item The prompt fragmentation of the nucleons hit by the projectile.
        This component has been predicted from elementary proton-nucleon
        interactions invoking isospin symmetry and the mean number of
        projectile collisions.
  \item The nuclear component which arises from intra-nuclear cascading
        and is partially accompanied by pion production.
\end{itemize}

At SPS energy the nuclear component extends up to $x_F$~=~-0.2. It is
well separated from the projectile fragmentation which feeds over
into the target hemisphere down to the same value of $x_F$~=~-0.2. The
target fragmentation in turn reaches from $x_F$~=~+0.2 down to 
$x_F$~=~\mbox{-1.2}. The latter range is well beyond the kinematic limit of elementary 
fragmentation with a diffractive peak close to and above $x_F$~=~-1.
This sharp diffractive structure is evidently diluted and smoothed by
quasi-elastic re-scattering of its low-momentum nucleons in
accordance with experimental results from other experiments.  

%
% ****************************** Section 15.5 ****************************
%
\subsection{A remark about ${\mathbf s}$-dependence}
\vspace{3mm}
\label{sec:sdep}

Even if the SPS energy range is, as shown above, a well suited region 
for the separation of the three components contributing to the 
measured net baryon density, it is interesting to look at the 
expected evolution with $\sqrt{s}$.

%
% ****************************** Section 15.5.1 ****************************
%
\subsubsection{The nuclear component}
\vspace{3mm}
\label{sec:sdep_nucl}

\begin{figure}[h]
  \begin{center}
  	\includegraphics[width=9.cm]{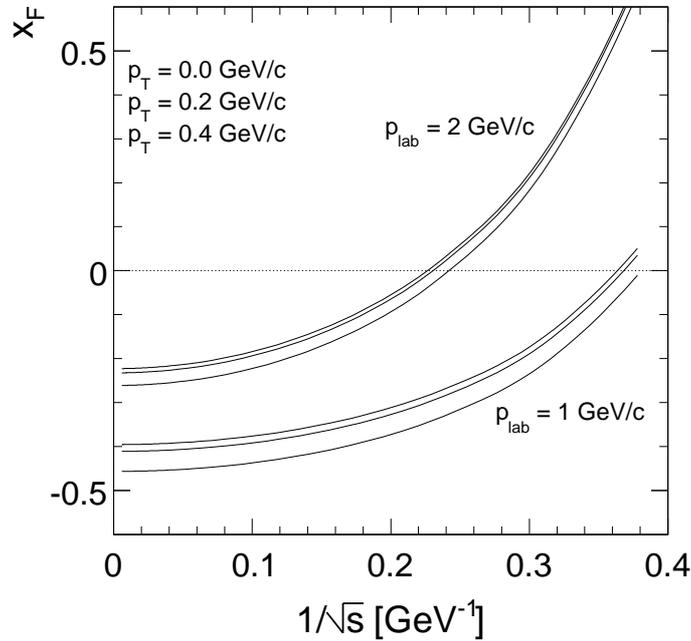}
 	\caption{$x_F$ as a function of 1/$\sqrt{s}$ for $p_{\textrm{lab}}$~=~1 and
              2~GeV/c, varying $p_T$ from 0 to 0.4~GeV/c}
  	 \label{fig:kin_sqs}
  \end{center}
\end{figure}

The study \cite{pc_survey} of backward proton production in p+C interactions
shows that only small changes in the momentum distributions are
to be expected by increasing $\sqrt{s}$ to RHIC energy and beyond,
or by decreasing it into the AGS energy region. Indeed one has
to move down to $\sqrt{s}$ below about 3~GeV in order to see
threshold effects drastically reducing the proton density. In
terms of the range in $x_F$ covered by the nuclear component,
however, kinematics will extend the upper limit of $\sim$~-0.25 at
SPS energy to higher values until $x_F$~=~0 is reached at $\sqrt{s} \sim$~4~GeV. 
This purely kinematic effect is presented in Fig.~\ref{fig:kin_sqs}
where $x_F$ is plotted as a function of 1/$\sqrt{s}$ for two values
of $p_{\textrm{lab}}$ and three values of $p_T$.

%       Fig.102 

This behaviour means that for given $x_F$ (or rapidity) the nuclear
contribution will increase with decreasing beam momentum. This
effect is of course also to be expected for peripheral heavy
ion collisions where the separation of prompt baryons from
the delayed nuclear cascade will become more important but
also more difficult with decreasing interaction energy.

%
% ****************************** Section 15.5.2 ****************************
%
\subsubsection{The target fragmentation}
\vspace{3mm}
\label{sec:sdep_targ}

As shown above prompt net baryons from target fragmentation are
well described by the superposition of elementary proton-nucleon
collisions. With increasing $\sqrt{s}$ this means that this
contribution will move back in $x_F$ due to increasing transparency
until at the highest ISR and at RHIC energies the central region
around $x_F$~=~0 will be only populated by pair produced baryons.
In the $x_F$ region below $\sim$~-0.3 the yields will not change
appreciably due to the approximate scaling behaviour.
Towards lower $\sqrt{s}$ the feed-over of the target fragmentation
into the projectile hemisphere will accordingly increase. Together
with the increasing nuclear component a clear separation will
therefore become more involved.

%
% ****************************** Section 15.5.3 ****************************
%
\subsubsection{The projectile component}
\vspace{3mm}
\label{sec:sdep_proj}

The evolution of the projectile component with collision energy
is of considerable interest as it carries unique information about
multiple hadronic collisions. Since its diffractive part is
connected - for minimum bias p+A collisions - to the fraction
of single projectile collisions which is as discussed above on
a level of 60\% for p+C interactions, there should be approximative
energy scaling with a progressive widening of the peak towards 
lower energies as measured in p+p interactions.
For the non-diffractive part the situation is however much less 
clear, in particular as there is no theoretical understanding of the
observed baryon number transfer and as there is only very limited
experimental information available on the required level of precision.  
It has been shown above that the net baryon density distribution
shifts downward by 0.15 units of $x_F$ in p+C collisions, the same
shift being observed for the neutrons. It has also been shown
that this baryon number transfer increases with centrality as
indicated by the dependence on the number of "grey" protons.
It has also in addition been indicated that resonance production 
and decay presents a very effective source of baryon transfer. 
If with increasing interaction energy the excitation of the projectile
proton in its multiple collisions also increases and therefore
the spectrum of produced resonances extends to higher masses,
the reduction of central net baryon density via transparency
should be reduced or even compensated. At lower $\sqrt{s}$ the
situation should become, as for the two other contributions, more 
obscure due to the increased mutual overlap of the different
components.

%
% ****************************** Section 15.6 ****************************
%
\subsection{Proton and net proton multiplicities}
\vspace{3mm}
\label{sec:total_yields}

Using the $p_T$  integrated density distributions $dn/dx_F$ extracted
above for the inclusive p+C interaction as well as for the different sub-components, 
the corresponding baryonic multiplicities
may now be determined by integration over $x_F$. In view of the statistical errors of the
$dn/dx_F$, Tables~\ref{tab:int_prot} and \ref{tab:int_aprot}, which are of order 1\% and 3\% for
p and $\overline{\textrm{p}}$, respectively, the statistical uncertainties of the total integration
are negligible with respect to the systematic errors given below in percent of the yields. The
integration limits in $x_F$ reach from -0.2 to 1 (projectile component, Fig.~\ref{fig:proj_net}), from -1.2
to 0.2 (target component, Figs.~\ref{fig:targ_net} and \ref{fig:targ_subtr_gaus}), and from -1.7 to -0.2
(nuclear component, Fig.~\ref{fig:targ_subtr_gaus}).
This results in the following numbers:

\begin{align}
   \textrm{inclusive proton multiplicity}         \qquad                \langle n_{\textrm{p}} \rangle &= 1.881  \pm 4.3\% \\
   \textrm{inclusive anti-proton multiplicity}    \qquad     \langle n_{\overline{\textrm{p}}} \rangle &= 0.052  \pm 6.0\% \\
   \textrm{multiplicity of pair produced protons} \qquad        \langle n_{\tilde{\textrm{p}}} \rangle &= 0.065  \pm 7.2\% \\
   \label{eq:tot_net} \textrm{inclusive net proton multiplicity}     \qquad \langle n_{\textrm{p}}^{\textrm{net}} \rangle &= 1.816  \pm 4.7\%
\end{align}

There are two predictions respectively for the net projectile component 
and for the expected net target component established from the elementary p+p collisions \cite{pp_proton}:

\begin{equation}
  \textrm{predicted net projectile component} \qquad  \langle n_{\textrm{p}}^{\textrm{proj,pred}} \rangle = 0.550  \pm 2.8\% 
\end{equation}

This assumes baryon number conservation and a negligible increase
in neutron and hyperon production.

\begin{equation}
  \textrm{predicted net target component}   \qquad    \langle n_{\textrm{p}}^{\textrm{targ,pred}} \rangle = 0.690  \pm 4.0\% 
  \label{eq:pntc}
\end{equation}

This uses isospin invariance and the measured number of 1.6
projectile collisions.

Subtracting the predicted target component from the total net
proton density in the projectile fragmentation region, the multiplicity of net projectile protons is measured as

\begin{equation}
  \textrm{measured net projectile component} \qquad  \langle n_{\textrm{p}}^{\textrm{proj,meas}} \rangle = 0.520  \pm 5.0\%
\end{equation}
which reproduces the expected multiplicity to within 5.5\% and
leaves room for some increase of the hyperon yields in p+C
collisions.

Subtracting the predicted target component from the total net
proton density in the backward region, the nuclear component
is measured as

\begin{equation}
  \textrm{measured nuclear component}    \qquad       \langle n^{\textrm{nucl,meas}} \rangle = 0.655  \pm 7.1\%
\end{equation}

A Gauss fit to the nuclear component in the symmetric regions
of the target-subtracted proton density yields the multiplicity

\begin{equation}
  \textrm{Gaussian nuclear component}     \qquad     \langle n^{\textrm{nucl,Gauss}} \rangle = 0.631  \pm 3.0\%
\end{equation}
which complies to within 3.6\% with the straight-forward subtraction.

Finally there is the modified target component obtained by subtraction
of the Gaussian nuclear component from the overall proton density
which smooths the diffractive peak in accordance with the Fermi
motion of the hit nucleons. This modified target component results
in the multiplicity

\begin{equation}
  \textrm{modified target component}      \qquad      \langle n^{\textrm{targ,mod}} \rangle = 0.670  \pm 5.3\% 
\end{equation}
which reproduces the predicted multiplicity (\ref{eq:pntc}) to within 2.9\%.

In conclusion it may be stated that the total measured net
proton multiplicity made up by the superposition of three components 
of comparable magnitude, namely a projectile multiplicity of 0.52, a target
multiplicity of 0.67 and a nuclear component of 0.65 units. These yields add up to 1.84 which is
within 1\% equal to the direct integration (\ref{eq:tot_net}). The multiplicities of anti-protons 
and pair produced protons are 0.052 and 0.065, respectively, corresponding to 2.9\% and
3.6\% of the total net proton yield. 

%
% ****************************** Section 16 ****************************
%
\section{Proton production in p+C collisions: ${\mathbf p_T}$ dependence}
\vspace{3mm}
\label{sec:pt_prot}

Following the study of $p_T$ integrated yields in the preceding
section, the double-differential proton cross sections will
now be studied for the three components of projectile fragmentation,
target fragmentation, and nuclear cascading, thus adding the 
transverse dimension to the experimental scrutiny. In order to 
comply with the discussion of the integrated proton yields
$dn/dx_F$, Sect.~\ref{sec:hadr_net}, the double differential cross sections 
will be used in their non-invariant form

\begin{equation}
  \frac{1}{p_T}\frac{d^2n}{dx_Fdp_T} = \frac{2\pi}{\sigma_{\textrm{inel}}}\frac{\sqrt{s}}{2}\frac{f(x_F,p_T)}{E} 
\end{equation}
see also (\ref{eq:int}). In a first step, the net proton cross sections
will be defined using the results on anti-protons, Sect.~\ref{sec:hadr}.
A detailed comparison of the overall backward cross sections
with the predicted contribution from target fragmentation
will reveal a necessary $p_T$ dependent modification of this
component which will then be employed to extract in turn the
projectile and nuclear components in their $p_T$ dependence.

%
% ****************************** Section 16.1 ****************************
%
\subsection{Pair produced protons and net proton density}
\vspace{3mm}
\label{sec:pt_pair}

\begin{figure}[h]
  \begin{center}
  	\includegraphics[width=9.5cm]{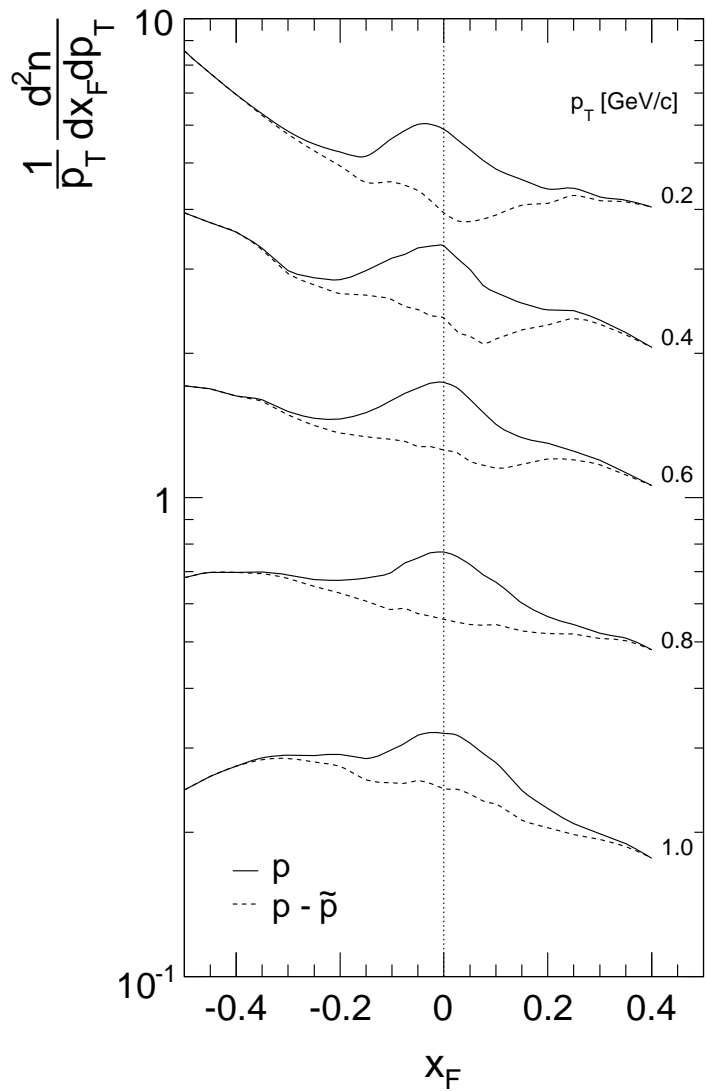}
 	\caption{Total and net proton density $\frac{1}{p_T}\frac{d^2n}{dx_Fdp_T}$
             as a function of $x_F$ for different $p_T$ values indicated in
             the panel. The distribution at $p_T$~=~0.2~GeV/c is multiplied by factor 
             of 1.2 for better separation}
  	 \label{fig:protnet}
  \end{center}
\end{figure}

Using the results of the discussion of the double differential
anti-proton yields and their separation into target and projectile
components in Sect.~\ref{sec:hadr_ptdist}, the yield of pair produced protons
may be estimated by multiplying the projectile component by the
isospin factor 1.6, maintaining the $p_T$ enhancement as parametrized
in (\ref{eq:pt_param}). The target component stays equal to the anti-proton
yield due to isospin symmetry. The total and the resulting net proton densities 
are shown in Fig.~\ref{fig:protnet} as a function of $x_F$ for different values of $p_T$.

%       Fig.103 

%
% ****************************** Section 16.2 ****************************
%
\subsection{Target component}
\vspace{3mm}
\label{sec:pt_targ}

\begin{figure}[h]
  \begin{center}
  	\includegraphics[width=15.5cm]{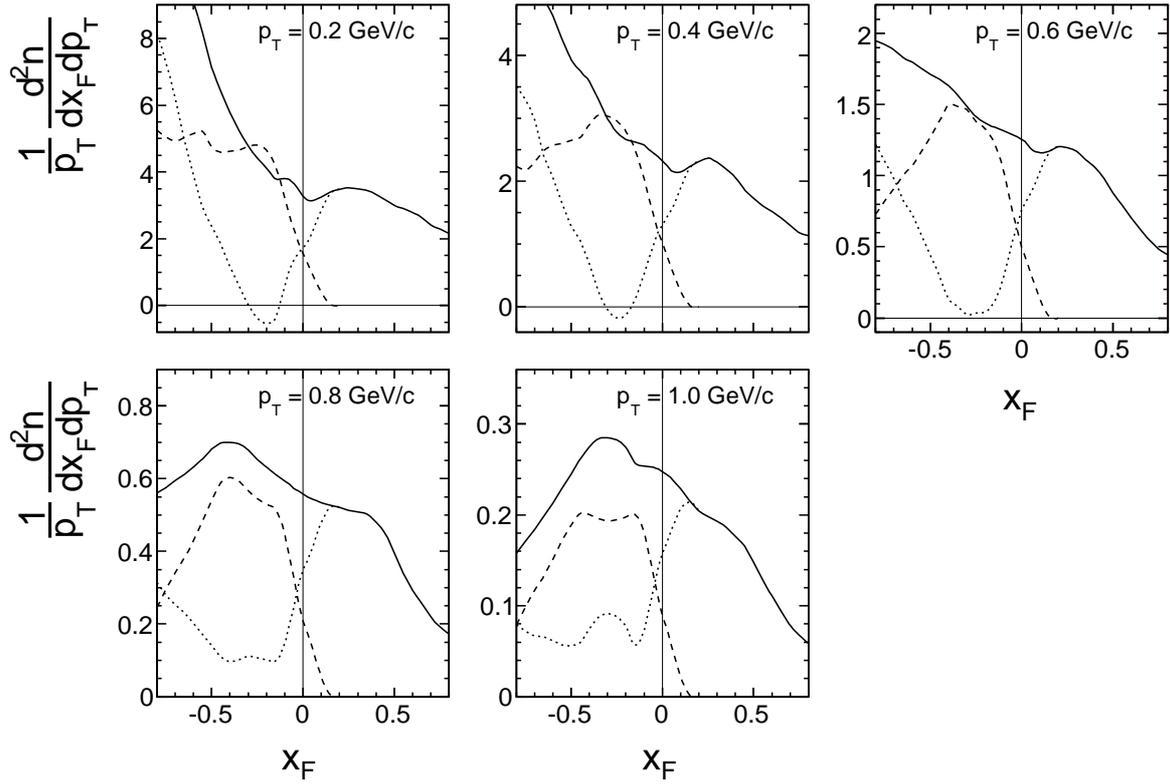}
 	\caption{Double differential net proton yield $\frac{1}{p_T}\frac{d^2n}{dx_Fdp_T}$
             as a function of $x_F$ for five different values of $p_T$. Full line:
             total yield, broken line: predicted target component, and
             dotted line: after subtraction of the target component}
  	 \label{fig:protnettarg}
  \end{center}
\end{figure}

As discussed in Sect.~\ref{sec:net_targ_pred} the target component of net proton
production in p+C interactions may be predicted from p+p collisions
using the number of projectile collisions $\langle \nu \rangle$~=~1.6, the target
overlap function $R^{c,\textrm{net}}_{\textrm{p}}$ ((\ref{eq:ratp}) and Fig.~\ref{fig:net_rat}), 
and the measured proton and neutron cross sections \cite{pp_proton}. The only additional assumption
to be made for the double differential yield is the equality of the
$p_T$ distributions for neutrons and protons, see \cite{pp_proton} for experimental
evidence. The result for the target component and its subtraction
from the overall proton yield is shown in Fig.~\ref{fig:protnettarg}.

%   Fig.104

\begin{figure}[h]
  \begin{center}
  	\includegraphics[width=12.5cm]{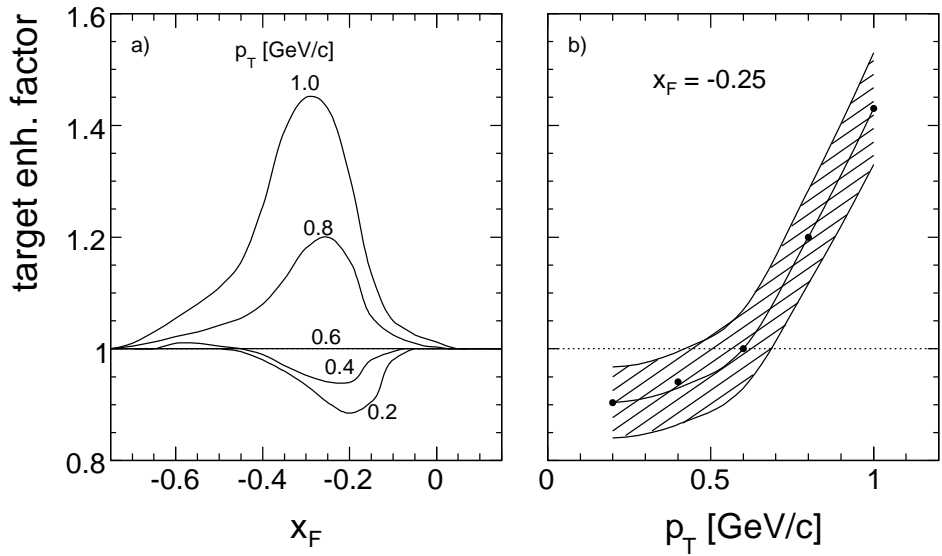}
 	\caption{Enhancement factor of the target component
             a) as a function of $x_F$ for $p_T$ values between 0.2 and
             1~GeV/c and b) as a function of $p_T$ for $x_F$~=~-0.25. The shaded
             region represents the error margins}
  	 \label{fig:targenh}
  \end{center}
\end{figure}

Compared to the same prediction and subtraction for the $p_T$ integrated
densities, Figs.~\ref{fig:targ_net} and \ref{fig:proj_net}, a $p_T$ dependence becomes visible. It is the
region of $x_F$ between -0.2 and -0.3 which is extremely sensitive to the
predicted target yield. At transverse momenta below the mean $p_T$ (Fig.~\ref{fig:meanpt_prot})
the prediction overshoots the total density by about 20\%, whereas for higher $p_T$ 
it falls low by up to 40\% at $p_T$~=~1~GeV/c. In this $x_F$
region neither the possible projectile feed-over nor the nuclear
component may explain, by their limited $x_F$ range, the observed $p_T$
dependence. Instead a yield suppression at low $p_T$ followed by an enhancement 
at high $p_T$ similar to the one found for anti-protons, Sect.~\ref{sec:hadr_ptdist}, 
has to be invoked also for the target contribution. This pattern of yield suppression and
enhancement is quantified in Fig.~\ref{fig:targenh}a as a function of $x_F$ for
the $p_T$ values shown in Fig.~\ref{fig:protnettarg}. Given the estimated systematic
error margins, Fig.~\ref{fig:targenh}b, the effect is significant for $p_T >$~0.7~GeV/c.
It should be noted that baryon number conservation imposes, for a yield increase in
this $p_T$ range, a corresponding yield suppression at low $p_T$.

%       Fig.105 

\begin{figure}[h]
  \begin{center}
  	\includegraphics[width=10.cm]{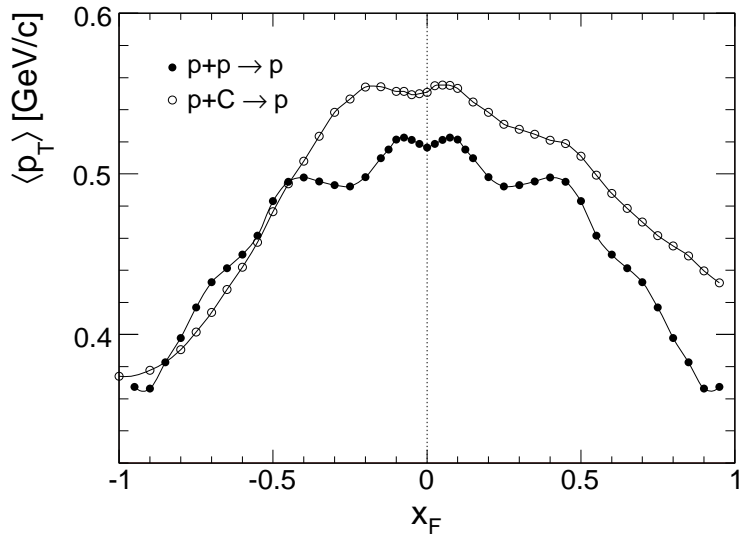}
 	\caption{Comparison of the mean $p_T$ values of protons in p+C
            (open circles) and p+p interactions (closed circles)
             as a function of $x_F$ from $x_F$~=~-1.0 to $x_F$~=~+0.95. The
             lines connecting the points are drawn to guide the eye}
  	 \label{fig:mpt_tot}
  \end{center}
\end{figure}

This phenomenon is also borne out by a comparison of the mean
$p_T$ values between p+C and p+p interactions shown in Fig.~\ref{fig:mpt_tot}.

%       Fig.106 

If the target component showed a $p_T$ dependence equivalent to
p+p interactions, the mean $p_T$ values should approach the ones for
p+p at $x_F$ around -0.2 to -0.3 where target fragmentation prevails.
The observed difference of 50~MeV/c in $\langle p_T \rangle$ is significant with
respect to the statistical errors of less than 1\%. The systematic errors are of
the same order as most error sources, or for certain the normalization error,
cancel in the mean value.
 
\begin{figure}[h]
  \begin{center}
  	\includegraphics[width=13.cm]{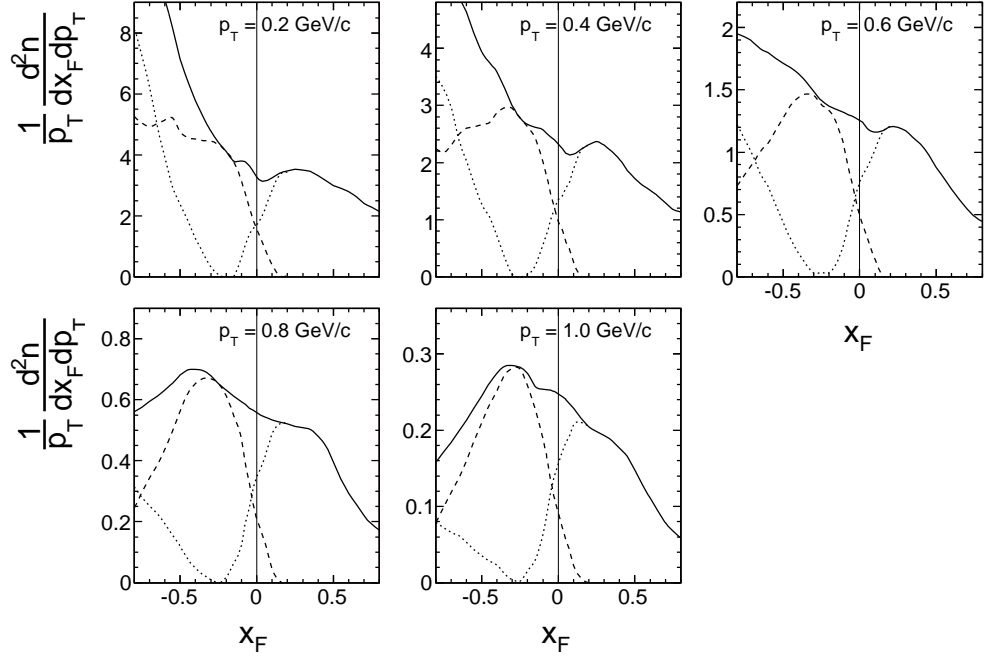}
 	\caption{Double differential net proton yield $\frac{1}{p_T}\frac{d^2n}{dx_Fdp_T}$
             as a function of $x_F$ for five values of $p_T$. Full line: total
             yield, broken line: predicted enhanced target component, dotted
             line: after subtraction of the enhanced target component}
  	 \label{fig:protnettargenh}
  \end{center}
\end{figure}

Applying this enhancement, the target component and its subtraction
from the total measured yield take the pattern shown in Fig.~\ref{fig:protnettargenh}.

%       Fig.107 

This subtraction leaves clearly defined and separated projectile
and nuclear components which will be discussed in the subsequent
sections

%
% ****************************** Section 16.3 ****************************
%
\subsection{The projectile component}
\vspace{3mm}
\label{sec:pt_proj}

\begin{figure}[h]
  \begin{center}
  	\includegraphics[width=8.cm]{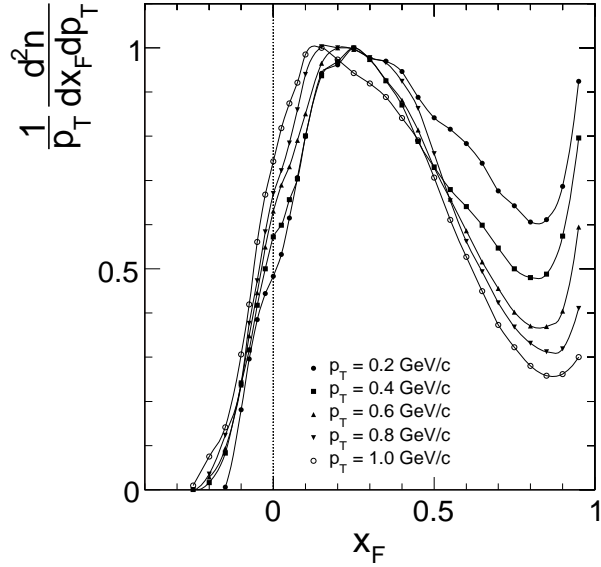}
 	\caption{Net proton density $\frac{1}{p_T}\frac{d^2n}{dx_Fdp_T}$ from
             projectile fragmentation as a function of $x_F$ normalized
             to its maximum value for five $p_T$ values between 0.2 and 1~GeV/c.
             The symbols and full lines correspond for $x_F >$~0.2 to the data interpolation \cite{site}.
             Below this $x_F$ value they represent the subtraction of the target component,
             derived from the p+p interaction \cite{site}, from the data interpolation. The systematic errors
             are 2.5 \% for $x_F >$~0.1 (p+p data) and increase to about 4.5 \% for $x_F <$~0}
  	 \label{fig:protprojnorm}
  \end{center}
\end{figure}

The hadronization of the projectile is already clearly visible
in the dotted line of Fig.~\ref{fig:protnettargenh}. It saturates the total yield at $x_F >$~0.2 
and comes down to zero at $x_F \sim$~-0.2 due to the limited range of the baryonic overlap function. 
In order to put this behaviour in perspective as far as the $x_F$ distributions 
for different transverse momenta are concerned, the projectile components normalized 
to their maximum densities are presented in Fig.~\ref{fig:protprojnorm}.

%      Fig.108 

Compared to the $p_T$ integrated distribution, Fig.~\ref{fig:proj_net}, there is
a clear dependence of baryon number transfer on $p_T$. The
maximum density shifts from $x_F$~=~0.3 at low $p_T$ to $x_F$~=~0.1 at
1~GeV/c. The density at $x_F \gtrsim$~0.6 corresponding to low mass
excitation is strongly reduced with increasing $p_T$ whereas
the transfer function is successively extending further into the backward hemisphere.

\begin{figure}[h]
  \begin{center}
  	\includegraphics[width=8.cm]{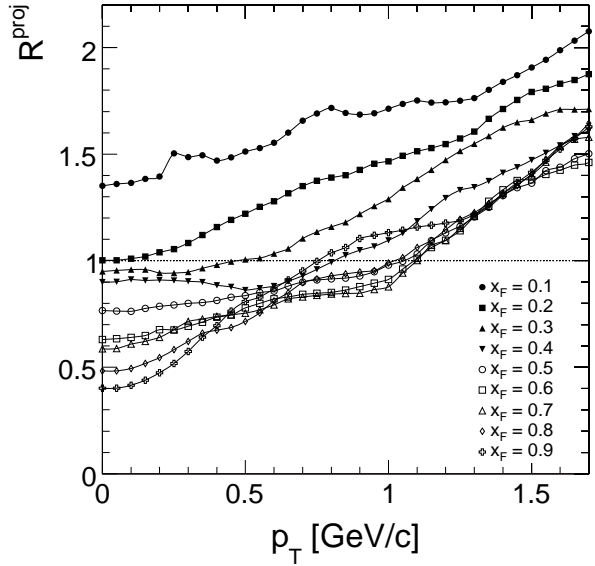}
 	\caption{Density ratio $R^{\textrm{proj}}$ of the projectile components
             of p+C and p+p interactions as a function of $p_T$ for different
             values of $x_F$ in steps of 0.05~GeV/c in $p_T$ as given by
             the data interpolation \cite{site}. The systematic uncertainties
             of the ratios correspond to 4.5 \%. They are to first order
             independent on $x_F$ and $p_T$. The lines are drawn to guide the eye}
  	 \label{fig:protproj_ptrat}
  \end{center}
\end{figure}

Further information concerning the $p_T$ dependence comes from
a direct comparison to p+p interactions. Extracting the
projectile component from the symmetric p+p collision using
the baryonic overlap function determined in Sect.~\ref{sec:two_comp}
the ratio

\begin{equation}
	R^{\textrm{proj}} = \left(\frac{1}{p_T}\frac{d^2n}{dx_Fdp_T}\right)^{\textrm{pC}}\bigg/\left(\frac{1}{p_T}\frac{d^2n}{dx_Fdp_T}\right)^{\textrm{pp}}
\end{equation}
may be obtained as shown in Fig.~\ref{fig:protproj_ptrat} as a function of $p_T$ for
different values of $x_F$.

%       Fig.109 

A clear increase with $p_T$ of the net proton yields is visible for all
values of $x_F$. The overall increase of the cross sections with decreasing $x_F$ is due 
to the general baryon number transfer in p+C interactions visible in the $p_T$ integrated distributions
of Fig.~\ref{fig:xf_ncd} ("minimum bias" compared to p+p). This can be normalized
out by multiplying $R^{\textrm{proj}}$ by the integrated density ratio between p+p and p+C collisions:

\begin{equation}
  R^{\textrm{proj}}_{\textrm{norm}} = R^{\textrm{proj}}\left(\frac{dn}{dx_F}\right)^{\textrm{pp}}\bigg/\left(\frac{dn}{dx_F}\right)^{\textrm{pC}}
\end{equation}

\begin{figure}[h]
  \begin{center}
  	\includegraphics[width=8.cm]{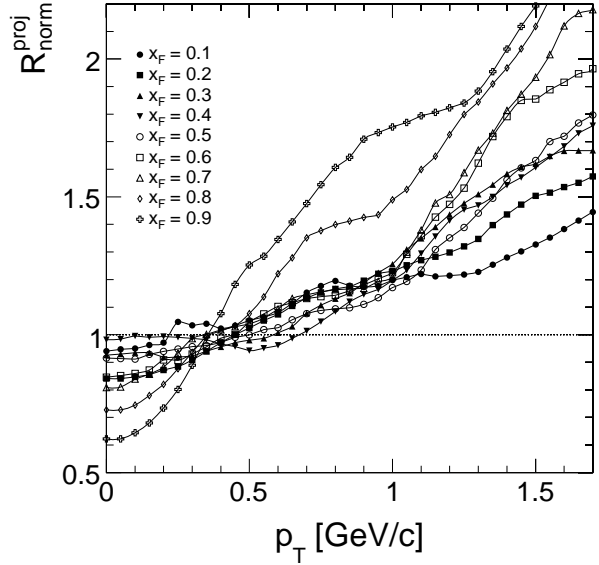}
 	\caption{Normalized density ratio $R^{\textrm{proj}}_{\textrm{norm}}$ as
             a function of $p_T$ for different values of $x_F$ in steps of 0.05~GeV/c in $p_T$ as given by
             the data interpolation \cite{site}. The systematic uncertainties
             of the ratios correspond to 4.5 \%. They are to first order
             independent on $x_F$ and $p_T$. The lines are drawn to guide the eye}
  	 \label{fig:protprojnorm_ptrat}
  \end{center}
\end{figure}

The resulting ratio $R^{\textrm{proj}}_{\textrm{norm}}$ is shown in Fig.~\ref{fig:protprojnorm_ptrat} again
as a function of $p_T$ for different values of $x_F$.

%      Fig.110 

The wide spread observed in Fig.~\ref{fig:protproj_ptrat} reduces to a narrow band of
$p_T$ enhancements which are about independent on $x_F$ except for the
$x_F$ range approaching diffraction. Again, as for $\overline{\textrm{p}}$ in Sect.~\ref{sec:hadr_ptdist} and 
for the target component in the preceding section, the normalized densities are reduced 
by 10-20\% at low $p_T$ and enhanced to factors of about 1.5 at $p_T$~=~1.5~GeV/c.

%
% ****************************** Section 16.4 ****************************
%
\subsection{The nuclear component}
\vspace{3mm}
\label{sec:pt_nucl}

\begin{figure}[h]
  \begin{center}
  	\includegraphics[width=12.5cm]{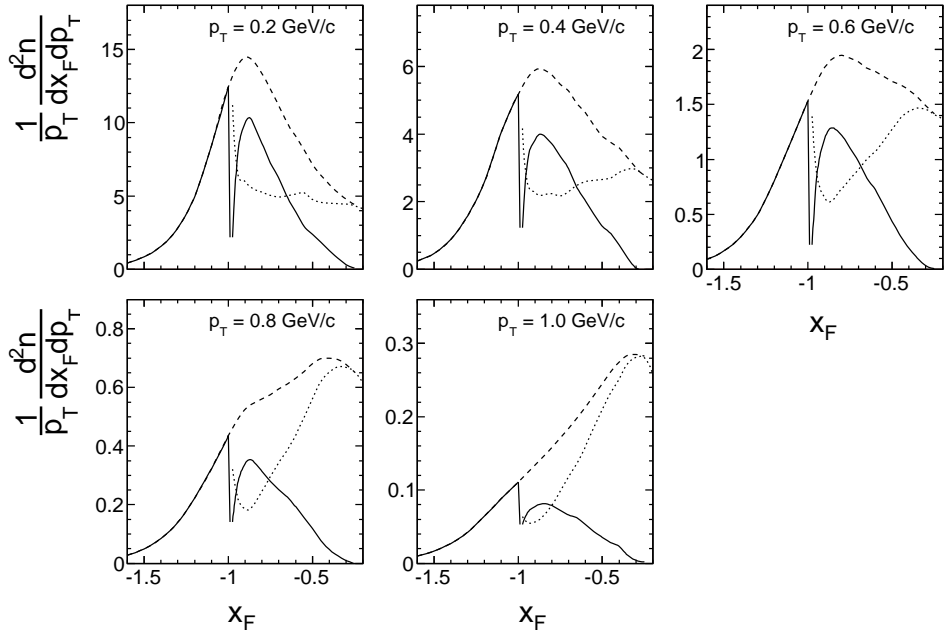}
 	\caption{Double differential net proton yield as a
             function of $x_F$ for five values of $p_T$. Broken line: total
             proton yield, dotted line: predicted target contribution,
             full line: target subtracted proton density}
  	 \label{fig:protnetnucl}
  \end{center}
\end{figure}

The onset of the contribution from nuclear cascading is already
visible in Fig.~\ref{fig:protnettargenh} by the dotted lines at $x_F <$~-0.25. This range is extended 
to the far backward hemisphere down to $x_F$~=~-1.6 in Fig.~\ref{fig:protnetnucl}. In this Figure 
the total proton density, the predicted target contribution and the target subtracted yield 
are shown as a function of $x_F$ for five values of $p_T$ between 0.2 and 1~GeV/c.

%       Fig.111 

\begin{figure}[h]
  \begin{center}
  	\includegraphics[width=12.5cm]{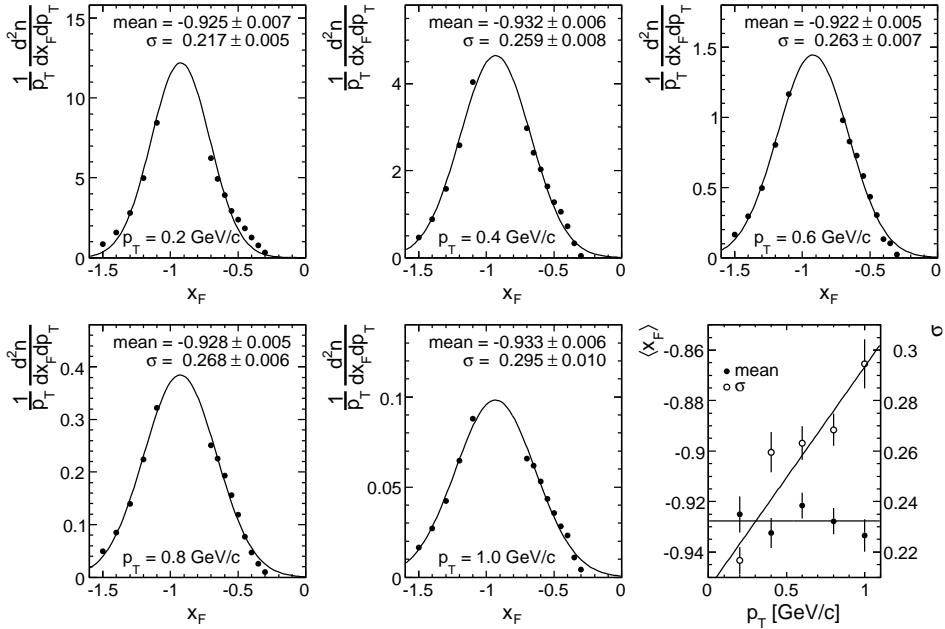}
 	\caption{Double differential, target subtracted proton
             yields as a function of $x_F$ for -1.6~$< x_F <$~-0.3 with the
             exclusion of the range -1.1~$< x_F <$~-0.8 for five values of
             $p_T$ between 0.2 and 1.0~GeV/c. Superimposed as full lines are
             Gaussian fits with mean values $\langle x_F \rangle$ and $\sigma$ indicated in each
             panel. The last panel shows plots of $\langle x_F \rangle$ (left scale) and
             $\sigma$ (right scale) as a function of $p_T$}
  	 \label{fig:nucl_gaus_sig}
  \end{center}
\end{figure}

As already discussed for the $p_T$ integrated densities in Sect.~\ref{sec:net_nucl} 
(Figs.~\ref{fig:net_prot} and \ref{fig:targ_subtr}), the presence of the diffractive peak from
proton fragmentation in the predicted target contribution creates
a sharp spike between $x_F \sim$~-0.9 and $x_F$~=~-1.0 which should be visible
in the total proton density distribution in this area. The absence
of such structure in the measured yield indicates, as argued in
Sect.~\ref{sec:net_nucl}, a smoothing of the predicted structure via quasi-elastic 
re-scattering of the diffrative protons inside the Carbon
nucleus. This re-scattering extends typically up to $\pm$0.2 units
of $x_F$ around $x_F$~=~-1 \cite{bell}. Beyond this range the target subtracted
yield shows a Gaussian behaviour as presented in Fig.~\ref{fig:nucl_gaus_sig}.

%       Fig.112 

\begin{figure}[h]
  \begin{center}
  	\includegraphics[width=9.5cm]{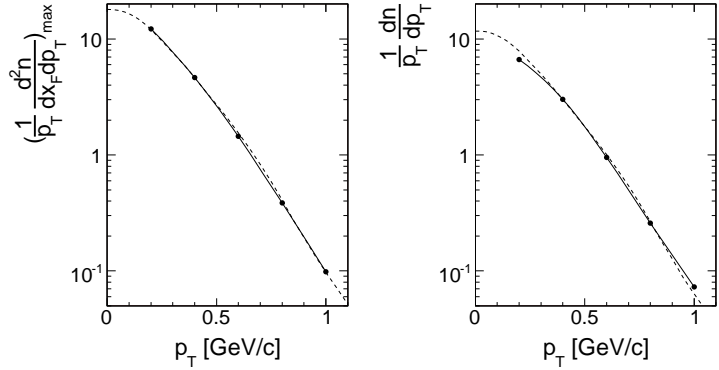}
 	\caption{a) Maximum proton density and b) yield
             $\frac{1}{p_T}\frac{dn}{dp_T}$ integrated over $x_F$ of the nuclear
             component as a function of $p_T$.
             The full lines represent data interpolations, the broken
             lines correspond to the $p_T$ dependence of the
             proton density in p+p interactions at $|x_F|$~=~0.95 \cite{pp_proton},
             normalized at $p_T$~=~0.4~GeV/c}
  	 \label{fig:mean_gaus}
  \end{center}
\end{figure}

The fit parameters show, within tight errors, a stable mean value
in $x_F$ between -0.92 and -0.93 in agreement with the $p_T$ integrated
fit (\ref{eq:nucl_gaus}). The rms deviation increases with $p_T$ 
from 0.22 to 0.3 units of $x_F$. The $p_T$ dependences of the maximum density and of the
yield integrated over $x_F$ are shown in Fig.~\ref{fig:mean_gaus}.

%       Fig.113 

It is interesting to observe that the $p_T$ dependence of the nuclear
component is not Gaussian and corresponds to the one measured in
the diffractive region of p+p interactions \cite{pp_proton}.

\begin{figure}[h]
  \begin{center}
  	\includegraphics[width=12cm]{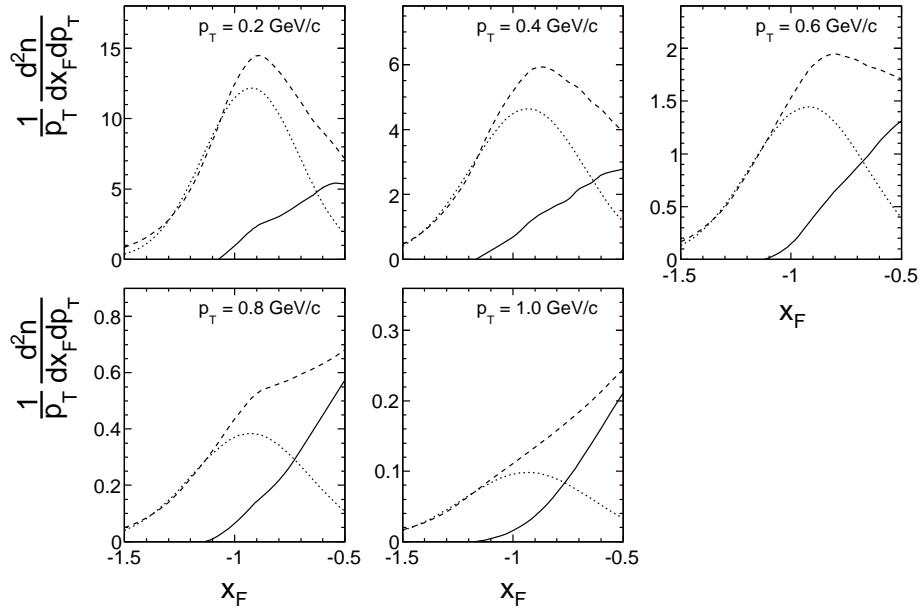}
 	\caption{Double differential proton yields as a function
             of $x_F$ for five values of $p_T$. Broken lines: total proton yield,
             dotted lines: Gaussian fit to the nuclear component, full
             lines: modified target component obtained by the subtraction
             of the Gaussian fits from the total proton density}
  	 \label{fig:prottarggaus}
  \end{center}
\end{figure}

Using the Gaussian fits as an estimator of the nuclear component
a modified target component may be obtained by subtracting these
fitted densities from the total proton yield as shown in Fig.~\ref{fig:prottarggaus}.

%       Fig.114 

The resulting modified target component extends now to $x_F$ values
below -1.0 and shows a smooth behaviour through the region of
proton diffraction in accordance with the $p_T$ integrated yield,
Sect.~\ref{sec:net_nucl}.

%
% ****************************** Section 17 ****************************
%
\section{Conclusion}
\vspace{3mm}
\label{sec:conclusion}

As part of a comprehensive study of hadronic production in
elementary and nuclear collisions at the SPS, new data from 
the NA49 experiment on proton, anti-proton, neutron and light ion production 
in minimum bias p+C interactions are presented. Making full use of the acceptance coverage and 
the particle identification capabilities of the NA49 detector, a wide phase space area 
from the far forward direction at Feynman $x_F$ of 0.95 to the far backward direction
down to $x_F$~=~-0.8 has been exploited. Using available data
from a Fermilab experiment the data coverage could be
further extended into the nuclear fragmentation region
down to $x_F \sim$~-2.0. In addition, deuteron and triton production
have been studied in the lab momentum range from 0.25 to 3~GeV/c 
making available there for the first time cross sections
in the low to medium $p_T$ region. In addition a limited amount of data with "grey" proton detection 
allows for a first look at the centrality dependence of baryon number transfer.

Given the complete phase space coverage of the combined data set, 
the main aim of this publication is the separation and
isolation of the three components of hadronization in p+C
collisions, namely projectile fragmentation, target fragmentation 
and nuclear cascading. This study has been conducted
both for $p_T$ integrated quantities and using double differential
cross sections to obtain a complete view of the $p_T$ dependence.
For this aim, the baryonic overlap functions from the projectile to the target regions 
and vice versa have been determined experimentally using both the elementary p+p 
and the asymmetric p+C interactions. This has been achieved in a completely model-independent 
way relying essentially on baryon number conservation and isospin symmetry. 
For anti-proton production the absence of a nuclear contribution has been shown 
and the superposition of the target and the projectile fragmentation has been established 
using the known number of projectile collisions
inside the Carbon nucleus. This allows for the definition
of the yield of pair produced protons and thereby of net proton densities.

In contrast to the a priory unknown projectile and nuclear components,
the target fragmentation occupies a special place as it should at
least to first order be describable by a superposition of single
nucleon fragmentations taking of course into account the number
of projectile collisions and isospin effects. This approach works
out well for $p_T$ integrated proton densities with the exception of the 
diffractive contribution contained in the elementary
interactions. This contribution is evidently smeared out by secondary, 
quasi-elastic interactions of the corresponding low
momentum protons in the nuclear rest system. The subtraction of
the thus predicted target component yields both the projectile
and the nuclear components. The former is proven to preserve baryon number 
combined with a sizeable amount of baryon transfer
of order 0.15 units of $x_F$. The latter turns out to have a Gaussian shape in 
$x_F$ centred at $x_F$~=~-0.92 with a substantial FWHM of 0.6 units far 
in excess of the narrow $x_F$ distribution centred
close to -1.0 which might be expected from Fermi motion alone.
In the case of the p+C collisions studied here, the three components have comparable 
total yields of 0.52, 0.67 and 0.65 net protons respectively for the projectile, target and nuclear
contributions. This sums up to 1.84 net protons in total, comparing to within 1\%
with the independent direct yield integration of 1.82 net protons (\ref{eq:tot_net}).

The study of double differential cross sections gives access to
the additional dimension of transverse momentum. All three components
show a distinct transverse activity which goes beyond the naive
expectation from elementary collisions or nuclear binding. The
projectile component features, both for protons and anti-protons,
a suppression by 10-20\% at low $p_T$ followed by an enhancement of
about 50\% at $p_T$~=~1.5~GeV/c. This pattern is rather $x_F$ independent
in the forward hemisphere. The target component shows a similar
behaviour, however with a distinct $x_F$ dependence centred at $x_F \sim$~-0.3.
The nuclear component finally has a wide, non-Gaussian $p_T$ dependence which goes 
far beyond the one expected from Fermi motion and
which is shown to be comparable to the one measured in the diffractive
region of p+p interactions.

As for the previous publications \cite{pp_pion,pp_kaon,pp_proton,pc_pion} the measured cross sections
and two-dimensional interpolation values are available on the Web Site \cite{site}.

%-------------- Acknowledgements -------------------
\section*{Acknowledgements}
\vspace{3mm}
This work was supported by
the Polish State Committee for Scientific Research (P03B00630),
the Polish National Science Centre (on the basis of decision no. DEC-2011/03/B/ST2/02634)
the Bulgarian National Science Fund (Ph-09/05),
the EU FP6 HRM Marie Curie Intra-European Fellowship Program,
the Hungarian Scientific Research Fund OTKA (T68506) and
the Hungarian OTKA/NKTH A08-77719 and A08-77815 grants.

\vspace{3cm}
%\clearpage

\end{document}